\documentclass[english]{bullsrsl}

\usepackage[latin1]{inputenc}
\usepackage[T1]{fontenc}

\usepackage{natbib}
\usepackage{xcolor}

\usepackage{graphicx}
\usepackage{amsmath}
\usepackage{booktabs}
\usepackage{rotating}

\DeclareRobustCommand{\ion}[2]{\textup{#1\,\textsc{\lowercase{#2}}}}
\newcommand*\fdg{\mbox{\rlap{\hskip1pt.}{°}}}
\newcommand*\farcs{\mbox{\rlap{\hskip1pt.}{$''$}}}

\begin{document} 

\title{A Detailed Spectroscopic Study of the\\
	   WN6ha\,+\,O5 Colliding-wind Binary WR\,25}

\author[affil={1,2}, corresponding]{Eric}{Gosset}
\author[affil={3,4}]{Roberto}{Gamen}
\author[affil=5]{Laurent}{Mahy}
\author[affil=6]{Nidia}{Morrell}
\author[affil=7]{Hugues}{Sana}
\author[affil=8,deceased]{Rodolfo~H.}{Barbá}

\affiliation[1]{Honorary Research Director F.R.S.-FNRS (Belgium)}
\affiliation[2]{Space sciences, Technologies and Astrophycics Research (STAR) Institute,
                Université de Liège, Allée du 6-Août, 19c, Bât. B5c, B--4000 Liège, Belgium}
\affiliation[3]{Instituto de Astrofísica de La Plata, CONICET-UNLP, 
               Paseo del Bosque S/N, B1900FWA La Plata, Argentina}
\affiliation[4]{Facultad de Ciencias Astronómicas y Geofísicas, 
               Universidad Nacional de La Plata, La Plata, B1900, Argentina}
\affiliation[5]{Royal Observatory of Belgium,
Avenue Circulaire/Ringlaan, 3, B--1180 Brussels, Belgium}
\affiliation[6]{Las Campanas Observatory, Carnegie Observatories, Casilla 601, La Serena, Chile}
\affiliation[7]{Institute of Astronomy, KU Leuven, Celestijnenlaan, 200D, B--3001 Leuven, Belgium}
\affiliation[8]{Departamento de Física y Astronomía, Universidad de la Serena, La Serena, Chile}

\correspondance{eric.gosset@uliege.be}

\maketitle

\begin{abstract}
Massive stars have a major dynamical as well as chemical impact on their environment in their host galaxy.
Despite their importance, many details of their life and evolution remain poorly constrained.
Observational determinations of their physical parameters are therefore required to better constrain evolutionary models.
In particular, the mass and the mass-loss rate---two key  parameters---are known to be more reliably determined through the study of binary stars.
In this context, the binary system WR\,25 ($\equiv$\,HD\,93162) whose dominant spectral type is that of a nitrogen sequence Wolf--Rayet star, is of particular interest.
It exhibits the highest known hydrogen content among the Milky Way Wolf--Rayet stars, as well as additional absorption lines.
It is therefore most probably an object still on the main-sequence or close to leaving it.
We acquired new high-resolution spectra of this extremely interesting object.
Following a detailed analysis, we present an improved version of the SB1 orbital solution for the motion of the WN component.
We confirm the detection of the previously reported probable O~companion and definitively prove that it is indeed gravitationally bound to the WN star.
A first SB2 solution is presented, and a mass-ratio WN/O of $2.02 \pm 0.36$ ($1\sigma$) is derived.
Following a spectral disentangling procedure, we identify the companion as an O5(f$^{+}$) star belonging to either the main-sequence (V) or giant (III) luminosity classes.
The spectral type of the companion, together with the above-mentioned mass ratio, points to a very massive primary WN star of at least 55\,$M_{\odot}$.
This confirms that WR\,25 belongs to the family of very massive WNLh objects, a very poorly populated class in the Milky Way.
We demonstrate that the absorption component in the P-Cygni profile of the \ion{He}{ii}\,$\lambda$4859+H$\beta$ line of WR\,25 is mainly due to the O-type companion, and we propose to adopt a WN6ha spectral type for the WN component of the system.
On the basis of the extracted disentangled spectra, we derive approximate physical parameters through a detailed CMFGEN analysis, given the difficulty in constraining both the extinction towards WR\,25 and the definitive brightness ratio between the components.
Finally, we propose the presence of a third component contributing to the spectrum of WR\,25.
\end{abstract}

\keywords{{\hskip-1pt}binary stars:spectroscopic, stars:early-type stars:Wolf--Rayet, star:individual:WR\,25}

\section{Introduction}\label{sec:intro}
Massive stars are very important objects in galaxies.
Through their strong winds and their associated high mass-loss rates, they exert a marked influence on their surroundings and are able to carve out and shape the interstellar medium, thereby affecting the local dynamical evolution.
In addition, they can modify the characteristics of local star formation \citep[see, e.g.,][and references therein]{2014MNRAS.444.2884L, 2018MNRAS.478.4142T}.
During their lifetime, and at its end, massive stars play a major role in the chemical enrichment of the interstellar medium.
Despite their importance, the physical parameters of these stars, including the most fundamental ones governing their evolution---namely the mass and the mass-loss rate (and its history)---remain poorly constrained.
The evolutionary details of massive stars are still to be understood.
The main-sequence O~stars are supposed to evolve towards a Wolf--Rayet stage, with or without going through a Luminous Blue Variable stage \citep[see][for discussions and references]{2016A&A...594A..94M, 2022A&A...657A...4M}.

It is therefore necessary to improve our theoretical understanding and to strengthen the constraints on evolutionary models.
This can only be achieved through an extensive observational effort aimed at reducing the uncertainties on stellar physical parameters.
One of the best ways to determine the physical parameters of massive stars is, as elsewhere in the Hertzsprung--Russell diagram, through the study of binary systems.
The minimum masses of both objects of a binary system can be derived from the Keplerian laws through the analysis of the orbital motion.
The determination of the inclination (through photometry and/or astrometry) then provides access to the true masses.
Photometric variations, as well as the total luminosity, could also place constraints on the stellar radii.
In the case of massive stars, the stellar winds are usually strong and radiatively driven \citep{2008A&ARv..16..209P}.
In massive binary systems, the winds of both components collide, and the collision zone emits strong X-rays.
The study of this phenomenon provides valuable information on the respective mass-loss rates, and in some cases, on the inclination \citep{1992ApJ...386..265S, 1976SvAL....2..138C, 1976SvA....20....2P, 1990ApJ...362..267L, 2011A&A...530A.119P}.
The detailed and comprehensive study of massive binary stars is therefore an important scientific objective.
Because massive stars are rare, the study of a unique, important object can be considered as a breakthrough in the field.
Interest in massive multiple stars has further increased in recent years following the long-awaited detection of the first Gravitational-Wave signal \citep[GW;][]{2016PhRvL.116f1102A}.
Indeed, a large fraction of the detected GW events have been attributed to mergers of massive binary black holes.

Binary systems can also be used to investigate the stellar evolution, although a caveat is of order here.
The evolution of a massive binary star does not necessarily correspond to the independent evolutions of two single stars.
This approximation may hold for very wide systems where the two stars evolve independently.
In closer systems, however, tidal forces could have some influence, and, even more dramatically, mass transfer can take place through Roche Lobe Overflow (RLOF).
\citet{2012Sci...337..444S, 2014ApJS..215...15S} presented the binary characteristics of Galactic \mbox{O-type} stars and revealed that O~binary systems are very common.
They further demonstrated that approximately 70\% of all stars born as O-type stars exchange mass with their companion during their lifetime and therefore undergo an evolution strongly affected by binary interaction:
binary interaction actually dominates the evolution of massive stars.
These results obtained for the Galactic O~stars have since been confirmed for early B~stars---which dominate the core-collapse supernovae progenitors \citep{2022A&A...658A..69B}---and at sub-solar metallicity \citep{2013A&A...550A.107S, 2025NatAs...9.1337S, 2015A&A...580A..93D, 2025A&A...698A..41V}.
This consequently has implications for the population of core-collapse supernovae.
For sufficiently young stars, mass transfer may not yet have taken place, offering the possibility, similarly to wide systems, to derive interesting constraints in a safe but cautious analysis.

In the 1990s, WR\,22 was presented by \citet{1996A&A...306..771R} as the first observational example of a WR\,+\,O~binary system, in which, by far, the most massive object was the WN star.
Although the existence of such objects had been anticipated by \citet{1995A&A...293..427C}, the detailed analysis of WR\,22 inaugurated the class of highly luminous objects containing an important amount of residual hydrogen and supposed to be core-hydrogen-burning objects disguised as WR stars \citep{1995LIACo..32..463R}.
These stars turned out to be massive stars still on the main-sequence \citep{1996A&A...306..771R, 1999A&A...347..127S}.
Since then, similar other objects have been identified in the Milky Way (such as WR\,29: \citealp{2009A&A...506.1269G}; WR\,20a: \citealp{2005A&A...432..985R}; WR\,21a: \citealp{2008MNRAS.389.1447N, 2016MNRAS.455.1275T, 2016A&A...590A.113G, 2022MNRAS.516.1149B}).
These objects, previously classified as WN7+abs, are now classified as late WN objects with residual hydrogen: WNLh.
Because they are very massive and still located on the main-sequence, they are particularly important to study.

Another object of this kind is the WN6ha\,+\,O~binary WR\,25 ($\equiv \mbox{HD\,93162}$, Hen\,3-478, \emph{Gaia}\,DR3\,5350357519345171200).
WR\,25 is a bright Wolf--Rayet star \citep[$V \approx 8.1$;][]{2001NewAR..45..135V} located in the region of the Carina Nebula south-west of $\eta$~Carinae.
This star is known as the WR star of the Milky Way with the highest hydrogen content \citep[see, e.g.,][]{2019A&A...625A..57H}.
WR\,25 was among the first massive stars recognised as an X-ray emitter with an abnormally high flux \citep[][using \emph{Einstein} observations]{1982ApJ...256..530S, 1979ApJ...234L..55S}.
This flux was usually attributed to the presence of a colliding-wind region in a binary system, although the object had not yet been identified as a true binary system at that time.
\citet{2006A&A...460..777G} were the first to report the detection of orbital motion and to definitively prove that WR\,25 was indeed a binary system.
They presented a preliminary orbital solution of type SB1 (i.e., a spectroscopic binary where the spectral signature of only one component is detected) for the WR component.
\citet{2008RMxAC..33...91G} subsequently reported new observations that revealed the first traces of the O~companion at a particular phase corresponding to the maximum radial-velocity separation.

In the present paper, we report on a new and more accurate orbital solution of type SB2 and further discuss its implications.
We also present a detailed analysis of the individual spectra and discuss the properties of the system.
A~preliminary account of this work can be found in \citet{gossetmorelia}.

The paper is organised as follows.
Section~\ref{sec:observations} details the spectroscopic observations and the data reduction procedures, including a particular method for the normalisation to the continuum.
Section~\ref{sec:wrspec} explores the spectrum of the WN star and analyses the various possibilities for adequately measuring the radial velocities (RVs) of the essential spectral lines in order to accurately derive the orbital motion.
A~reassessment of the period is then performed in Sect.~\ref{sec:period}, and a new improved SB1 orbital solution is deduced from the carefully selected measurements in Sect.~\ref{sec:orbsol}.
Section~\ref{sec:ospec} investigates all possible detections of the O-type companion in the WR\,25 spectrum and prepares the derivation of the SB2 orbital solution, which requires the use of spectral disentangling techniques applied in Sect.~\ref{sec:disentangling}.
Section~\ref{sec:discussion} presents additional considerations and derived results, whereas Sect.~\ref{sec:CMFGEN} presents a first CMFGEN analysis of the disentangled spectra.
The conclusions are presented in Sect.~\ref{sec:conclusion}.

\section{Observations and Data Reduction}\label{sec:observations}
Since the discovery paper by \citet{2006A&A...460..777G}, we have organised several observing campaigns to improve the phase coverage and the signal-to-noise (S/N) ratio of the spectroscopic data.
These different campaigns have provided various data sets, which are described below.
The main effort took place in data set~I.

\subsection{Data set I}\label{ssec:observations_DS1}
Data set~I consists of spectra obtained during the FEROS observation campaigns carried out between 2006 and 2009.
It forms the backbone of the present study.

\subsubsection{Instrumental configuration}%
\label{sssec:observations_DS1_instrconf}
The Fiber-fed Extended Range Echelle Optical Spectrograph  \citep[FEROS;][]{1999Msngr..95....8K} is a two-fiber bench-mounted cross-dispersed echelle spectrograph currently installed at the MPG/ESO 2.2\,m telescope at ESO La Silla (Chile).
The two fibers have a 2$''$ aperture and are separated by 2.9$'$ on the sky.
The science target is observed through one fiber, while the sky contribution is recorded simultaneously through the other fiber.
The instrument covers 39 orders in a fixed cross-dispersed format, providing an effective coverage from 3700\,{\AA} to 9200\,{\AA}.
The resolving power is around 48\,000.
The detector is an EEV $\mbox{2K}\times\mbox{4K}$ CCD with a square pixel size of 15\,{\textmu}m.
All observations used the 225\,Kps-fast, low-gain read-out mode and the $1\times1$ binning.
FEROS is equipped with an atmospheric dispersion corrector (ADC) that can be inserted into the light path to reduce the losses due to atmospheric refraction.
However, the ADC significantly reduces transmission below 4000\,{\AA} and, given the interest of this domain, we usually did not use the ADC.

ESO has implemented a detailed Calibration Plan that provides sufficient support to calibrate the night observations \citep{Pritchard05}.
These calibrations consist of five daily bias frames and one dark exposure allowing to estimate the bias level, the read-out noise, the dark current, and to detect CCD cosmetic defects as well as hot pixels.
Flat-field exposures were also acquired (10 exposures of 2.4\,s each) to correct for the blaze function and pixel-to-pixel sensitivity variations.
The flat-field lamp consisted of two bulbs.
However, residual fringing above 6500\,{\AA} was not fully corrected using these data.
Wavelength calibration (WLC) was performed via the acquisition of lamp spectra producing emission lines.
Two lamp types were available: ThArNe and ThAr+Ne.
In practice, a series of two frames with exposure times of 3, 15 and 30\,s are obtained on a daily basis with each lamp.
In addition, Solar and flux standard-star spectra are also acquired and the linearity of the detector is regularly monitored.
Usually, observers acquire additional flat-field and WLC frames in the afternoon prior to the night in order to initiate the data reduction system (DRS) and to further improve the quality of the reduction procedure.
In most cases, we also acquired an O-type star spectrum close to the WR\,25 exposure.
Its usefulness will be clarified in the following sections.                

The twenty-seven WR\,25 spectra constituting data set~I were collected between March 01, 2006 and March 31, 2009.
The exposure time typically ranged between 23 and 30\,min.
The ADC was generally not used, except on three occasions.
In 12 cases, a 60-s ThAr WLC has been obtained immediately after the WR\,25 observation, using the ThAr+Ne lamp.
The detailed journal of the observations is provided in Table~\ref{tabA1}.

\subsubsection{Data basic reduction}\label{sssec:observations_DS1_reduction}
Data reduction was performed using a modified version of the FEROS DRS working under the eponymous context in ESO-MIDAS.
For each night, a master bias frame was created by averaging the individual bias frames with the appropriate binning and read-out mode.
The master bias was used to check the homogeneity of the read-out noise and the cosmic defects of the detector throughout the campaign.
However, the master bias frame is not used to subtract the bias level.
Instead, this latter was determined using the overscan on each individual frame.
During the data reduction process, detected bad columns were replaced over their full length by the average of the two adjacent columns.
This procedure usually provides a satisfactory approximation, except for columns 1299 and 1300, which both suffer from more severe cosmetic defects.

When available, dark frames were bias-subtracted and used to monitor the dark current.
The contribution of the latter was found to be negligible and it has been neglected throughout the data reduction.
A~daily master flat-field frame was created by averaging all 2.4-s frames available for a given night (i.e., taken both before and after the observing night).
It was used to detect the location of orders on the detector and to determine the blaze function, as well as the relative pixel-to-pixel response.
The order localisation step uses a cross-order profile, which is required to accurately trace the order positions by cross-correlating, starting from an initial guess solution at the centre of each order.
The adopted cross-order profile and guess starting values were determined using the June 2006 campaign data \citep{2009A&A...501..291S}.
They were found to be suitable for the reduction of the whole WR\,25 data set.

The instrument's diffuse light was estimated for each frame by fitting a 2D polynomial using the zero inter-order regions.
Prior to this step, residuals from bad column corrections and individual bad pixels were removed using additional filtering.
The spectra from the SKY and OBJECT fibers are then extracted using the order locations determined on the master flat-field frame.
To reduce contamination arising from a narrower inter-order separation in the red part of the spectrum, we used different extraction window widths for different order ranges.
A~window width of 13, 15, and 17 pixels has been used for order ranges 35--39, 32--34, and 1--31, respectively \citep[as in][]{2006MNRAS.371...67S}.
The same procedure is applied to both calibration and nighttime observations.

The wavelength calibration proceeds through two alternative approaches, depending on whether an attached 60\,s ThAr+Ne frame is actually available.
When such a frame is available, it is used with the specific extraction windows described above;
otherwise, a series of $3\times2$ ThArNe exposures is used.
The full procedure is described in \citet{2009A&A...501..291S}.
Different integration time exposures are used to calibrate different spectral orders:
frames with individual integration times of 3, 15, and 30\,s are used to calibrate orders 30--39, 10--29, and 1--9, respectively.
In all cases, saturated lines are excluded from the wavelength calibration solution.
Finally, a two-pass $3 \sigma$ clipping is applied to reject remaining discordant lines.
The resulting WLC residuals are systematically better than 0.003\,{\AA} when using the ThArNe series.
Comparable performance is obtained with ThAr+Ne exposures, although values as high as 0.006\,{\AA} have occasionally been reached.
For the observation at $\mbox{HJD} = \mbox{2,454,092.827}$, the attached ThAr+Ne exposure failed to provide an acceptable WLC and the ThArNe series was used instead.

Finally, the blaze function and the WLC are applied to nighttime observations.
During this process, the final spectra are remapped onto a linear wavelength scale with a step size of 0.03\,{\AA}.
This corresponds to sampling the resolution element with the Nyquist frequency at 3000\,{\AA} and with twice that sampling rate at 6000\,{\AA}.
Because of the long time base of our observations, and because of the different WLC approaches, the \ion{Na}{i} $\lambda$5896 interstellar line in the spectrum of WR\,25 was used to check the consistency of our wavelength calibration.
The adopted line presents multiple components.
Although the four main ones are fitted simultaneously (see below), the solution is tested on the basis of the strongest red-shifted component, which is indeed less affected by the uncertainty of the deblending.
The average velocity measured for this component is $4.014 \pm 0.097$\,km\,s$^{-1}$;
the maximal deviation from the average value is 0.127\,km\,s$^{-1}$.
The dispersion of the measurements is in perfect agreement with the measurement uncertainties obtained from the multi-Gaussian fitting.
No systematic uncertainties seem to affect the calibration, which is therefore fully consistent throughout the entire duration of the data set~I campaign.

\subsubsection{Normalisation to the continuum}%
\label{sssec:observations_DS1_normal}
Even after all these reduction steps, the continuum of the spectra still presents oscillations due to the global response function of the spectrograph.
The latter may vary from one epoch to another, due to imperfect corrections of the blaze function, and to variations in the flux of the flat-field lamps depending on their temperature.
The accurate normalisation of the WR spectra to the continuum is a very difficult step.
To verify the consistency of our results, we used a twofold approach.
A~first possibility consists in fitting a (relatively) low-degree polynomial to the mean flux value over a set of selected narrow windows.
The observed spectrum is then approximated order by order and normalised by dividing the spectrum by the resulting local polynomial.
Although this technique is usually good, the true spectrum of a WR star generates some problems.
Indeed, the broad underlying lines and their wings are usually mistakenly identified as part of the continuum.
The zones affected by these lines are thus clearly modified, generally leading to a weakening of the emission lines.\bigskip

The second possible approach consists in determining a relative response by using a continuum star obtained as close as possible (both in time and on the sky) to the science observation.
This star, preferentially an O-type star, is reduced in exactly the same way as the science exposure.
However, this is not always possible, and sometimes the continuum star was observed several hours apart and/or under quite different airmass or observing conditions.
On other occasions, no suitable continuum star could be found, so that the approach is not universally applicable.
The quality of the response curve can be well tested by merging the individual orders of the continuum star.
However, even if this merging is excellent for the continuum star (i.e., better than a fraction of a percent), the transposition to the science object (here WR\,25) is not necessarily as good.
When successful, the division by the response function derived from the continuum reference star leaves a linear trend in the data that results from the temperature difference between the two objects.
An additional division by a straight line could still be necessary.
This method could also be more efficient and has the major advantage of preserving the broad spectral lines.
Its main benefit is to better handle broad line regions with little or no continuum spanning several consecutive orders.
The possibility of having two stars (science and continuum) of the same brightness greatly improves the quality of the normalisation.
In the following, this procedure is referred to as the \emph{global} normalisation by opposition to the first approach, which is termed \emph{local} normalisation.\bigskip

\begin{figure}
	\centering
	\includegraphics[width=0.45\textwidth]{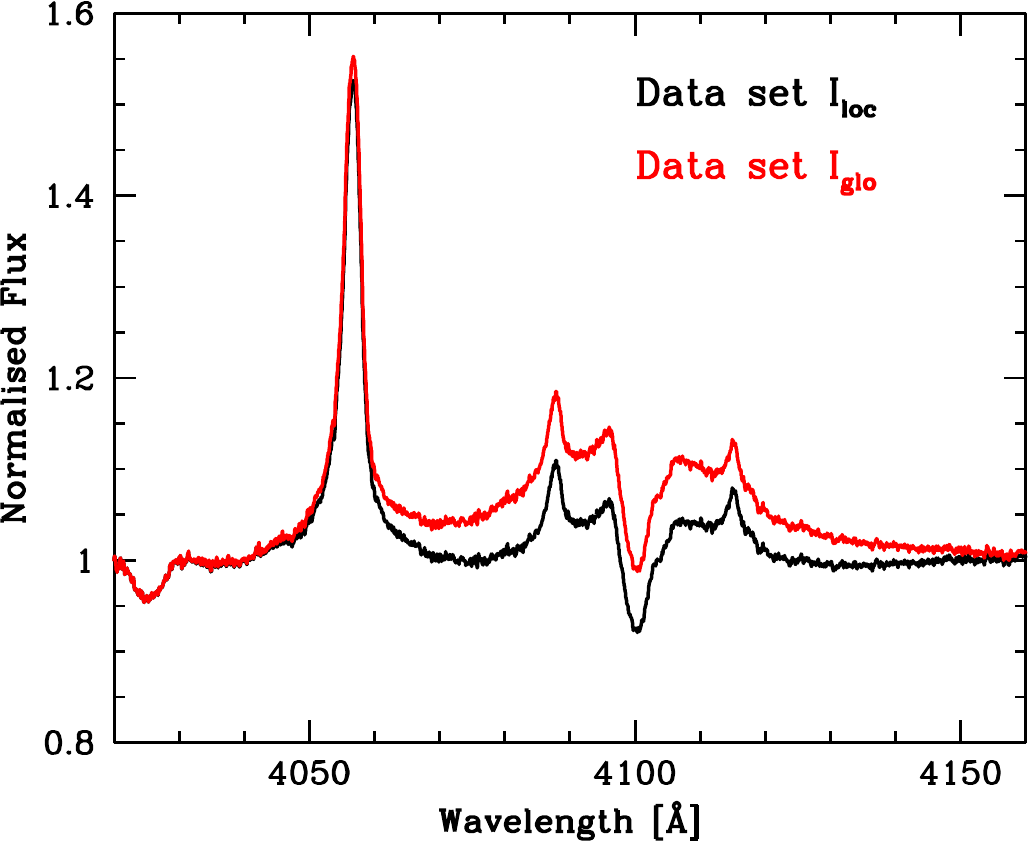}
	\hspace{1cm}
	\includegraphics[width=0.45\textwidth]{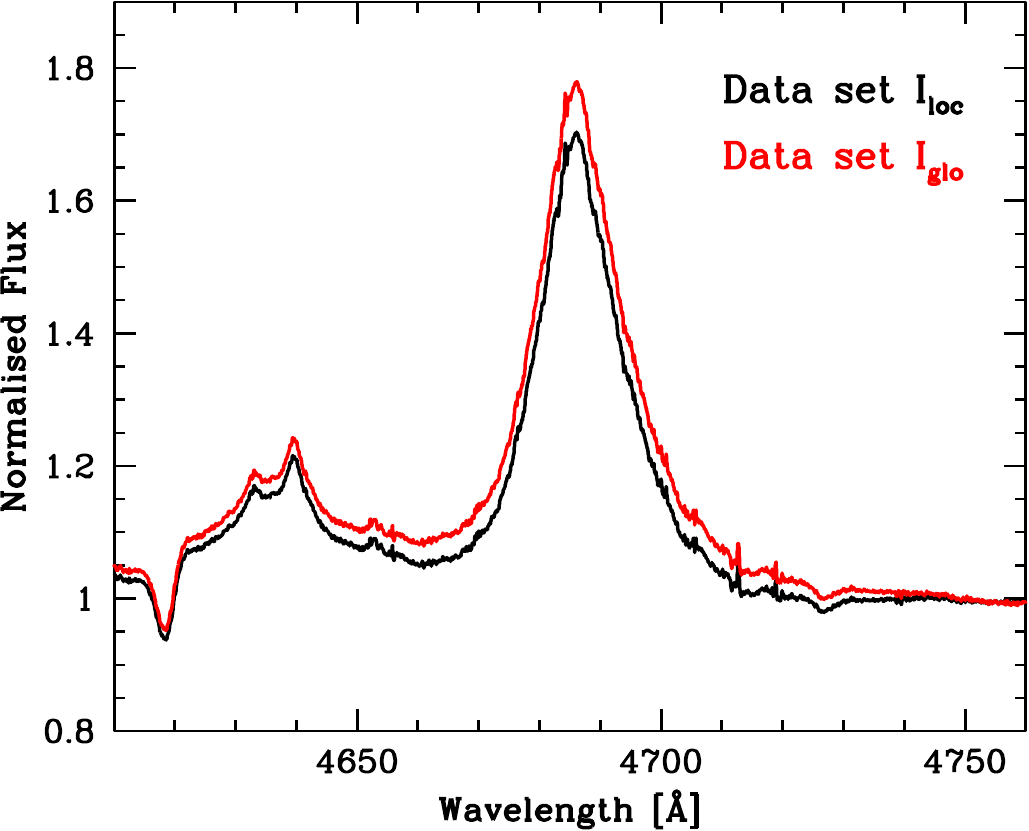}
	\vskip1ex
	
	\begin{minipage}{12cm}
		\caption{%
			Illustration of the spectrum of WR\,25 and of the dissimilarities between the normalisations to the continuum in the local (black) and in the global (red) reduction mode.
			\emph{(Left)} Region of the \ion{N}{iv}\,$\lambda$4058 and the \ion{He}{ii}\,$\lambda$4100+H$\delta$ lines.
			\emph{(Right)} Region between the \ion{N}{v}\,$\lambda$4619 transition and the \ion{He}{ii}\,$\lambda$4686 line.%
			\label{fig:fglocglo}}
	\end{minipage}
\end{figure}

\begin{figure*}
	\centering
	\includegraphics[width=0.8\textwidth]{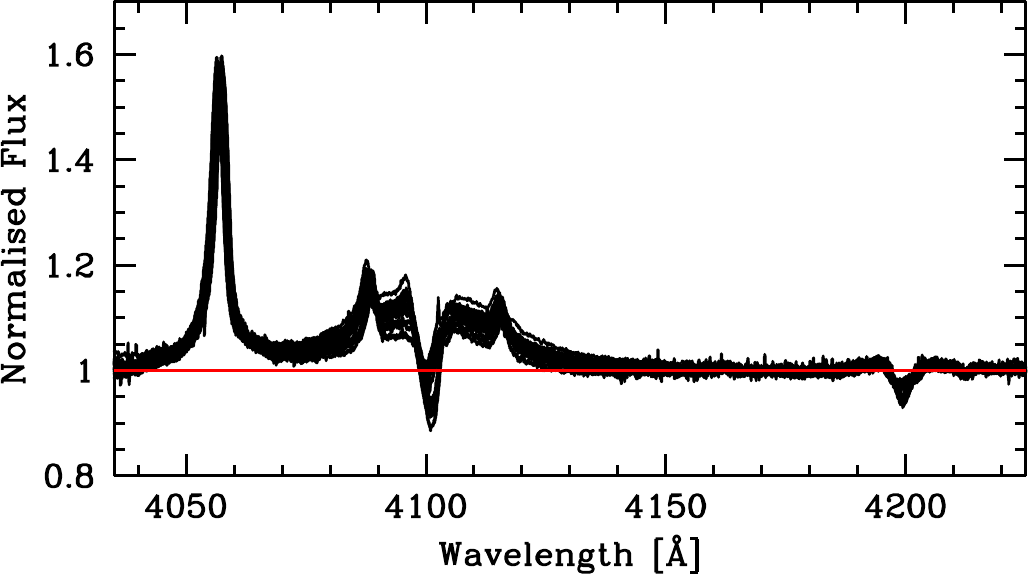}
	\vskip1ex
	
	\begin{minipage}{12cm}
		\caption{%
			The 21 spectra of the I$_{\text{glo}}$ data set overplotted to illustrate the stability of the global reduction procedure in this spectral region.%
			\label{fig:figplot21}}
	\end{minipage}
\end{figure*}

Generally, after order merging, both approaches provide similar results in relatively straightforward spectral regions, with dispersions over the continuum well below one percent.
In some more difficult regions, the global approach may be better suited to detecting very extended spectral emission regions.
Therefore, for data set~I, we have produced two versions of the reduced spectra:
I$_{\text{loc}}$ and I$_{\text{glo}}$, for the local and global procedures, respectively.
A~comparison of the dissimilarities between the results from the two methods used to define the continuum is illustrated in Fig.~\ref{fig:fglocglo} where we present selected wavelength domains for which the difficulty to trace the continuum is evident.
%
To demonstrate the stability of the reduction obtained with the global approach, Fig.~\ref{fig:figplot21} displays a selected portion of the WR\,25 spectrum for all 21 spectra included in I$_{\text{glo}}$.
Indeed, only 21 of the 27 spectra from I$_{\text{loc}}$ had a suitable reference O-type star available for the global normalisation procedure.
\vspace{-1ex} 

\subsection{Data set II}\label{ssec:observations_DS2}
Data set~II consists of a single observation obtained at a critical phase for which very few observations are available (see below).
The spectroscopic observation was acquired in February 2007 with the Magellan Inamori Kyoceros Echelle (MIKE) high-throughput double-echelle spectrograph mounted on the Magellan~II Clay Telescope located at Las Campanas Observatory.
The MIKE spectrograph has two arms:
a blue one (3350--5000\,{\AA}) and a red one (4500--9500\,{\AA}).
The two $\mbox{2K}\times\mbox{2K}$ detectors were both used with a $2\times2$ binning and in slow read-out mode.
The reading noise was about 3.5\,e$^{-}$.
The resolving power was approximately 33\,000.
The spectrum was processed with \textsc{IRAF} standard echelle routines.
Wavelength calibration was performed with ThAr lamp exposures taken before or after the stellar spectrum at the same telescope position.
Continuum normalisation was carried out in the local mode, by polynomial fitting.
The details and the journal of the observations are provided in Table~\ref{tabA2}.
\vspace{-1ex} 

\subsection{Data set III}\label{ssec:observations_DS3}
The third data set consists exclusively of FEROS spectra previously analysed by \citet{2006A&A...460..777G}.
Some 15 spectra acquired between 2001 and 2004 were found to be of sufficient quality to be included in our sample.
The first ones were still acquired with the ESO 1.5-m telescope, whereas the remaining ones (obtained after 2002) were acquired with the MPG/ESO 2.2-m telescope.
The reduction procedure has been described in \citet{2006A&A...460..777G}.
The normalisation to the continuum was performed in local mode only as no suitable continuum-star observations were acquired during the same nights.
The details and the journal of the observations are provided in Table~\ref{tabA3}.

\subsection{Data set IV}\label{ssec:observations_DS4}
The fourth data set consists of spectra acquired in 2006 and 2007 with the 2.5-m Irénée du Pont Telescope at Las Campanas Observatory, Chile.
One of these spectra was used to illustrate the detection of the companion reported in \citet{2008RMxAC..33...91G}.
The telescope was equipped with an echelle spectrograph providing spectra in the range $\sim 3700$\,{\AA} to 7000\,{\AA} with a resolving power of about 40\,000.
The detector was a TeK5 2K by 2K CCD with a pixel size of 24\,{\textmu}m.
We used a 1 by 4 pixel slit and the typical S/N ratio was at least 100.
Some 10 spectra were acquired in 2006 and two (acquired during the same night) in 2007.
Spectra were processed following the same standard procedure described above (see Sect.~\ref{ssec:observations_DS2}).
The normalisation to the continuum was performed in local mode only as no suitable continuum-star observations were acquired during the same nights.
The details and the journal of the observations are provided in Table~\ref{tabA4}.

\section{The WR Spectrum and Related RV Measurements}%
\label{sec:wrspec}
The visible spectrum of WR\,25 is dominated by the \ion{He}{ii}\,$\lambda$4686 line and the \ion{He}{ii}\,$\lambda$6560+{\hskip0pt}H$\alpha$ blend.
The \ion{N}{iv}\,$\lambda$4058 line also stands out.
This is precisely the line that was used to reveal the binarity of the system \citep{2006A&A...460..777G}.
As expected for a WN object, several other nitrogen lines are detected:
the \ion{N}{v}\,$\lambda\lambda$4604-4619 doublet, the \ion{N}{iv} lines around 6220 and 7110\,{\AA} and others.
Numerous \ion{N}{iii} lines are also present.
The \ion{Si}{iv}\,$\lambda \lambda$4089-4116 doublet is observed in emission and is blended with the \ion{He}{ii}\,$\lambda$4100+H$\delta$ lines.

The main objective of the work described in this section is to measure the RVs of the WR star in order to derive its orbital motion.
We therefore focus on spectral lines that are expected to best trace the true motion of the star.
Lines such as \ion{He}{ii}\,$\lambda$4686 and \ion{He}{ii}\,$\lambda$6560+H$\alpha$ are too broad and too complex, and may further contain (variable) contributions from the wind-collision region.
As a result, these features do not provide sufficiently reliable RV measurements.
Lines that could be hampered by the presence of the companion (e.g., Balmer lines) should also be avoided.
The \ion{N}{iv}\,$\lambda$4058 is certainly much more suitable and reliable to use.
In the following, we measured the positions and RVs of a series of lines in each individual spectrum and carefully examined their behaviour.
The line positions were estimated on the basis of single- or multiple-Gaussian fitting and the method is detailed below for each line individually.

\subsection{Data set I}%
\label{ssec:wrspec_DS1}

\subsubsection{\texorpdfstring{The \ion{N}{iv}\,$\lambda$4058 line}{The N IV lambda4058 line}}%
\begin{figure*}
	\centering
	\includegraphics[width=0.45\textwidth]{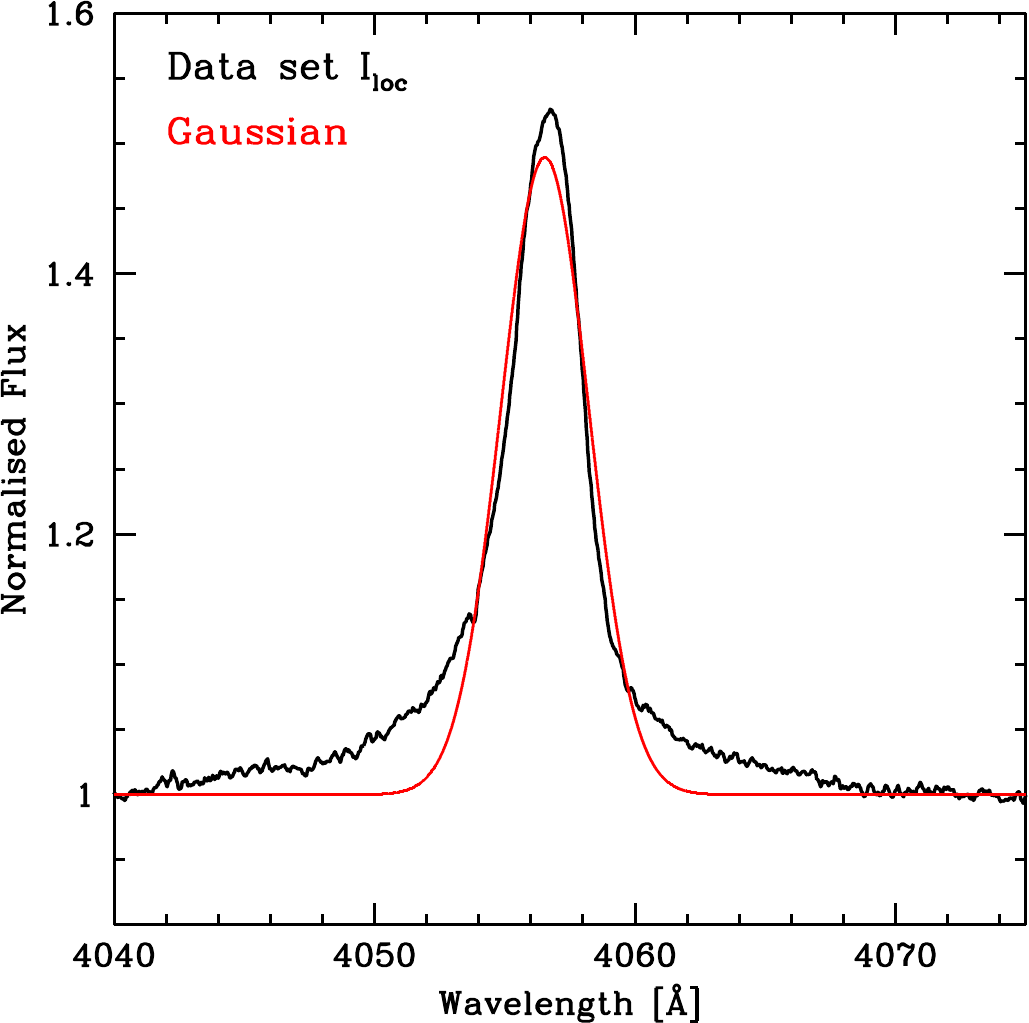}
	\hspace{1cm}
	\includegraphics[width=0.45\textwidth]{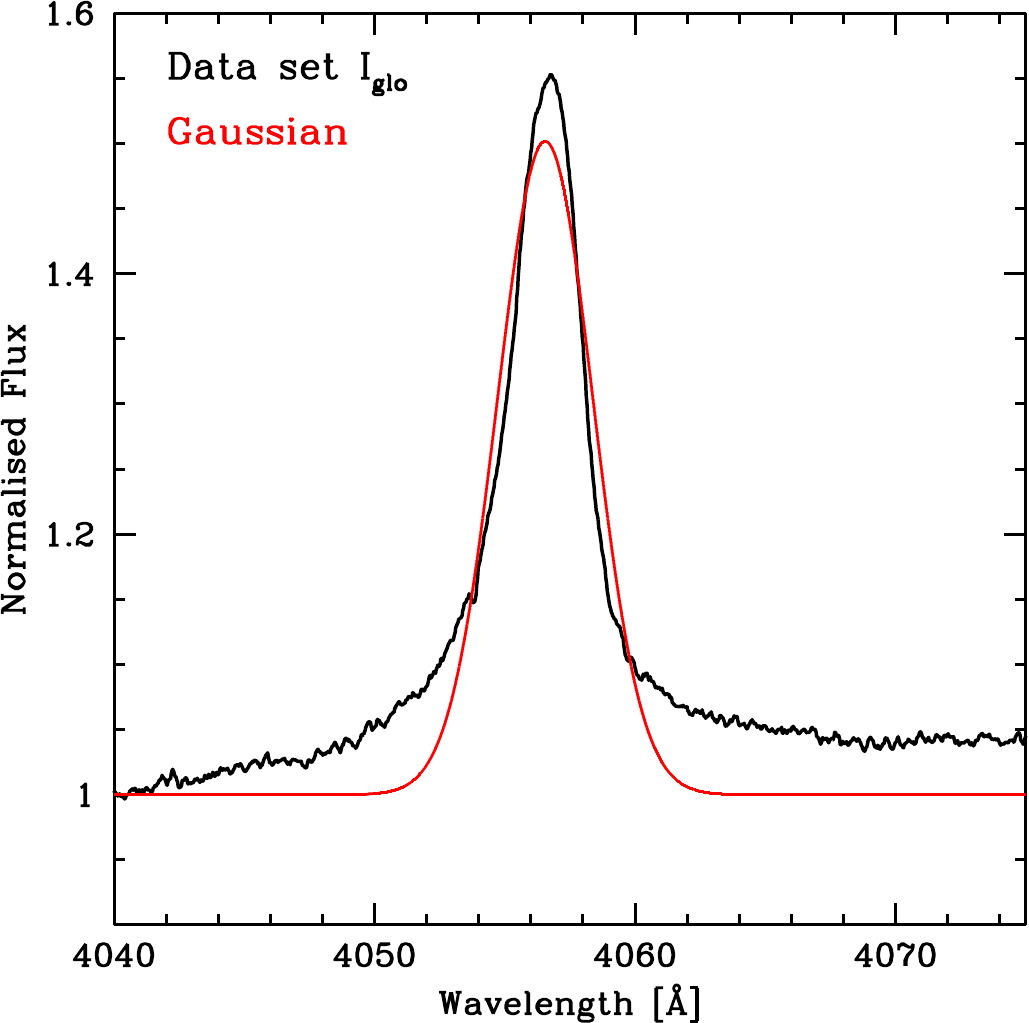}
	\vskip1ex
	
	\begin{minipage}{12cm}
		\caption{%
			Single-Gaussian fit to the \ion{N}{iv}\,$\lambda$4058 line in WR\,25 illustrated for a spectrum acquired at HJD\,2,453,868.628.
			The black curve represents the observed spectrum, whereas the red curve corresponds to the fitted profile.
			\emph{(Left)} Spectrum reduced using the local normalisation procedure.
			\emph{(Right)} Same spectrum reduced using the global normalisation procedure.%
			\label{fig:fgfitlg4058g1}}
	\end{minipage}
\end{figure*}

\label{sssec:wrspec_DS1_4058}
A~major and particularly important line in the spectra of WN stars is \ion{N}{iv}\,$\lambda$4058, essentially as a result of its strength and relative narrowness.
It arises from a highly ionised element and is expected to form deep within the stellar wind.
This line is often used to derive radial velocities \citep{2006A&A...460..777G}.
We adopted a typical laboratory wavelength $\lambda_{0} = 4057.90$\,{\AA} for this line, as proposed by \citet{1977ApJ...214..759C}.
We first fitted this line with a single Gaussian curve.
An example is shown in Fig.~\ref{fig:fgfitlg4058g1}, illustrating the best fits for both data sets I$_{\text{loc}}$ and I$_{\text{glo}}$, and for the same spectrum acquired at HJD\,2,453,868.628.
Although a single Gaussian does not provide an entirely satisfactory representation of the line profile, it nevertheless provides some valuable estimate of the RV.
The observed profile presents excess flux, particularly in the blue wing, although the broad part of the line is detected on both wings.
Further to the red, a contribution from the blend of the \ion{He}{ii}\,$\lambda$4100+H$\delta$ wings is also visible in the I$_{\text{glo}}$ data.

\begin{figure*}
	\centering
	\includegraphics[width=0.45\textwidth]{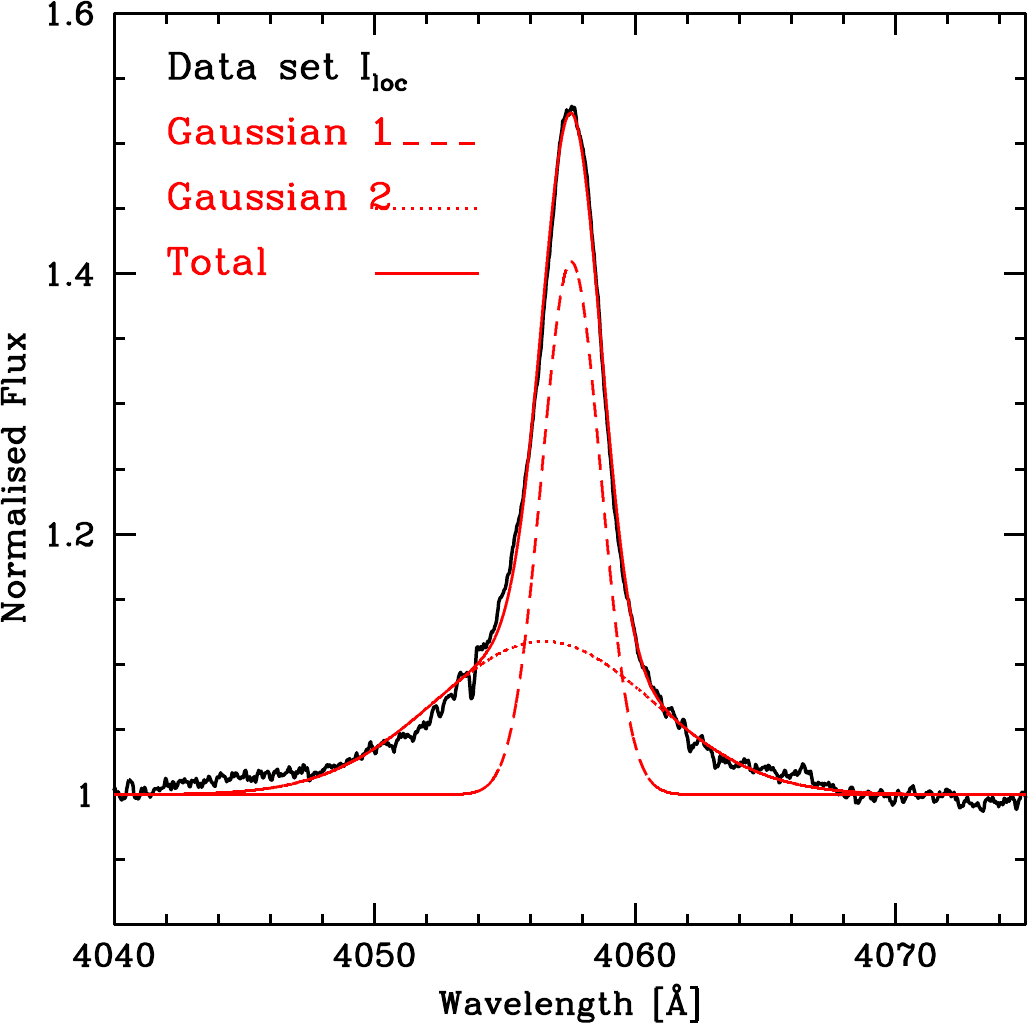}
	\hspace{1cm}
	\includegraphics[width=0.45\textwidth]{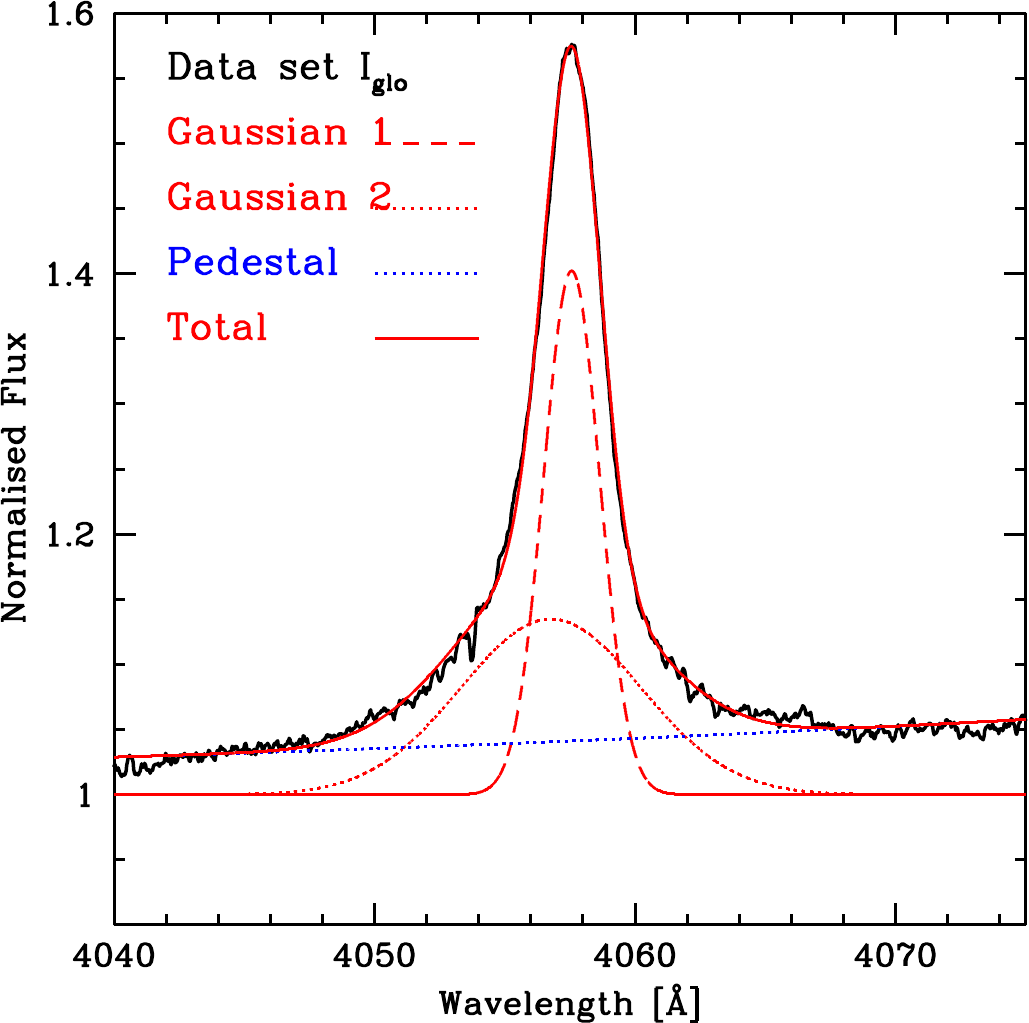}
	\vskip1ex
	
	\begin{minipage}{12cm}
		\caption{%
			Two-Gaussian fits to the \ion{N}{iv}\,$\lambda$4058 line in WR\,25 illustrated for a spectrum acquired at HJD\,2,453,901.587.
			The black curve represents the observed spectrum, whereas the red curve corresponds to the fitted profile.
			\emph{(Left)} Spectrum reduced using the local normalisation procedure.
			\emph{(Right)} Same spectrum reduced using the global normalisation procedure.
			This latter fit required the introduction of an additional pedestal component, shown in blue.
			Gaussian~1 was adopted for the RV measurement.%
			\label{fig:fgfitlg4058gm}}
	\end{minipage}
\end{figure*}
\begin{figure*}
	\centering
	\includegraphics[width=0.45\textwidth]{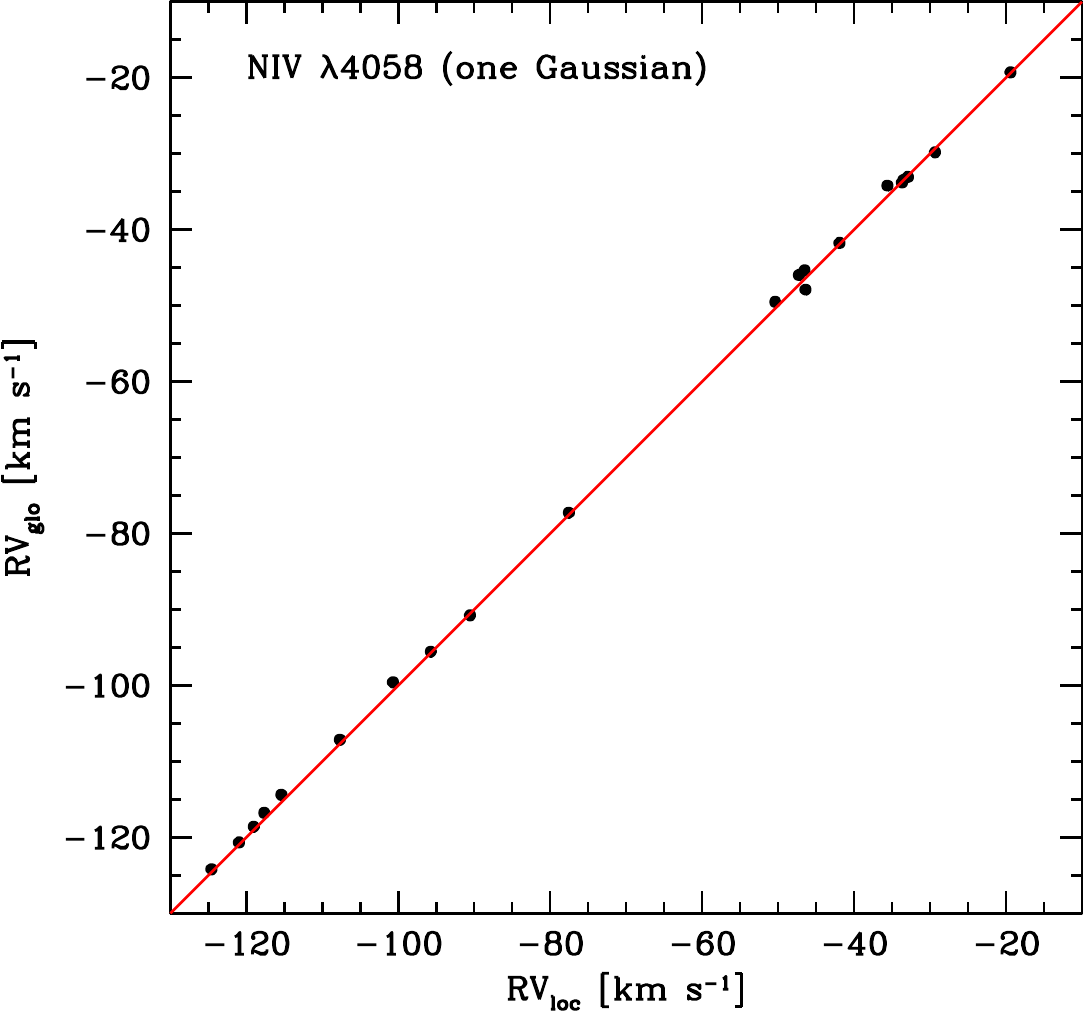}
	\hspace{1cm}
	\includegraphics[width=0.45\textwidth]{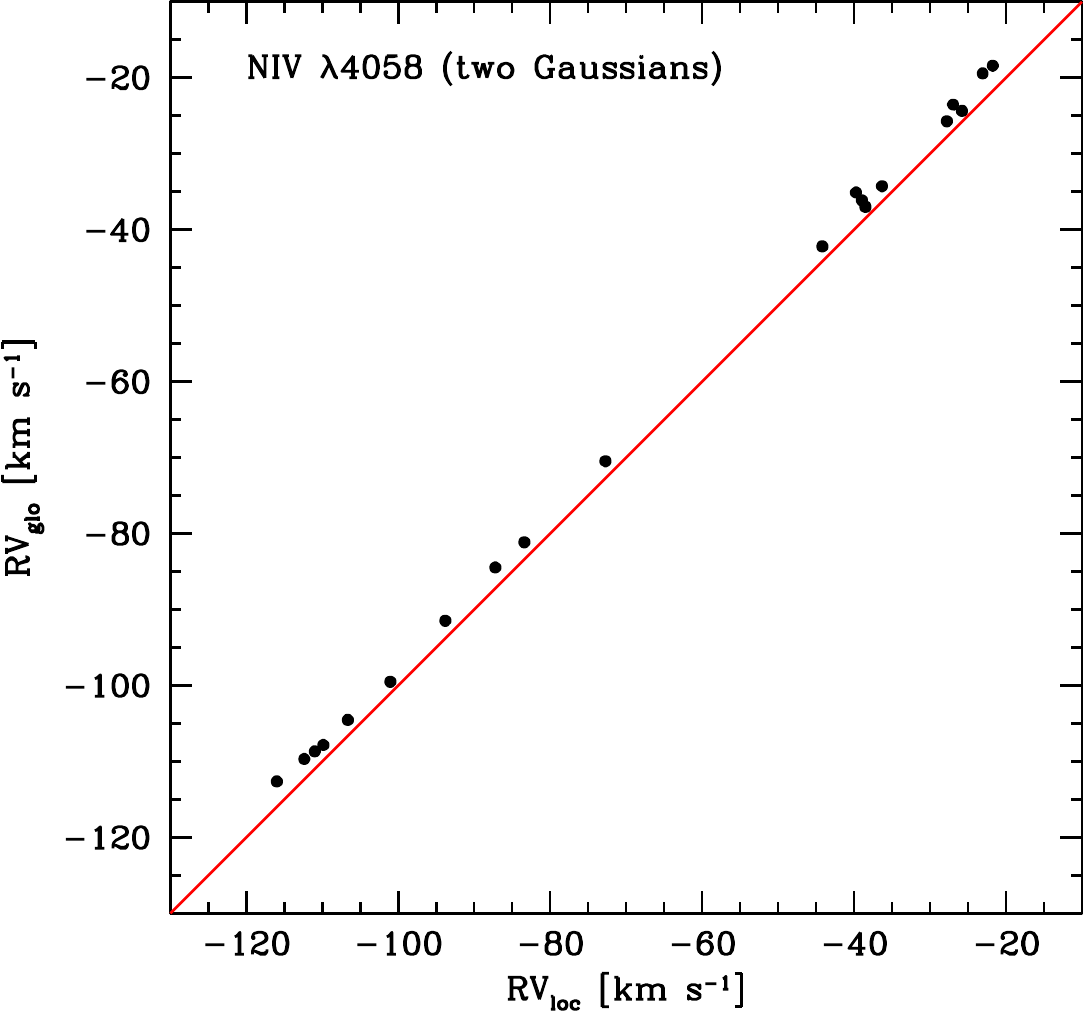}
	\vspace{0.5cm} \\
	\mbox{}\hspace{1.5mm}\includegraphics[width=0.432\textwidth]{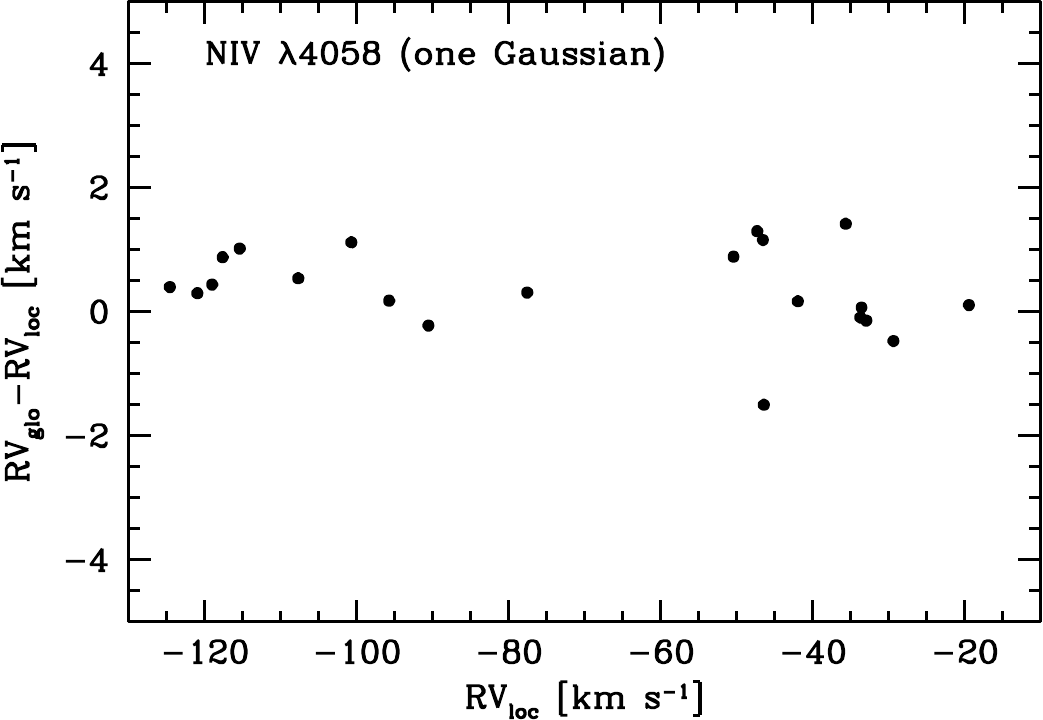}
	\hspace{13mm}
	\includegraphics[width=0.432\textwidth]{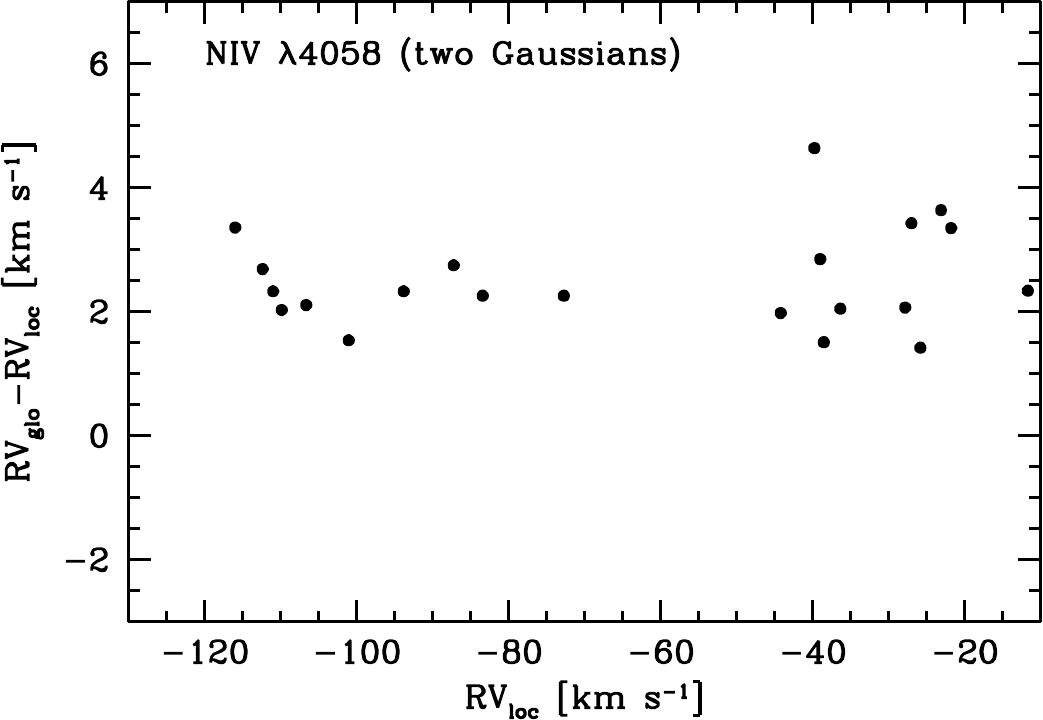}
	\vskip1ex
	
	\begin{minipage}{12cm}
		\caption{%
			Comparison between the RVs derived from the globally normalised spectra and those measured from the locally normalised spectra.
			\emph{(Upper left)} Results obtained from single-Gaussian fits.
			\emph{(Upper right)} Results obtained from two-Gaussian fits, adopting Gaussian 1 for the RV measurements.
			The red line indicates the perfect one-to-one relation.
			The agreement is particularly good, especially when allowing for a constant offset between the two determinations.
			\emph{(Lower panels)} Same comparison, with the ordinate now representing the differences between the two RV measurements.%
			\label{fig:fgcorr4058lg}}
	\end{minipage}
\end{figure*}

We also performed two-Gaussian fits, including an additional trend component in the case of the globally normalised spectra.
In the following, this additional component is referred to as the {\emph{pedestal}}.
Examples of these fits are shown in Fig.~\ref{fig:fgfitlg4058gm} for the spectrum acquired at HJD\,2,453,901.587.
They demonstrate that the two-Gaussian fits provide a significantly better representation of the profile.
The primary Gaussian (Gaussian~1) reproduces the position of the main emission peak more accurately.
Only very minor residuals remain.
However, the two fitting approaches do not necessarily yield compatible absolute RV values.
Figure~\ref{fig:fgcorr4058lg} compares the RVs obtained from the dominant component (the single Gaussian in the one-component fit and Gaussian~1 in the two-Gaussian fit) for the two kinds of continuum-normalisation methods (local versus global).
The two types of reduction clearly have no significant impact on the RV measurement in the single-Gaussian case, whereas a small but systematic effect is present for the two-Gaussian fit.
To first order, the two RV determinations differ only by a constant, but the slope of the relation is preserved.
This means that the absolute positions (and therefore the absolute RVs) of the emission line are determined only up to an arbitrary constant, whereas the relative positions---and thus the relative variations of the positions and consequently of the RVs---are fully determined.
This behaviour appears to be a general property of the WR spectra:
absolute RV measurement cannot be reliably obtained.
The absolute wavelengths of blends of emission lines remain unknown, as they may depend on the physical conditions in the line-formation zone.
In addition, the line profiles can be strongly asymmetric, and the relation between the centroid of the line (as derived, for example, by a Gaussian fit) and the absolute zero-velocity position of the line remains unknown.
The position of the line could also be affected by systematic effects.
Since we preferentially measure narrow lines expected to trace the orbital motion more closely, these lines are likely formed close to the stellar core, and occultation effects could thus induce line shifts.
As a consequence, the orbital motion can be accurately determined, but not the systemic velocity.
If we accept this conclusion, the precise choice of laboratory rest wavelengths becomes less critical.
However, it demonstrates that we can reach a good precision on the orbital motion provided other spectroscopic lines agree with the observed motion.
The RVs derived from the \ion{N}{iv}\,$\lambda$4058 line are listed in Appendix~\ref{sec:appB}.

Figure~\ref{fig:fgcorr4058ll} compares the RVs obtained from the single-Gaussian fits with those measured from the two-Gaussian fits, adopting Gaussian~1 for the RV measurement.
We restrict the comparison to the I$_{\text{loc}}$ data set.
Here again, the two RV sets essentially differ only by a single offset.
\begin{figure}
	\centering
	\includegraphics[width=0.45\textwidth]{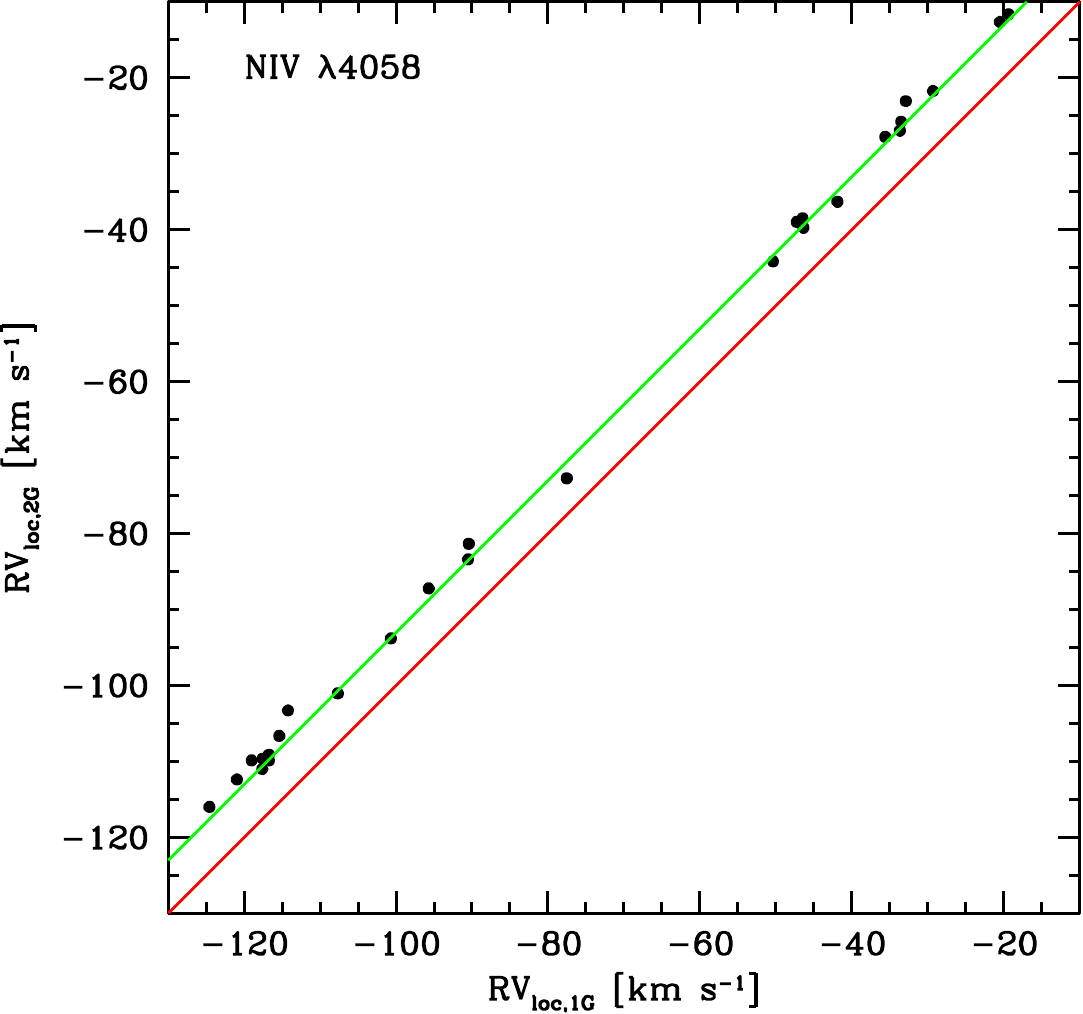}
	\hspace{1cm}
	\includegraphics[width=0.45\textwidth]{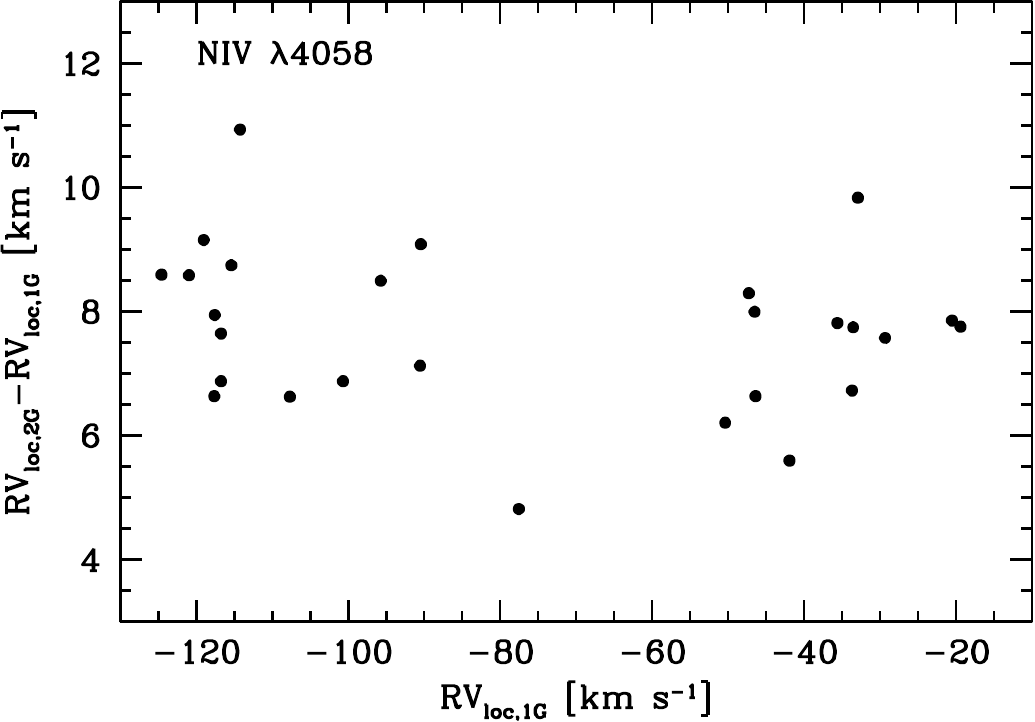}
	\vskip1ex
	
	\begin{minipage}{12cm}
		\caption{%
			\emph{(Left)} Comparison between the RVs derived from the single-Gaussian fits (1G) and those associated with Gaussian 1 from the two-Gaussian fits (2G);
			both measurements correspond to the locally normalised spectra.
			The red line indicates the one-to-one relation, whereas the green line is an arbitrary line drawn parallel to it.
			This illustrates that the two RV sets differ by a mere constant.
			\emph{(Right)} RV differences for the same data set.%
			\label{fig:fgcorr4058ll}}
	\end{minipage}
\end{figure}

\begin{figure}
	\centering
	\includegraphics[width=0.45\textwidth]{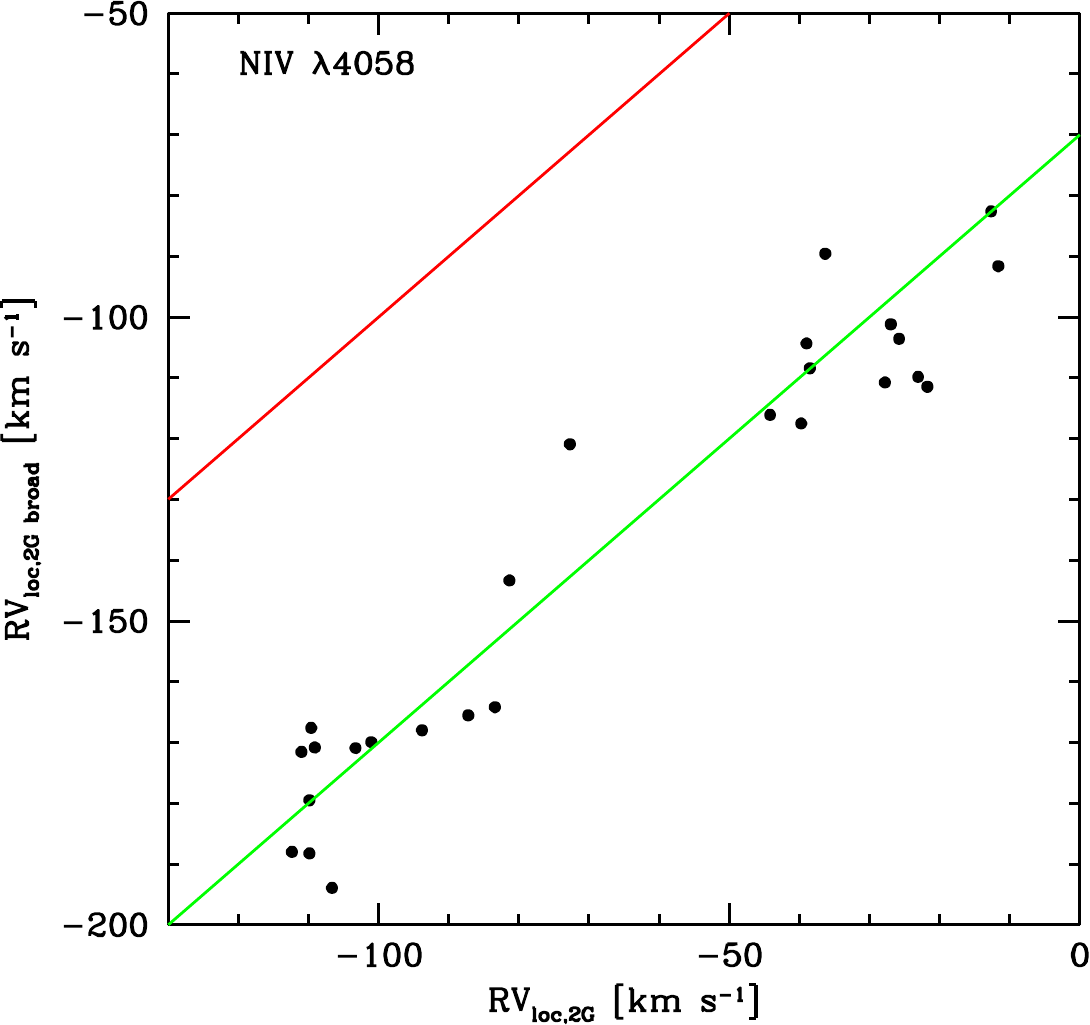}
	\vskip1ex
	
	\begin{minipage}{12cm}
		\caption{%
			Comparison between the RVs associated with Gaussian~1 and those associated with Gaussian~2, the latter representing the broader component of the profile. Despite the observed dispersion, both components vary essentially together and differ primarily by a constant offset, confirming that the combination of two Gaussians provides an adequate representation of the line profile.%
			\label{fig:fgcorr4058add}}
	\end{minipage}
\end{figure}

Figure~\ref{fig:fgcorr4058add} compares the RVs associated with Gaussian~1 and Gaussian~2 (i.e., respectively the major and the weaker).
Although some dispersion is present, the separation between the two Gaussian components remains approximately constant to first order.
Thus, to first order, the two Gaussians are considered necessary for a satisfactory representation of the physical line profile.
However, they vary essentially in concert, and the relative motion of the star can be reliably associated with the dominant component, Gaussian~1.
The same conclusion is reached when using the globally normalised data set.

Since some spectra are not included in the I$_{\text{glo}}$ data set, we decided to adopt the RV measurements derived from the two-Gaussian fits to the I$_{\text{loc}}$ data set, using the RVs associated with the dominant component (Gaussian~1).

\subsubsection{\texorpdfstring{The \ion{N}{v}\,$\lambda \lambda$4604-4619 doublet}{The N V lambda lambda 4604-4619 doublet}}%
\label{sssec:wrspec_DS1_4604}
This doublet consists primarily of the \ion{N}{v}\,$\lambda$4604 transition, which presents a P-Cygni profile, and the \ion{N}{v}\,$\lambda$4619 transition, which is observed mainly in absorption.
In numerous objects, both lines present P-Cygni profiles.
However, in the case of WR\,25, the second line is seen essentially in absorption and with a very weak emission, if any.
The two lines of the doublet are not blended with each other.
To the red, a broad blend of several lines is present (see Fig.~\ref{fig:fglocglo}) and the \ion{N}{iii}\,$\lambda \lambda$4634-4641 triplet emission is rather close and induces a marked slope in the apparent continuum.
The contamination is significant even in the locally normalised version of the spectra.
It is therefore necessary to include a pedestal component in the fitting procedure.
We only used the local version because the corresponding fit is much more stable.

\begin{figure}[t]
	\centering
	\includegraphics[width=0.45\textwidth]{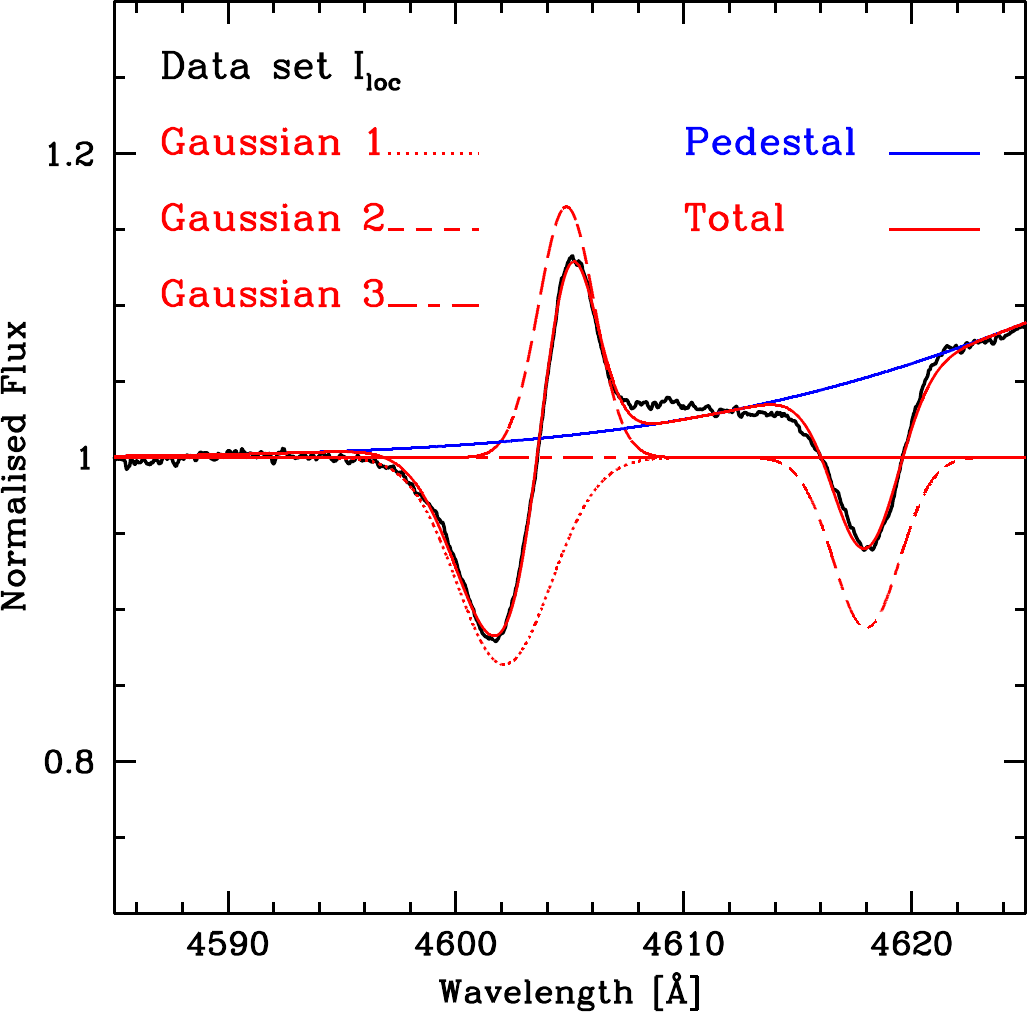}
	\vskip1ex
	
	\begin{minipage}{12cm}
		\caption{%
			Determination of an RV associated with the \ion{N}{v}\,$\lambda \lambda$4604-4619 doublet for a spectrum acquired at HJD\,2,454,912.516.
			Three Gaussian components were used to fit the spectral region:
			one for the absorption of the \ion{N}{v}\,$\lambda$4604 transition (Gaussian~1), one for the corresponding emission component (Gaussian~2), and one accounting for the absorption of the \ion{N}{v}\,$\lambda$4619 transition (Gaussian 3).
			The surrounding spectral region further required the introduction of a pedestal component (shown in blue).
			The position of Gaussian~2 has been adopted as the reference RV.%
			\label{fig:fgfitl4604}}
	\end{minipage}
\end{figure}

\begin{figure}
	\centering
	\includegraphics[width=0.45\textwidth]{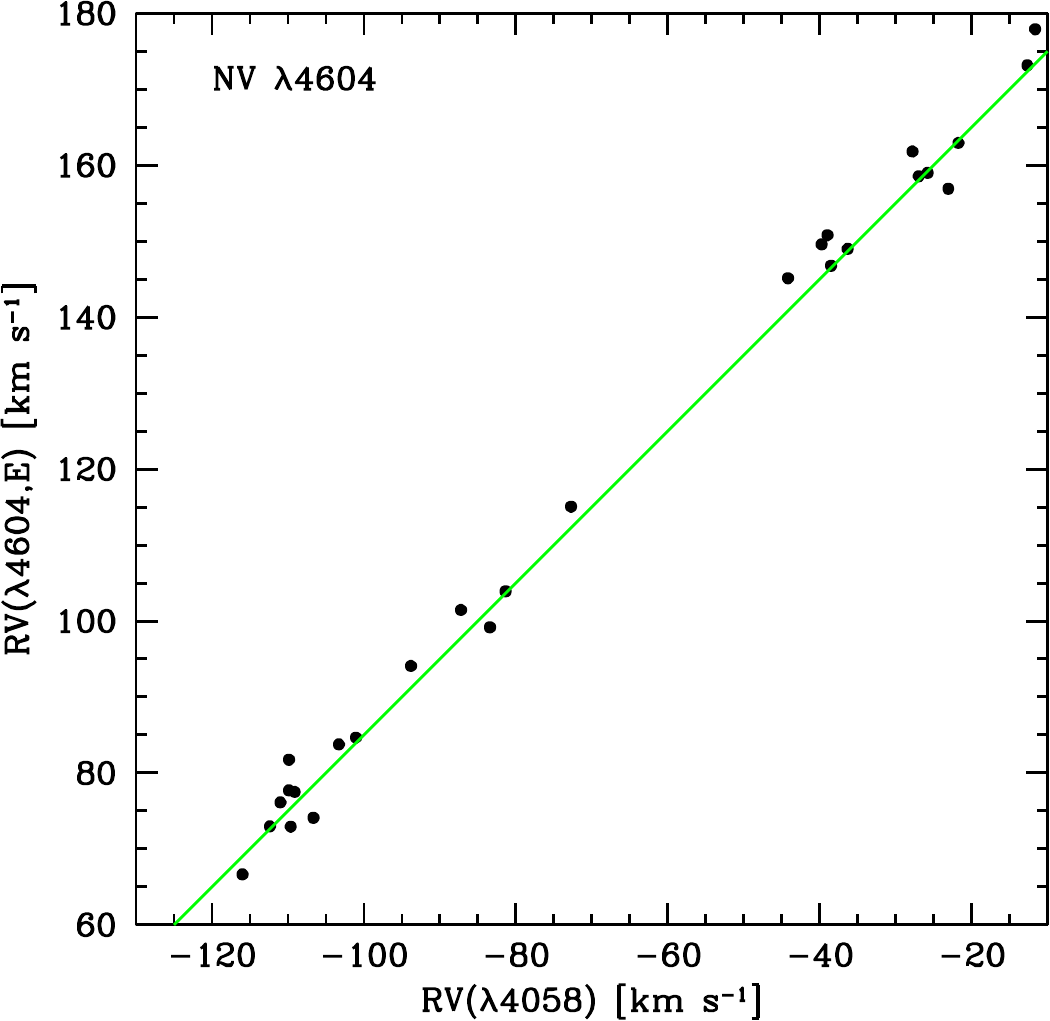}
	\hspace{1cm}
	\includegraphics[width=0.45\textwidth]{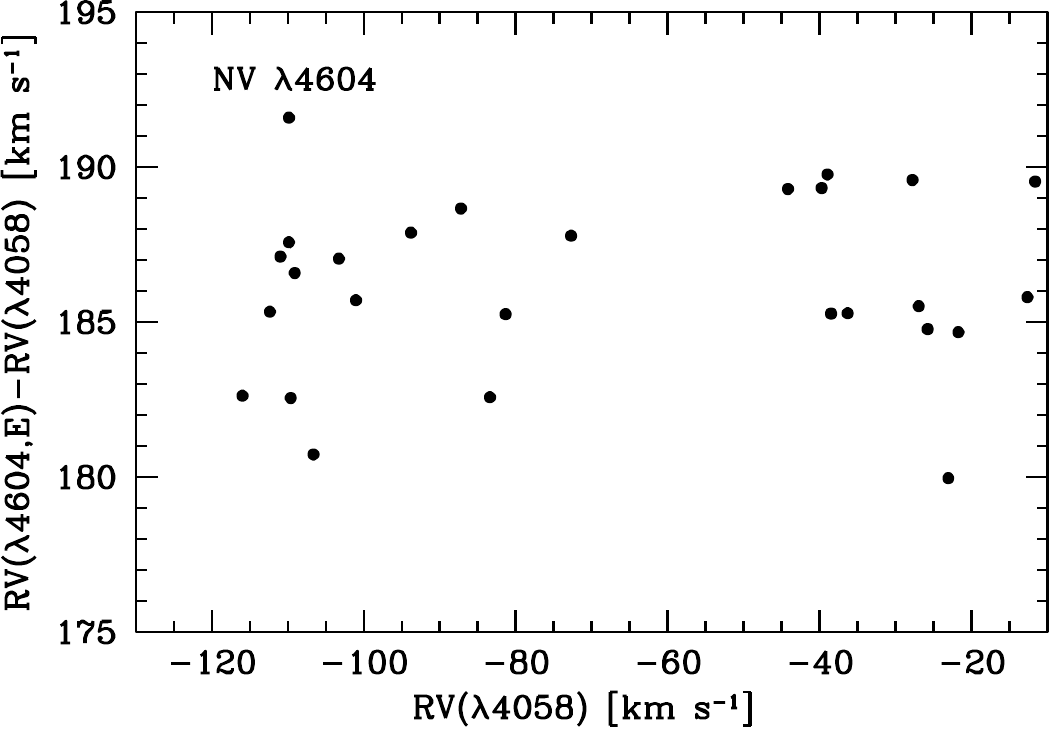}
	\vskip1ex
	
	\begin{minipage}{12cm}
		\caption{%
			\emph{(Left)} Comparison between the RVs associated with Gaussian~2 in Fig.~\ref{fig:fgfitl4604} and those derived from the \ion{N}{iv}\,$\lambda$4058 line.
			The one-to-one correlation line lies outside the plotted range.
			The green line is an arbitrary line parallel to it.
			The large---but essentially constant---difference between the RVs of the two data sets is not unexpected owing to the particular choice of the profile component adopted as the RV reference.
			\emph{(Right)} Same comparison shown in terms of RV differences.%
			\label{fig:fgcorr4604}}
	\end{minipage}
\end{figure}

The spectral region has been fitted with three Gaussians together with a pedestal.
Here and throughout the following analysis, the parameters of each Gaussian (position, amplitude, and width) were allowed to vary freely and independently.
For this spectral region, the RVs associated with the \ion{N}{v}\,$\lambda$4604 line were adopted, but the \ion{N}{v}\,$\lambda$4619 line was fitted simultaneously to improve the stability of the procedure.
Gaussian~1 was assigned to the absorption part of the \ion{N}{v}\,$\lambda$4604 transition profile, Gaussian~2 to the emission part and Gaussian~3 to the absorption of the other line.
An example of the fit is illustrated in Fig.~\ref{fig:fgfitl4604}.
Because the slopes of the P-Cygni profile differ strongly on the blue side, on the red side, and between the absorption and emission components, Gaussians~1 and 2 strongly interact;
we nevertheless consider that this remains the best way to proceed.
In contrast to the case described in the previous section, several possible reference points can be adopted here.
We investigated the position of the absorption component (Gaussian~1), the position of the emission component (Gaussian~2), and the wavelength corresponding to the intersection of the total fitted profile with the normalised flux level of unity.
The position derived from the absorption turned out to be noisier, whereas the two other options yielded quite equivalent results, differing mainly by a constant offset.
We therefore decided to report the RV value corresponding to the emission component (Gaussian~2).
As before, the derived RVs are not intended to provide absolute velocity measurements.
In all cases, we adopted $\lambda_{0} = 4603.73$\,{\AA} for the laboratory wavelength.
%
Figure~\ref{fig:fgcorr4604} compares the RVs derived from the \ion{N}{iv}\,$\lambda$4058 line with those obtained here.
The two RV series are in good agreement despite a huge---and expected---difference that is essentially constant.
Both lines therefore tend to trace the same relative motion.

\subsubsection{\texorpdfstring{The \ion{N}{iv}\,$\lambda$5737 line}{The N IV lambda 5737 line}}%
\begin{figure}
	\centering
	\includegraphics[width=0.45\textwidth]{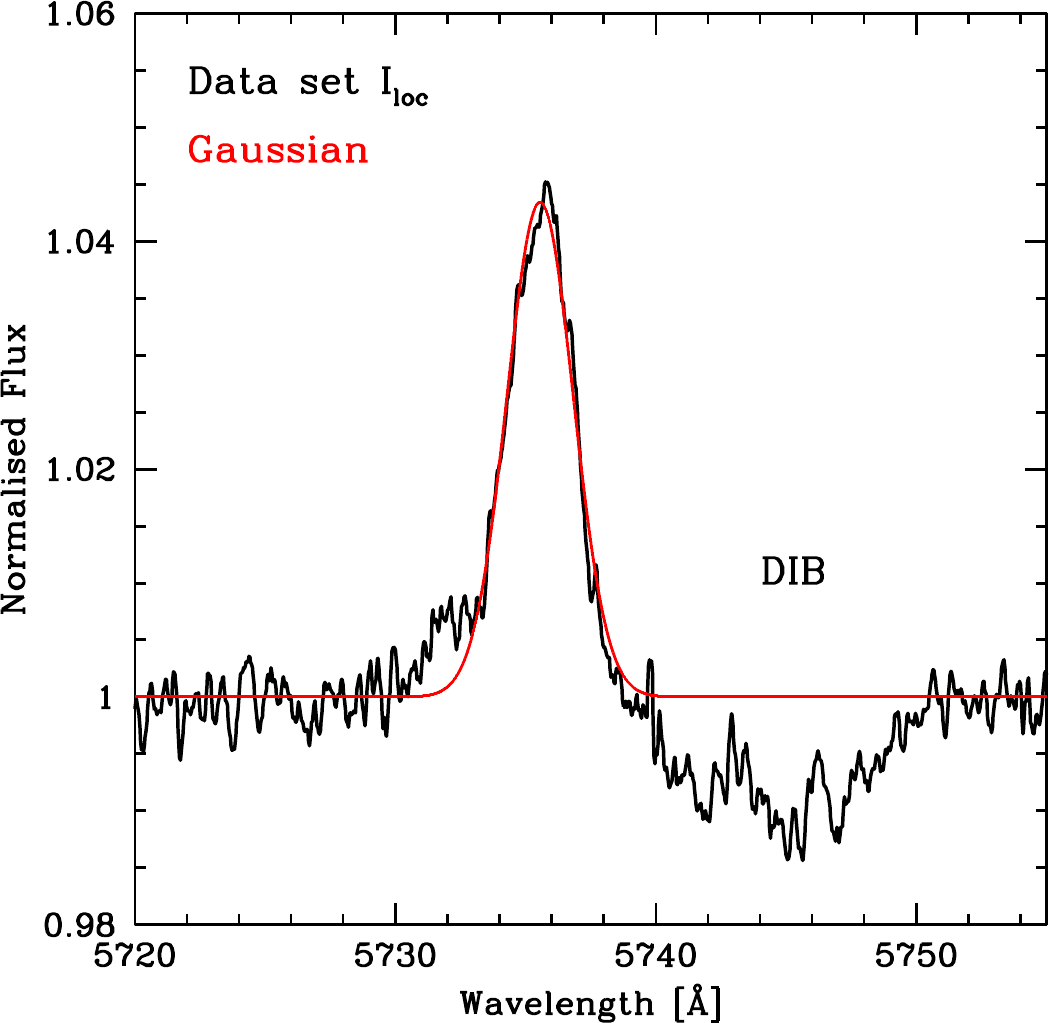}
	\vskip1ex
	
	\begin{minipage}{12cm}
		\caption{%
			Example of the RV determination based on the fit of the \ion{N}{iv}\,$\lambda$5737 line.
			The spectrum shown here was acquired at HJD\,2,453,864.483.%
			\label{fig:fgfitl5737}}
	\end{minipage}
\end{figure}

\begin{figure}
	\centering
	\includegraphics[width=0.45\textwidth]{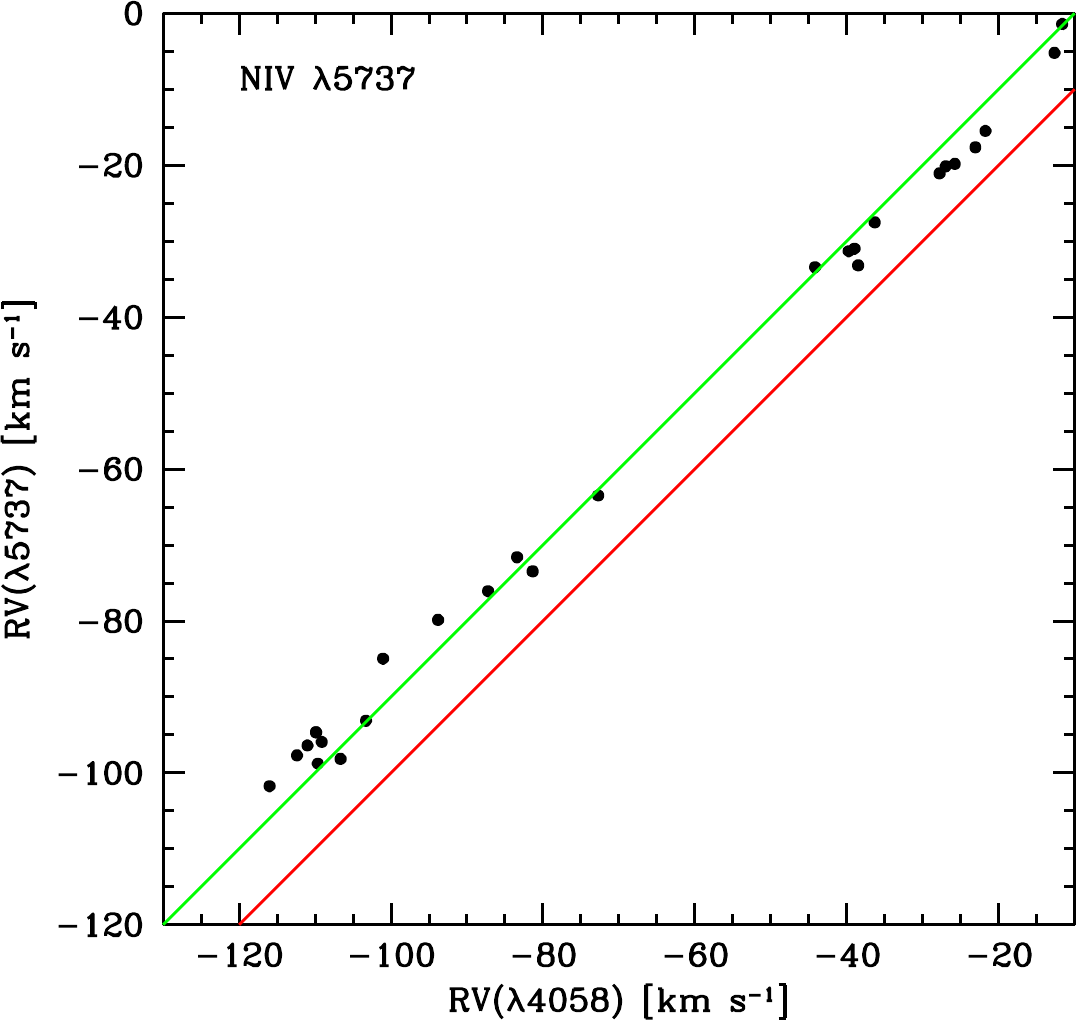}
	\vskip1ex
	
	\begin{minipage}{12cm}
		\caption{%
			Comparison between the RVs derived from the \ion{N}{iv}\,$\lambda$5737 and \ion{N}{iv}\,$\lambda$4058 lines.
			The red line represents the one-to-one correlation, whereas the green line is drawn parallel to it.%
			\label{fig:fgcorr5737}}
	\end{minipage}
\end{figure}

\label{sssec:wrspec_DS1_5737}
The \ion{N}{iv}\,$\lambda$5737 line is quite isolated, and the continuum in its vicinity is the same in data sets I$_{\text{glo}}$ and I$_{\text{loc}}$.
Therefore, the measured positions of the lines in both the globally and locally normalised data sets are strongly similar and, in practice, almost identical.
Although the line is comparatively isolated, a diffuse interstellar band (DIB) at $\lambda$5747.81 is present near its red side.
We therefore adopted a single-Gaussian fit for this line, as illustrated in Fig.~\ref{fig:fgfitl5737}.
The shape of the DIB is too complex to justify the introduction of an additional fitting component, which would introduce unwanted arbitrariness in the process.
The adopted laboratory wavelength for this line is $\lambda_{0} = 5736.93$\,{\AA}.
The Gaussian profile provides a good representation of the observed line shape.
A~comparison between the RVs derived from this transition and those obtained from \ion{N}{iv}\,$\lambda$4058 is presented in Fig.~\ref{fig:fgcorr5737}.
Once again, the two RV series differ primarily by a constant offset, while exhibiting very similar slopes.
An extremely slight difference may still be present, but it remains below the level of dispersion of the measurements.
Here, the offset is very small, since both compared lines are defined through the position of their main emission peak.
The \ion{N}{iv}\,$\lambda$5737 line is therefore a good indicator of the relative orbital motion.
It also originates from a highly ionised species.
\vspace{-1ex} 

\subsubsection{\texorpdfstring{The \ion{N}{v}\,$\lambda$4945 line}{The N V lambda4945 line}}%
\begin{figure}
	\centering
	\includegraphics[width=0.45\textwidth]{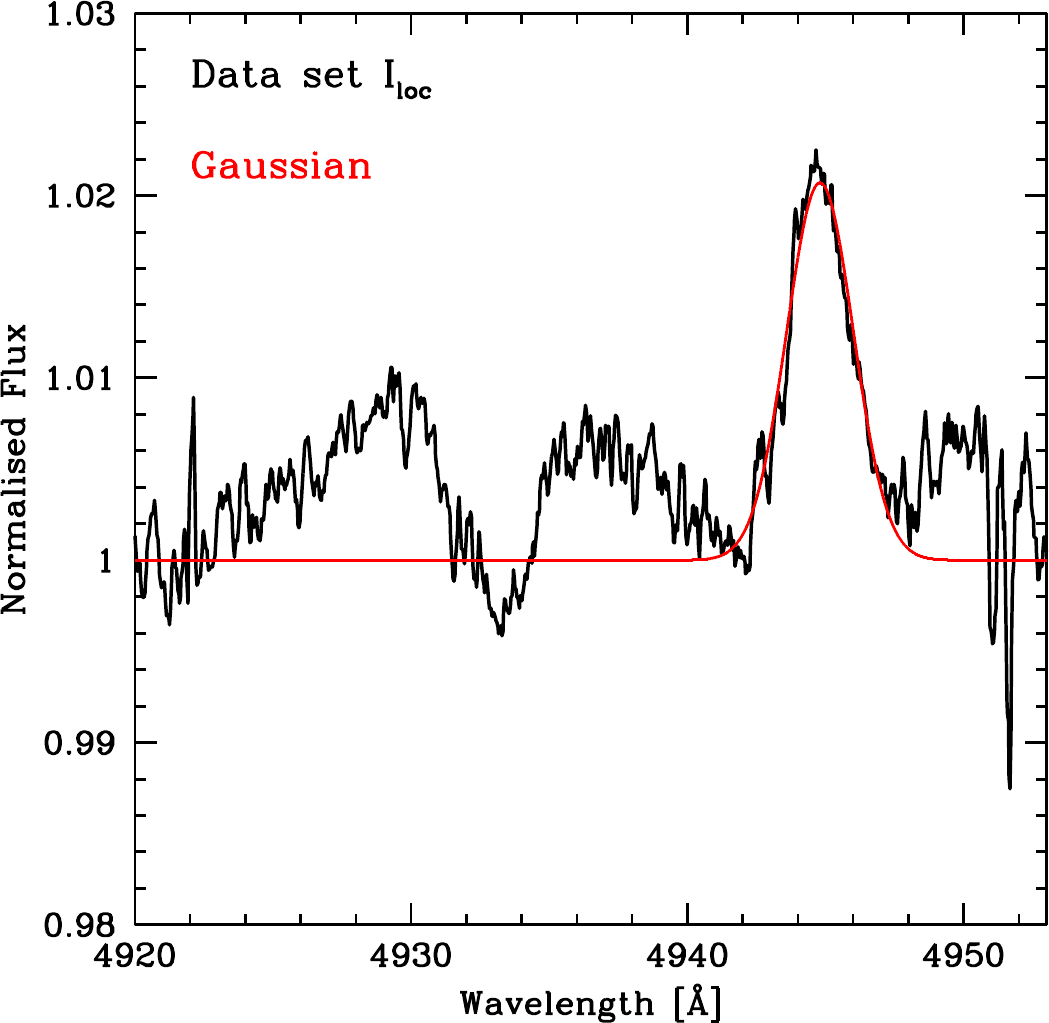}
	\vskip1ex
	
	\begin{minipage}{12cm}
		\caption{%
			Same as Fig.~\ref{fig:fgfitl5737}, but for the
			\ion{N}{v}\,$\lambda$4945 line.
			The spectrum was acquired at HJD\,2,453,897.469.%
			\label{fig:fgfitl4945}}
	\end{minipage}
\end{figure}

\begin{figure}
	\centering
	\includegraphics[width=0.45\textwidth]{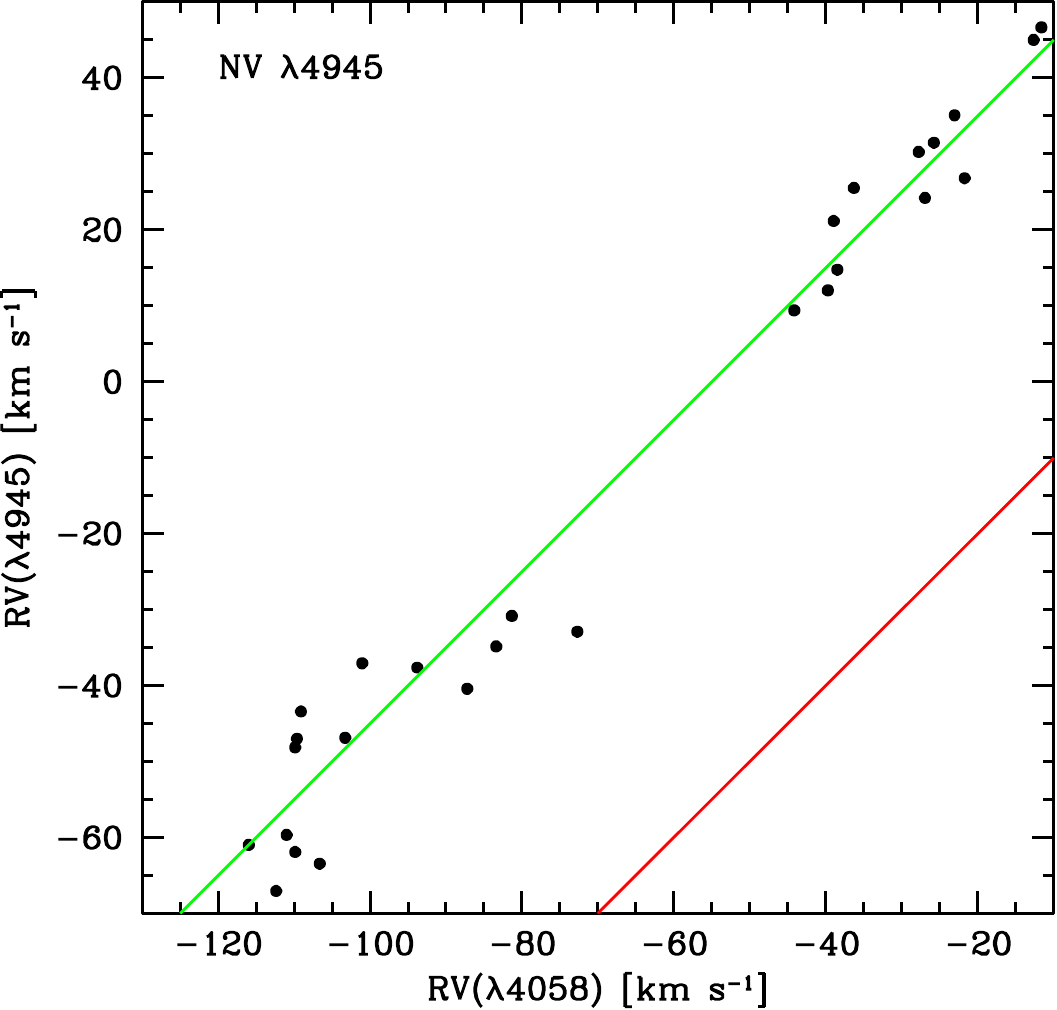}
	\vskip1ex
	
	\begin{minipage}{12cm}
		\caption{%
			Same as Fig.~\ref{fig:fgcorr5737}, but for the \ion{N}{v}\,$\lambda$4945 line.%
			\label{fig:fgcorr4945}}
	\end{minipage}
\end{figure}

\label{sssec:wrspec_DS1_4945}
The \ion{N}{v}\,$\lambda$4945 feature appears as a single emission line, composed of several blended triplets in a complex spectral region.
It is relatively faint, reaching only slightly more than 2{\%} above the continuum level.
In general, we fitted it with a single Gaussian profile, as illustrated in Fig.~\ref{fig:fgfitl4945}.
%
Despite the complexity of the surrounding spectral environment, the deduced RVs appear to be useful.
We adopted a laboratory wavelength of $\lambda_{0} = 4944.56$\,{\AA}.
Figure~\ref{fig:fgcorr4945} compares the RVs derived from  \ion{N}{v}\,$\lambda$4945 with those obtained from \ion{N}{iv}\,$\lambda$4058.
The green line drawn parallel to the red one-to-one relation, indicates that the two RV series merely differ by a constant offset.
This spectral line may therefore also be regarded as an interesting tracer of the orbital motion.
%
\vspace{-1ex} 

\subsubsection{\texorpdfstring{The \ion{N}{iv}\,$\lambda \lambda$6212-6215-6220 triplet}{The N IV lambda lambda 6212-6215-6220 triplet}}%
\begin{figure}
	\centering
	\includegraphics[width=0.45\textwidth]{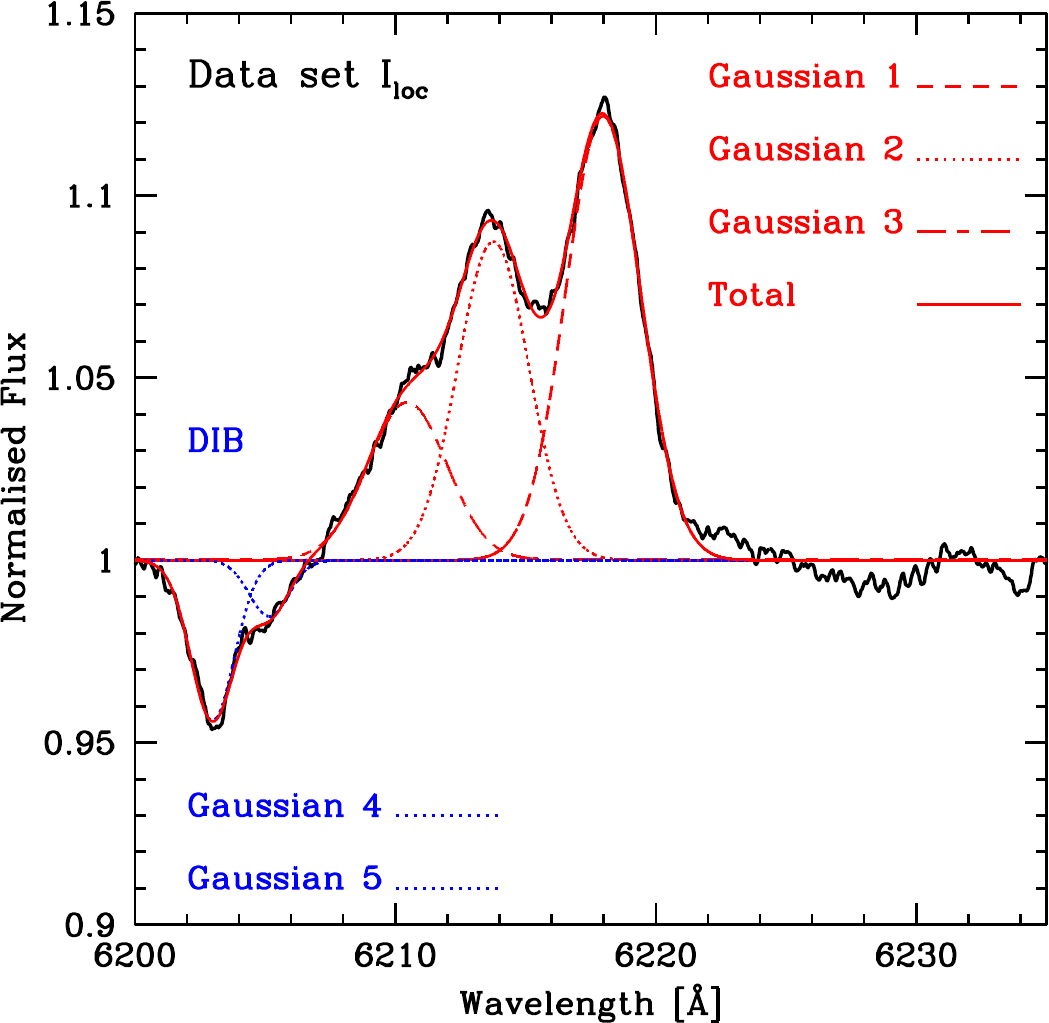}
	\vskip1ex
	
	\begin{minipage}{12cm}
		\caption{%
			Illustration of the RV determination based on the \ion{N}{iv}\,$\lambda \lambda$6212-6215-6220 triplet.
			The three Gaussian components fitted to the emission lines are shown in red and the two Gaussians that account for the neighbouring DIB feature are shown in blue.
			This spectrum was acquired at HJD\,2,454,918.588.
			Gaussian~1 was adopted as the RV reference.%
			\label{fig:fgfitl6220}}
	\end{minipage}
\end{figure}

\begin{figure}
	\centering
	\includegraphics[width=0.45\textwidth]{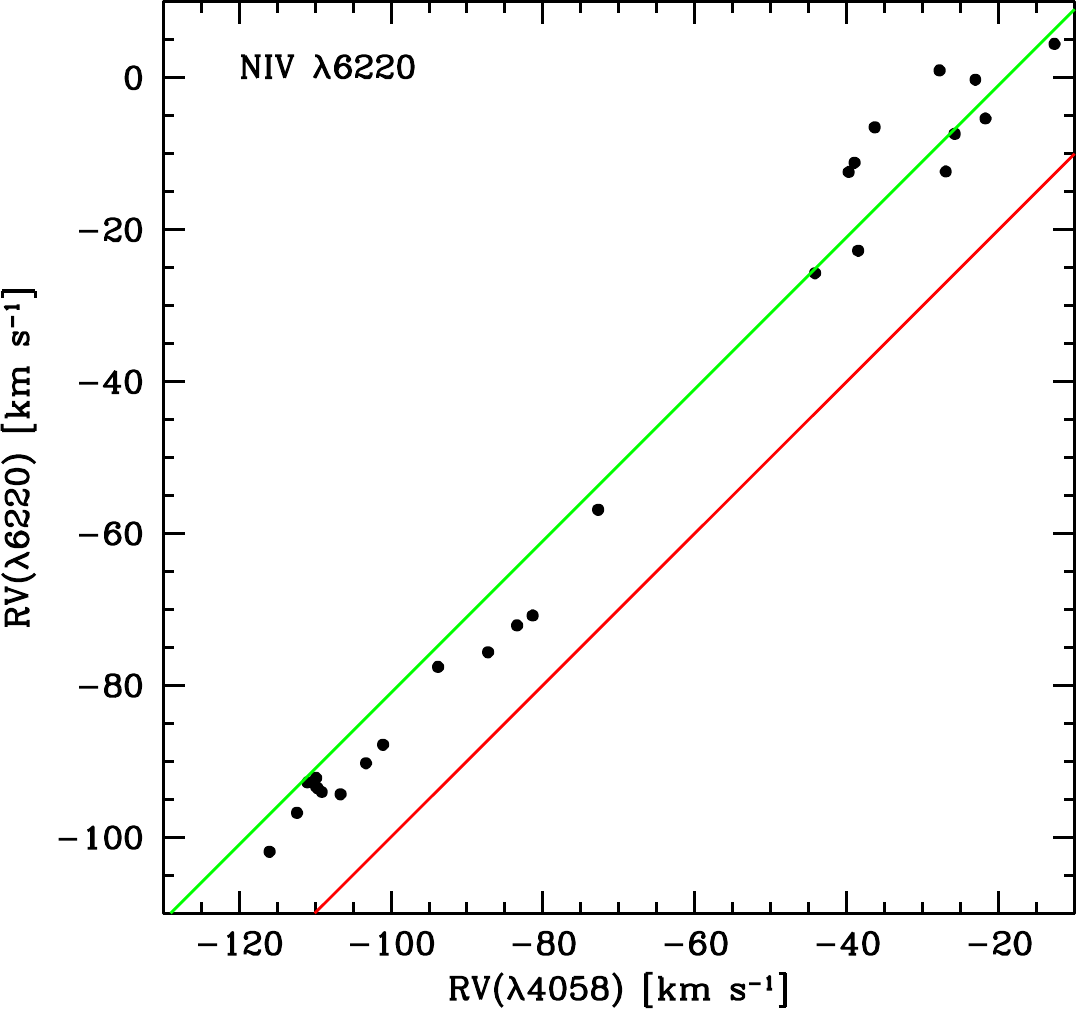}
	\vskip1ex
	
	\begin{minipage}{12cm}
		\caption{%
			Same as Fig.~\ref{fig:fgcorr5737}, but for Gaussian~1 from Fig.~\ref{fig:fgfitl6220}.%
			\label{fig:fgcorr6220}}
	\end{minipage}
\end{figure}

\label{sssec:wrspec_DS1_6220}
A~blend of three lines is present in this spectral region.
The dominant component of this triplet in WR\,25 is the feature located near 6220\,{\AA} (see Fig.~\ref{fig:fgfitl6220}).
A~DIB is also present in the blue wing of the blend.
To determine an RV for this group of lines, we decided to use the dominant component, for which a laboratory wavelength of $\lambda_{0} = 6219.89$\,{\AA} was adopted.
To obtain the position of this line, the full region was fitted using three Gaussian components (Gaussians 1--3);
the DIB could be adequately represented by two additional Gaussian components (Gaussians~4 and 5).
These five Gaussians were fitted simultaneously, without imposing constraints on any of their parameters, including those associated with the DIBs.
The position of Gaussian~1 was adopted as the RV reference for this multiplet.
The RV variations derived from the \ion{N}{iv}\,$\lambda$6220 feature and the \ion{N}{iv}\,$\lambda$4058 line can, to first order, be represented by a straight line relation, although the results in this case are somewhat noisier.
As in the previous figures, this behaviour is represented in Fig.~\ref{fig:fgcorr6220} by the green line drawn strictly parallel to the red one-to-one relation.
The residual deviations may either correspond to higher-order effects that could possibly be ignored, or simply reflect noise.
This spectral feature can also be considered as a useful tracer of the orbital motion.

\subsubsection{\texorpdfstring{The \ion{N}{iv} multiplet around 7100--7120\,{\AA}}{The N IV multiplet around 7100-7120 Angstroem}}%
\label{sssec:wrspec_DS1_7110}
\begin{figure}
	\centering
	\includegraphics[width=0.45\textwidth]{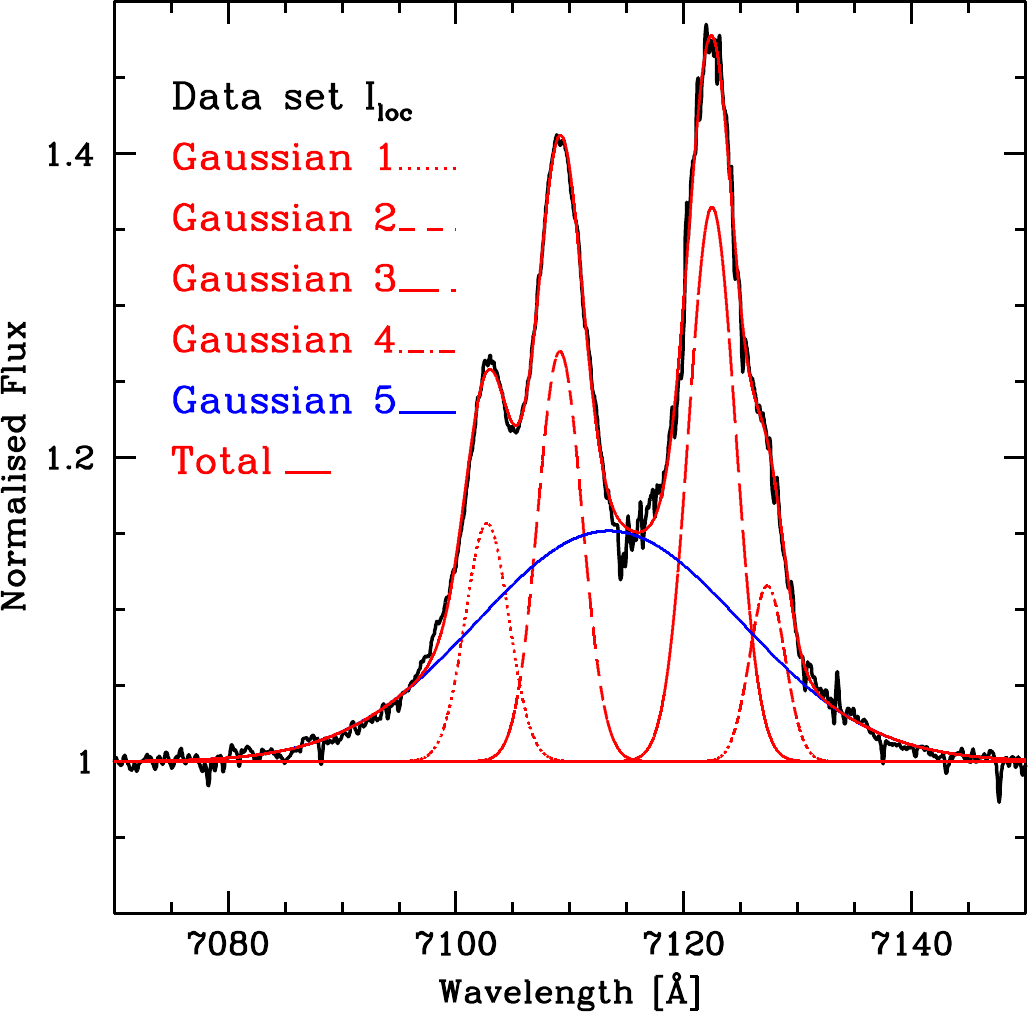}
	\vskip1ex
	
	\begin{minipage}{12cm}
		\caption{%
			Illustration of the RV determination based on the \ion{N}{iv} transitions in the 7100--7120\,{\AA} spectral region.
			Fitting this purely emission-line blend requires five Gaussian components: four narrow Gaussians (shown in red) together with one broader (in blue).
			This spectrum was acquired at HJD\,2,453,837.576.
			The Gaussian~2 was adopted as the RV reference.%
			\label{fig:fgfitl7110}}
	\end{minipage}
\end{figure}
The emission multiplet in this spectral region is rather complex, although it presents neither absorption components nor strong neighbouring DIBs (see Fig.~\ref{fig:fgfitl7110}).
It appears as the blend of four apparent emission lines, the reddest one being visible only as a small bump in the red wing of the dominant component.
This weak line is essentially associated with the transition at 7127.25\,{\AA} whereas the dominant one corresponds to the transition at 7122.98\,{\AA}.
On the far blue side, the transition at 7103.24\,{\AA} is present.
The second-strongest line arises from two fully blended transitions at 7109.35 and 7111.28\,{\AA}.
This blended feature was selected as the reference line for precise RV measurements.
We adopted a laboratory wavelength of $\lambda_{0} = 7109.99$\,{\AA} for this unresolved pair and refer to the feature hereafter as the 7110 line.
The dominant line was deliberately avoided because it seems to be affected by very weak, most likely telluric, absorption lines.
Moreover, at certain orbital phases, it exhibits a very strange behaviour characterised by additional absorption components.
By contrast, this effect does not appear to affect the selected 7110 line, which is therefore better suited for the RV measurements.
Additional wings are visible on both sides of the full blend, requiring the inclusion of a broad Gaussian to the fit.
In addition to this broad Gaussian, four narrower Gaussians were used to reproduce the complete profile.
For each epoch, the RV associated with Gaussian~2 was adopted as the reference value.

\begin{figure}
	\centering
	\includegraphics[width=0.45\textwidth]{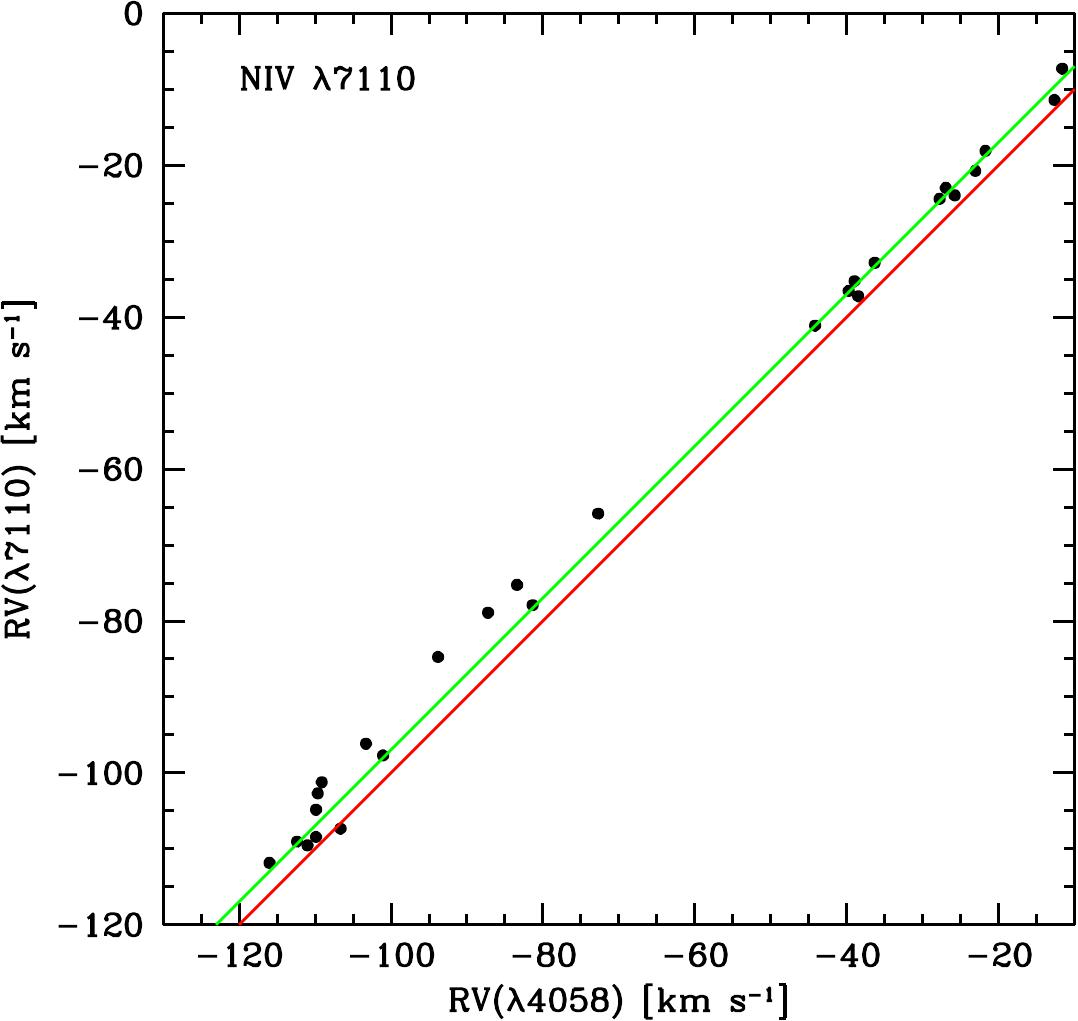}
	\vskip1ex
	
	\begin{minipage}{12cm}
		\caption{%
			Same as Fig.~\ref{fig:fgcorr5737}, but for Gaussian~2 from Fig.~\ref{fig:fgfitl7110}.%
			\label{fig:fgcorr7110}}
	\end{minipage}
\end{figure}

Figure~\ref{fig:fgcorr7110} compares the RVs derived from this feature with those obtained from the \ion{N}{iv}\,$\lambda$4058 line.
The agreement is very good, despite a small systematic offset.
This line thus constitutes an excellent tracer of the orbital motion.

\subsubsection{\texorpdfstring{The \ion{N}{iii}\,$\lambda$4905 line}{The N III lambda 4905 line}}%
\begin{figure}
	\centering
	\includegraphics[width=0.45\textwidth]{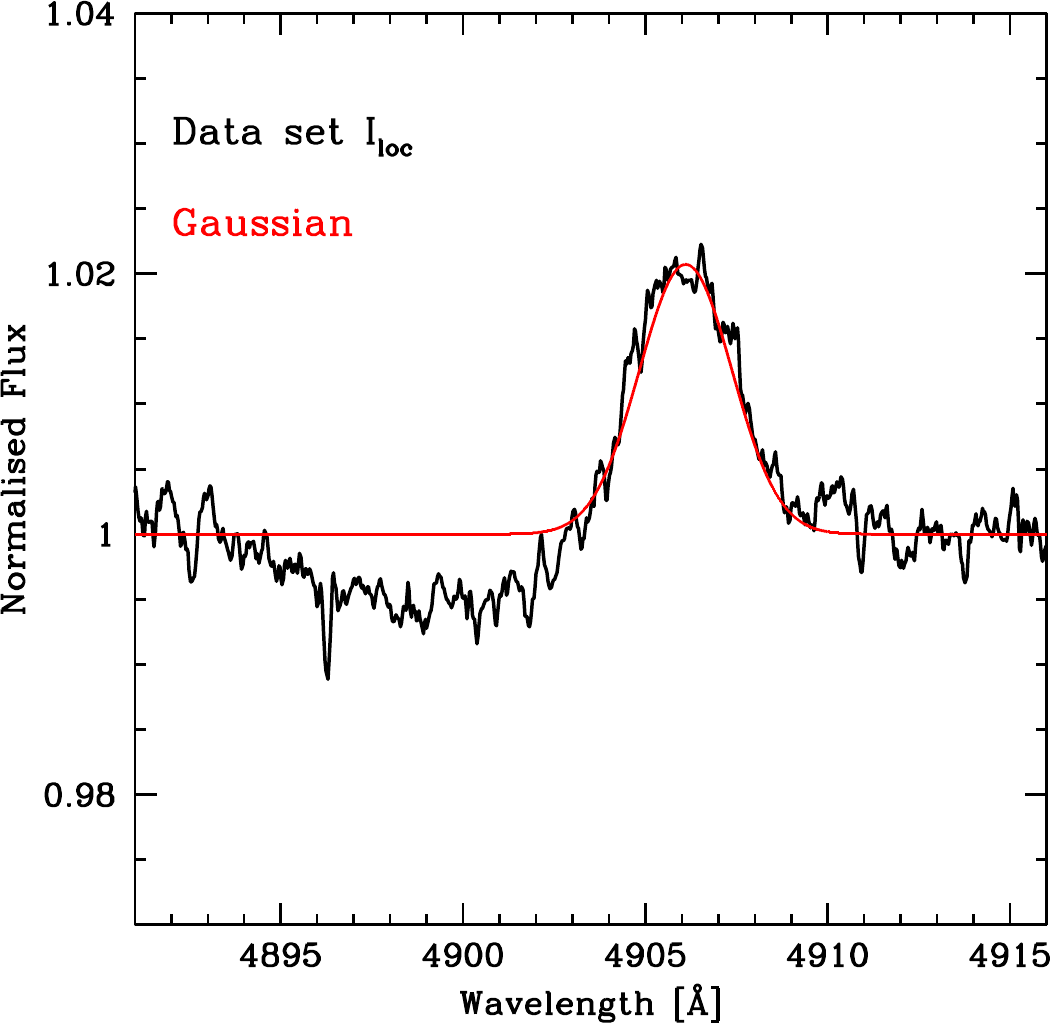}
	\vskip1ex
	
	\begin{minipage}{12cm}
		\caption{%
			Same as Fig.~\ref{fig:fgfitl5737}, but for the \ion{N}{iii}\,$\lambda$4905 line.
			The spectrum was acquired at HJD\,2,453,897.469.%
			\label{fig:fgfitl4905}}
	\end{minipage}
\end{figure}

\begin{figure}
	\centering
	\includegraphics[width=0.45\textwidth]{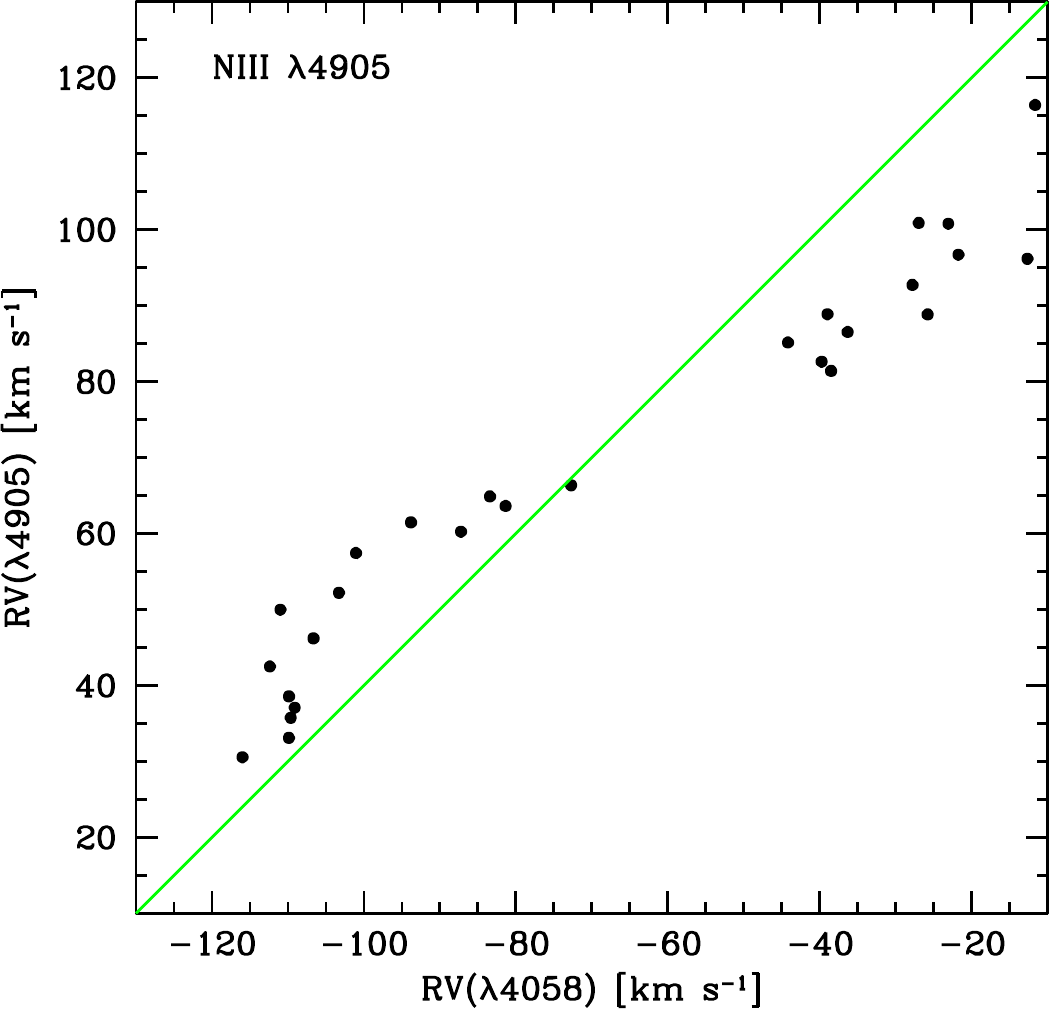}
	\vskip1ex
	
	\begin{minipage}{12cm}
		\caption{%
			Comparison between the RVs determined from the \ion{N}{iii}\,$\lambda$4905 feature and those obtained from the \ion{N}{iv}\,$\lambda$4058 line.
			The two sets of RVs are clearly inconsistent.%
			\label{fig:fgcorr4905}}
	\end{minipage}
\end{figure}

\label{sssec:wrspec_DS1_4905}
A~weak emission line is present near 4905\,{\AA} and we will refer to it accordingly.
Inspection of the NIST Atomic Spectra Database (\url{https://www.list.gov/pml/atomic-spectra-database}) suggests the existence of a line at $\lambda_{0} = 4904.78$\,{\AA}, part of a triplet whose two bluer components are not detected.
The spectral region is shown in Fig.~\ref{fig:fgfitl4905}, where it appears that the feature can be well fitted by a single Gaussian profile.
%
The continuum on the blue side of the line is somewhat depressed in flux, which may complicate the placement of the continuum.
This depression probably results from the presence of some DIBs and the onset of the flux rise due to the far red wing of \ion{He}{ii}\,$\lambda$4859+H$\beta$.
We performed RV measurements for all individual spectra of the I$_{\text{loc}}$ data set.
More recently, we noted that the University of Kentucky Atomic Line List database \citep[][\url{https://linelist.pa.uky.edu}]{2018Galax...6...63V}, lists an \ion{N}{iii} triplet at 4907.04, 4907.30, and 4907.30\,{\AA}.
This blend may therefore also contribute, partially or even predominantly, to the observed feature.
This would not pose a problem a priori, provided that both contributors do present the same relative variations, since our work focuses on relative motions.
%
Figure~\ref{fig:fgcorr4905} compares the RVs derived from this feature with those obtained from the \ion{N}{iv}\,$\lambda$4058 line.
In addition to a global offset, which could be partly due to the \ion{N}{iii} triplet just mentioned, the line clearly shows a significant dependency of the RV shift on the measured RV itself.
This behaviour may be linked to the continuum depression mentioned above, although no definitive interpretation is currently available.
In any case, this feature does not provide reliable RVs and does most probably not trace the orbital motion of the WR component.
We therefore excluded it from further orbital analysis.

\subsubsection{\texorpdfstring{The \ion{N}{iii}\,$\lambda\lambda$4634-4641 triplet}{The N III lambda lambda 4634-4641 triplet}}%
\begin{figure}
	\centering
	\includegraphics[width=0.45\textwidth]{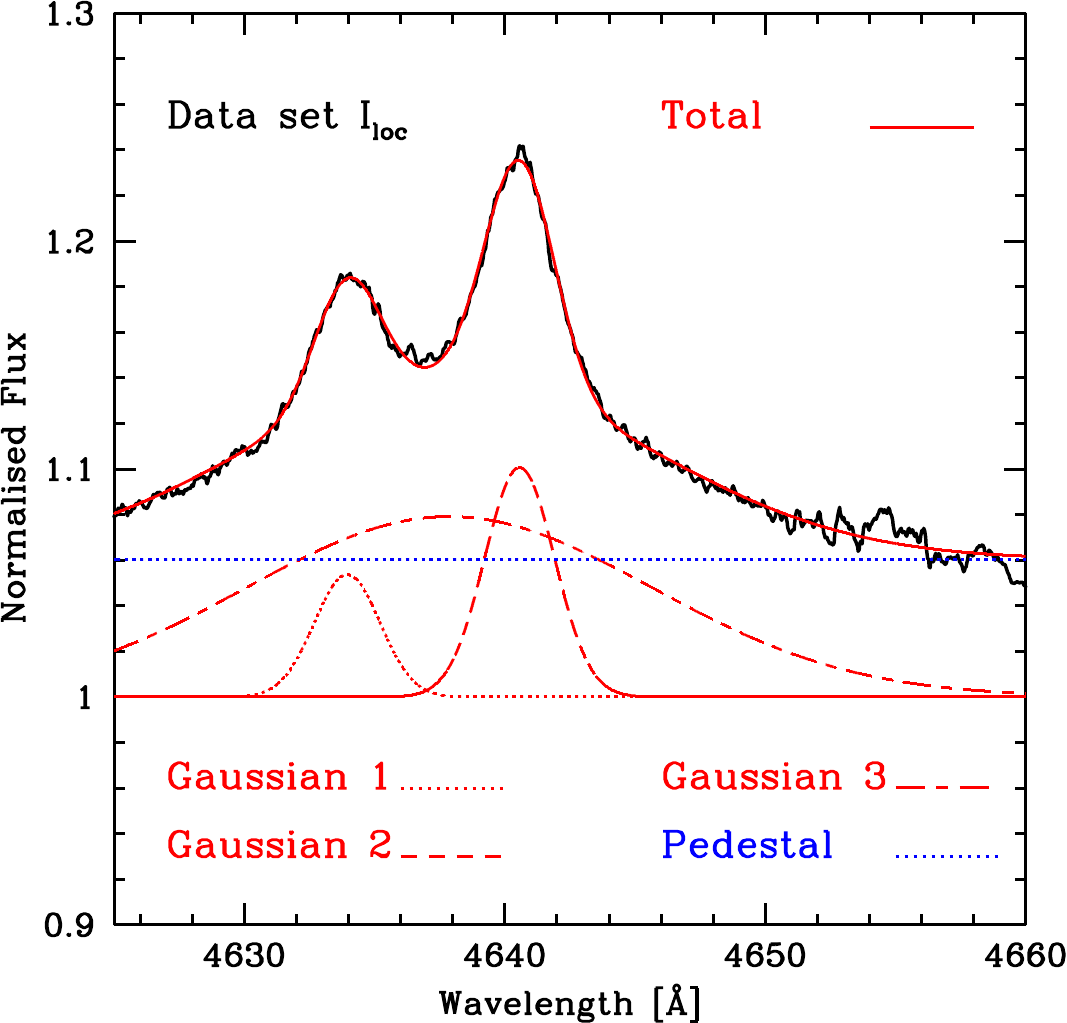}
	\vskip1ex
	
	\begin{minipage}{12cm}
		\caption{%
			RV measurements based on the well-known \ion{N}{iii}\,$\lambda \lambda$4634-4641 triplet.
			The spectrum was acquired at HJD\,2,453,918.499.
			Reproducing the full profile of the spectral region requires three Gaussians:
			two narrow and one broad.
			The measurement shown here yields essentially the same results when using the globally normalised data set instead of the locally normalised one.%
			\label{fig:fgfitl4641}}
	\end{minipage}
\end{figure}

\begin{figure}
	\centering
	\includegraphics[width=0.45\textwidth]{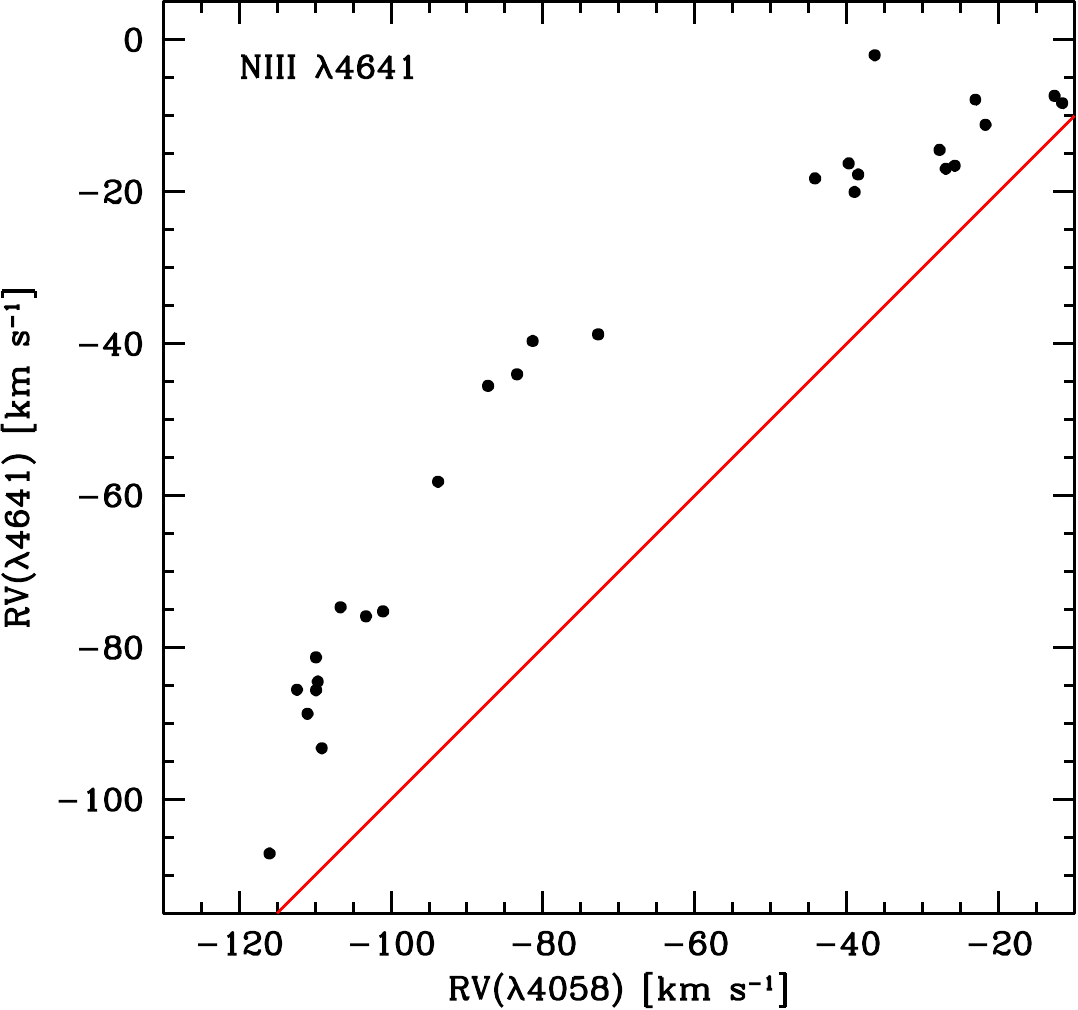}
	\vskip1ex
	
	\begin{minipage}{12cm}
		\caption{%
			Comparison between the RVs derived from Gaussian~2 in Fig.~\ref{fig:fgfitl4641} and those obtained from the \ion{N}{iv}\,$\lambda$4058 transition.
			The two RV sets are clearly inconsistent, with particularly large discrepancies.%
			\label{fig:fgcorr4641}}
	\end{minipage}
\end{figure}

\label{sssec:wrspec_DS1_4641}
This \ion{N}{iii} triplet transition is well known to be present in the spectra of certain O~stars displaying the ``f'' classification suffix, as well as in WN stars, where it is usually strong.
The observed profile consists of the two main emission peaks:
one corresponding to the transition at 4634.12\,{\AA}, and the other to the transition at 4640.64\,{\AA} blended with the weaker line at 4641.85\,{\AA}.
Figure~\ref{fig:fgfitl4641} illustrates this spectral region.
%
The usual lines are clearly visible.
We fitted the profile using two narrow Gaussians corresponding to the two apparent emission peaks.
It was immediately clear that an additional broad Gaussian was required to represent the excess emission in the surrounding region.
We thus fitted three Gaussians simultaneously.
We considered the position of Gaussian~2 to provide the best representative reference point and therefore adopted it for the RV measurements, using the arbitrarily chosen laboratory wavelength $\lambda_{0} = 4640.64$\,{\AA}.
The resulting RVs are compared in Fig.~\ref{fig:fgcorr4641} with those derived from the \ion{N}{iv}\,$\lambda$4058 line.
The total disagreement between the two RV sets is evident, with differences occasionally reaching 30\,km\,s$^{-1}$, well above the typical random dispersion (noise), which is estimated to be lower than 10\,km\,s$^{-1}$.
This transition should therefore not be used to trace the orbital motion of the WN star.
The line being bright, the observed discrepancy clearly reflects a systematic effect.
A~straightforward explanation for this phenomenon is given in Sect.~\ref{ssec:ospec_anom4641}.

\subsubsection{\texorpdfstring{The \ion{Si}{iv}\,$\lambda \lambda$4089-4116 lines}{The Si IV lambda lambda 4089-4116 lines}}%
\begin{figure}
	\centering
	\includegraphics[width=0.45\textwidth]{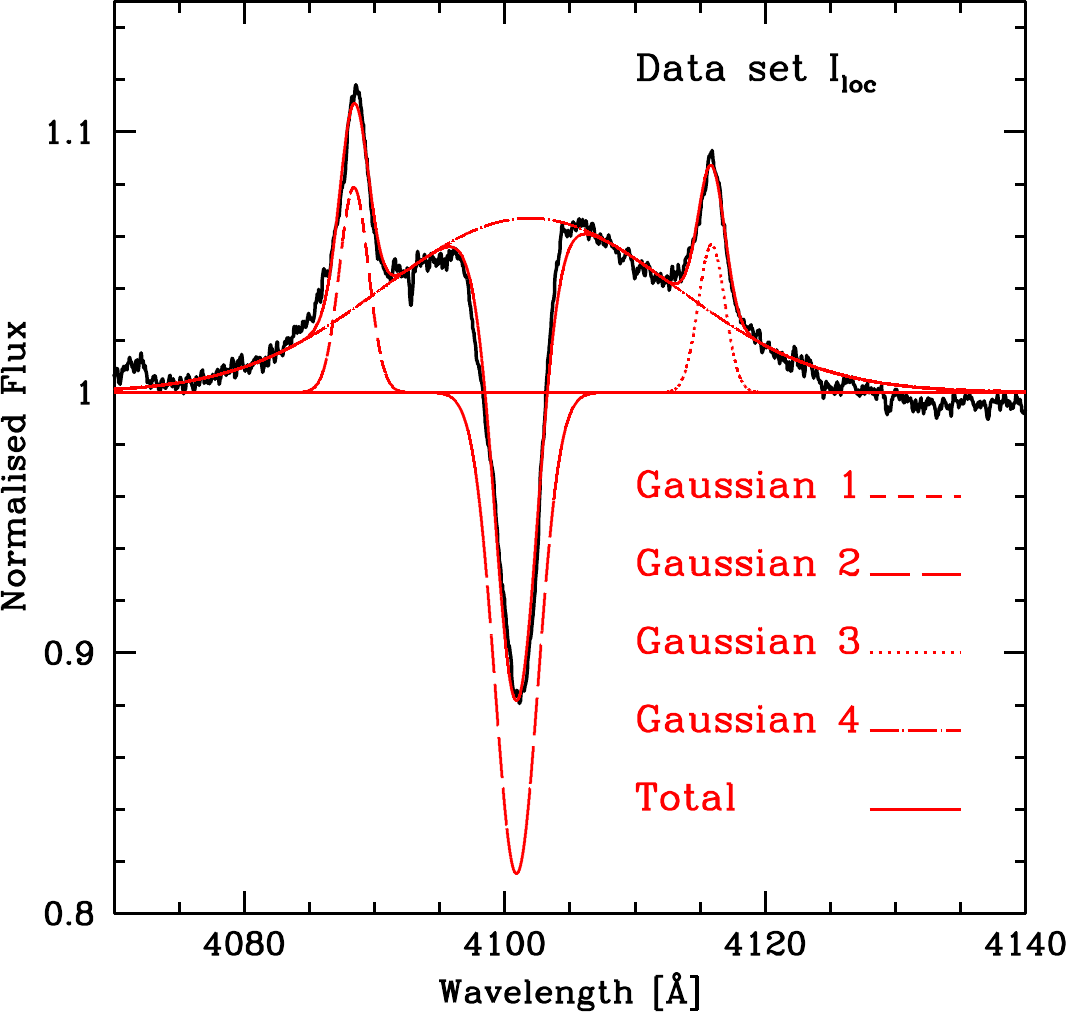}
	\vskip1ex
	
	\begin{minipage}{12cm}
		\caption{%
			Illustration of the RV measurement of the \ion{Si}{iv}\,$\lambda$4089 line and of the surrounding blend consisting of the \ion{Si}{iv}\,$\lambda$4116 line and the \ion{He}{ii}\,$\lambda$4100+H$\delta$ feature.
			Four Gaussians are required to reach a good total fit:
			Gaussians~1 and 3 (relatively narrow) reproduce the two silicon emission lines, Gaussian~2 represents a slightly broader absorption for the \ion{He}{ii}\,$\lambda$4100+H$\delta$ transitions and Gaussian~4 accounts for its much broader emission component.
			Fitting the full spectral region yields a very robust RV determination for the \ion{Si}{iv}\,$\lambda$4089 line.
			The situation is more complicated for the \ion{Si}{iv}\,$\lambda$4116 line (see text).
			The spectrum shown here was acquired at HJD\,2,453,833.528.%
			\label{fig:fgfitl4089}}
	\end{minipage}
\end{figure}

\begin{figure}
	\centering
	\includegraphics[width=0.45\textwidth]{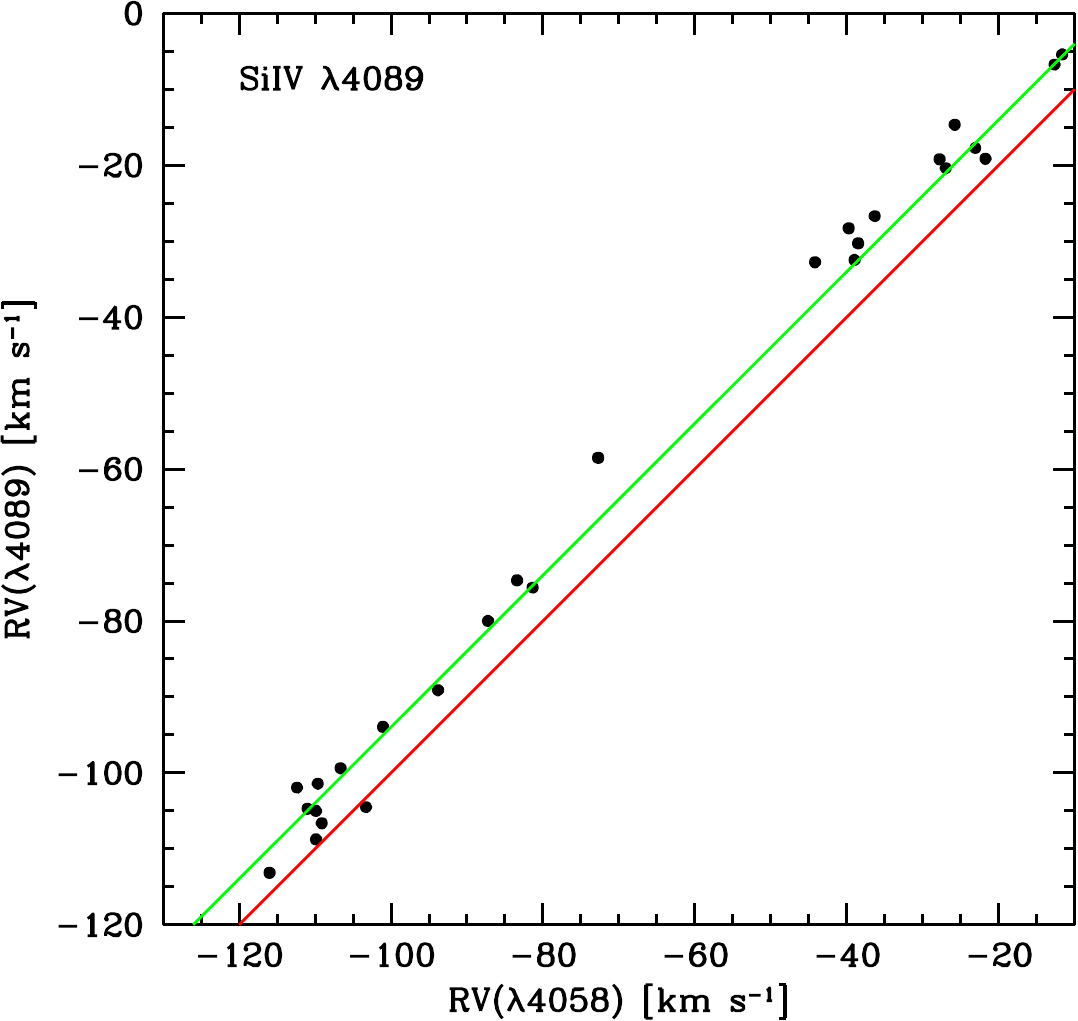}
	\vskip1ex
	
	\begin{minipage}{12cm}
		\caption{%
			Comparison between the RVs derived from Gaussian~1 in Fig.~\ref{fig:fgfitl4089} and those obtained from the \ion{N}{iv}\,$\lambda$4058 transition.
			The two RV sets are nicely consistent, differing only by a small constant offset.%
			\label{fig:fgcorr4089}}
	\end{minipage}
\end{figure}

\begin{figure}
	\centering
	\includegraphics[width=0.45\textwidth]{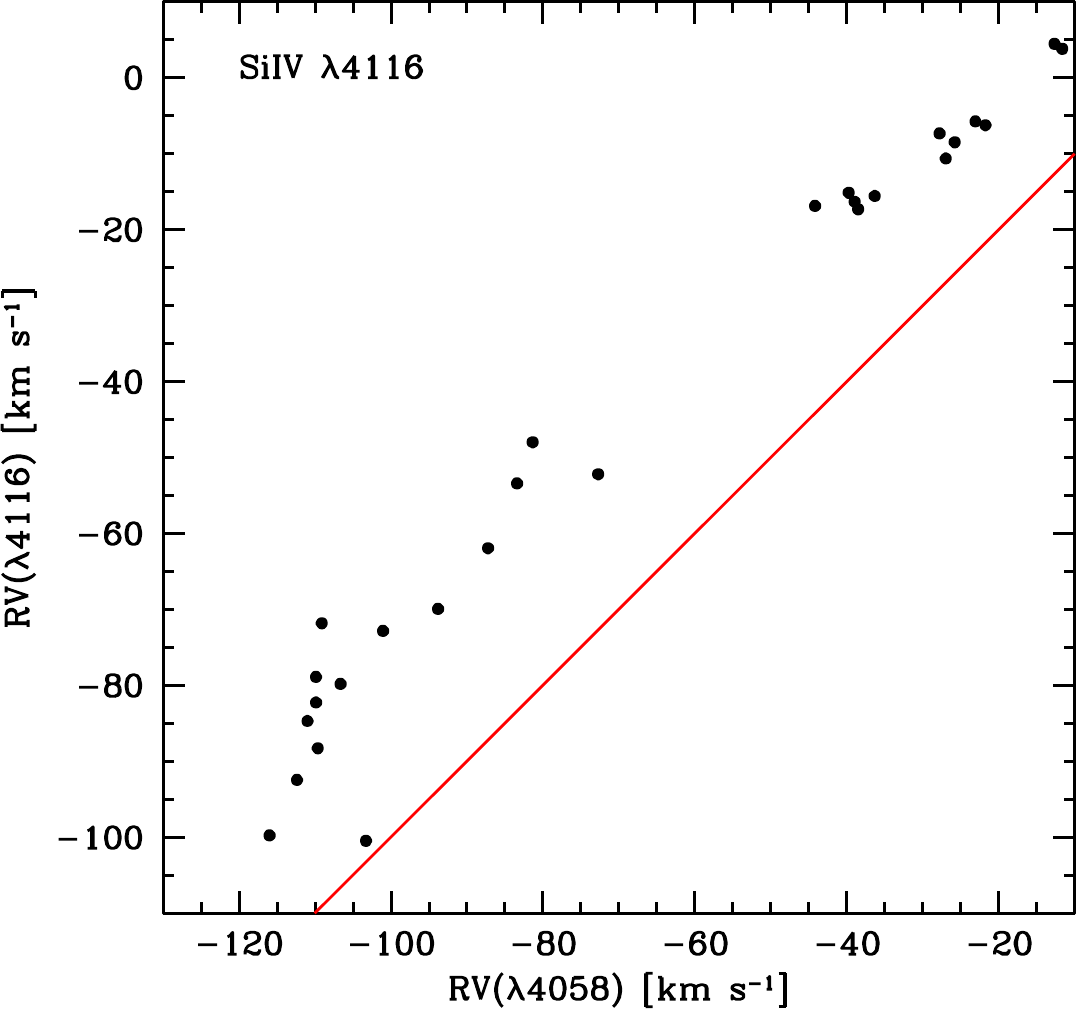}
	\vskip1ex
	
	\begin{minipage}{12cm}
		\caption{%
			Comparison between the RVs derived from Gaussian~3 (\ion{Si}{iv}\,$\lambda$4116) in Fig.~\ref{fig:fgfitl4089} and those obtained from the \ion{N}{iv}\,$\lambda$4058 transition.
			The measurement of this line may require additional Gaussians (see text).
			The two RV sets are not fully consistent in this case.%
			\label{fig:fgcorr4116}}
	\end{minipage}
\end{figure}

\label{sssec:wrspec_DS1_4089}
The silicon doublet is a well-known feature in the spectra of late WN stars.
The two lines correspond to transitions located at 4088.862\,{\AA} and 4116.103\,{\AA}.
They are well separated and should, in principle, be of interest for RV measurements.
However, the spectral region is affected by other lines, notably \ion{He}{ii}\,$\lambda$4100+H$\delta$ and the \ion{N}{iii}\,$\lambda \lambda$4097-4103 doublet.
Figure~\ref{fig:fgfitl4089} offers a view of this spectral region.
%
In the adopted fit, Gaussian~1 and Gaussian~3 correspond to the silicon lines, whereas the broad Gaussian~4 represents the Balmer-line emission.
Gaussian~2 accounts for an absorption component of the transition.
The nitrogen lines are not visible in the figure because they are much weaker.
The RV measurement for \ion{Si}{iv}\,$\lambda$4089 can be performed using four Gaussians.
Additional Gaussians are occasionally required for RV measurements of the \ion{Si}{iv}\,$\lambda$4116 line, depending on orbital phase.
This complexity is partly due to the interaction between the structure of the  \ion{He}{ii}\,$\lambda$4100+H$\delta$ line and the nitrogen lines, which affect the broad component.
Furthermore, at very particular phases, the red wing of the Balmer profile becomes more complex, as will be detailed in Sect.~\ref{ssec:ospec_detect}.
%
Figure~\ref{fig:fgcorr4089} compares the RVs derived from the \ion{Si}{iv}\,$\lambda$4089 with those obtained from \ion{N}{iv}\,$\lambda$4058.
Apart from a small offset, the two lines clearly exhibit the same behaviour.
We therefore consider \ion{Si}{iv}\,$\lambda$4089 to be a reliable tracer of the orbital motion and we will retain it as a basic indicator.
In contrast, the \ion{Si}{iv}\,$\lambda$4116 measurements are much more corrupted, most likely because the fitting procedure requires a larger number of Gaussians.
The corresponding RV comparison is shown in Fig.~\ref{fig:fgcorr4116}.
We decided to exclude this line from the list of reliable orbital tracers.

\subsubsection{\texorpdfstring{The \ion{N}{iii}\,$\lambda \lambda$4511-4515 multiplet}{The N III lambda lambda 4511-4515 multiplet}}%
\label{sssec:wrspec_DS1_4511}
The two dominant emission lines of this \ion{N}{iii} multiplet are clearly present in our WR\,25 spectra.
They are actually part of a broader blend composed of the two dominant lines together with extended surrounding wings.
The feature typically reaches a peak intensity of about 4{\%} above the continuum level.
We attempted to derive RV measurements from these lines;
however, the blue wing is affected by a strong nearby DIB.
As a result, our efforts to obtain reliable RVs from these transitions were unsuccessful.
Direct measurements (by cursor) of the line positions yield RVs that vary similarly to those derived from the \ion{N}{iv}\,$\lambda$4058 line, although the data are dispersed around this trend, as expected.
Consequently, these RVs cannot be used to improve our knowledge of the orbital motion.
However, the blend moves in phase with the \ion{N}{iv} line and must therefore be associated with the WN star, which is not unexpected.

\subsubsection{\texorpdfstring{The \ion{N}{iii}\,$\lambda$4379 line}{The N III lambda 4379 line}}%
\label{sssec:wrspec_DS1_4379}
This emission line is actually a tight triplet and reaches an intensity of approximately 3.5{\%} above the continuum.
The profile was fitted either with a single Gaussian or with the combination of one narrow and one broad Gaussian.
This non-uniform fitting procedure inevitably introduces additional dispersion into the RV measurements.
The resulting RVs were compared with those obtained from the \ion{N}{iv}\,$\lambda$5737 line.
Despite some scatter, the \ion{N}{iii}\,$\lambda$4379 feature clearly follows the motion of the WN star and can be securely attributed to it.
However, the observed level of dispersion indicates that this line is not sufficiently reliable to serve as an efficient tracer of the orbital motion.

\subsubsection{\texorpdfstring{The \ion{N}{iii}\,$\lambda$4321 line}{The N III lambda 4321 line}}%
\label{sssec:wrspec_DS1_4321}
An emission line is detected in the blue wing of the \ion{He}{ii}\,$\lambda$4339+H$\gamma$ line.
The corresponding fit was performed using a single Gaussian, together with a second Gaussian accounting for the proximity of the broad feature.
(The spectral region is illustrated in the \ion{He}{ii}\,$\lambda$4339+H$\gamma$ panel of Fig.~\ref{fig:fgoextr}---see Sect.~\ref{ssec:ospec_detect}.)
The RVs derived from this fit exhibit clear changes reminiscent of the orbital motion of the WN star.
This line must therefore be associated with the WN star.
We tentatively identify it with the \ion{N}{iii}\,$\lambda$4321 transition and label it accordingly.
This identification should nevertheless be treated with caution (see Appendix~\ref{sec:appE}).
Because of the substantial scatter in the RV distribution, we do not consider this line to provide a sufficiently reliable tracer of the orbital motion, despite its clear association with the WN star.

\subsubsection{\texorpdfstring{Two unknown lines on both sides of H$\beta$}{Two unknown lines on both sides of Hbeta}}%
\begin{figure}
	\centering
	\includegraphics[width=0.45\textwidth]{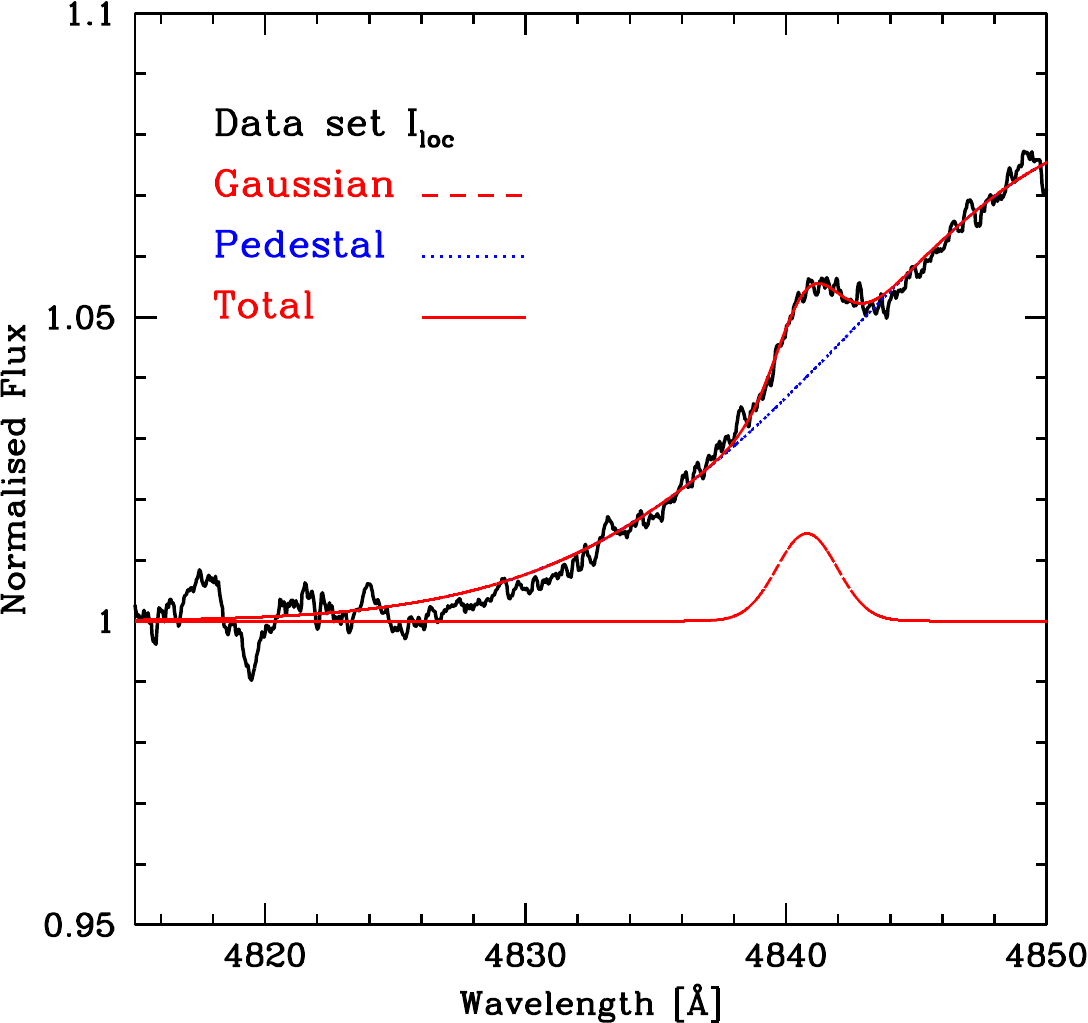}
	\vskip1ex
	
	\begin{minipage}{12cm}
		\caption{%
			Example of the fit applied to the unidentified line on the blue side of H$\beta$ (near 4841\,{\AA}).
			A~single Gaussian is sufficient to reproduce the line profile but a pedestal is necessary to account for the influence of the neighbouring \ion{He}{ii}\,$\lambda$4859+H$\beta$ feature.
			The spectrum shown here was obtained at HJD\,2,454,918.588.%
			\label{fig:fgfitl4841}}
	\end{minipage}
\end{figure}

\label{sssec:wrspec_DS1_4841}
Two weak emission features are present on either side of the \ion{He}{ii}\,$\lambda$4859+H$\beta$ line.
To the best of our knowledge, these lines have not previously been reported in the spectrum of a WR star.
The one on the blue side is located near 4841\,{\AA} and reaches approximately 1.5{\%} above the continuum, whereas that on the red side is situated near 4883\,{\AA} and reaches only about 1{\%} above the continuum.
We measured the relative RVs for both lines, as the laboratory wavelengths are unknown.
Figure~\ref{fig:fgfitl4841} illustrates the Gaussian fit to the line near 4841\,{\AA} which requires a pedestal to mitigate the influence of the neighbouring \ion{He}{ii}\,$\lambda$4859+H$\beta$ line.
%
These unidentified lines approximately follow the RV variations of the WN object.
It is tempting to propose the following \ion{N}{iii} transitions as identifications for these lines 
(see the University of Kentucky Atomic Line List database \citep{2018Galax...6...63V}, accessible at \url{https://linelist.pa.uky.edu}):%
\begin{align*}
\mathrm{1s^{2}\,2s\,2p\,(^{3}P^{o})\,4f~^{4}D_{5/2}}
&- \mathrm{1s^{2}\,2s\,2p\,(^{3}P^{o})\,5d~^{4}D^{o}_{3/2}}
&\text{at 4841.87\,{\AA},}\\
\makebox[0pt][l]{\hspace*{-57.5mm}and}\\
\mathrm{1s^{2}\,2s\,2p\,(^{3}P^{o})\,3p~^{4}D_{5/2}}
&- \mathrm{1s^{2}\,2s\,2p\,(^{3}P^{o})\,3d~^{4}F^{o}_{3/2}}
&\mbox{at 4881.78\,{\AA},}\\
\mathrm{1s^{2}\,2s\,2p\,(^{3}P^{o})\,3p~^{4}D_{7/2}}
&- \mathrm{1s^{2}\,2s\,2p\,(^{3}P^{o})\,3d~^{4}F^{o}_{7/2}}
&\mbox{at 4884.14\,{\AA}.}
\end{align*}
The latter feature may additionally be affected by the presence of a DIB at 4880.1\,{\AA}.
These identifications remain highly tentative and require further confirmation.
In particular, other transitions in multiplets that these lines also belong to are not detected, which may argue against the proposed identifications.
It is nevertheless noteworthy that these features are also observed in weak emission in the spectra of WR\,24 and WR\,78 (see also Sect.~\ref{ssec:ospec_hbeta}).

\subsubsection{Carbon lines}\label{sssec:wrspec_DS1_carbon}
\begin{figure}
	\centering
	\includegraphics[width=0.45\textwidth]{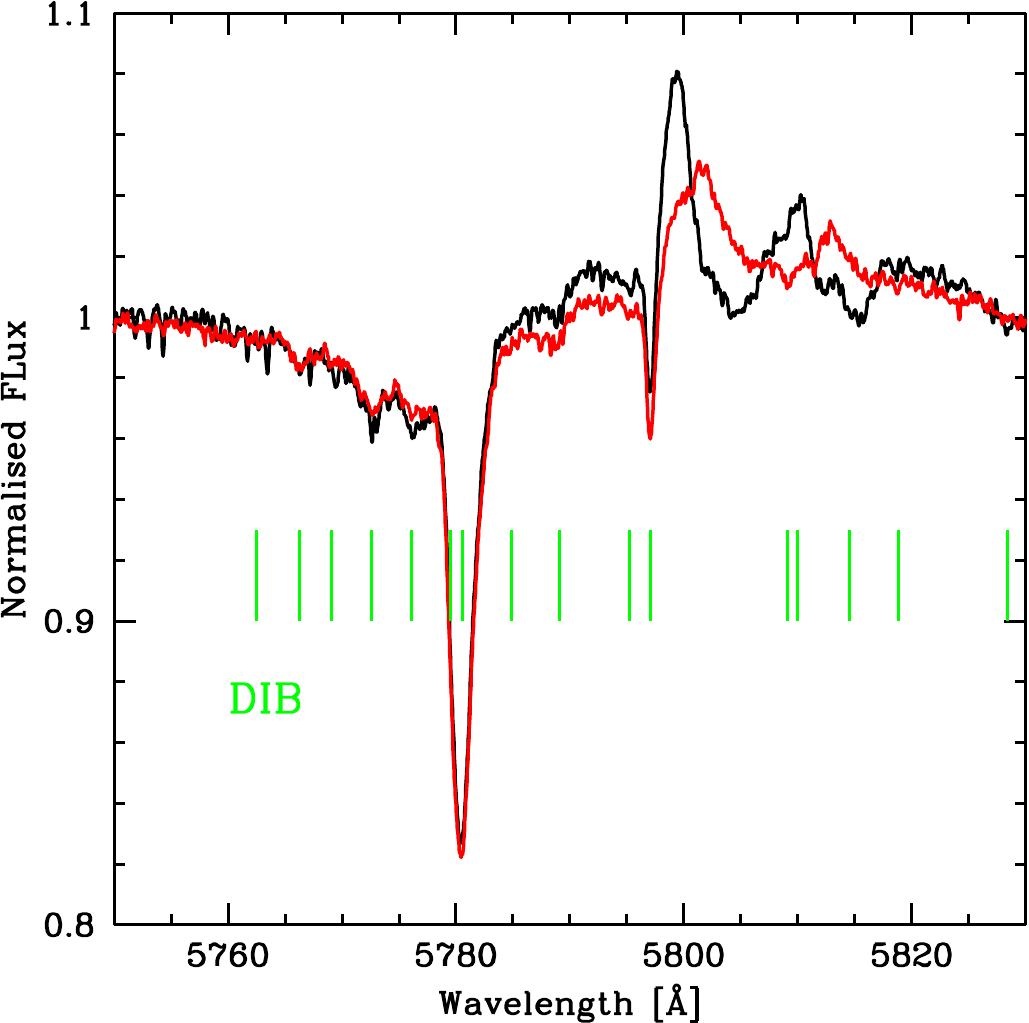}
	\vskip1ex
	
	\begin{minipage}{12cm}
		\caption{%
			Spectral region around the \ion{C}{iv}\,$\lambda \lambda$5801-5812 doublet for two spectra acquired at HJD\,2,453,800.671 (red) and HJD\,2,454,916.541 (black).
			At the epoch corresponding to the black spectrum, the WN star was shifted towards the blue, whereas at the epoch of the red spectrum, it was shifted to the red.
			The carbon lines thus follow the motion of the WN star.
			However, the large number of DIBs present in this spectral region prevents these features from being used as reliable tracers of the WN orbital motion.%
			\label{fig:fgcarbon5800}}
	\end{minipage}
\end{figure}

The \ion{C}{iv}\,$\lambda \lambda$5801-5812 doublet, although rather faint, is clearly present in the spectrum of WR\,25, as can be seen in Fig.~\ref{fig:fgcarbon5800}, where the region is shown at two different epochs.
At the epoch corresponding to the spectrum plotted in red, the WN star has a relative shift to the red (according to the behaviour of the \ion{N}{iv}\,$\lambda$4058 line as described in Sect.~\ref{sssec:wrspec_DS1_4058}).
The \ion{C}{iv} lines therefore clearly belong to the WN star.
Unfortunately, they are rather faint and the spectral region moreover contains a plethora of DIBs, preventing any reliable RV measurements without the use of a special dedicated analysis method, which would inevitably introduce additional uncertainties.
We therefore chose not to include this doublet in our selected set of RV tracers.

The \ion{C}{iii}\,$\lambda$5696 line is also detectable, although it is very faint, and it is hard to associate it with either component of the binary system.
This line will be addressed further in Appendix~\ref{sec:appE}.

\subsection{The other data sets (II, III and IV)}
\label{ssec:wrspec_DS234}
An analysis similar to that described in Sect.~\ref{ssec:wrspec_DS1} was also carried out for the remaining data sets.
We will not describe them here in detail.
The spectrum from data set~II was reduced only using the local normalisation procedure and is slightly noisier than the spectra from data set~I.
For data set~III, all the lines were carefully remeasured using the local normalisation approach, as no simultaneously observed O-type star was available as reference for the continuum determination.
We also measured the positions of the \ion{N}{iv}\,$\lambda$4058 line in the spectra from data set~IV, since this line is crucial for the determination of the orbital period (values are provided in Appendix~\ref{sec:appB}).
An extended version of Fig.~\ref{fig:fgcorr4058add}, including data sets II and III, is likewise presented in Appendix~\ref{sec:appB}.

\subsection{The radial velocities}
\label{ssec:wrspec_radvel}
We finally selected seven lines that are expected to follow the orbital motion of the WN star with the best accuracy.
These lines are
\ion{N}{iv}\,$\lambda$4058,
\ion{N}{v}\,$\lambda$4604,
\ion{N}{iv}\,$\lambda$5737,
\ion{N}{v}\,$\lambda$4945,
\ion{N}{iv}\,$\lambda$6220,
\ion{N}{iv}\,$\lambda$7110,
and \ion{Si}{iv}\,$\lambda$4089.
It should be emphasised that we do not have absolute positions for these lines, as different features were measured using Gaussians fitted to profiles that were not necessarily symmetric.
Moreover, this lack of absolute value makes the measurements effectively independent of the adopted zero velocity (laboratory) wavelength.
We therefore focus exclusively on the relative motion.
To improve the precision of the RV determinations, we computed the mean RV from these seven transitions after subtracting the individual zero-point offsets. The resulting modified RVs constitute what we refer to hereafter as the ``gold sample''.
The absolute velocities measured on the dominant Gaussian fitted to the \ion{N}{iv}\,$\lambda$4058 line together with the gold sample RVs derived from the seven selected spectroscopic lines are listed in Appendices~\ref{sec:appB} and \ref{sec:appC}, respectively.

\section{Period Reassessment}\label{sec:period}
The first and only determination of the orbital period of the WR\,25 system dates back to the discovery paper of \citet{2006A&A...460..777G}, who reported a period of $207.85 \pm 0.02$\,d.
The new data presented here are of higher precision;
it is therefore natural to attempt a refinement of this value.
Although the precision of the RV measurements is important, the time span covered by them is even more so.
Data sets~I to IV cover the interval from HJD\,2,452,037 to HJD\,2,454,921, and comprise 27 measurements from data set~I, one from data set~II, 15 from data set~III and 12 from data set~IV.
We considered it useful to extend this combined data set with the \ion{N}{iv}\,$\lambda$4058 RV measurements reported by \citet{2006A&A...460..777G}.
We therefore complemented our combined data set (I+II+III+IV) with RVs from the high-resolution spectra (i.e., labelled HRS) published by \citet{2006A&A...460..777G}, from which we removed the FEROS data already included in data set~III.
The latter are already part of the combined data set and were thus remeasured in the present work (see Sect.~\ref{ssec:observations_DS3}).
The RVs from spectra not included in data set~III were not remeasured, but directly taken from Table~2 of \citet{2006A&A...460..777G}.
These additional data are thus made up of previously determined RVs and could therefore belong to a slightly different reference system.
The first such observation dates back to HJD\,2,450,505; the additional data set adds 35 measurements.
Although these data are of lower quality and were measured using a different method, they extend the observed interval, increase the number of measurements by 35, and could thus improve upon the previous result.
We redetermined the period on the basis of this complete data set, which contains 90 measurements distributed over a time span of about 4416\,d.
We first analysed the time-series with a Fourier-type technique;
the peaks in the Power Spectrum have a typical width of $\delta \nu = 2.3\times 10^{-4}$\,d$^{-1}$.
We used the Heck--Manfroid--Mersch (HMM) method, which is particularly suited for the Fourier analysis of unevenly sampled data \citep{1985A&AS...59...63H, 2001MNRAS.327..435G}.
There is one peak standing out in the periodogram;
the corresponding frequency $\nu$ is given in Table~\ref{tabperiod} together with the associated period.
\begin{table}
	\centering
	\begin{minipage}{104mm}
		\caption{%
			Values of the frequencies $\nu$ corresponding to the largest peak in the periodogram as a function of the number of harmonics included in the considered method.
			The corresponding periods $P$ are also given.
			The last line corresponds to a harmonic content identical to that of a Keplerian motion.%
			\label{tabperiod}}
	\end{minipage}
	\vskip1ex
	
	\begin{tabular}{l@{\hspace{18mm}}c@{\hspace{15mm}}c}
		\toprule
		\textbf{Periodogram}
		& $\mathbf{\boldsymbol\nu}$\,[\textbf{d}$\mathbf{^{-1}}$]
		& \textbf{\textit{P}\,[d]} \\
		\midrule
		$\nu$                      & $0.47697 \times 10^{-2}$ & 209.660 \\
		$\nu + 2\nu$               & $0.48159 \times 10^{-2}$ & 207.646 \\ 
		$\nu + 2\nu + 3\nu$        & $0.48205 \times 10^{-2}$ & 207.447 \\ 
		$\nu + 2\nu + 3\nu + 4\nu$ & $0.48162 \times 10^{-2}$ & 207.633 \\
		Keplerian                  & $0.48161 \times 10^{-2}$ & 207.638 \\
		\bottomrule
	\end{tabular}
\end{table} 

This value definitively confirms the existence of the period previously reported by \citet{2006A&A...460..777G}.
Since the RV curve of WR\,25 is known to be eccentric, the position of the peak could be slightly inaccurate due to an interaction between the exact shape of the curve and the actual temporal distribution of the data points.
Therefore, we need to estimate the period with a method that is less sensitive to the shape of the curve.
The inclusion of harmonics allows us to reduce the dependence on the shape (and thus on the true harmonic content) of the curve.
We used the formalism of the multifrequency extension of the HMM method, as introduced in the appendix of \citet[][equations (A13)--(A19)]{2001MNRAS.327..435G} to construct a super-periodogram.
This super-periodogram gathers within each peak the power of the trial fundamental frequency as well as that of the related harmonics.
The positions of the main peak in the periodogram as a function of the number of harmonics are reported in Table~\ref{tabperiod}.
Increasing in the number of included harmonics leads to a form of convergence of the derived period value.
We also used a version of the periodogram based on the harmonic content of a Keplerian orbit, similarly to \citet{2025A&A...693A.124G}.
We finally adopted the value $P = 207.638$\,d.
The determination of the uncertainty on the period is not an easy task.
We can obtain a very conservative upper limit by adopting a value equal to one-tenth of the natural width of the peaks;
this corresponds to $\sigma_{\nu} = 2.3 \times 10^{-5}$\,d$^{-1}$, which in turn corresponds to $\sigma_{P} = 0.97$\,d.
We could also use the analytical formula of \citet{1971AJ.....76..544L}.
In this case, we obtain $\sigma_{\nu} = 1.3 \times 10^{-6}$\,d$^{-1}$, i.e.\ $\sigma_{P} = 0.06$\,d.
This analytical value is often considered to be slightly underestimated.
If we consider the three values of the period for the super-Fourier with two, three and four harmonics, their convergence suggests an uncertainty of $\sigma_{P} = 0.11$\,d.
The correct value should be somewhere between 0.06 and 0.11.
It is interesting to note that the value associated with the inspection of the objective function for the Keplerian fit gives an uncertainty of $\sigma_{P} = 0.093$\,d, which is the value we finally adopt.

Since the shape of the RV curve could be problematic, we also applied the method of \citet{1978A&A....63..125R}, which is expected to be more robust with respect to the exact shape, although it makes locating the extremum of the statistic more difficult.
The value obtained with the Renson method is about 207.6\,d, which provides further support for our Fourier analysis.
As an additional check, we also performed a detailed analysis of the I+II+III+IV data sets ($N = 55$) which yielded results in good agreement with the value obtained from the $N = 90$ data set.
We therefore finally adopted the value $P = 207.638$\,d with $\sigma = 0.093$\,d.

\section{The New SB1 Solution for the WN Star}\label{sec:orbsol}

\subsection{The SB1 model}%
\begin{figure}
	\centering
	\includegraphics[width=0.6\textwidth]{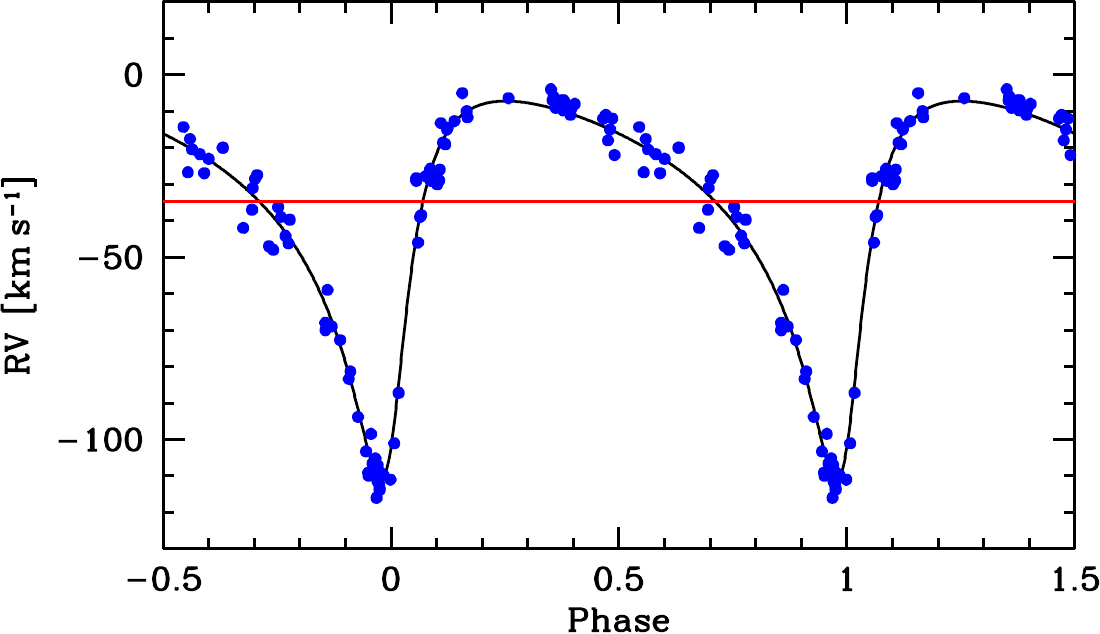}
	\vskip1ex
	
	\begin{minipage}{12cm}
		\caption{%
			RV curve of WR\,25.
			The blue points represent the 90 individual RVs associated with the position of Gaussian~1 fitted to the \ion{N}{iv}\,$\lambda$4058 line.
			The phase was computed on the basis of the orbital elements given in the third column of Table~\ref{tabelement}.
			The solid black curve corresponds to the fitted Keplerian solution derived from these elements;
			the red line represents the apparent systemic velocity associated with the spectroscopic line considered.%
			\label{fig:RVorbit1}}
	\end{minipage}
\end{figure}

\begin{figure}
	\centering
	\includegraphics[width=0.6\textwidth]{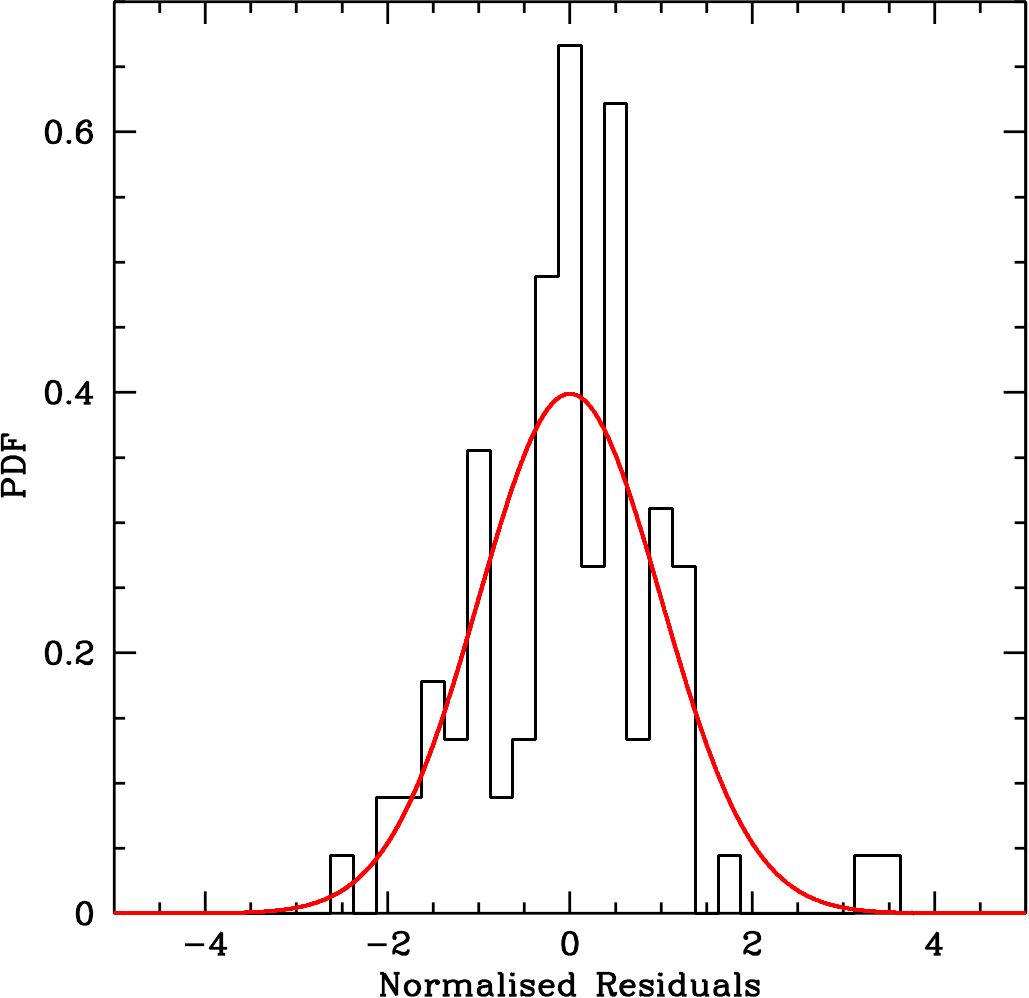}
	\vskip1ex
	
	\begin{minipage}{12cm}
		\caption{%
			Histogram of the residuals of the fit shown in Fig.~\ref{fig:RVorbit1}.
			The residuals have been normalised by the $\sigma_{\text{fit}} = 4.86$\,km\,s$^{-1}$.
			The histogram counts have been normalised so that they can be directly compared with the expected probability density function (PDF), assumed here to be a Gaussian.%
			\label{fig:pltresiduals}}
	\end{minipage}
\end{figure}

\label{ssec:orbsol_sb1model}
The period having been determined, we can now perform a fit of the Keplerian motion of the WN star in the SB1 mode.
First, we performed the fit on the basis of the general data set of 90 RVs associated with the \ion{N}{iv}\,$\lambda$4058 line.
The results of this first fit are given in Table~\ref{tabelement}.
\begin{table}[t]
\centering
\begin{minipage}{154mm}
\caption{%
Orbital elements derived from the fit of a Keplerian model to the two selected data sets of the present work, compared with the previous results reported in \citet{2006A&A...460..777G}.%
\label{tabelement}}
\end{minipage}
\vskip1ex

\begin{tabular}{lccc}
\toprule
$\mathbf{Parameter}$
  & \textbf{Value (Gamen et al.)}
    & \textbf{Value (\ion{N}{iv}\,$\boldsymbol\lambda$4058)}
      & \textbf{Value (``gold sample'')} \\
\midrule
$P$ [d] & 207.85 (0.02) & 207.638 (0.093) & 207.638 (fixed) \\
$e$ & 0.50 (0.02) & 0.570 (0.011) & 0.595 (0.013) \\
$\omega_{\text{WN}}$ [degree] & 215 (3) & 214.05 (2.21) & 212.04 (1.59) \\
$T_{0}$ (2,400,000+) & 51598.0 (1.0) & 51184.33 (5.06/0.59) & 51184.08 (0.46) \\
$K_{\text{WN}}$ [km\,s$^{-1}$] & 44.0 (2.0) & 52.83 (0.82) & 53.18 (0.82) \\
$\gamma_{\text{WN}}$ [km\,s$^{-1}$] & --34.6 (0.5) & --34.63 (0.59) & 0.00 (0.60) \\
$\rho$($\omega_{\text{WN}},T_{0}$) & --- & 0.7 & 0.8 \\
$\sigma_{\text{fit}}$ [km\,s$^{-1}$] & 4.7 & 4.86 & 3.39 \\
$N$ & 124 & 90 & 43 \\
$f$($M$) [$M_{\odot}$] & 1.2 (0.2) & 1.765 (0.097) & 1.687 (0.098) \\
$\phi$ (lower conj.) & --- & 0.039 & 0.038 \\
$\phi$ (upper conj.) & --- & 0.843 & 0.859 \\
$\phi$ (min.\ RV) & --- & 0.978 & 0.981 \\
\bottomrule\addlinespace[\belowrulesep]
\multicolumn{4}{p{150mm}}{\footnotesize
{\bf{Notes:}} the listed parameters are the period $P$, the eccentricity $e$, the argument of periastron $\omega_{\text{WN}}$ associated with the WN star, the time of passage at periastron $T_{0}$, the semi-amplitude $K_{\text{WN}}$, and the apparent systemic velocity $\gamma_{\text{WN}}$.
Additional information is also provided, such as the correlation coefficient $\rho$ between $\omega_{\text{WN}}$ and $T_{0}$, the $\sigma$ of the fit, the number of data points in the analysed time-series, the mass function $f(M)$, as well as the phases $\phi$ of the two conjunctions and of the minimum RV for the WN (i.e., phase of maximum RV separation).
The figures in parentheses give the $1\sigma$ uncertainties on the parameters as estimated from the fit.
For $T_{0}$, two values are presented in the third column:
one corresponding to the fit with the period being adjusted and the other for the period kept fixed.}
\end{tabular}
\vspace*{2ex}
\end{table} 
For this data set, the period is not fixed but is naturally identical to the adopted value, since we selected the period obtained from the Keplerian model.
Compared to the preliminary solution of \citet{2006A&A...460..777G}, the eccentricity is slightly but significantly larger.
The corresponding solution is compared with the RV data in Fig.~\ref{fig:RVorbit1}.
The data points are rather well distributed in phase, and the new data provided in the present work strongly alleviate, and indeed solve, the uncertainty problem around the minimum previously reported by \citet{2006A&A...460..777G}.
The typical dispersion of the residuals can be characterised by $\sigma_{\text{fit}} = 4.86$\,km\,s$^{-1}$, which is quite satisfactory for a WN star.
As a final check, we renormalised the individual residuals corresponding to each data point by $\sigma_{\text{fit}}$ and plotted the histogram of these normalised residuals in Fig.~\ref{fig:pltresiduals}.
We compared them with the Gaussian probability density function (PDF) expected for Gaussian behaviour (with $\sigma = 1$).
The histogram is consistent with the expected Gaussian distribution, and we therefore detect no significant anomaly.

\begin{figure}
	\centering
	\includegraphics[width=0.6\textwidth]{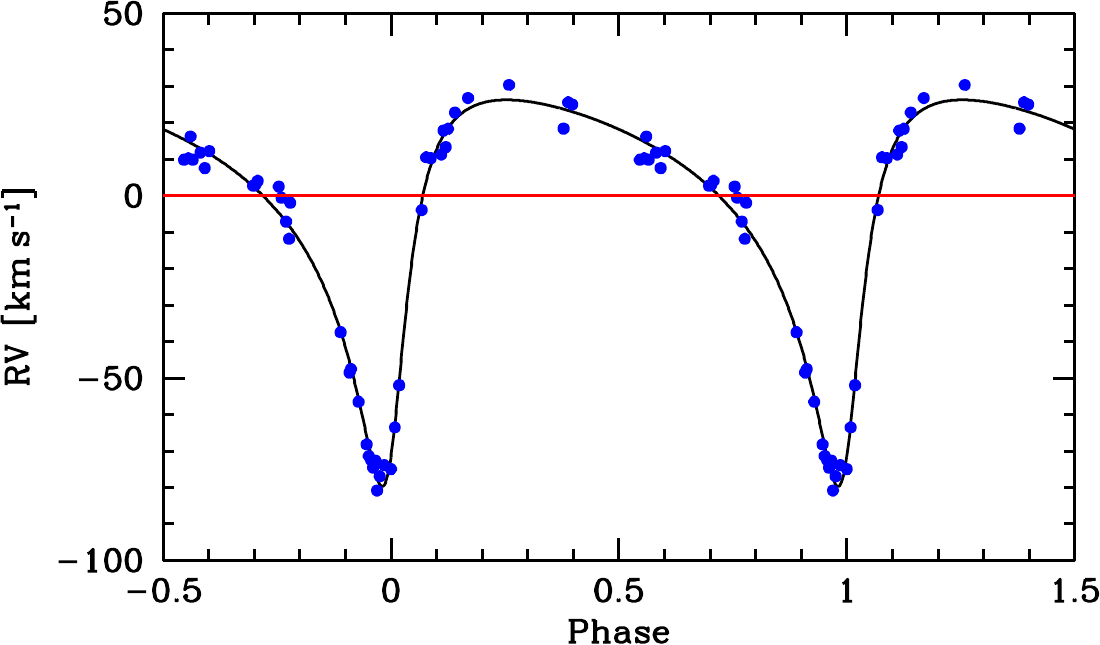}
	\vskip1ex
	
	\begin{minipage}{12cm}
		\caption{%
			Final RV curve of WR\,25.
			The blue points represent the 43 individual RVs obtained from the combination of the seven lines that best trace the orbital motion (``gold sample'').
			The phase is computed using the final elements given in the fourth column of Table~\ref{tabelement}.
			The solid black curve corresponds to the fitted Keplerian solution derived from these elements.
			The red line represents the artificial systemic velocity, which is zero by construction.%
			\label{fig:RVorbit2}}
	\end{minipage}
\end{figure}

Secondly, we analysed the gold data set of 43 RVs obtained from the combination of the seven lines exhibiting a well-defined and coherent behaviour.
For this fit, we fixed the period to the adopted value.
The resulting fitted parameters are given in Table~\ref{tabelement}, and the fit is illustrated in Fig.~\ref{fig:RVorbit2}.
As expected, this fit produces smaller residuals.
Both fits reported in Table~\ref{tabelement} provide orbital parameters that are fully compatible with each other.
Each value lies well within three times the uncertainty associated with the corresponding parameter in the other solution.
Due to the general compatibility of the two solutions, we decided to adopt the second one as the preferred solution.
The systemic velocity is here equal to zero by construction of the gold data set.
However, the zero value reported was not fixed during the fit, which allows us to estimate the associated uncertainty.
It should be noted that the time of periastron passage ($T_{0}$) has a smaller uncertainty in this second fit because the period was fixed and because the period and $T_{0}$ parameters are correlated.
The second uncertainty on $T_{0}$ in the third column refers to the uncertainty expected when the period is fixed.
The ephemerides adopted here were used to determine the phases of the individual observations; these phases are reported in Appendix~\ref{sec:appA}.
For a better understanding of the following discussion, we also report in Table~\ref{tabelement} the SB1 mass function, as well as the phases corresponding to the lower conjunction (the WN star being closer than the centre of mass) and upper conjunction (the WN star being behind the centre of mass).
\vspace{-2ex} 

\subsection{Soft X-ray eclipse and validation}
\label{ssec:orbsol_xrayval}
In the absence of confirmed photometric variations, it is difficult to validate the spectroscopic orbital solution, despite the early claim by \citet{1994ASIC..436..101G} of a possible period longer than 200\,d.
In contrast, the X-ray domain may provide additional support.
WR\,25 was first recognised as variable in the X-ray domain by \citet{2006A&A...445.1093P}, although this preceded the report of the discovery of the periodic spectral variations \citep{2006A&A...460..777G}.
This latter work provided confirmation and triggered subsequent X-ray monitoring of WR\,25, leading to several published analyses \citep{2014ApJ...788...84P, 2019MNRAS.487.2624A, 2021ApJ...915..114P}.
Based on \emph{Swift} data, \citet{2019MNRAS.487.2624A} performed a time-series analysis of the X-ray flux and found a period of \(207.8 \pm 3.4\)\,d, thereby independently confirming the period of the system.
A~similar analysis based on \emph{XMM-Newton} data provided a value of \(208.3 \pm 2.2\)\,d.
These results definitively confirm the period value, although they are not sufficiently precise to discriminate between the period reported by \citet{2006A&A...460..777G} and the value proposed here.
In any case, this difference between the two values is not significant.

We retrieved the fluxes from the \emph{Swift} satellite as measured by \citet[][see their Tables~2 and 3]{2019MNRAS.487.2624A}.
The observed \emph{Swift} spectra, binned in phase, were fitted by these authors in the energy range 0.5--10.0\,keV using a two-temperature thermal plasma model with a free absorbing column in the foreground.
The latter consists of a free local column density, to which is added the interstellar column density fixed at \(N_{\text{H}}(\mbox{ISM}) = 3.7 \times 10^{21}\)\,cm$^{-2}$ (equivalent hydrogen-column density).
Solar chemical abundances from \citet{1989GeCoA..53..197A} were adopted.
The plasma temperatures were fixed at 0.268\,keV and 2.75\,keV for all fits.
In their Table~3, \citet{2019MNRAS.487.2624A} present the observed fluxes, the fluxes corrected for interstellar extinction, and the intrinsic fluxes, corrected for both the interstellar extinction and the assumed local wind absorption.
 
\begin{figure}
	\centering
	\includegraphics[width=0.6\textwidth]{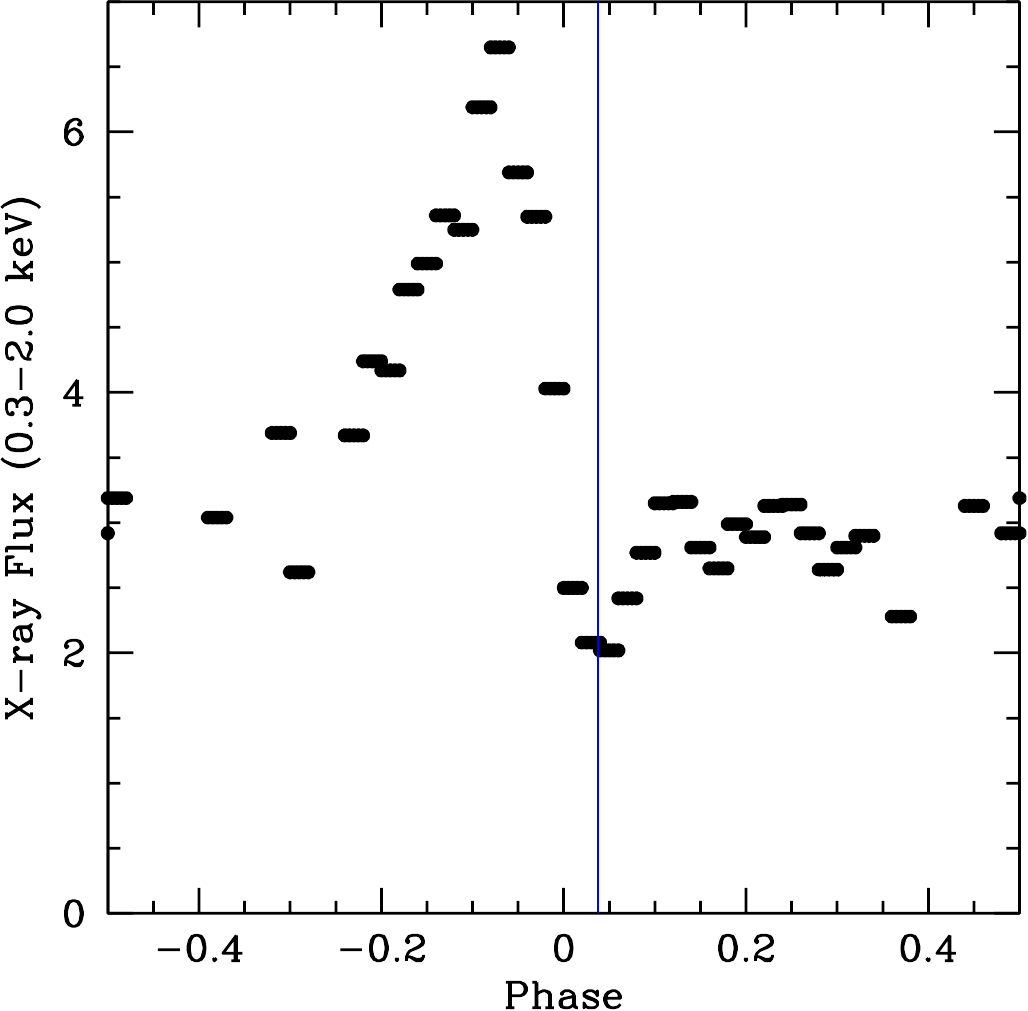}
	\vskip1ex
	
	\begin{minipage}{12cm}
		\caption{%
			Run of the soft X-ray flux of WR\,25 as a function of phase (see Table~\ref{tabelement}).
			The fluxes plotted here are corrected for extinction by the interstellar medium and are expressed in units of $10^{-12}$\,erg\,cm$^{-2}$\,s$^{-1}$.
			The measured fluxes are taken from \citet{2019MNRAS.487.2624A}.
			The soft band corresponds to the 0.3--2.0\,keV energy range.
			The data points are horizontally elongated to illustrate the phase uncertainty of 0.02 (see text).
			The blue vertical line indicates the time of conjunction with the WN star in front according to the ephemerides presented here.%
			\label{fig:fgXsoft}}
	\end{minipage}
\end{figure}

Unfortunately, the work of \citet{2019MNRAS.487.2624A} was performed using phase binning based on the ephemerides of \citet{2006A&A...460..777G}.
The slight modification of the period proposed here implies a corresponding shift in phase.
Even assuming that the grouping of the various exposures remains unchanged, the \emph{Swift} data span 15 orbital cycles of the binary system, which leads to a phase uncertainty of about 0.02 solely due to the revised period.
In Fig.~\ref{fig:fgXsoft}, we show the run of the X-ray flux corrected for interstellar extinction in the soft band (0.3--2.0\,keV) as a function of phase.
The data points are horizontally elongated to illustrate this uncertainty of 0.02.
Although our discussion is based on the flux corrected for interstellar extinction (taken from \citealt{2019MNRAS.487.2624A}), the conclusions remain unchanged if the observed fluxes are considered instead.
The blue vertical line indicates the position of the conjunction with the WN star in front (see Table~\ref{tabelement}).
The observed soft X-ray eclipse confirms and therefore supports our ephemerides.
However, due to the possible phase error, the observations also remain generally
consistent with the earlier orbital solution of \citet{2006A&A...460..777G}.
A full reanalysis of the X-ray data to resolve this ambiguity is, however, beyond the scope of this paper.

The eccentricity reported by \citet{2006A&A...460..777G} could be slightly underestimated, as first suggested by \citet{2021ApJ...915..114P}.
This is not surprising given the lack of good spectroscopic data around phase 0.98 in that early study.
The eccentricity derived here ($e = 0.595$) implies a conjunction phase of 0.038 whereas an eccentricity of $\sim 0.5$ would imply a conjunction phase of $\sim 0.05$.
The conjunction therefore occurs 7.9\,d after periastron instead of 10.5\,d, respectively.
If the time of conjunction in Fig.~2 of \citet{2021ApJ...915..114P} is shifted accordingly, the agreement between the conjunction and the soft X-ray eclipse appears to improve somewhat.

\section{The O-Component Detection and Motion}\label{sec:ospec}

\subsection{The detection of the O component}\label{ssec:ospec_detect}
Now that the SB1 orbital solution of the WN star in WR\,25 has been determined, it is time to investigate the orbit of the companion.
Shortly after the paper by \citet{2006A&A...460..777G} which reported the discovery of the binarity of WR\,25, we had the opportunity to complement our data set with a spectrum acquired in May 2006 that observed the star at the epoch of maximal radial-velocity separation, a phase not previously covered by our high- and medium-resolution spectra.
The detection of the companion was illustrated in Fig.~1 of \citet{2008RMxAC..33...91G} on the basis of the \ion{He}{ii}\,$\lambda$4542 line, indicating that the companion is an O-type star.
This spectrum is included in data set~IV.
\begin{sidewaysfigure}
	\centering
	\includegraphics[width=0.27\textwidth]{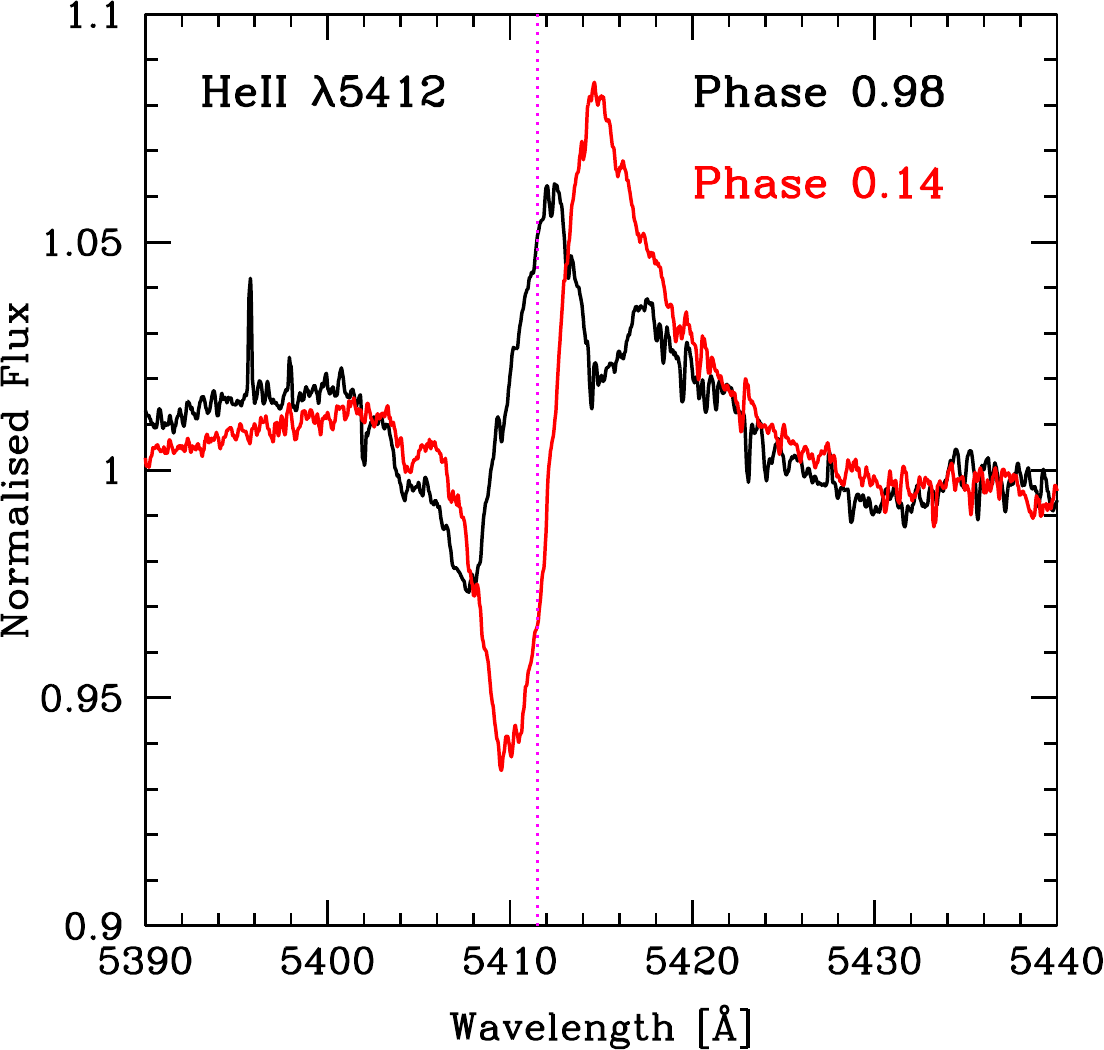}
	\hspace{5mm}
	\includegraphics[width=0.27\textwidth]{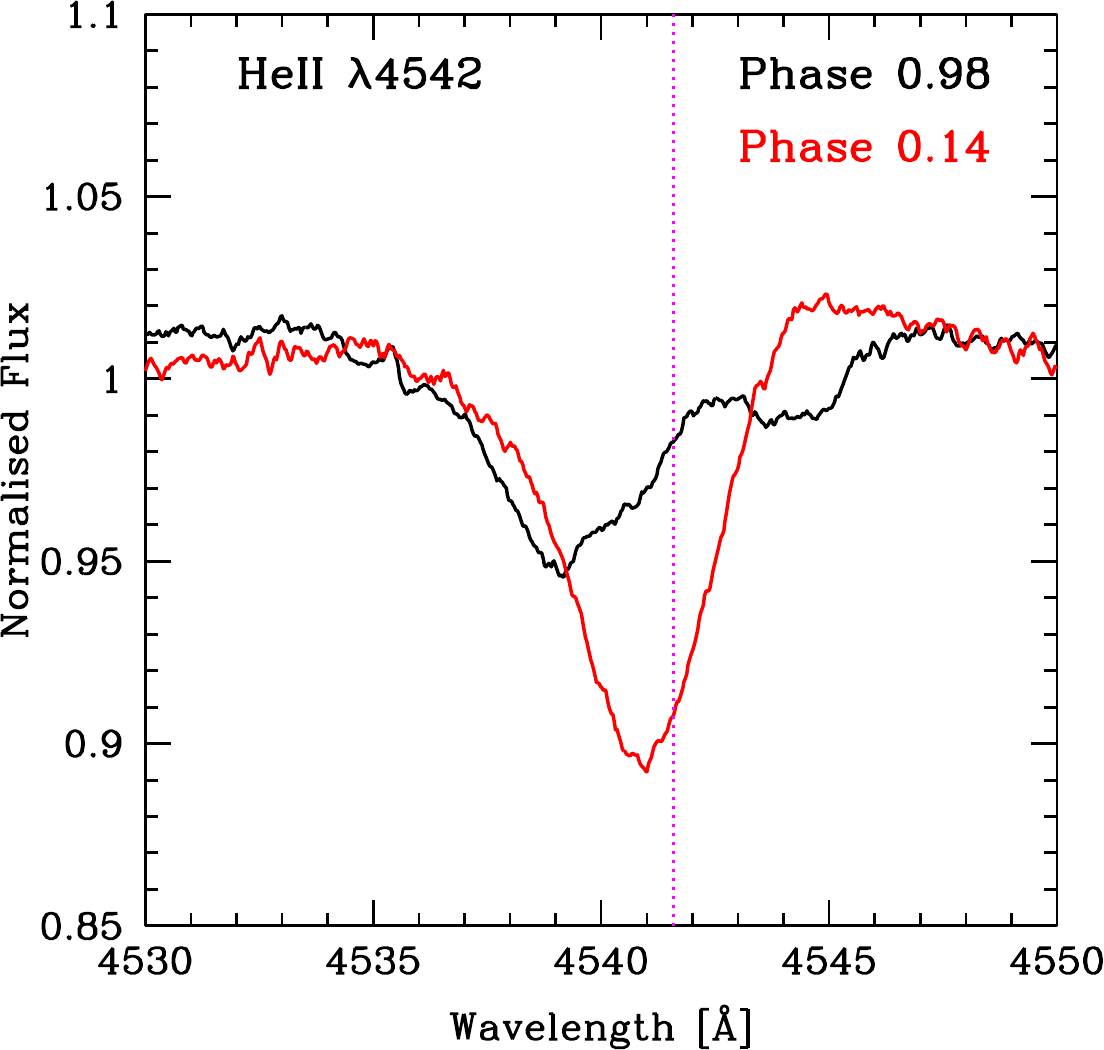}
	\hspace{5mm}
	\includegraphics[width=0.27\textwidth]{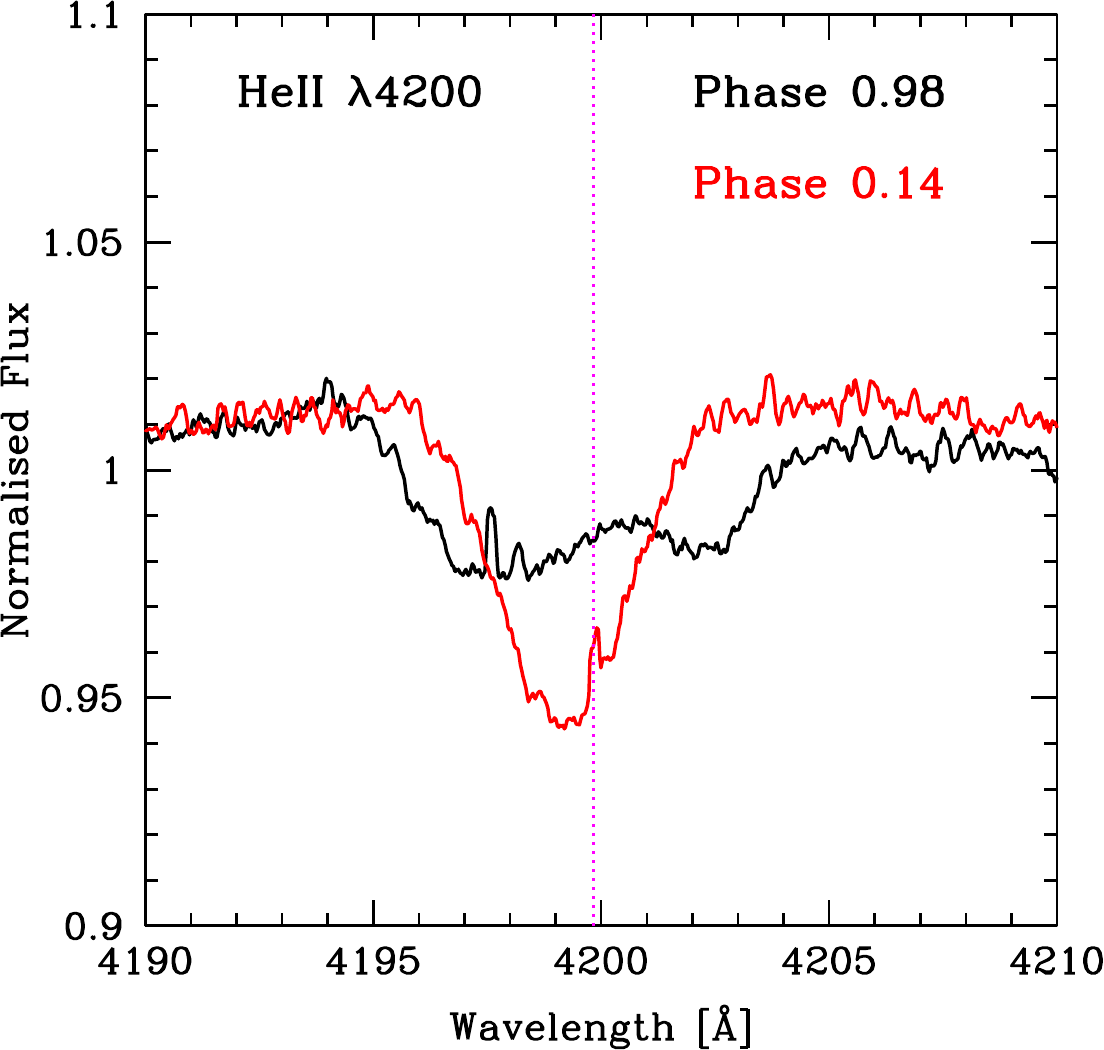} 
	\\
	\hskip-3mm\includegraphics[width=0.263\textwidth]{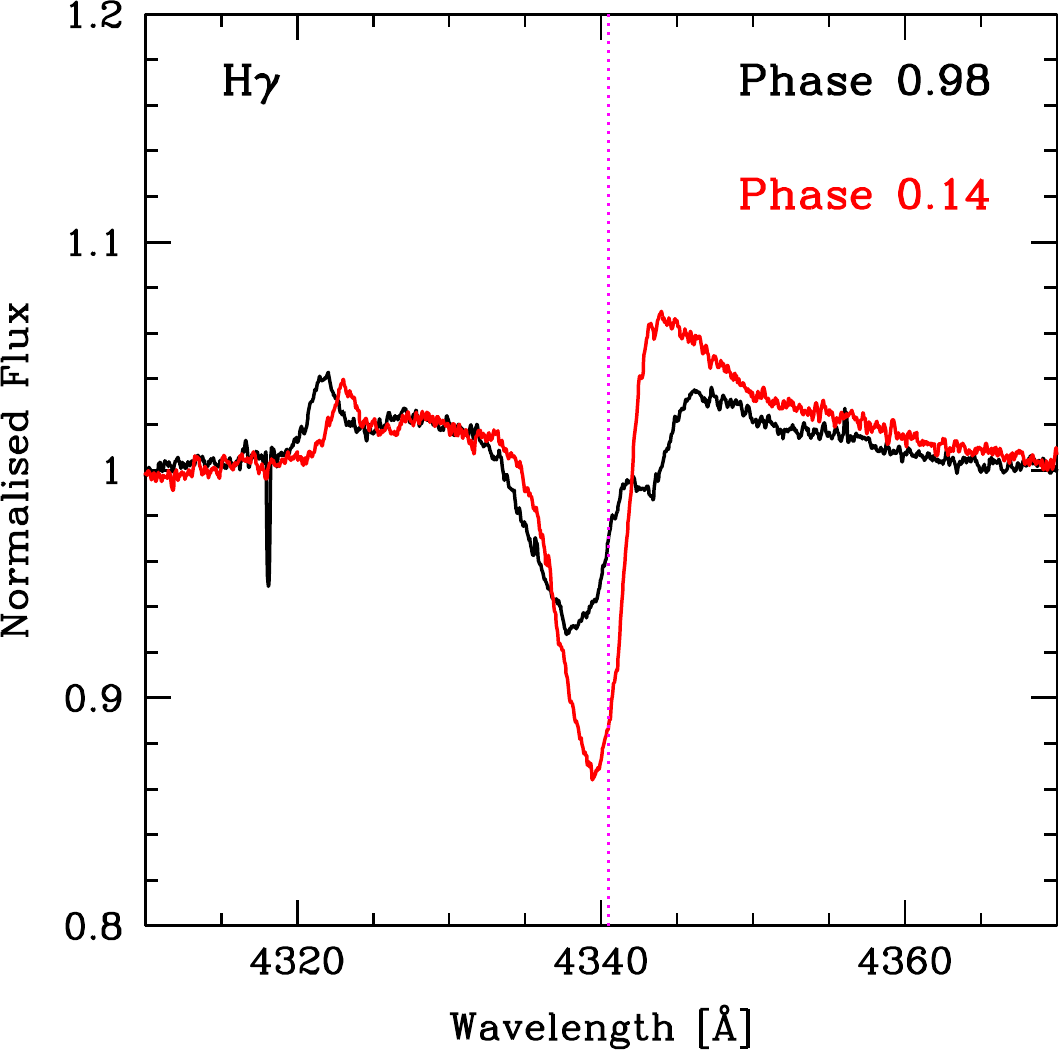}
	\hspace{5mm}
	\hskip3mm\includegraphics[width=0.263\textwidth]{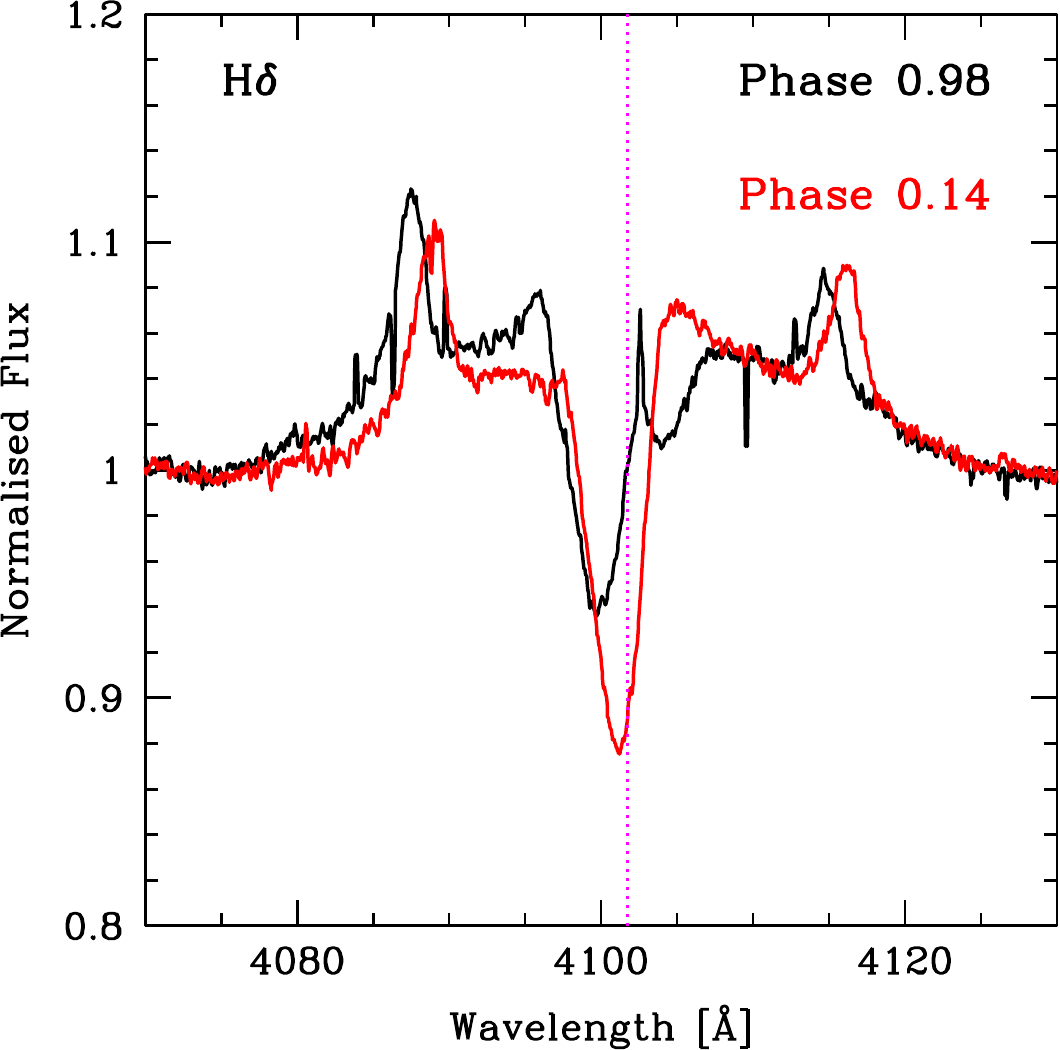}
	\hspace{5mm}
	\includegraphics[width=0.27\textwidth]{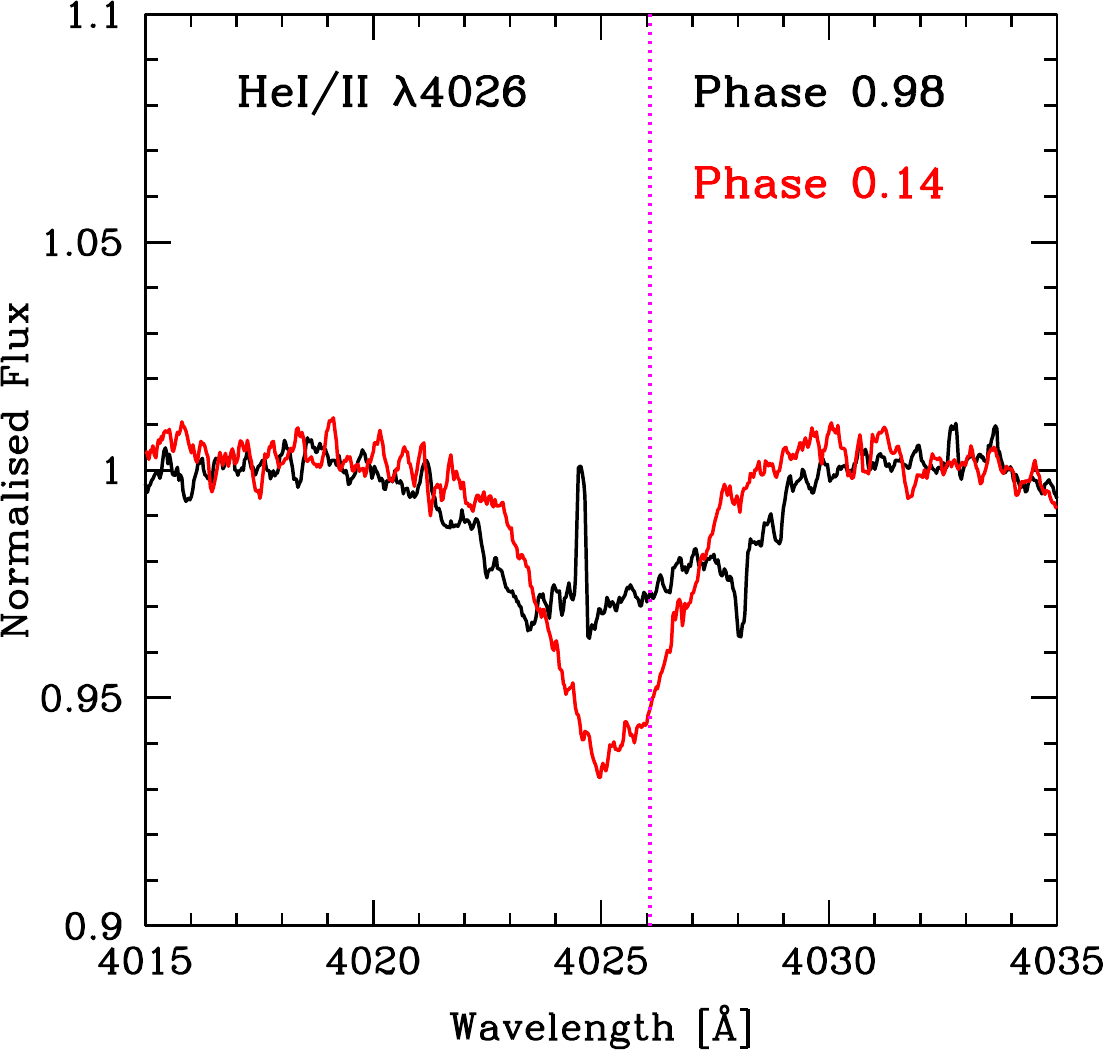}
	\vskip1ex
	
	\begin{minipage}{16cm}
		\caption{%
			Comparison of the shape and position of selected spectroscopic lines at two orbital phases: 0.14 (red) and 0.98 (black).
			These phases correspond to the maximum RV separation (phase 0.98, WN star shifted towards the blue) and to the approximate opposite extreme separation (phase 0.14,WN star shifted towards the red).
			The panels show the \ion{He}{ii}\,$\lambda$5412, \ion{He}{ii}\,$\lambda$4542, \ion{He}{ii}\,$\lambda$4200, H$\gamma$, H$\delta$, and \ion{He}{i-ii}\,$\lambda$4026 lines.%
			\label{fig:fgoextr}}
	\end{minipage}
\end{sidewaysfigure}
%
To further illustrate the detection of the companion, we show in Fig.~\ref{fig:fgoextr} two spectra: one displayed in black (phase 0.98), where the WN component is shifted towards the blue and the \mbox{O-star} component towards the red, corresponding to the phase of maximum radial-velocity separation;
and a second spectrum displayed in red, acquired at phase 0.14, which approximately corresponds to the opposite configuration.
The actual extremum occurs at phase 0.256.
Both spectra are part of data set~I$_{\text{glo}}$.
Six different wavelength regions are shown, each illustrating different spectroscopic lines.
At phase 0.98, the \ion{He}{ii}\,$\lambda$4542 line clearly exhibits an additional weaker component shifted towards the red that can be associated with the O~companion, thereby confirming the results of \citet{2008RMxAC..33...91G}.
As the system evolves towards the phase 0.14 configuration, the \mbox{O-star} component shifts towards the blue and the WN absorption towards the red.
The relative motion of the WN star can be seen in the short-wavelength wing of the WN absorption line.
At phase 0.14, both absorption lines have merged.
A~very similar behaviour is observed for the \ion{He}{ii}\,$\lambda$4200 transition.
The \ion{He}{ii}\,$\lambda$5412 line exhibits the same behaviour, although the situation is more complex due to the effect of a marked P-Cygni profile in the WN spectrum:
at phase 0.98, the \mbox{O-star} absorption component is superimposed on the
emission component of this WN P-Cygni profile.

In the panel of Fig.~\ref{fig:fgoextr} showing the behaviour of the H$\gamma$ line, the blue-to-red shift of the WN star is visible in the unrelated \ion{N}{iii}\,$\lambda$4321 line, which has no contribution from the O~star.
The same shift is also visible in the short-wavelength wing of the central absorption line and in the overall displacement of the centre of this same line.
The case of H$\delta$ is also similar, although the displacement of the short-wavelength wing of the central absorption is almost invisible, probably due to the additional presence of the \ion{N}{iii} lines.
The spectroscopic line \ion{He}{i}\,$\lambda$4026 (with a contribution also from \ion{He}{ii}) is not deblended at phase 0.98, and thus never deblends.
The line is simply broader at this phase.
The strengths (equivalent widths) of the absorption components associated with the WN star and the O~one are here very similar.
The \ion{He}{i}\,$\lambda$5876 line (not shown here) also never becomes deblended.
The depth of the WN absorption at phase 0.14 is always greater than at phase 0.98, but the difference is very roughly comparable to the contribution of the O~star.
This suggests that we are essentially observing a simple merging of the absorption components of the two objects.
Consequently, at phase 0.14, the contribution of the O~star to the total absorption is not negligible for these lines.

From the inspection of all these lines, it is clear that the spectral components of the two objects can be separated only for the \ion{He}{ii} and \ion{H}{} lines and even then only over a narrow range of phases around 0.98.
This indicates that we should search for other spectroscopic lines of the O~star that are significantly narrower and for which the influence of the WN component is much more limited.

\begin{figure}
	\centering
	\includegraphics[width=0.6\textwidth]{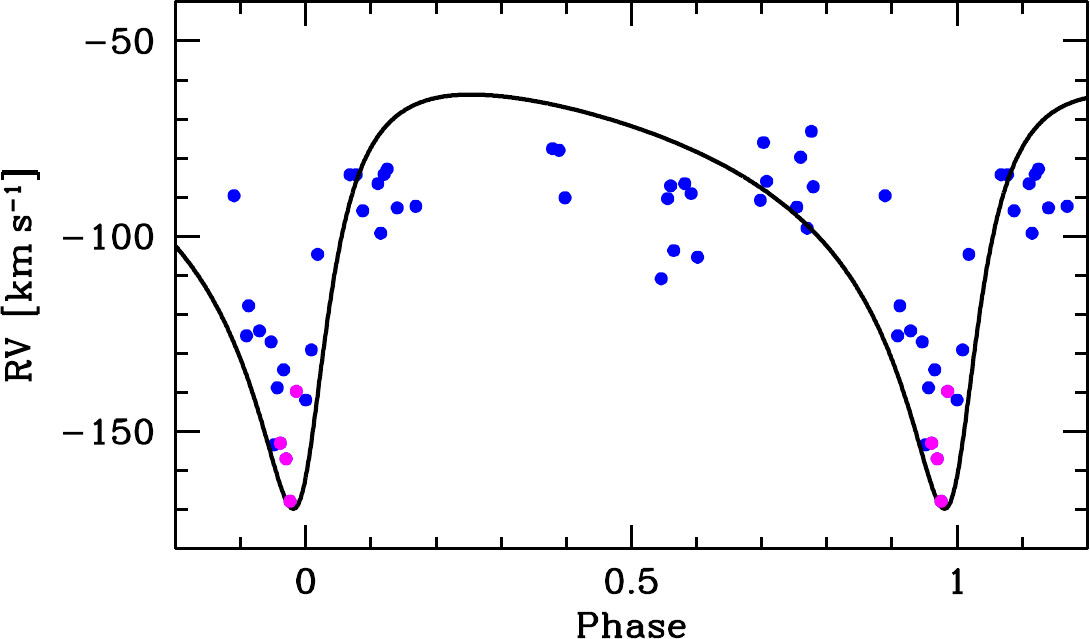}
	\vskip1ex
	
	\begin{minipage}{12cm}
		\caption{%
			Phase diagram showing the RVs associated to the main absorption component of the H$\gamma$ profile.
			The blue points represent the RVs for the phases where only one absorption component is visible, whereas the few magenta points correspond to phases where two absorption components are visible and can be separated (the RVs associated solely with the O~star are not shown here).
			The continuous line represents the adopted WN orbital solution from Sect.~\ref{ssec:orbsol_sb1model} (see Fig.~\ref{fig:RVorbit2}), with the zero systemic velocity shifted by an arbitrary $-90$\,km\,s$^{-1}$ for illustration purposes.%
			\label{fig:fgrvhgamabs}}
	\end{minipage}
\end{figure}

In order to further illustrate the behaviour of the absorption components in the above-mentioned transitions, we focus on the case of H$\gamma$.
The absorption lines in H$\gamma$ were fitted with one or two Gaussians.
The resulting RVs are displayed in the phase diagram shown in Fig.~\ref{fig:fgrvhgamabs}.
The majority of the data points (in blue) correspond to a single-Gaussian fit, whereas a very few data points (in magenta) correspond to a simultaneous fit with two Gaussians, one for the WN component and one for the \mbox{O-star} component (not shown here).
The magenta points associated with the WN component are concentrated at the most negative velocities when the two lines are separated.
In contrast, the blue points at phases between approximately 0.5 and 0.8 are located at less negative velocities.
This indicates that the absorption component broadly follows the motion of the WN star, although not perfectly.
This provides definitive evidence that WR\,25 contains absorption lines associated with both stars.
The same phenomenon is observed in the other lines discussed in the present section.
At phases between 0.75 and 0.92, the RVs are shifted towards less negative velocities.
This is an effect of the blending between the WN absorption line and the line of the O~star.
At phases between 0.1 and 0.7, the RVs do not change dramatically.
If an orbital fit is performed on these data, an eccentricity of 0.7 is obtained, which is not consistent with the orbital motion derived for the WN component.
Two effects could play a role:
either the WN absorption line does not perfectly follow the orbital motion because of distortions in the line-formation region, or the absorption component of the O~star in H$\gamma$ shifts the apparent position of the blended absorption profile.
From an observational point of view, it is difficult, if not impossible, to disentangle these two effects.
We nevertheless consider that the RVs at phases just below 0.98 strongly suggest that the second effect is indeed present.

\subsection{\texorpdfstring{The particular case of H$\boldsymbol\beta$}{The particular case of Hbeta}}%
\label{ssec:ospec_hbeta}
\begin{figure}
	\centering
	\includegraphics[width=0.6\textwidth]{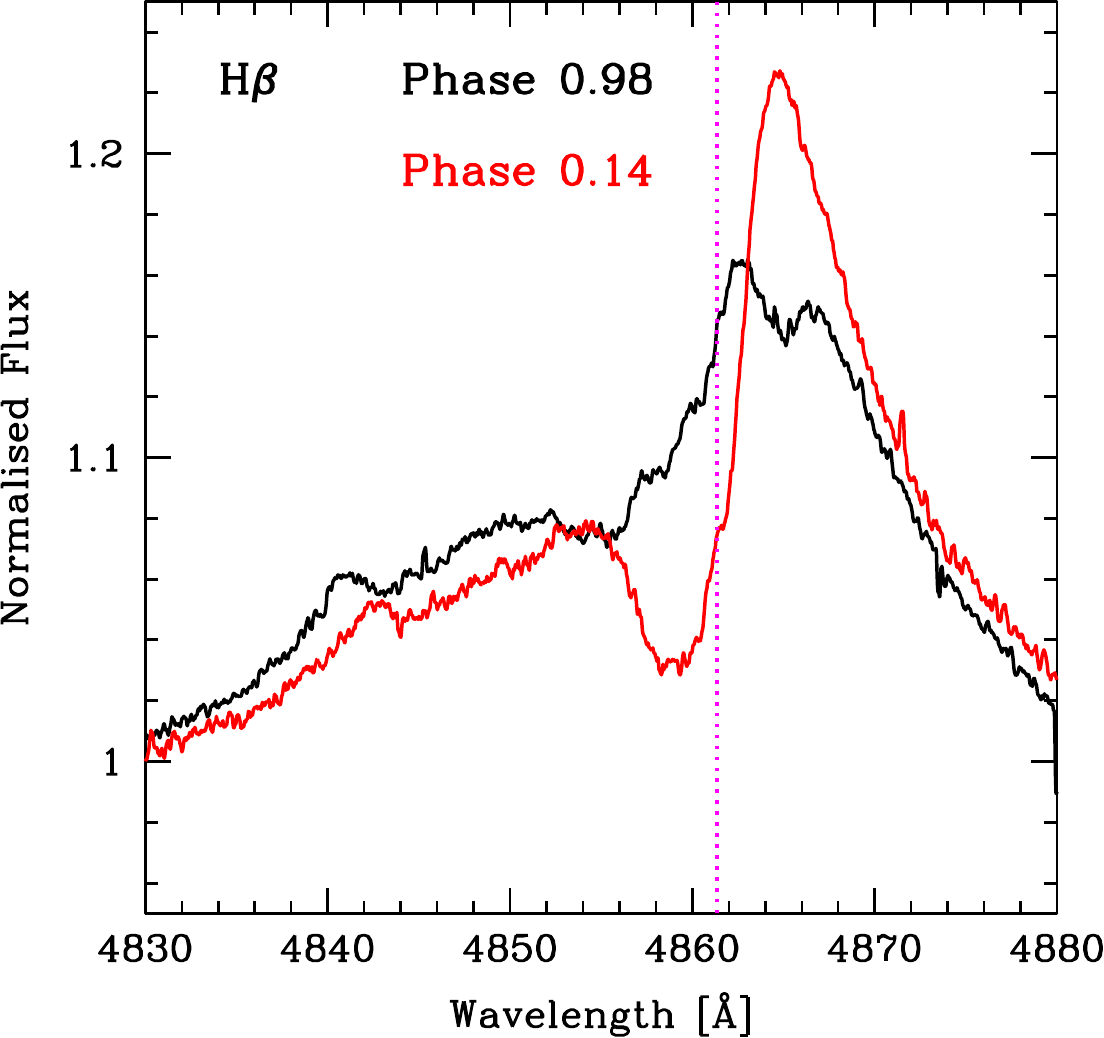}
	\vskip1ex
	
	\begin{minipage}{12cm}
		\caption{%
			Illustration of the \ion{He}{ii}\,$\lambda$4859+H$\beta$ profile in WR\,25 at the two usual extreme phases, 0.98 (black) and 0.14 (red).
			The dotted magenta line indicates the position of the laboratory wavelength of this spectral feature.
			The small emission line to the blue side of \ion{He}{ii}\,$\lambda$4859+H$\beta$ is attributed (see Sect.~\ref{sssec:wrspec_DS1_4841}) to the transition \ion{N}{iii}\,$\lambda$4842.%
			\label{fig:fgextrpluhbet}}
	\end{minipage}
\end{figure}
Similarly to what was done in Sect.~\ref{ssec:ospec_detect}, we show in Fig.~\ref{fig:fgextrpluhbet} the spectrum of WR\,25 in the vicinity of the \ion{He}{ii}\,$\lambda$4859+H$\beta$ line at phases 0.14 and 0.98.
At phase 0.14, the WN spectrum is close to its maximum shift towards the red, as can be seen from the \ion{N}{iii}\,$\lambda$4842 line.
The spectrum is then characterised by a P-Cygni profile.
The presence of such a marked absorption component has been reported in several stars by \citet{2011MNRAS.416.1311C}, prompting them to introduce the composite spectral type O2.5f$^{*}$/WN6, for which WR\,25 was selected as the archetype.
The main and sole defining criterion for this spectral type is the presence in H$\beta$ of a P-Cygni profile with a strong absorption component \citep[see Table~2 of][]{2011MNRAS.416.1311C}.

\begin{figure}
	\centering
	\includegraphics[width=0.6\textwidth]{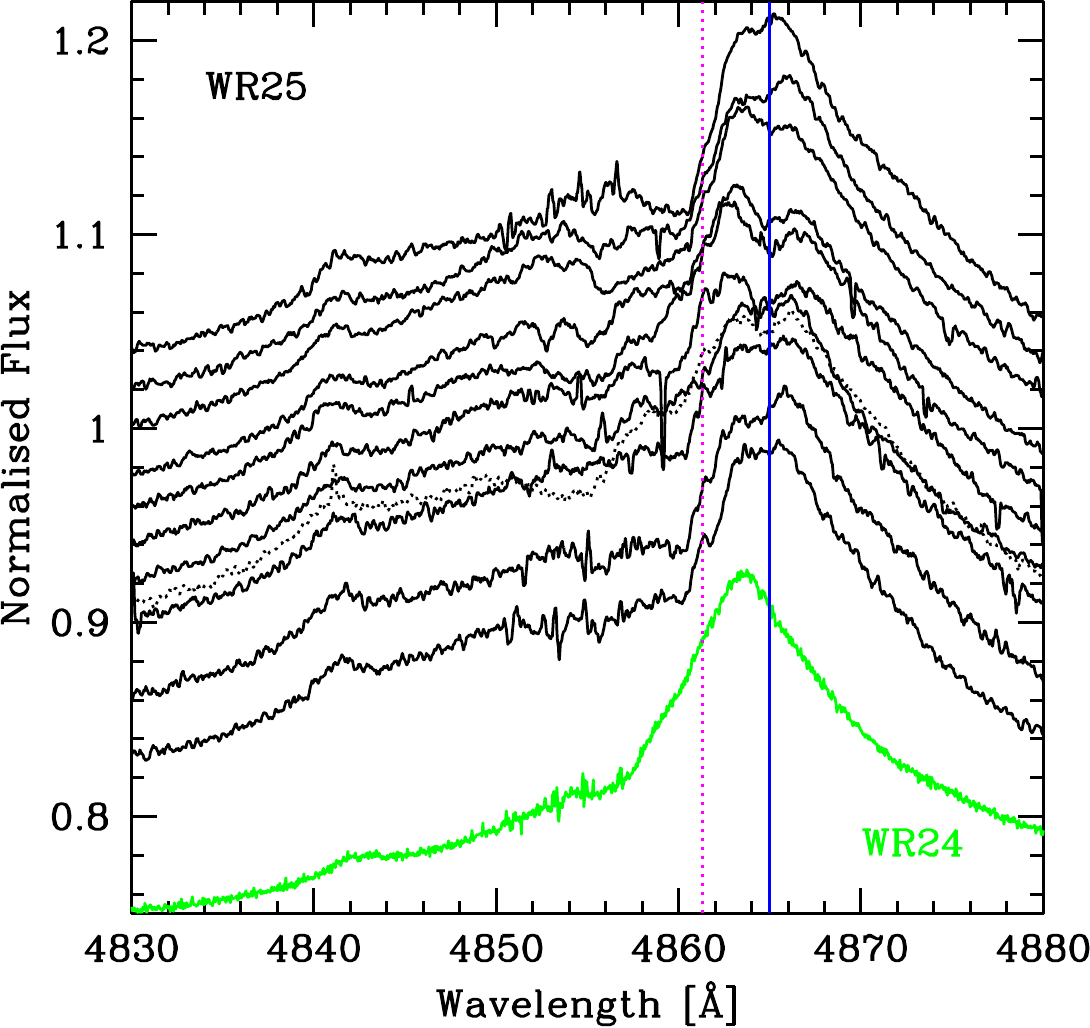}
	\vskip1ex
	
	\begin{minipage}{12cm}
		\caption{%
			Selected WR\,25 spectra (black) at phases close to 0.98.
			For illustrative purposes, the spectra have been vertically shifted relative to the continuum level at 1 by an amount equal to twice the difference between phase zero and the phase of the corresponding spectrum.
			From \emph{top} to \emph{bottom}, the phases are 0.018, 0.009, 0.000, 0.986, 0.976, 0.966, 0.956, 0.952 (dotted), 0.947, 0.929, and 0.912.
			The magenta vertical line indicates the laboratory wavelength, whereas the blue vertical line marks the maximum redward excursion of the O~component.
			An arbitrarily rescaled spectrum of WR\,24 (green) is also shown to facilitate comparison between the two stars.%
			\label{fig:fghbetamax}}
	\end{minipage}
\end{figure}

At phase 0.98, when the WN star is at its maximum shift towards the blue, and the O~star is expected to be at its maximum shift towards the red, the above-mentioned absorption component is no longer prominent.
In addition, the emission component of the global P-Cygni profile is strongly attenuated.
This effect is actually caused by the H$\beta$ absorption of the O~star.
At this phase, the absorption feature is located almost exactly at the centre of the emission component, producing the observed double-subpeak structure and the associated flux deficit.
We note that the area between the red and black lines around 4864\,{\AA} is similar, but of opposite sign, to the area between the same two lines at the position of the absorption feature near 4859\,{\AA}.
It is tempting to attribute the absorption component observed at phase 0.14 to the O~component.
However, this interpretation would require a velocity shift substantially larger than that observed for the \ion{He}{ii}\,$\lambda$4200 line.
To investigate this issue further, we illustrate the behaviour of the \ion{He}{ii}\,$\lambda$4859+H$\beta$ line near phase 0.98 in Fig.~\ref{fig:fghbetamax}.
Several selected spectra of WR\,25 are shown for phases between 0.91 and 0.02 (or 1.02).
None of these spectra exhibits a pronounced P-Cygni absorption.
Clearly, the region between 4850 and 4860\,{\AA} shows significant variability, but none of the phases exhibits the P-Cygni absorption visible at the other phases (as illustrated in this paper by phase 0.14).
Around the dotted magenta line, a small level of variability is also present, but this is essentially due to an imperfect subtraction of the narrow nebular H$\beta$ line, which is ubiquitous in the Carina nebula region.
The spectrum at phase 0.95 (dotted line) differs markedly from that at phase 0.96 in the region around 4850--4855\,{\AA}.
This provides additional evidence for variability in this region, since the two spectra were acquired only one day apart.

As indicated by the blue vertical line, the absorption component superimposed on the P-Cygni profile is located approximately at the position of this line around phase 0.98.
In contrast, before and after this phase, the absorption shifts towards the blue (left-hand side), thereby tracing the RV motion of the O~star.
For comparison, we also show in Fig.~\ref{fig:fghbetamax} an arbitrarily rescaled spectrum of the \ion{He}{ii}\,$\lambda$4859+H$\beta$ profile of WR\,24 (in green).
At these phases, the \ion{He}{ii}\,$\lambda$4859+H$\beta$ profile of WR\,25 is very similar to that of WR\,24, which has the spectral type WN6ha \citep{2011MNRAS.416.1311C}.

\begin{figure}
	\centering
	\includegraphics[width=0.6\textwidth]{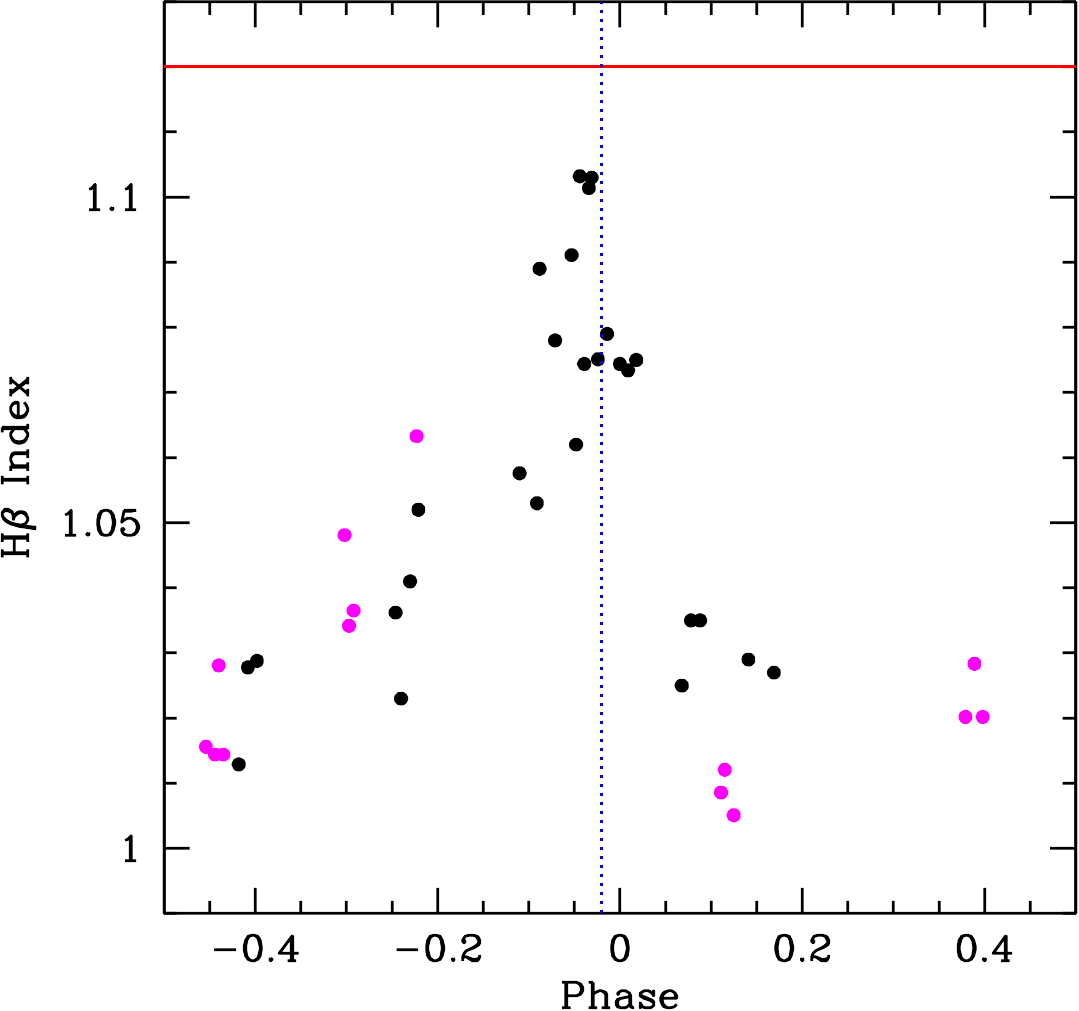}
	\vskip1ex
	
	\begin{minipage}{12cm}
		\caption{%
			Run of the H$\beta$ index as a function of orbital phase.
			The black points correspond to the spectra from the I$_{\text{loc}}$ data set, while magenta points represent data set~III.
			The blue dotted vertical line indicates the phase of minimum radial velocity of the WN star.
			The horizontal red line marks the maximum flux level possible in the region around 4860\,{\AA}.%
			\label{fig:fghbetaabs}}
	\end{minipage}
\end{figure}

The variability observed between 4850 and 4860\,{\AA} at phases around 0.98 is most probably due to instabilities in the wind of the WN component.
The origin of the P-Cygni absorption component visible between phases 0.14 and 0.80 is more difficult to assess.
It appears to be
related to the absorption line of the O~component, although further modelling is required to confirm this interpretation.
In any case, the pronounced absorption component observed in WR\,25 is only visible at certain phases.
To illustrate this further, we define an H$\beta$ index corresponding to the minimum normalised flux in the vicinity of 4860\,{\AA}.
The run of this index as a function of phase is shown in Fig.~\ref{fig:fghbetaabs}.
The phase dependence of the absorption component is striking.
Even if the effect is not entirely caused by the spectrum of the O~star itself, it is certainly linked to the presence of the \mbox{O-star} companion and therefore to the binary nature of the system.
Consequently, the selection of WR\,25 as the archetype of the O2.5f$^{*}$/WN6 spectral type, defined by the presence of a strong absorption component in the \ion{He}{ii}\,$\lambda$4859+H$\beta$ profile, appears somewhat unfortunate.
\vspace{-2ex}

\subsection{\texorpdfstring{The anomaly of \ion{N}{iii}\,$\boldsymbol\lambda$4641}{The anomaly of N III lambda 4641}}%
\begin{figure}
	\centering
	\includegraphics[width=0.6\textwidth]{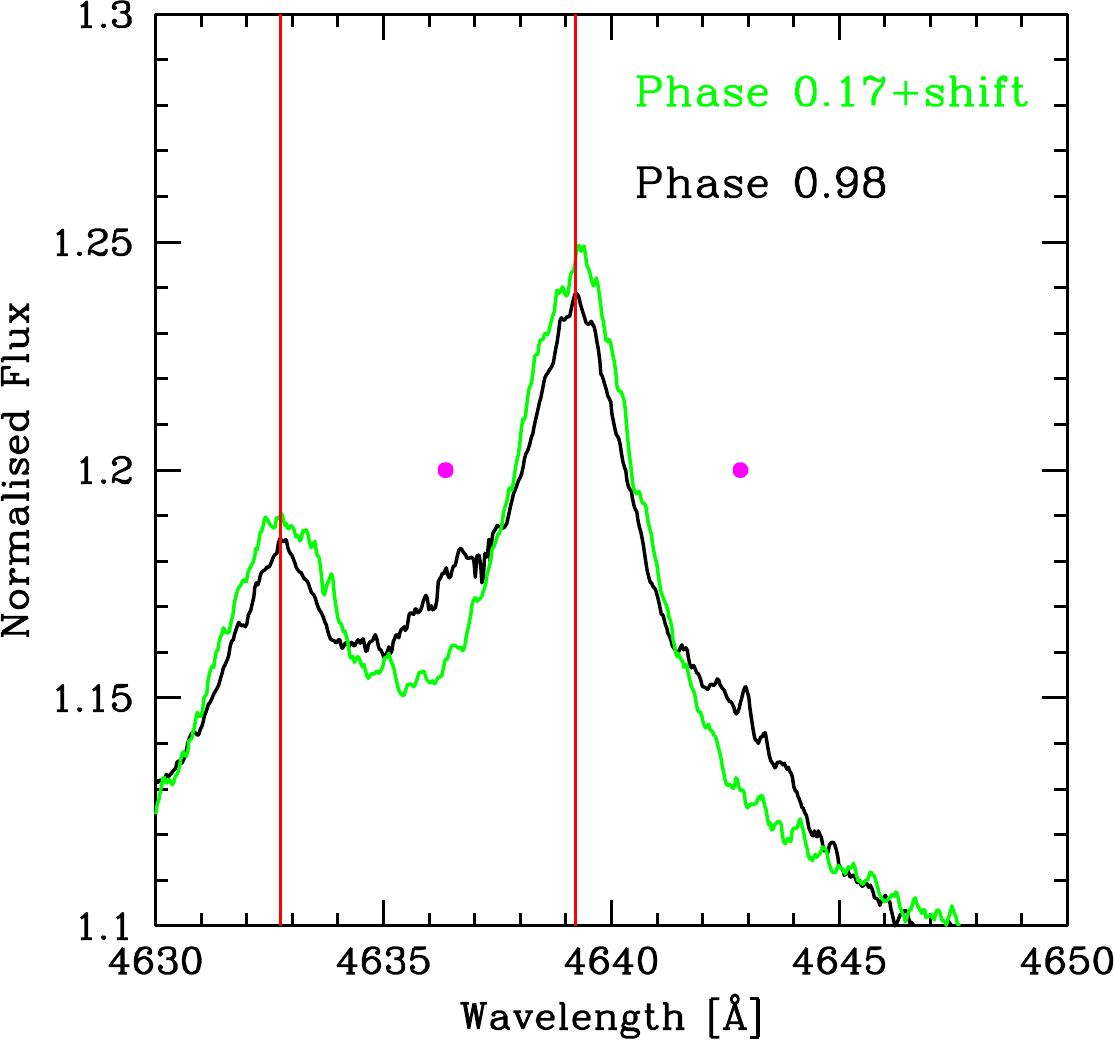}
	\vskip1ex
	
	\begin{minipage}{12cm}
		\caption{%
			Spectrum of WR\,25 in the region of the \ion{N}{iii}\,$\lambda \lambda$4634-4641 triplet at phase 0.98 (black) with the spectrum at phase 0.17 (green) superimposed after being shifted so that the two apparent emission lines of the triplet coincide.
			The positions of the two transitions are marked by red vertical lines.
			The two magenta dots indicate the position of the \mbox{O-star} contribution at phase 0.98.%
			\label{fig:fgpftglo4641}}
	\end{minipage}
\end{figure}

\label{ssec:ospec_anom4641}
In this section, we further investigate the anomaly reported in Sect.~\ref{sssec:wrspec_DS1_4641}.
Figure~\ref{fig:fgpftglo4641} shows the spectrum of WR\,25, in the region of the \ion{N}{iii}\,$\lambda \lambda$4634-4641 triplet at phase 0.98 (black).
The positions of the two apparent lines formed by the triplet are marked by red vertical lines.
We also plot a spectrum obtained at phase 0.17 (green), shifted in RV so that the two apparent lines associated with the WN component are superimposed on those of the spectrum at phase 0.98.
This shift is a strict minimum;
the true shift could be slightly larger.
Both spectra shown in Fig.~\ref{fig:fgpftglo4641} are taken from data set~I$_{\text{glo}}$.
The two apparent emission lines in the black spectrum are weaker than those in the green spectrum.
Conversely, the black spectrum  exhibits excess flux relative to the green spectrum at two positions marked by magenta dots.
The explanation is straightforward.
The excesses are due to an additional emission from the O~star, and are clearly visible and separated from the WN component at phase 0.98.
At phase 0.17, however, the same O~star emission contributes directly to the WN features, which explains why the green spectrum is somewhat brighter at the positions marked by the red vertical lines.
Therefore, the O~star exhibits this triplet in emission and should at least be classified as ``((f))'' \citep{1971ApJS...23..257W, 1996LIACo..33....1V}.
The observed displacement suggests a very rough lower limit of about 80\,km\,s$^{-1}$ for the semi-amplitude $K_{\text{O}}$.
Throughout the orbital cycle, the emissions from the two stars combine in different ways, and this hampers the correct determination of the RVs in a phase-dependent manner.
This provides a natural explanation for the anomaly reported in Sect.~\ref{sssec:wrspec_DS1_4641}.
At phase 0.98, the \ion{N}{iii} triplet emission lines of the WN and O~components are well separated and the measurement of the RV of the WN component is essentially unbiased and agrees well with the RV derived from the \ion{N}{iv}\,$\lambda$4058 line (see Fig.~\ref{fig:fgcorr4641});
at phase 0.14, and approximately until phase 0.3, the emission lines of both objects occur at nearly the same wavelengths, so the position of the blended feature is not significantly affected by the O-star contribution.
Beyond phase 0.3, the \mbox{O-star} emission lines progressively separate from the WN features while moving towards the red.
As this occurs, the \mbox{O-star} emission lines pull the centroid of the fitted Gaussian redwards, producing a bias towards larger RVs, which is precisely the behaviour illustrated in Fig.~\ref{fig:fgcorr4641}.

These results are particularly interesting because they raise the possibility that information on the orbit of the O~component could be extracted through spectral-disentangling techniques (see Sect.~\ref{sec:disentangling}).

\subsection{\texorpdfstring{The \ion{He}{i}\,$\boldsymbol\lambda$4471 line}{The He I lambda 4471 line}}%
\begin{figure}[t]
	\centering
	\includegraphics[width=0.6\textwidth]{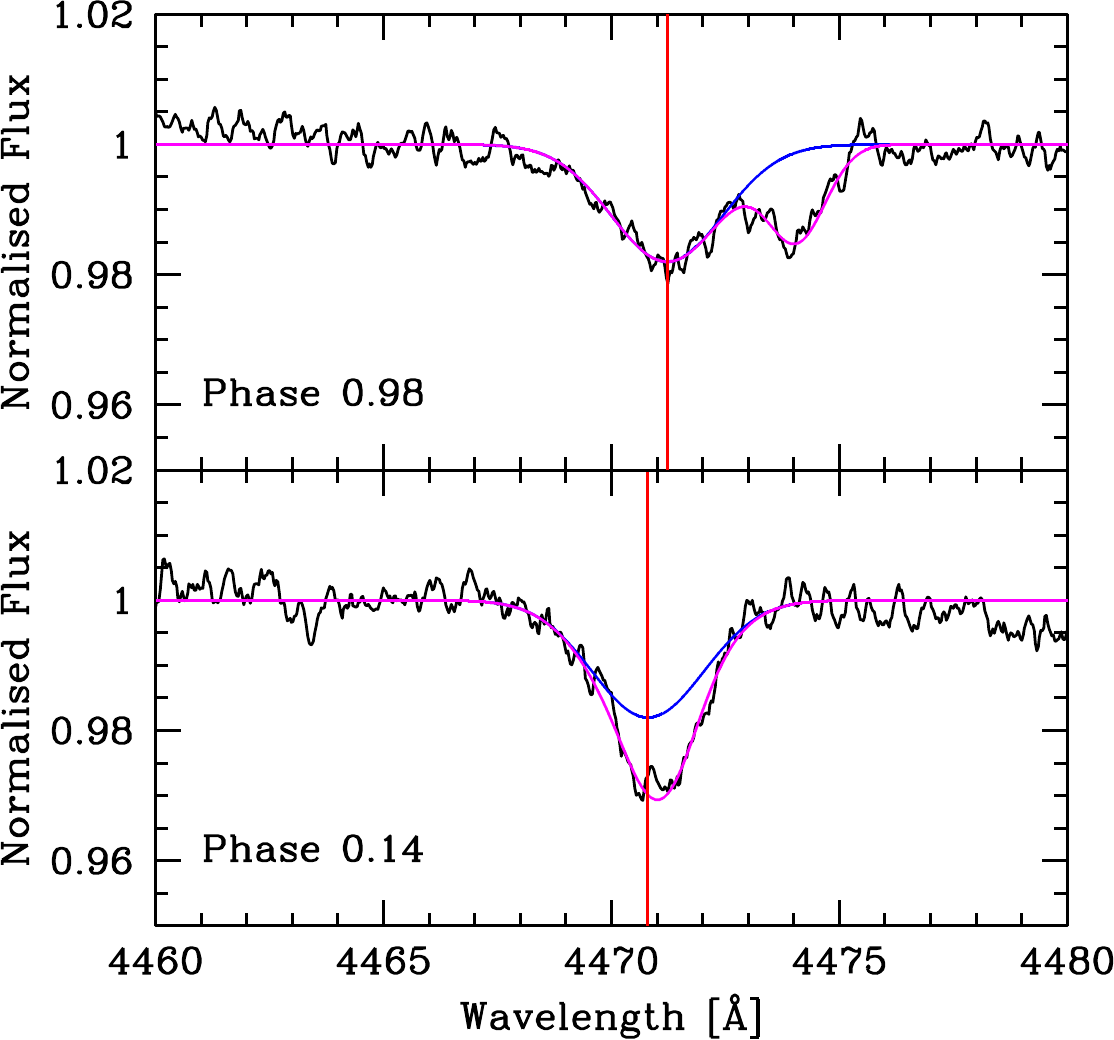}
	\vskip1ex
	
	\begin{minipage}{12cm}
		\caption{%
			Spectrum of WR\,25 in the region of the \ion{He}{i}\,$\lambda$4471 line at phase 0.98 (\emph{upper panel}) and phase 0.14 (\emph{lower panel}). 
			The blue curves show the single-Gaussian fit, while the magenta lines correspond to the two-Gaussian fit.
			The red vertical lines mark the position of the component attributed to the WN component, highlighting its limited radial-velocity variation.%
			\label{fig:fg4471extr}}
	\end{minipage}
\end{figure}

\label{ssec:ospec4471}
In order to improve our knowledge of the orbital motion of the O~component, we searched for additional spectral lines that could provide further constraints.
In particular, we looked for lines for which the detrimental effect of the WN component might be reduced.
Although weak, \ion{He}{i}\,$\lambda$4471 could be of interest.
Figure~\ref{fig:fg4471extr} shows this line at two extremal orbital configurations, where it is always observed in absorption.
%
In the upper panel, contributions from both the WN and O~components are clearly visible.
At phase 0.98, the two absorption features are sufficiently separated to allow a
simultaneous fit.
The blue curve represents the Gaussian fitted to the left-hand component alone, to be associated with the WN star, whereas the magenta curve represents the fit obtained with two Gaussians and, by comparison, highlights the contribution of the O~star.
At phase 0.98 (more precisely 0.976), the RV of the \mbox{O-star} component is 171.5\,km\,s$^{-1}$, in good agreement with the velocity derived at the same phase from, e.g., the \ion{He}{ii}\,$\lambda$4542 line (observed at 4544.2\,{\AA}, corresponding to a velocity of 172.3\,km\,s$^{-1}$---see Fig.~\ref{fig:fgoextr}).
However, two other spectra at phases 0.970 and 0.986 favour a velocity closer to 180\,km\,s$^{-1}$.
We therefore adopt a velocity of 176\,km\,s$^{-1}$ at phase 0.98.
This choice is admittedly somewhat arbitrary.
However, 5\,km\,s$^{-1}$ corresponds to 0.075\,{\AA}, i.e., roughly one resolution element (see the end of Sect.~\ref{sssec:observations_DS1_reduction}), and a substantially better precision can therefore not be expected from a single isolated line.
Inspection of Fig.~\ref{fig:fg4471extr} suggests that this level of accuracy is not achieved in practice.
The RV associated with the WN component is $-19.5$\,km\,s$^{-1}$ (marked by the red vertical line), a result that remains to be explained.
This velocity is clearly shifted with respect to the actual RV associated with the WN component.
Indeed, at phase 0.98, the WN star should in principle be shifted by more than 70\,km\,s$^{-1}$ towards the blue relative to the systemic velocity.
The \ion{He}{i} line could potentially be formed in the expanding wind, or it may arise from an unrelated source---an intruder.
The only nearby source reported for WR\,25 is situated 0\farcs 79 away and has a magnitude difference $\Delta V$ of about 5.8\,mag \citep{2014ApJS..211...10S}.
It is unlikely that the line can be attributed to this faint object.
This object is not part of the \emph{Gaia} DR3 catalogue \citep{2016A&A...595A...1G, 2021A&A...649A...1G, 2023A&A...674A...1G} and no other nearby star is listed in this catalogue in the immediate vicinity of WR\,25.
This point will be discussed further below.

\begin{figure}[t]
	\centering
	\includegraphics[width=0.6\textwidth]{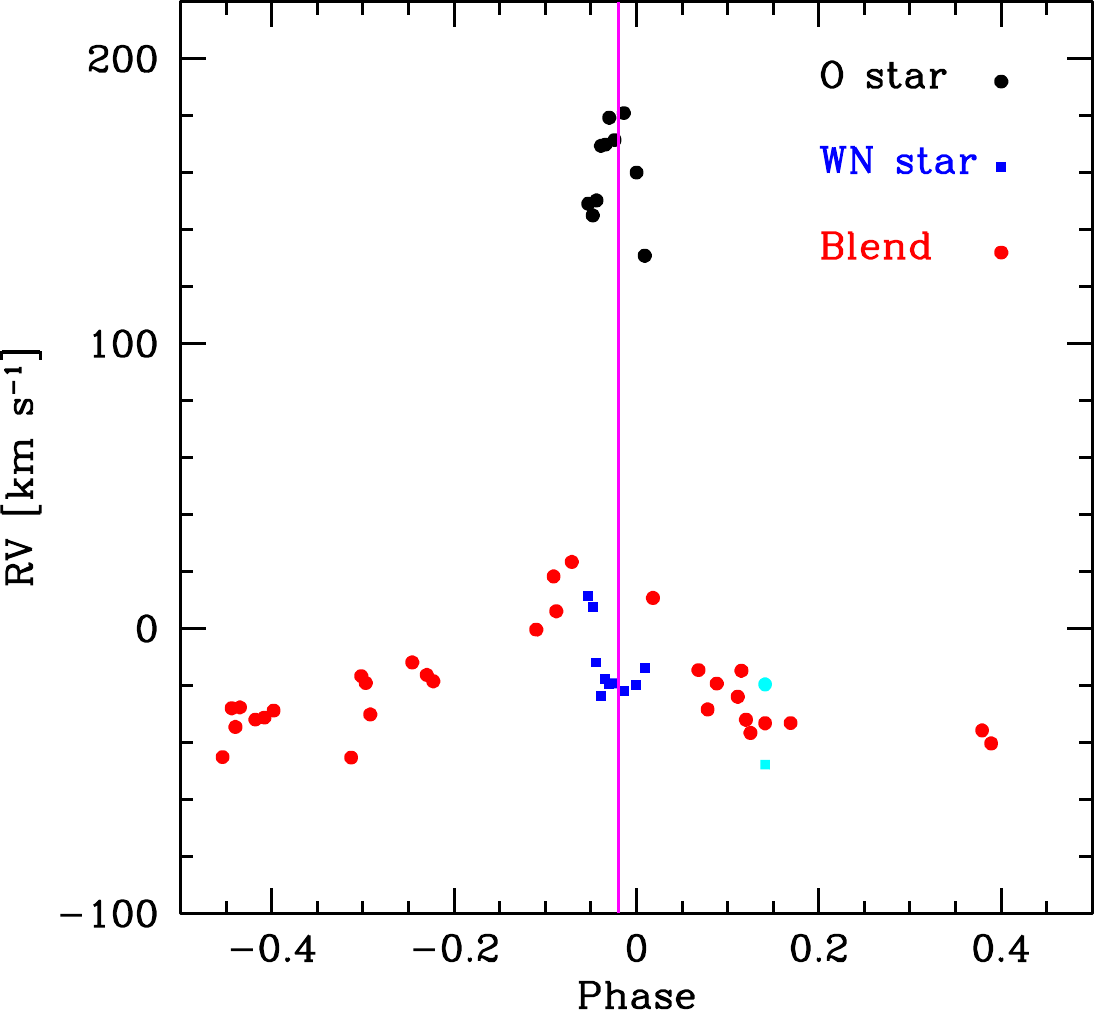}
	\vskip1ex
	
	\begin{minipage}{12cm}
		\caption{%
			Radial velocities derived from the \ion{He}{i}\,$\lambda$4471 line as a function of orbital phase.
			The different symbols and colours correspond to the O~component (black circles), the WN-star component (blue squares), and the blends (red circles).
			The cyan circle and cyan square indicate the RVs measured for the O- and WN-star components, respectively, depicted in the lower panel of Fig.~\ref{fig:fg4471extr} (see also text).
			The vertical line marks phase 0.98.%
			\label{fig:fgrv4471general}}
	\end{minipage}
\end{figure}

At phase 0.14, the two lines are clearly blended.
A~tentative two-component fit is shown in the lower panel of Fig.~\ref{fig:fg4471extr}.
However, this particular fit is not secure, and other solutions are possible.
No other spectrum obtained at these phases clearly reveals two separate components.
Although the \mbox{O-star} component shifted towards the blue is reasonably consistent with the expected orbital motion of the O~star, the component attributed to the WN star also appears to be shifted towards the blue, which is certainly puzzling.
It is therefore possible that the assignment of the two components is not correct.
In any case, this feature does not appear to follow the orbital motion of the WN star.
To further illustrate the intricacy of the problem,  Fig.~\ref{fig:fgrv4471general} shows the measured RVs as a function of phase.
Around phase 0.98, the RVs of the WN and O~components can be measured separately, whereas at other phases, only the global position of the blended profile can be reported.
As the spectra approach the phase interval where the two components become resolvable, the position of the blend is progressively displaced towards more positive RVs.
This occurs because at these phases the O~component is on the edge of deblending.
The WN component itself appears to exhibit only limited motion.
For the data point at phase 0.14, we report both the RV of the blend and the two RVs derived from the tentative fit shown in the lower panel of Fig.~\ref{fig:fg4471extr}.
These two latter measurements are indicated by two cyan symbols in Fig.~\ref{fig:fgrv4471general}.
Studying this \ion{He}{i}\,$\lambda$4471 line would probably benefit from analysis using a spectral-disentangling technique (see Sect.~\ref{sec:disentangling}).
Indeed, the equivalent width of the O~component is not too faint compared to the supposed WN contribution, and its influence on the equivalent width of the blended profile is potentially favourable for disentangling.

At phase 0.98, the \ion{He}{i}\,$\lambda$4471 line associated with the O~component is measured at a velocity of 176\,km\,s$^{-1}$.
At phase 0.14, the two Gaussians fitted to the blended profile are located at velocities of $-27.6$\,km\,s$^{-1}$ (the probable O~component) and $-47.7$\,km\,s$^{-1}$.
The resulting velocity differences are therefore 203.6\,km\,s$^{-1}$ and 223.7\,km\,s$^{-1}$, respectively.
Phase 0.14 does not correspond exactly to the extremum, which occurs at phase 0.256.
Using the shape of the WN-star orbit derived in Sect.~\ref{ssec:orbsol_sb1model}, we estimate that the passage from phase 0.14 to phase 0.256 requires an increase of 5.5{\%} in the measured velocity range to recover the full RV excursion of the O~star,  which should be a good estimator of $2K$.
From the velocity differences quoted above, we obtain a preliminary estimate of the \mbox{O-star's} semi-amplitude, $K_{\text{O}}$, in the range 107--108\,km\,s$^{-1}$.
Under the alternative---less likely---identification of the two components in the blended profile, the corresponding value would be 118\,km\,s$^{-1}$.
This latter value may be regarded as an approximate upper limit.

\begin{figure}
	\centering
	\includegraphics[width=0.6\textwidth]{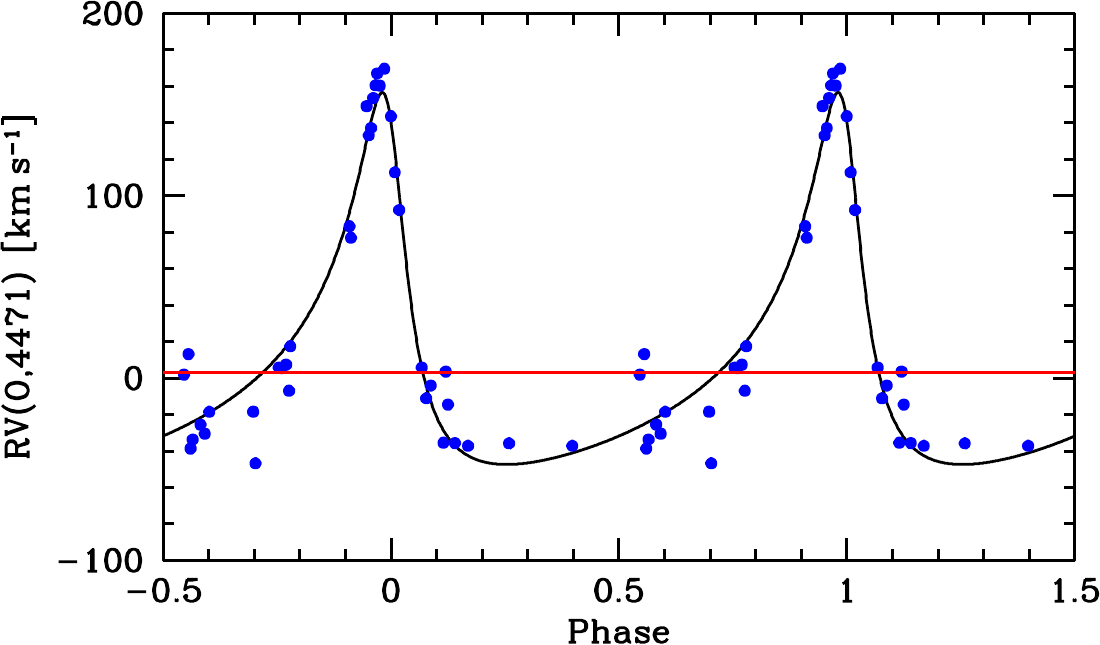}
	\vskip1ex
	
	\begin{minipage}{12cm}
		\caption{%
			Preliminary RV curve of the O~component as a function of orbital phase.
			The blue points represent the individual RV measurements obtained after removing the \ion{He}{i}\,$\lambda$4471 component assumed to be constant in RV and not associated with the O~star.
			An orbital fit was performed (see Sect.~\ref{ssec:ospec4471}), with parameters $K_{\text{O}}$ and $\gamma_{\text{O}}$ treated as free parameters while all other orbital elements were fixed to the values of the SB1 solution.
			The solid black curve shows the resulting Keplerian fit, and the red horizontal line indicates the fitted systemic velocity.%
			\label{fig:figpreorb4471}}
	\end{minipage}
\end{figure}

Since the main \ion{He}{i}\,$\lambda$4471 spectral line does not seem to follow the motion of the WN component and remains nearly stationary, we explored an alternative approach.
As a working hypothesis, we assumed that this feature is constant in wavelength.
We adopted a central wavelength of 4471.1\,{\AA}, corresponding approximately to the midpoint of the positions shown in Fig.~\ref{fig:fg4471extr}.
We also adopted the same depth (0.018 in continuum units) and width ($\sigma = 1.2$\,{\AA}) as in that figure.
This component was then subtracted from all the spectra in data sets~I, II and III.
In principle, this procedure leaves only the absorption feature associated with the O~component.
We then fitted a Gaussian profile to this residual feature in each spectrum and derived the corresponding RV.
Because the line is very weak, the procedure could not be applied successfully to all spectra (depending on the S/N ratio).
The successfully determined RVs are listed in Appendix~\ref{sec:appD}.
These RVs were plotted in a phase diagram and fitted for the parameters $K_{\text{O}}$ and $\gamma_{\text{O}}$, while the other orbital parameters were fixed to the values given in Table~\ref{tabelement}.
The resulting fit is shown in Fig.~\ref{fig:figpreorb4471}.
We derived \mbox{$K_{\text{O}} = 102.6\,(3.8)$\,km\,s$^{-1}$} and \mbox{$\gamma_{\text{O}} = 3.2\,(3.0)$\,km\,s$^{-1}$}.
We recall that, throughout this paper, uncertainties quoted in parentheses correspond to $1\sigma$ errors.
The parameters derived for the O~star are not very precise.
The line itself is very faint (with a typical depth of only 0.013 in continuum units), and the assumption that the subtracted primary component is strictly constant remains a working hypothesis that may introduce an unknown systematic bias.
A~more detailed analysis will be presented in the section devoted to spectral disentangling (Sect.~\ref{sec:disentangling}).

\subsection{\texorpdfstring{The \ion{N}{iv} triplet around 5200\,{\AA}}{The N IV triplet around 5200 Angstroem}}
\begin{figure}
	\centering
	\includegraphics[width=0.6\textwidth]{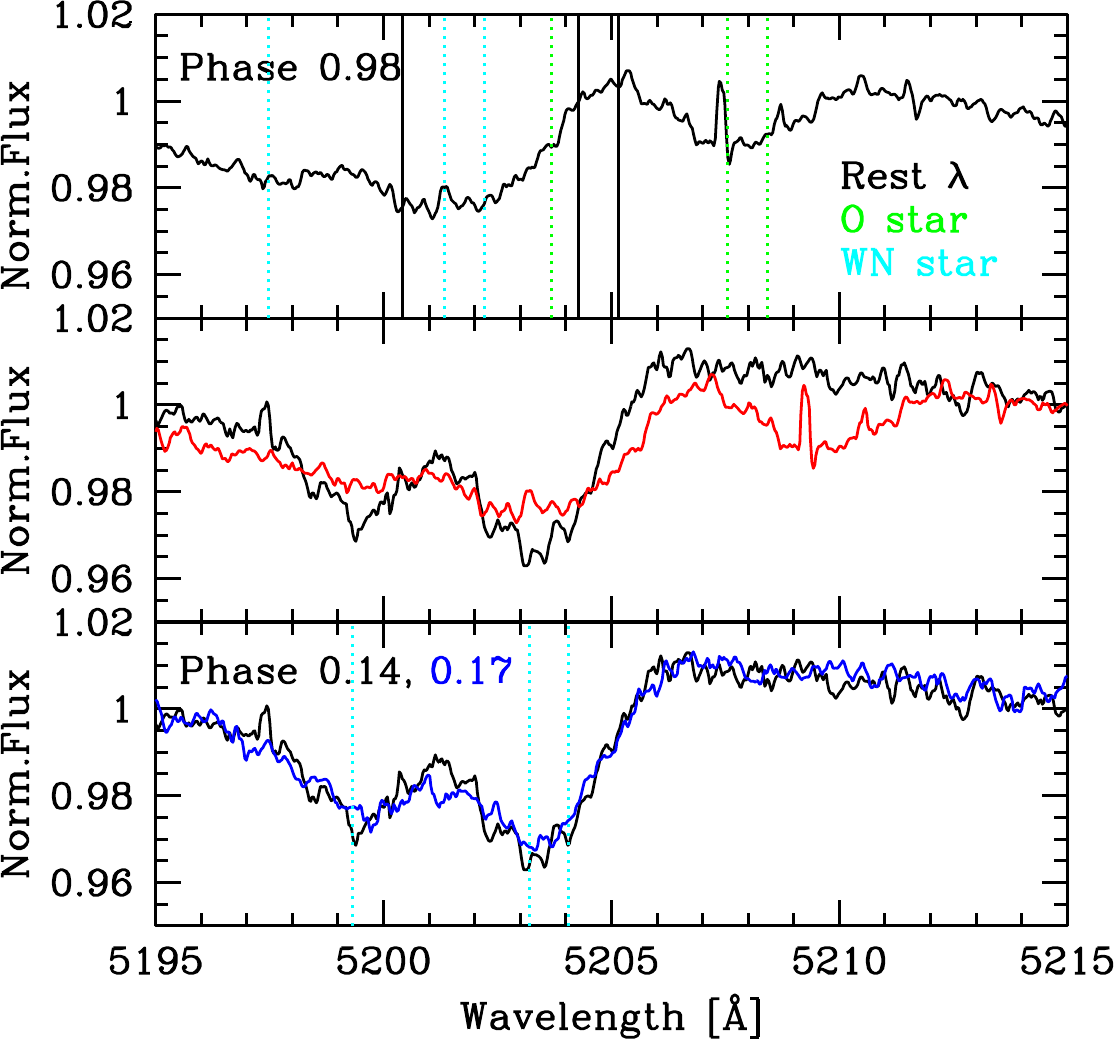}
	\vskip1ex
	
	\begin{minipage}{12cm}
		\caption{%
			Spectrum of WR\,25 in the region of the \ion{N}{iv} triplet.
			The \emph{upper panel} shows the spectrum at phase 0.98, while the \emph{lower panel} presents spectra obtained at phases 0.14 (black) and 0.17 (blue).
			In the upper panel, the identified absorption features are labelled, and the rest wavelengths of the triplet are indicated by the solid vertical black lines.
			The green dotted lines mark the expected positions of the O~component, whereas the cyan dotted lines indicate those of the WN component.
			In the lower panel, the cyan lines identify the WN component blended with the \mbox{O-star} contribution.
			The \emph{middle panel} compares the spectrum at phase 0.14 with that at phase 0.98 shifted redwards by twice the value of $K_{\text{WN}}$.
			\label{fig:fg5204extr}}
	\end{minipage}
\end{figure}

The \ion{N}{iv} ion exhibits a triplet with laboratory wavelengths of 5200.41\,{\AA}, 5204.29\,{\AA}, and 5205.15\,{\AA}.
The latter two transitions are closely spaced and remain unresolved in all our spectra.
We therefore adopt a rest wavelength of 5204.55\,{\AA} for this blended pair.
Figure~\ref{fig:fg5204extr} shows the spectrum of WR\,25 in the region of this triplet.
%
In the upper panel, corresponding to phase 0.98, we identify the various observed absorption features.
The positions of the triplet at rest are indicated by solid vertical black lines, while their positions shifted to a radial velocity of 188.3\,km\,s$^{-1}$, corresponding to the O~star, are marked by green dotted lines.
At this phase, the 5200.41\,{\AA} transition of the O~component is blended with another feature, namely the 5204.55\,{\AA} pair associated with the WN component.
The velocity of the WN component can be estimated from the 5200.41\,{\AA} transition and is approximately $-169.2$\,km\,s$^{-1}$.
The lower panel of Fig.~\ref{fig:fg5204extr} shows spectra obtained at phases 0.14 and 0.17, which are very similar.
At these phases, the lines of the WN and the O~components appear to be fully blended.
In the middle panel, we reproduce the observed spectrum at phase 0.14 together with that obtained at phase 0.98 after shifting it redwards by 106.35\,km\,s$^{-1}$ (i.e.\ by $2\,K_{\text{WN}}$).
This comparison clearly demonstrates that the \ion{N}{iv} triplet follows the orbital motion of the WN component along the orbit, unlike the \ion{He}{i}\,$\lambda$4471 line.
However, the triplet remains displaced towards the blue with respect to its laboratory wavelengths.
The \mbox{O-star} lines are again separated from those of the WN component around phase 0.98, whereas at phases 0.14/0.17, the blended lines are noticeably deeper.
This provides further evidence that the \mbox{O-star} lines contribute to the observed spectrum at all phases and supports the idea that a spectral-disentangling analysis could prove beneficial.

Between phases 0.98 and 0.14/0.17, the \mbox{O-star} lines shift by approximately 4.35\,{\AA} towards shorter wavelengths.
Applying the same correction as in the previous section, we obtain a rough estimate of $K_{\text{O}}$ in the range 125--130\,km\,s$^{-1}$ from this shift.

\subsection{The directly observed motion of the O component}
\begin{figure}
	\centering
	\includegraphics[width=0.6\textwidth]{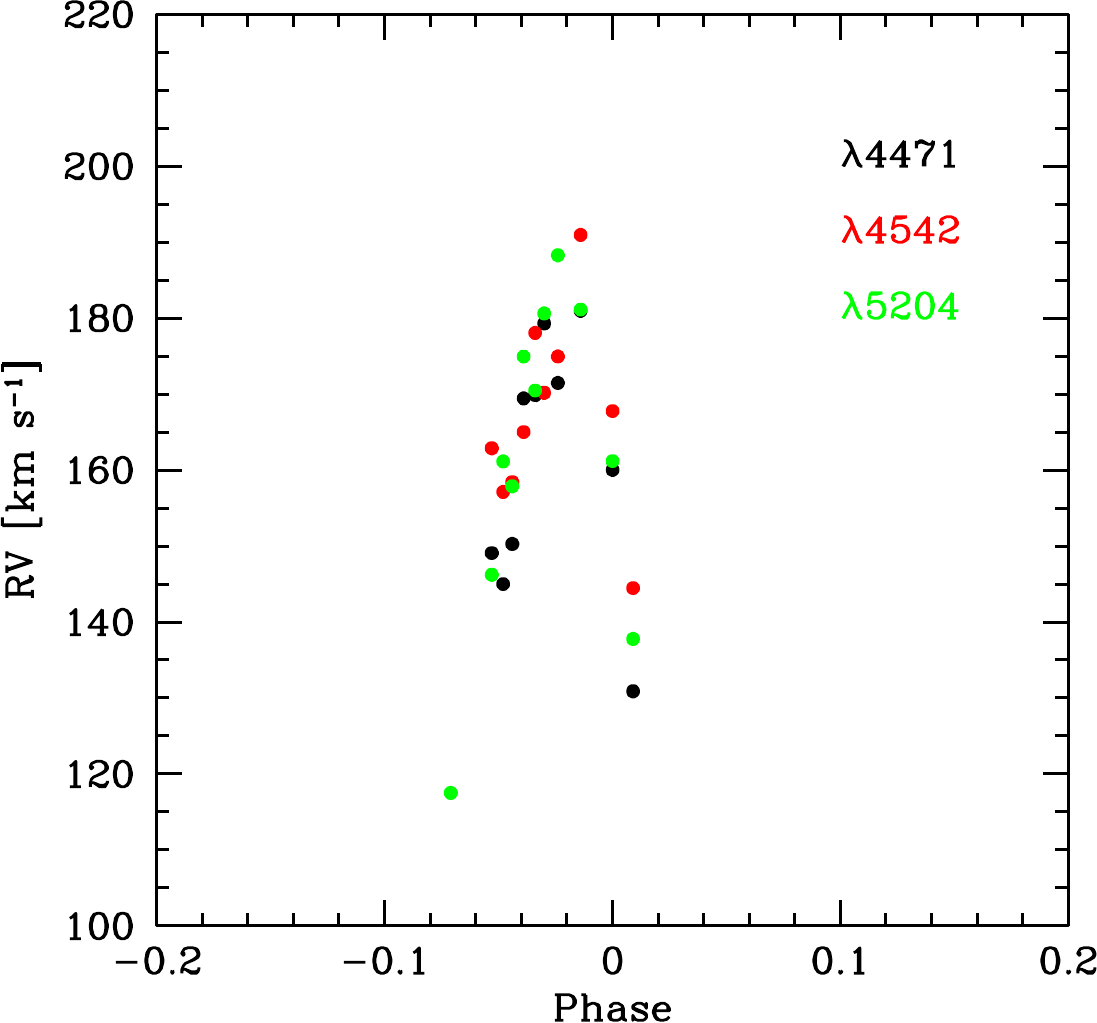}
	\vskip1ex
	
	\begin{minipage}{12cm}
		\caption{%
			Radial velocities associated with the \mbox{O-star} component as a function of orbital phase.
			Black dots correspond to measurements from the \ion{He}{i}\,$\lambda$4471 line, red dots to those from the \ion{He}{ii}\,$\lambda$4542 line, and green dots to the \ion{N}{iv}\,$\lambda$5204.55 blend.
			The data points highlight the RV motion of the O~star in the vicinity of the phase interval corresponding to the maximum velocity separation between the two components.%
			\label{fig:fgtoptopfin}}
	\end{minipage}
\end{figure}

\begin{figure}[t]
	\centering
	\includegraphics[width=0.6\textwidth]{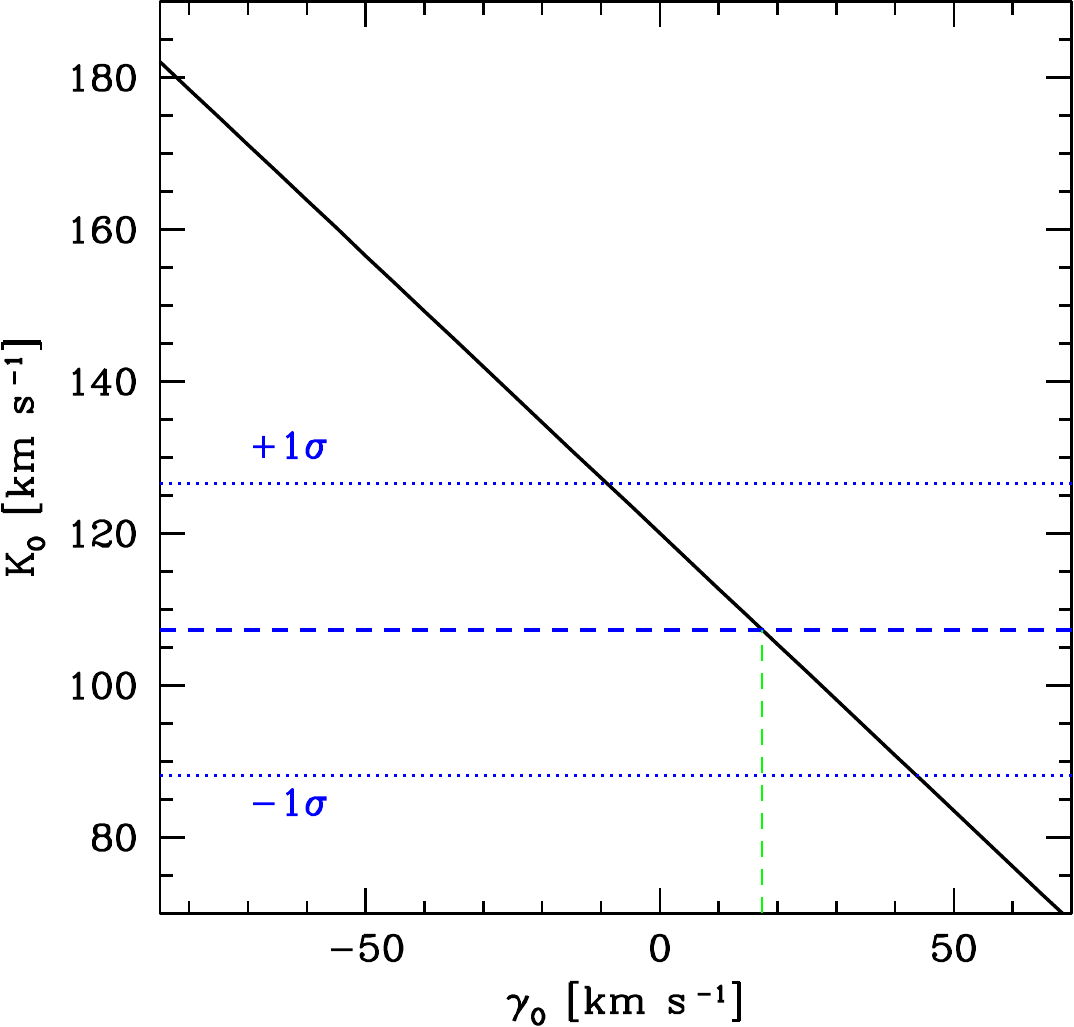}
	\vskip1ex
	
	\begin{minipage}{12cm}
		\caption{%
			Fitted values of the semi-amplitude $K_{\text{O}}$ as a function of the assumed systemic velocity $\gamma_{\text{O}}$.
			The formal uncertainty on the fitted parameter is comparable to the thickness of the black curve.
			The fit is based on the RV measurements shown in Fig.~\ref{fig:fgtoptopfin}.
			The dashed horizontal blue line indicates the value of $K_{\text{O}}$ derived in Sect.~\ref{sec:disentangling}, while the dotted blue lines delineate the corresponding $\pm 1\sigma$ interval.
			The green line shows the determination of $\gamma_{\text{O}}$ obtained by projecting the relation for the adopted value $K_{\text{O}} = 107.36$\,km\,s$^{-1}$.%
			\label{fig:figKgamma}}
	\end{minipage}
\end{figure}

We systematically measured the RVs of the O~component using three spectral features:
\ion{He}{ii}\,$\lambda$4542, \ion{He}{i}\,$\lambda$4471, and the \ion{N}{iv} triplet around 5200\,{\AA}.
These lines were selected following the inspection described in Sect.~\ref{sec:ospec}.
Using multi-Gaussian fitting similarly to the method used in Sect.~\ref{sec:wrspec}, we directly measured the RVs of the \ion{He}{ii}\,$\lambda$4542 and \ion{He}{i}\,$\lambda$4471 lines by assigning one Gaussian to the O~component.
For the \ion{N}{iv} triplet, we measured the position of the blend formed by the \ion{N}{iv}\,$\lambda$5204.29 and \ion{N}{iv}\,$\lambda$5205.15 transitions, adopting a laboratory wavelength of 5204.55\,{\AA} for the blend.
Naturally, such measurements are possible only for spectra acquired near phase 0.98.
The resulting RVs are listed in Appendix~\ref{sec:appD}.
These velocities can be regarded as having an absolute character, in the sense that they may represent the actual dynamical motion of the O~component, unlike the velocities derived from the WN-star lines.
%
Figure~\ref{fig:fgtoptopfin} shows the securely measured RVs associated with the O~star.
The three spectral features yield mutually consistent velocities.
In this phase interval, corresponding to the region of maximum radial velocity, the RVs are relatively well constrained.
However, this covers only a very limited range in phase, and no constraint is available on the opposite side of the systemic velocity.
Nevertheless, these RV measurements around the velocity extremum provide definitive evidence that the O~star is gravitationally bound to the WN component.
The overall shape of the RV curve derived from the WN star (see Sect.~\ref{ssec:orbsol_sb1model}) can therefore reasonably be assumed to apply to the O star as well.
We can thus fit an orbital solution to these measurements, taking $\gamma_{\text{O}}$ and $K_{\text{O}}$ as the only free parameters.
Unfortunately, because all of the available velocities sample only one side of the orbit, the two parameters are extremely, indeed nearly perfectly, correlated.
Consequently, the fit provides little hope of determining them independently and with useful precision.
Figure~\ref{fig:figKgamma} shows the results of a series of fits in which $\gamma_{\text{O}}$ was fixed at a range of assumed values.
The resulting semi-amplitude $K_{\text{O}}$, the only fitted parameter, is plotted as a function of the adopted $\gamma_{\text{O}}$, and the correlation is evident.
Therefore, apart from the measurement obtained from the spectrum acquired at phase 0.14 (see Sect.~\ref{ssec:ospec4471}), only a spectral-disentangling analysis is likely to provide a significantly improved determination of the orbital solution of the O~component.

\section{Spectral Disentangling and Individual Spectra}%
\label{sec:disentangling}
Up to this point, we have investigated the orbital motion through the analysis and fitting of measured RVs.
We have shown that, when the WN component reaches its maximum positive velocity, the spectral lines of the O~component become blended with those of the WN star.
Since the \mbox{O-star} lines are systematically weaker, their contribution is difficult to isolate in such circumstances.
This is precisely the type of situation in which a spectral-disentangling technique can provide a significant advantage.
The application of such a technique is the subject of the present section.

\subsection{Disentangling and derived semi-amplitudes}%
\label{ssec:disentanglingforK}
Thus, in order to refine the values of $K_{\text{WN}}$ and $K_{\text{O}}$, and to extract the individual spectral contributions of the two components, we applied a spectral-disentangling code to the spectra belonging to data set~I$_{\text{glo}}$.
The code used here is the same as that described by \citet{2022A&A...664A.159M} and is based on the Fourier approach \citep{1995A&AS..114..393H, 2004ASPC..318..107I}.

\begin{figure*}
	\centering
	\includegraphics[width=0.9\textwidth]{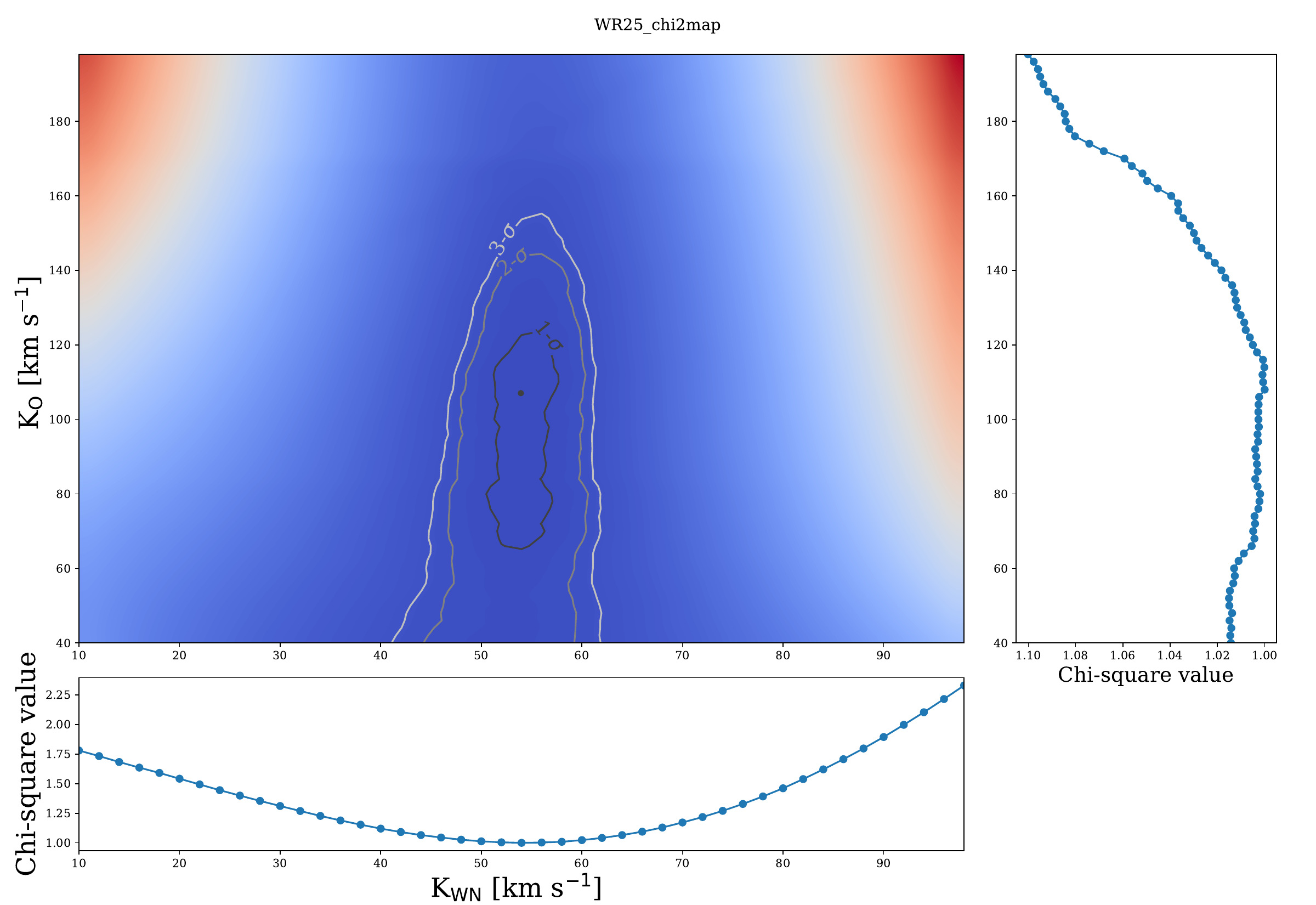}
	\vskip1ex
	
	\begin{minipage}{12cm}
		\caption{%
			Two-dimensional reduced $\chi^2$ map obtained from the Fourier disentangling as a function of the parameters $K_{\text{WN}}$ and $K_{\text{O}}$.
			The global minimum is marked by a black dot, and contour levels corresponding to $1\sigma$, $2\sigma$, and $3\sigma$ are shown.
			The \emph{bottom panel} shows the one-dimensional variation of $\chi^{2}$ around the minimum as a function of $K_{\text{O}}$ while the \emph{right-hand panel} shows the corresponding one-dimensional variation as a function of $K_{\text{WN}}$.%
			\label{fig:chicarremap}}
	\end{minipage}
\end{figure*}

As in the successful applications reported by \citet{2021A&A...651A.119F} and \citet{2024Sci...384..214F}, we reduced the number of free parameters by fixing the orbital period $P$, the eccentricity $e$, the argument of periastron $\omega_{\text{WN}}$, and the reference time $T_{0}$.
Only $K_{\text{WN}}$ and $K_{\text{O}}$ were allowed to vary. The apparent systemic velocities of the two components were excluded from the optimisation process.
This grid search was designed to refine the determination of the RV semi-amplitudes of both components.
The procedure was applied to four selected spectral features:
\ion{N}{iv}\,$\lambda$4058, representative of the WN component, and \ion{He}{ii}\,$\lambda$4200, \ion{He}{ii}\,$\lambda$4542 and the \ion{N}{iv} blend slightly above 5200\,{\AA}, all of which exhibit a non-negligible contribution from the O~component.
The resulting $\chi^{2}$ map is shown in Fig.~\ref{fig:chicarremap}.
%
We compared the results obtained with the Fourier spectral-disentangling technique with those derived using the shift-and-add method \citep{2022A&A...665A.148S}, another spectral disentangling technique.
The two approaches yield consistent results.
Combining them leads to $K_{\text{WN}} = 53.82\,(0.99)$\,km\,s$^{-1}$ and $K_{\text{O}} = 107.36\,(19.24)$\,km\,s$^{-1}$.
Figure~\ref{fig:chicarremap} shows that the determination of $K_{\text{WN}}$ is well behaved, with a marginal distribution that is close to symmetric.
In contrast, the determination of $K_{\text{O}}$ is more challenging, and the minimum along this axis is noticeably asymmetric.
However, the overall two-dimensional $\chi^{2}$ map indicates that the two parameters are not strongly correlated.
This suggests that the observed asymmetry arises primarily from the blending discussed above when the \mbox{O-star} component is shifted towards the blue, which constitutes a significant limitation of the analysis.
The uncertainty of 19.24\,km\,s$^{-1}$ on $K_{\text{O}}$ is therefore not strictly symmetric and could be more appropriately expressed as asymmetric confidence intervals, namely $\sigma^{+} = 11.5$\,km\,s$^{-1}$ and $\sigma^{-} = 27.4$\,km\,s$^{-1}$.

The derived value of $K_{\text{WN}}$ is in excellent agreement with that reported in Table~\ref{tabelement}, providing further support for the robustness of our results.
If the value $K_{\text{O}} = 107.36$\,km\,s$^{-1}$ is projected onto Fig.~\ref{fig:figKgamma}, it corresponds to a possible systemic velocity $\gamma_{\text{O}}$ of 17.4\,km\,s$^{-1}$.
At the position of WR\,25, the radial velocity projected along the line of sight due to Galactic rotation is expected to be approximately $-11$\,km\,s$^{-1}$.
These two values are not fully consistent and would imply an intrinsic radial motion of WR\,25 of a few tens of km\,s$^{-1}$.
The uncertainties on the determination of $\gamma_{\text{O}}$ are, however, huge, and no firm conclusion can be drawn from this apparent discrepancy.

\subsection{Individual spectra}\label{ssec:disentanglingspectra}
We therefore fixed the values $K_{\text{WN}}$ and $K_{\text{O}}$, and re-applied the spectral disentangling procedure to extract the individual spectra of both components over the useful wavelength range covered by the observations.

\begin{sidewaysfigure}
	\centering
	\includegraphics[width=\textwidth]{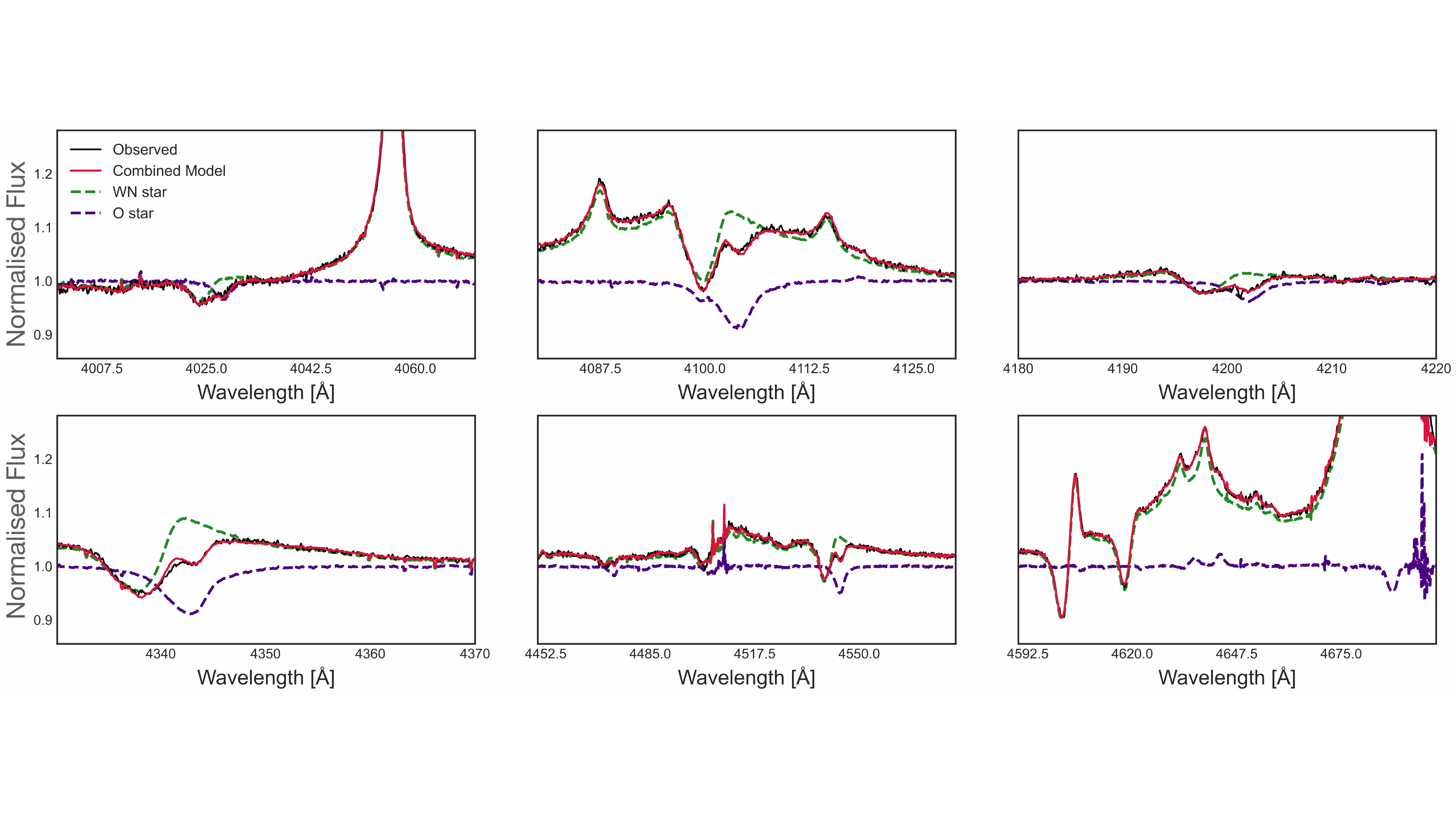}
	\vskip-10ex
	
	\begin{minipage}{16cm}
		\caption{%
			Combination (red) of the two disentangled spectra (green for the WN star and blue for the O~star) shifted according to their respective radial velocities and summed, compared with the observed spectrum (black) obtained at HJD\,2,454,084.750.%
			\label{fig:figdisentangling}}
	\end{minipage}
\end{sidewaysfigure}

Depending on the disentangling technique employed, the continua of the individual components may be lost during the process (see \citealp{2017A&A...607A..96M} for details).
It is therefore necessary to correct the disentangled spectra for this effect.
This correction differs from a standard normalisation: rather than dividing the spectra by the continuum, we force the pseudo-continuum to unity.
Since the sum of the continua of the O~star and the WN star must reproduce the continuum of the observed spectrum, the correction applied to the \mbox{O-star} continuum is directly correlated to that applied to the WN-star continuum.
We apply this procedure to both objects.
%
Figure~\ref{fig:figdisentangling} illustrates the quality of the reconstruction by comparing the sum of the two disentangled spectra shifted according to their respective radial velocities with an observed spectrum obtained at HJD\,2,454,084.750.
%
A~key limitation of the disentangling analysis of WR~25 is that the \mbox{O-star} contribution cannot be fully separated from the WN profile when its lines are shifted towards the blue.
As a result, the blue wings of the \mbox{O-star} lines cannot be completely reconstructed, producing artificial asymmetries in the disentangled profiles.
In addition, because the brightness ratio of the two stars cannot be constrained through eclipses in the light curve of WR~25, it cannot be determined from spectroscopy alone.
As a first approximation, we therefore arbitrarily adopted equal brightness contributions for the two components, i.e.\ a 0.50/0.50 ratio.
This assumption naturally affects the absolute equivalent widths of the lines, but not their ratios.
To recover the contribution of the spectral features to the global spectrum of WR\,25, the equivalent widths measured in the disentangled spectra must therefore be multiplied by a factor of 1/2.
%
Examples of the disentangled spectra are shown in Fig.~\ref{fig:figspecOdis} and in Appendix~\ref{sec:appE}.
\begin{figure}
	\centering
	\includegraphics[width=0.8\textwidth]{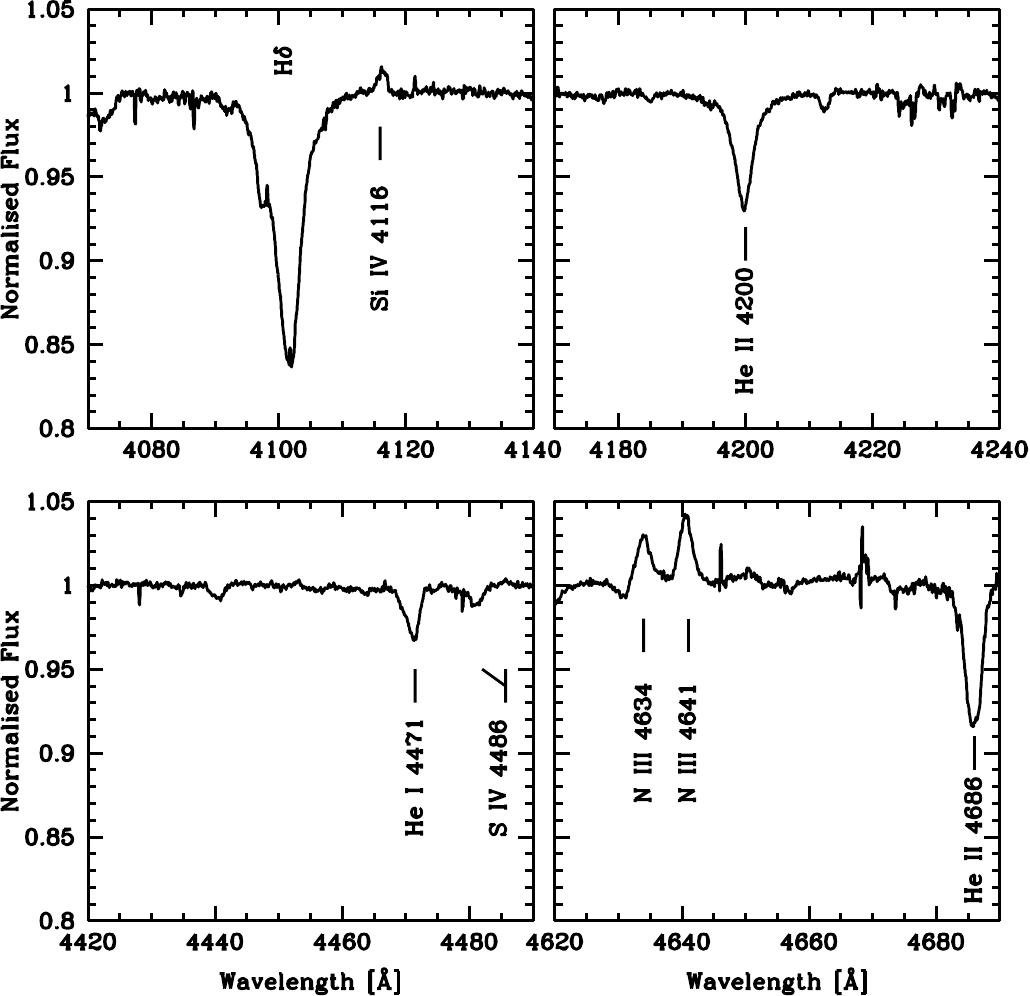}
	\vskip1ex
	
	\begin{minipage}{12cm}
		\caption{%
			Disentangled spectrum of the O~component in several spectral regions.
			These regions correspond to the \ion{He}{ii}\,$\lambda$4100+H$\delta$ line and the \ion{Si}{iv}\,$\lambda \lambda$4089-4116 doublet (\emph{upper left}; the first line is not visible), the \ion{He}{ii}\,$\lambda$4200 line (\emph{upper right}), the \ion{He}{i}\,$\lambda$4471 line (\emph{lower left}), and the triplet \ion{N}{iii}\,$\lambda \lambda$4634-4641 together with the \ion{He}{ii}\,$\lambda$4686 line (\emph{lower right}).%
			\label{fig:figspecOdis}}
	\end{minipage}
\end{figure}
The secondary component displays a typical O-type star spectrum.
The ratio  
\[
W = \log_{10}(\text{EW(\ion{He}{I})}/\text{EW(\ion{He}{ii})}) 
= -0.54~(0.12)
\]
derived from the equivalent widths of the \ion{He}{i}\,$\lambda$4471 and \ion{He}{ii}\,$\lambda$4542 lines corresponds to a spectral type O5, with a possible uncertainty of one subclass (O4 to O5.5; \citealp{1971ApJ...170..325C, 1973ApJ...179..181C, 1988A&AS...76..427M, 2018A&A...616A.135M}).
It is noteworthy that the \ion{Si}{iv}\,$\lambda$4116 line appears in emission, whilst the \ion{Si}{iv}\,$\lambda$4089 transition is absent (see Fig.~\ref{fig:figspecOdis}).
The presence of \ion{Si}{iv}\,$\lambda$4116 emission in the spectrum of the O~component explains the anomaly discussed in Sect.~\ref{sssec:wrspec_DS1_4089} and illustrated in Fig.~\ref{fig:fgcorr4116}.
Figure~\ref{fig:figspecOdis} also shows that the \ion{N}{iii}\,$\lambda \lambda$4634-41 lines are clearly in emission, supporting the addition of a suffix ``f'' in parentheses to the spectral classification.

In the vicinity of spectral type O5, the EW of \ion{He}{ii}\,$\lambda$4542 is relatively constant and largely independent of whether the star belongs to the main-sequence or giant luminosity class.
From the EWs published by \citet{1988A&AS...76..427M} for six stars (namely HD\,168112, HD\,93204, HD\,152233, HD\,157857, HD\,167633, and HD\,303308), we derive a typical EW of 705\,(60)\,m{\AA} for \ion{He}{ii}\,$\lambda$4542.
Because of the lack of a precise brightness ratio, we do not have access to absolute EWs.
However, the stable behaviour of the \ion{He}{ii}\,$\lambda$4542 line makes it a suitable reference line.
Around spectral type O5, the ratio of \ion{He}{ii}\,$\lambda$4542 to \ion{He}{ii}\,$\lambda$4686 is close to unity, even for stars classified as ``((f))'' (see the EWs of the six O~stars listed above; \citealp{1988A&AS...76..427M}).
In contrast, stars carrying the ``(f)'' qualifier generally exhibit a \ion{He}{ii}\,$\lambda$4686 line with a lower EW.
From the data of \citet{1988A&AS...76..427M}, we find that the ``(f)'' stars among the six objects are characterised by ratio below approximately 0.85.
In some cases, the absorption line may even be completely filled in by emission or present a weak emission (e.g., the O5III(f) star HD\,97253, Fig.~11 of \citealp{2014ApJS..211...10S}).
In WR\,25, the O5 component displays a clear \ion{He}{ii}\,$\lambda$4686 absorption line, but its EW amounts to only about 76{\%} of that of \ion{He}{ii}\,$\lambda$4542, and its depth is also noticeably smaller.
Such a ratio is more characteristic of ``(f)'' stars \citep[][]{1988A&AS...76..427M}.
Indeed, the relative strengths of the two lines closely resemble those observed in the O5III(f) star HD\,168112 \citep[see Fig.~5 of][]{2011ApJS..193...24S}.
Following the classification scheme of \citet{1990PASP..102..379W} and \citet{2016ApJS..224....4M}, while taking into account the similarity with HD\,168112, we assign the qualifier ``(f)'' to the companion.
Admittedly, the star lies very close to the boundary separating the ``(f)'' and ``((f))'' classes.
A~caveat is warranted.
Since the publication of \citet{1988A&AS...76..427M}, several of the reference stars have been recognised as binaries.
This is notably the case for HD\,168112 \citep{2023MNRAS.525.6084P, 2024A&A...687A.106B}, which is now classified as O4.5IV((f))\,+\,O5.5V(n)((f)).
These two components closely bracket the O5((f)) spectral type.
Consequently, the original ``(f)'' classification has effectively become ``((f))''.
In their Sect.~3.2, \citet{2023MNRAS.525.6084P} review the history of the revision, illustrating the borderline nature of such classifications.
A~more recent study by \citet{2018A&A...616A.135M} suggests to place the boundary between the ``((f))'' and ``(f)'' classes at $\mbox{EW(\ion{He}{ii}\,$\lambda$4686)} = 600$\,m{\AA}.
Using the representative EW of \ion{He}{ii}\,$\lambda$4542 quoted above, the threshold of \citet{2018A&A...616A.135M} corresponds to a ratio of 0.85.
The observed value of 0.76 therefore clearly favours our ``(f)'' qualifier for the O5 component of WR\,25.
The presence of \ion{Si}{iv}\,$\lambda$4116 in emission provides observational support for adding a ``+'' suffix, yielding the classification O5(f$^{+}$).
The ``f$^{+}$'' notation has been in use since \citet[][see also \citet{1990PASP..102..379W}, \citet{1996LIACo..33....1V}, and references therein]{1971ApJS...23..257W}.
Although it was originally associated with hot O-type supergiants, being an observational evidence, it has occasionally been applied to hot objects that are non supergiants \citep[see, e.g.,][]{2001ASPC..242..217W, 2006ApJ...638..409H, 2006A&A...456.1131M}.
\citet{2011ApJS..193...24S} regard this ``+'' suffix as non-essential and recommend that it be considered obsolete.
In conclusion, we assign the spectral type O5(f$^+$) to the companion, although the ``+'' suffix may be omitted if one follows the recommendation of \citet{2011ApJS..193...24S}.

The \ion{He}{i}\,$\lambda$4471 line in Fig.~\ref{fig:figspecOdis} exhibits a slight asymmetry, which is a consequence of the blending effect described above.
A~few other lines also present this artefact.

The spectrum of the WN star shows a clear emission of the \ion{N}{iv}\,$\lambda$4058 line, and a well-marked P-Cygni profile at the \ion{N}{v}\,$\lambda$4604 transition.
WR\,25 is reputed to be a weak-lined WN star.
It has been classified as WN6ha according to the criteria of \citet{1996MNRAS.281..163S}, but \citet{2011MNRAS.416.1311C} suggested that it should rather be classified O2.5If$^{*}$/WN6 on the basis of various criteria involving \ion{N}{} lines, as well as the presence of a marked absorption component in the \ion{He}{ii}\,$\lambda$4859+H$\beta$ P-Cygni profile.
The uncertainties surrounding the classification as either an Of/WN or a WN star were discussed in Sect.~3 of \citet{2011MNRAS.416.1311C}.
If we use the luminosities published by \citet{2023MNRAS.521..585C}, we find that the ratio L(4058)/L(4630) (where 4630 denotes the \ion{N}{v}--\ion{N}{iii} complex) is 0.52 for WR\,25, 0.48 for WR\,24 (WN6ha), and 0.63 for the classical WN6 archetype WR\,67 (classified as WN6o according to \citealp{1996MNRAS.281..163S}).
WR\,25 therefore lies approximately midway between two WN6 stars.
If we consider the height (normalised to the continuum) of the peak of \ion{N}{iv}\,$\lambda$4058 (see Figs.~\ref{fig:fglocglo} and \ref{fig:fgfitlg4058gm}), it clearly exceeds that of the 4630 complex by at least a factor of 2.
We therefore conclude that WR\,25 is not very different from WR\,24 in this respect.
In our high-resolution spectra, the height of the emission component of the \ion{N}{v}\,$\lambda$4604 line (see Fig.~\ref{fig:fgfitl4604}) is between about 0.2 and 0.85 times that of the \ion{N}{iii}\,$\lambda \lambda$4634-4641 triplet (see Fig.~\ref{fig:fgfitl4641}).
This is not sufficiently discriminating, and a comparison with Table~2 of \citet{2011MNRAS.416.1311C} does not rule out a pure WN6 spectral type.
Therefore, the determining clue for classifying WR\,25 as an Of/WN star, on which we must rely following the prescription of \citet{2011MNRAS.416.1311C}, is the presence of a marked absorption component in the \ion{He}{ii}\,$\lambda$4859+H$\beta$ profile, as can be seen in Fig.~1 of \citet{2011MNRAS.416.1311C}.
This feature is also visible in our spectrum of WR\,25, e.g., at phase 0.14 (see Fig.~\ref{fig:fgextrpluhbet}).
\citet{2011MNRAS.416.1311C} emphasised that this H$\beta$-based criterion is not entirely secure and requires high-resolution spectra to support such a classification.

\begin{figure}[t]
	\centering
	\includegraphics[width=0.6\textwidth]{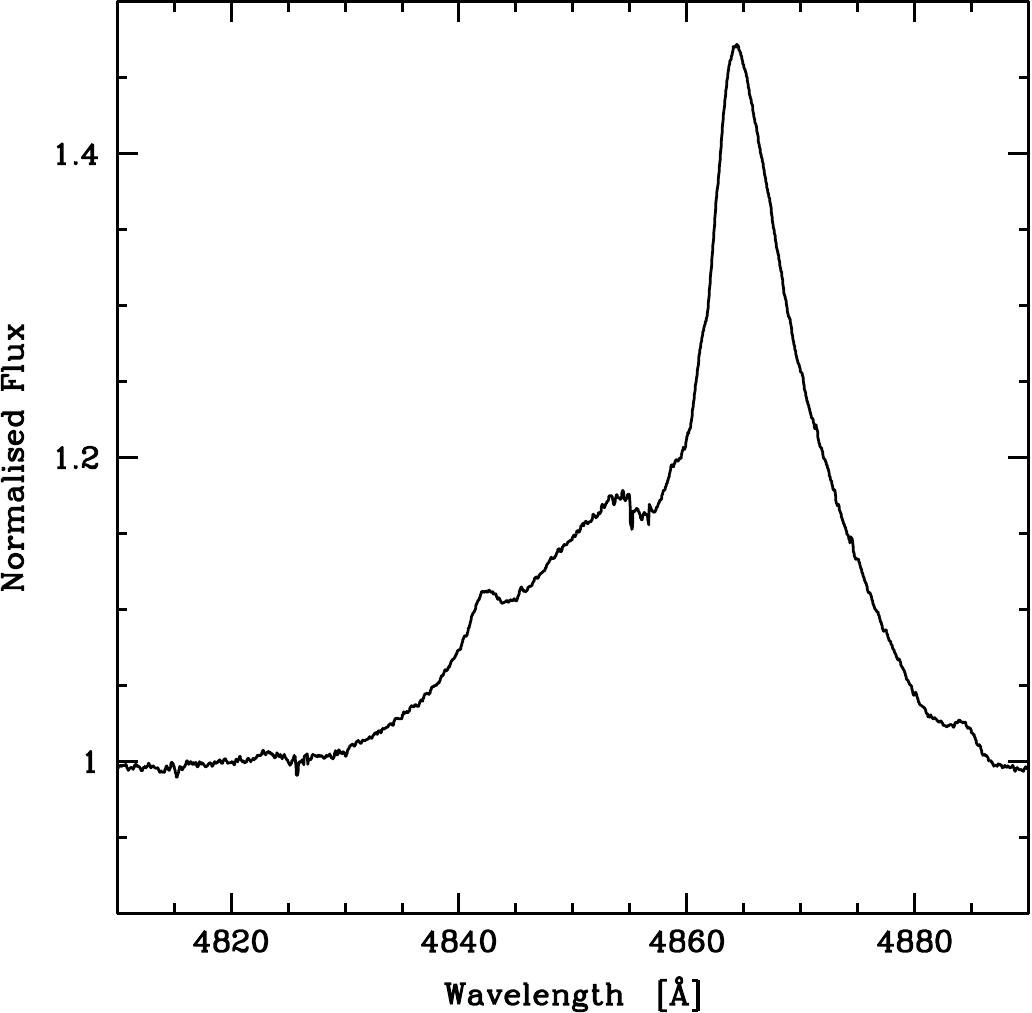}
	\vskip1ex
	
	\begin{minipage}{12cm}
		\caption{%
			Disentangled spectrum of the WN component in the region of the \ion{He}{ii}\,$\lambda$4859+H$\beta$ line.
			The absence of any strong absorption component in the P-Cygni profile is conspicuous.
			It is interesting to note that the small emission lines \ion{N}{iii}\,$\lambda$4842 and \ion{N}{iii}\,$\lambda \lambda$4882-4884 are correctly attributed to the WN component by the disentangling procedure.%
			\label{fig:fighbetadis}}
	\end{minipage}
\end{figure}

In Sect.~\ref{ssec:ospec_hbeta}, we suggested that this absorption component could, in the case of WR\,25, be due to the presence of the O~companion.
Indeed, the disentangled spectrum of the WN component shows no strong absorption in the \ion{He}{ii}\,$\lambda$4859+H$\beta$ line (see Fig.~\ref{fig:fighbetadis}).
Only a weak absorption may be present around 4855--4856\,{\AA}.
The absolute value of its equivalent width is smaller than that of the neighbouring \ion{N}{iii}\,$\lambda$4842 line.
Expressed in terms of the H$\beta$ index, this absorption component would correspond to a minimum value of 1.08 in the global H$\beta$ profile (see Fig.~\ref{fig:fghbetaabs}).
Therefore, at some orbital phases, the prominent absorption observed in the H$\beta$ profile in WR\,25 is absent.
The component intrinsic to the WN star is very weak and occurs in a region of the profile that is sensitive to wind fluctuations (see Sect.~\ref{ssec:ospec_hbeta}).
In fact, the main contribution to the absorption usually observed in the H$\beta$ line of WR\,25 is predominantly due to the absorption associated with the O~star and is not intrinsic to the WN component.
Consequently, we advocate classifying the WR component in WR\,25 as WN6ha, similarly to WR\,24.

The weak-line character of WR\,25 could perhaps be due to the dilution caused by the presence of the O~companion.
We selected the three strong emission lines \ion{N}{iv}\,$\lambda$4058, \ion{He}{ii}\,$\lambda$4686, and \ion{He}{ii}\,$\lambda$6560+H$\alpha$.
Based on the spectra from the dataset I$_{\text{glo}}$, we derive typical line  heights of 0.58, 0.75 and 0.78, respectively.
Performing the same analysis on a spectrum of WR\,24 yields values of 0.78, 1.39 and 1.60, respectively.
From these three lines, we estimate the dilution factors 0.74, 0.54, and 0.49, respectively.
We thus derive brightness ratios ranging roughly between 0.5/0.5 and 0.26/0.74.
This domain of ratios could be correct, but it is not sufficiently precise to allow further analysis.
It is also very sensitive to the possible presence of a third object.
Another caveat is that WR\,24 is not necessarily the best comparison object.
Additional observations will be necessary to investigate this issue further.
Assuming that the entire dilution is due to the O~component, we derive a magnitude difference between the two objects of up to 1.1\,mag.

\subsection{Dilution of the O5 star lines}
\label{ssec:disentanglingdilution}%
In the vicinity of spectral type O5, the EW of the \ion{He}{ii}\,$\lambda$4542 line is rather constant.
We estimated it above to be 705\,(60)\,m{\AA}.
If we restrict ourselves to HD\,157857 and HD\,303308 (which are not known to be binaries at the time of writing), we obtain a value of 725\,m{\AA}, well within the error bar.
The EW of this line in the disentangled spectrum of the O5 component is 360\,m{\AA}, i.e.\ 180\,m{\AA} as observed in the WR\,25 spectrum.
Given that the disentangled spectra have been corrected for the dilution factor (0.5), the measured EW is twice that measured in the observed spectra.
Therefore, the EW of \ion{He}{ii}\,$\lambda$4542 must be divided by a factor of 2.
The dilution factor, mainly due to the presence of the WN, is thus 0.26, which leads to an estimated brightness ratio of 0.26/0.74.

The same exercise can be performed with H$\gamma$.
Considering the EW of H$\gamma$ of approximately spectral type O5, we can deduce from the works of \citet{1973ApJ...179..161C}, \citet{1974MNRAS.166..203B}, and \citet{1993A&AS..101..599C} a typical EW of about 2000--2200\,m{\AA}.
The EW observed for the O~star component in the spectrum of WR\,25 is 457\,m{\AA}, leading to brightness ratios in the range 0.20/0.80 to 0.23/0.77, in good agreement with the above results derived from the \ion{He}{ii} line.

\section{Additional Considerations}\label{sec:discussion}

\subsection{Brightness ratio, absolute magnitude and luminosity}%
\label{ssec:brightnessratio}
We adopt an apparent magnitude $V = 8.1$\,mag.\ for WR\,25, as suggested by \citet{1968MNRAS.138..109S} (with the correction proposed by \citealp{1983ApJ...264..126M}), \citet{1969MNRAS.143..273F}, \citet{1993AJ....105..980M}, \citet{1982AJ.....87.1012F}, and \citet{2013AJ....145...44Z}.
The value of 8.8 mag.\ \citep{2002yCat.2237....0D} is probably erroneous and is not compatible with the \emph{Gaia} value $G = 7.808$\,mag.\ \citep{2021A&A...649A...1G, 2023A&A...674A...1G}.
The $B$ magnitude of WR\,25 is 8.55 \citep{2013AJ....145...44Z} and a small or negative $B - V$ implies a very small value of $G - V$ (see the transformation in Fig.~15 of \citealp{2016A&A...595A...7C}).
This underlines the incompatibility of the $V = 8.8$\,mag.\ value.
Alternatively, a $B - V$ of 0.45 (8.55--8.1) translates into $G - V =  \mbox{0.13--0.25}$, yielding $V \approx 8.0$, in better agreement.
WR\,25 is not expected to follow exactly the characteristics of the stars used to establish the relations of \citet{2016A&A...595A...7C}.
Adopting a distance of 2.19\,kpc for WR\,25 (see the discussion in Sect.~\ref{ssec:semimajoraxis}), we derive a distance modulus $\mbox{DM} = 11.7$ (or 11.9 considering the ALS~III distance of 2.37--2.40\,kpc).
We therefore obtain
\[
M_{V} +  A_{V} = -3.6 \quad \text{(or $-3.8$)}.
\]
The evaluation of the absolute magnitude requires knowledge of the extinction.
Unfortunately, this is hampered by the fact that the extinction inferred from the blue/UV range and that inferred from the red/NIR range are not mutually compatible.
A~large uncertainty remains, depending on whether one assumes a normal extinction law and thus the presence of a red/NIR excess, or an anomalous extinction law.
A~discussion of this issue is provided by \citet{1995A&A...293..427C}.
From the analyses of \citet{1995A&A...293..427C} and \citet{1993A&A...274..397H, 2019A&A...625A..57H}, we derive a range of $A_{V}$ values between 1.9 and 3.4.
Consequently, the absolute magnitude of WR\,25, $M_V$, lies between $-5.5$ and $-7.2$.
This is a rigorous range of possible values, and the correct value, which remains unknown, is not necessarily close to the midpoint of the interval.

The reddening correction might be smaller if we use the $H$- and $K$-band magnitudes ($H = 5.970$\,mag.\ and $K = 5.721$\,mag.).
The intrinsic $H-K$ colour is taken from \citet{2006A&A...457..637M} to be $(H-K)_{0} = -0.10$.
It is interesting to note that the value taken from \citet{2006A&A...457..637M} is independent of the actual spectral type of the O~star.
However, the use of \mbox{O-star} synthetic colours is a possible approximation.
Synthetic colours for WN stars might be more appropriate, but no intrinsic $(H-K)_{0}$ for a WN star is available in the literature.
We can instead refer to the analysis of WR\,25 by \citet{2006A&A...457.1015H}, updated in \citet{2019A&A...625A..57H}, where the PoWR model characterised by $T_{*} = 50100$\,K and a transformed radius $R_{\text{t}} = 1.5$ is associated with this star.
This model is characterised by $(H-K)_{0} = -0.025$.
One should also be cautious with this value because the fit of \citet{2019A&A...625A..57H} does not take into account the dilution effects due to the presence of the O5 companion.
In addition, the temperature we attribute to the WN component in Sect.~\ref{sec:CMFGEN} is lower than that advocated by these authors.
We then derive $E(H-K) = 0.35$ or 0.27 (adopting either the value from \citet{2006A&A...457..637M} or that from \citet{2019A&A...625A..57H}, respectively).
From \citet{2005ApJ...619..931I}, $A_{H} \approx 2.8 \times E(H-K)$, where 2.8 is the mean of the values 2.6 and 3.0 provided by \citet{2009ApJ...696.1407N} and \citet{1999PASP..111...63F}, respectively.
Using this relation and $\mbox{DM} = 11.7$, the absolute $H$ band magnitude of the whole WR\,25 system is $M_{H} = -6.71~(0.10)$ or $-6.50~(0.10)$, respectively.

For an assumed brightness ratio and luminosity class of the O~star, we can compute the expected total $M_V$ of WR\,25.
The results are presented in Table~\ref{taabs}.
\begin{table}
	\centering
	\begin{minipage}{96mm}
		\caption{%
			Absolute magnitude $M_{V}$ of WR\,25 derived as a function of an unknown brightness ratio and from the three luminosity classes attributable to the O~companion.
			Possible values are reported in bold (see text).%
			\label{taabs}}
	\end{minipage}
	\vskip1ex
	
	\begin{tabular}{cccc}
		\toprule
		\textbf{Brightness}
		& \textbf{O5V}
		& \textbf{O5III}
		& \textbf{O5I} \\
		\textbf{ratio}
		& \textbf{\emph{M\textsubscript{V}} = --{\hskip0.25mm}5.21}
		& \textbf{\emph{M\textsubscript{V}} = --{\hskip0.25mm}5.85}
		& \textbf{\emph{M\textsubscript{V}} = --{\hskip0.25mm}6.33}\\
		\midrule
		0.5/0.5 & $-5.96$ & $-6.59$ & $-7.0$8 \\
		0.4/0.6 & $\mathbf{-6.20}$ & $\mathbf{-6.83}$ & $-7.32$ \\
		0.3/0.7 & $\mathbf{-6.52}$ & $\mathbf{-7.15}$ & $-7.64$ \\
		0.2/0.8 & $\mathbf{-6.96}$ & $-7.59$ & $-8.08$ \\
		0.1/0.9 & $-7.71$ & $-8.34$ & $-8.83$ \\
		\bottomrule
	\end{tabular}
\end{table}
Clearly, there is no room for an O5I companion in WR\,25;
only a main-sequence star or a giant is compatible with the observations.
This conclusion reinforces the one drawn from the absence of the line \ion{N}{iii}\,$\lambda$4514 (see Appendix~\ref{sec:appE}).
The brightness ratio 0.1/0.9 must also be rejected.
The lines of the O~star are rather faint, making a brightness ratio of 0.5/0.5 very unlikely.
Likewise, a brightness ratio of 0.2/0.8 is incompatible with an O5III companion.
The acceptable values are highlighted in bold in Table~\ref{taabs}.
The constraint is not particularly strong, and it would be useful to have an independent way of estimating the brightness ratio.
In particular, the possible intruder has not been taken into account in these computations.

Assuming a brightness ratio of 0.26/0.74 as derived in Sect.~\ref{ssec:disentanglingdilution}, a negligible effect of any intruder, and adopting the absolute $V$-band magnitude of WR\,25, we obtain absolute magnitudes of $M_{V}^{\text{prim}} = -5.93 \pm 0.76$ for the WN primary and $M_{V}^{\text{seco}} = -4.80 \pm 0.76$ for the O5 secondary.
These values are obtained by converting the range of possible absolute magnitudes discussed above.
In this context, the quoted uncertainty of 0.76 does not represent a standard deviation associated with a statistical uncertainty, but rather the extent of the allowed range.
Bolometric corrections were computed using the prescription of \citet{2006A&A...457..637M}, based on the effective temperatures determined with the atmosphere code described in Sect.~\ref{sec:CMFGEN}.
We obtain $BC_{V}^{\text{prim}} = -3.99 \pm 0.13$ and $BC_{V}^{\text{seco}} = -3.74 \pm 0.14$ for the primary and secondary, respectively.
Applying these corrections yields luminosities of $\log(L/L_{\odot}) = 5.87 \pm 0.35$ and $\log(L/L_{\odot}) = 5.31 \pm 0.35$ for the primary and secondary, respectively.
Again, these values represent rigorous ranges of possible luminosities.
Referring to \citet{2019A&A...625A..57H} and their Table~1, we note that the typical $V$-band bolometric correction for WN6h stars is $BC_{V}^{\text{prim}} = -4.2$; this is also the value attributed to WR\,25.
Using this alternative bolometric correction, we derive $\log(L/L_{\odot}) = 5.95 \pm 0.35$ for the primary, fully consistent with the previous estimate.
Any more precise guess remains necessarily only prospective.
Alternatively, to further constrain the discussion, one may adopt the anomalous extinction scenario advocated by \citet{1995A&A...293..427C}.
We therefore consider $A_{V} = 3.4$ and $M_{V} = -7.00$ as the most likely values for the extinction and the absolute magnitude of the system.
In this case, we get $M_{V}^{\text{prim}} = -6.67$ for the primary, which yields $\log(L/L_{\odot}) = 6.16$ for $BC_{V}^{\text{prim}} = -3.99$, and $\log(L/L_{\odot}) = 6.24$ for $BC_{V}^{\text{prim}} = -4.2$.
For the secondary, we derive $M_V^{\text{seco}} = -5.54$, corresponding to $\log(L/L_{\odot}) = 5.61$.

Using instead the absolute magnitude of WR\,25 derived from the $H$ band, and assuming the same brightness ratio, we obtain $M_{H}^{\text{prim}} = -6.38 \pm 0.10$ and $M_{H}^{\text{seco}} = -5.25 \pm 0.10$ for the WN primary and the O5 secondary, respectively.
The corresponding bolometric corrections are $BC_{H}^{\text{prim}} = -4.80 \pm 0.14$ and $BC_{H}^{\text{seco}} = -4.54 \pm 0.15$.
This leads to luminosities of $\log(L/L_{\odot}) =  6.37 \pm 0.07$ for the primary and $\log(L/L_{\odot}) = 5.81 \pm 0.07$ for the secondary.
Alternatively, if we adopt the bolometric correction associated with WR\,25 ($BC_{H}^{\text{prim}} = -4.95$), from de model of \citet{2019A&A...625A..57H}, we obtain $\log(L/L_{\odot}) = 6.43$ for the primary.

Although the uncertainty on the bolometric correction is not particularly severe, the luminosities derived from the $V$ and $H$ bands differ, depending on the adopted extinction.
It is therefore essential to better constrain the extinction towards WR\,25 and to obtain a more accurate determination of the brightness ratio between the two components.

An additional issue is the unknown effect of the possible intruder.
We illustrate this in the framework of the adopted value $M_{V} = -7.00$.
Since the brightness ratio 0.26/0.74 is estimated from the dilution of the lines of the O5 star, the possible intruder has no effect on the luminosity derived for this component, and we can therefore conclude that $\log(L_{\text{O5}}/L_{\odot}) = 5.61$.
In contrast, the 74{\%} of the brightness not attributed to the O5 star includes the combined contributions of the WN component and the intruder.
To further fix ideas, let us assume that the intruder contributes half the visible brightness of the O5.
The corresponding brightness ratios would then be $\mbox{intruder/O5/WN} = 0.13/0.26/0.61$.
In this case, we obtain $M^{\text{WN}}_{V} = -6.46$ and $\log(L_{\text{WN}}/L_{\odot}) = 6.08$ or 6.16, depending on the adopted bolometric correction.
The dependence of the results on the properties of the intruder is therefore not negligible.

\subsection{Mass ratio and masses}
\label{ssec:massratio}
Following the determination of $K_{\text{WN}}$ and $K_{\text{O}}$ in Sect.~\ref{sec:disentangling}, we can compute the mass ratio of the system from the ratio of the semi-amplitudes:
\[
q = \frac{M_{\text{WN}}}{M_{\text{O}}}
  = \frac{K_{\text{O}}}{K_{\text{WN}}}
  = 2.02~(0.36).
\]
This value is not particularly precise, but the uncertainty on $q$ is dominated by the uncertainty on $K_{\text{O}}$.
From this mass ratio and the mass function from the SB1 solution (see Table~\ref{tabelement}), we obtain
\begin{equation}
M_{\text{O}} \sin^3 i = 
f(M) \times (1+q)^2 = 15.39 \, (2.35) \, M_{\odot}, 
\label{eqmtwo}
\end{equation}
where $i$ is the unknown inclination of the system.
At this stage, we must make a strong assumption.
If the companion is a canonical O~star obeying the calibration of \citet{2005A&A...436.1049M}, its mass can be taken as either $M_{\text{O5V}} = 37.3\,M_{\odot}$ or $M_{\text{O5III}} = 41.5\,M_{\odot}$.
Using Eq.~(\ref{eqmtwo}), we derive inclinations of $i = 48\fdg 1$ or $i = 45\fdg 9$, respectively.
Such values are not unexpected given the apparent absence of eclipses \citep[see][]{gossetmorelia}.
Assuming an O5V (or O5III) classification, we derive approximate masses for the WN component of $M_{\text{WN}} = 75.3\,M_{\odot}$ (or 83.8\,$M_{\odot}$), respectively.
These estimates remain highly uncertain.
The masses provided by the calibration of \citet{2005A&A...436.1049M} carry uncertainties of 35--50{\%}, suggesting that it would be misleading to quote formal error bars.
In any case, we consider it unlikely that the O~companion has a mass below $27\,M_{\odot}$ which implies that the WN component has at least a mass of $55\,M_{\odot}$.
An O5-star mass of $27\,M_{\odot}$ corresponds to an upper limit on the inclination of 56°.
This minimum mass of $27\,M_{\odot}$ for the O5 star is motivated by the masses determined for the components of HD\,168112, as well as the accurately measured  masses of the secondaries in HD\,150136 and 9~Sgr \citep{2023MNRAS.525.6084P, 2018A&A...616A..75M, 2021A&A...651A.119F}.
These results confirm that WR\,25 is a very massive WNLh star belonging to the same family as WR\,22 \citep{1996A&A...306..771R, 1999A&A...347..127S}, WR\,29 \citep{2009A&A...506.1269G}, both components of WR\,20a \citep{2005A&A...432..985R}, and WR21a \citep{2016MNRAS.455.1275T, 2022MNRAS.516.1149B}.

\subsection{Semi-major axes}\label{ssec:semimajoraxis}
Based on the previously determined $K$, $P$, and $e$, we can compute the projected semi-major axes.
We obtain
\begin{align*}
a_{_{\text{WN}}} \, \sin i & = 0.816 \, (0.023) \, \text{AU} \\
a_{_{\text{O}}} \, \sin i & = 1.647 \, (0.298) \, \text{AU} \\
a_{_{\text{tot}}} \, \sin i & = (a_{_{\text{WN}}} + 
a_{_{\text{O}}}) \, \sin i  = 2.463 \, (0.299) \, \text{AU}.
\end{align*}
The deprojected semi-major axis of the system can be estimated, assuming the inclinations derived above, to lie between 3.31 and 3.43\,AU, and certainly above 2.97\,AU.
At the \emph{Gaia} \citep{2016A&A...595A...1G, 2021A&A...649A...1G, 2023A&A...674A...1G} DR3 distance of 2.19\,(0.10)\,kpc (RUWE is 1.16; \citealp{2021AJ....161..147B}), the total projected semi-major axis corresponds to an angular separation on the sky of about 1.1--1.5\,mas.
On the basis of the same parallax but adopting a more recently derived zero-point and a prior better suited to hot stars, \citet[ALS III,][]{2025MNRAS.543...63P} derived alternative distances of $2.37^{+0.30}_{-0.24}$\,kpc (median distance) or $2.40^{+0.30}_{-0.24}$\,kpc (mean distance), which, if adopted, would lead to a 9{\%} decrease of the separation on the sky, well within the uncertainties.
These estimated separations are at the limit of what can be resolved using optical/IR long-baseline interferometry and could perhaps be observed, provided the close field is not too crowded.

\section{CMFGEN Analysis}\label{sec:CMFGEN}
Now that the individual contributions for both components have been separated, we can perform an atmosphere-modelling analysis.
We applied a brightness ratio of 0.26/0.74 to correct the individual spectra for the dilution effect.
This value was inspired by the work described in Sect.~\ref{ssec:disentanglingdilution}, but remains uncertain.
We will refer to the resulting spectra as the \emph{modified disentangled spectra}.
To estimate the effective temperature ($T_{\text{eff}}$) and the surface gravity ($\log\,g$) of the O~star, we used the CMFGEN atmosphere code \citep{1998ApJ...496..407H}.
CMFGEN computes non-LTE (Local Thermodynamic Equilibrium) models of massive-star atmospheres, including winds and line blanketing.
The hydrodynamical structure consists of a hydrostatic photosphere and a wind.
The trans-sonic velocity profile is assumed to follow a $\beta$-velocity law.
The spectroscopic level populations are calculated through the rate equations.
A~super-level approach is used to reduce the size of the problem (and thus the computing time).
To ensure good modelling of our observed spectra, we included the following elements in the calculations:
H, He, C, N, O, Ne, Mg, Al, Si, S, Ar, Ca, Fe, and Ni.
Solar abundances from \citet{2010Ap&SS.328..179G} were adopted for all elements unless otherwise stated.
Once the atmospheric structure was obtained, a formal solution of the radiative transfer equation was computed.

\subsection{The O star}%
\label{ssec:Ostar}

\subsubsection{Projected rotational velocities and macroturbulence}%
\label{sssec:vsini}
The projected rotational velocity ($v\,\sin i$) and the macroturbulent velocity ($v_{\text{mac}}$) were determined independently using the \texttt{iacob-broad} package of \citet{2014A&A...562A.135S}.
Given the spectral type of the O~star, \ion{He}{i} lines are present in the spectrum but remain weak.
Despite the asymmetry introduced by the spectral disentangling procedure, we selected the \ion{He}{i}~$\lambda$4471 line, together with the \ion{He}{ii}~$\lambda$4200 line, to estimate the $v\,\sin i$ and $v_{\text{mac}}$.
The \ion{He}{ii} lines are not ideally suited for determining these parameters, but given the faintness of the other \ion{He}{i} lines and the difficulty of identifying strong metal lines in the spectra, they provide a useful alternative.
The values of $v\,\sin i$ and $v_{\text{mac}}$ were then kept fixed in the atmosphere analysis.

\subsubsection{Effective temperature and surface gravity}%
\label{sssec:teff}
The effective temperature, $T_{\text{eff}}$, and surface gravity, $\log g$, were estimated from a grid of synthetic CMFGEN spectra.
The grid was constructed using the same stellar parameters as those adopted in the evolutionary tracks of \citet{2011A&A...530A.115B}.
We adopted a constant step in $\log g$ of $\Delta \log g = 0.1$.
Appropriate values of $T_{\text{eff}}$ and $\log(L/L_{\odot})$ were used to build the models, together with mass-loss rate prescriptions from \citet{2000A&A...362..295V, 2001A&A...369..574V} for solar metallicity, scaled by a factor of three.
We used the empirical terminal wind-speed prescription provided by \citet{2024A&A...688A.105H}.
The exponent $\beta$ of the velocity law was set to 1.0 \citep{2004A&A...415..349R}, and the clumping filling factor was adopted as $f_{\text{cl}} = 0.1$.

The final grid covers non-uniformly the ranges $10,000\,\mbox{K} < T_{\text{eff}} < 50,000\,\mbox{K}$ and $2.6 < \log(L/L_{\odot}) < 6.0$.
For each model, a depth-dependent microturbulent velocity was used to compute the emergent spectrum.
The microturbulence varies from 2, 5, 7, 10, 12, 15, 17, or 20\,km\,s$^{-1}$ at the photosphere to 10{\%} of the terminal wind velocity at the top of the atmosphere.
The final grid at Galactic metallicity comprises 4138 synthetic spectra.

The effective temperature was constrained through the ionisation balance between \ion{He}{i} and \ion{He}{ii} lines.
We used the
\ion{He}{i/ii}\,$\lambda$4026,
\ion{He}{i}\,$\lambda$4143,
\ion{He}{i}\,$\lambda$4388,
\ion{He}{i}\,$\lambda$4471,
\ion{He}{ii}\,$\lambda$4200,
\ion{He}{ii}\,$\lambda$4542, and
\ion{He}{ii}\,$\lambda$5412 lines.
The surface gravity was determined from the wings of the H$\delta$ and H$\gamma$ lines, as these are less sensitive to wind effects than H$\beta$ and H$\alpha$.
The quality of the fit was quantified by means of a $\chi^2$ analysis applied to the spectral lines mentioned above.
The $\chi^{2}$ value was computed for each model in the grid and linearly interpolated between grid points with steps of $\Delta \, T_{\text{eff}} = 100$\,K and $\Delta \log\,g = 0.01$~(cgs).
The 1$\sigma$ uncertainties on the two parameters were estimated from $\Delta \chi^{2} = 2.30$ (for two degrees of freedom; \citealp{Press2007}).

\begin{sidewaysfigure}
	\centering
	\includegraphics[width=0.78\textwidth]{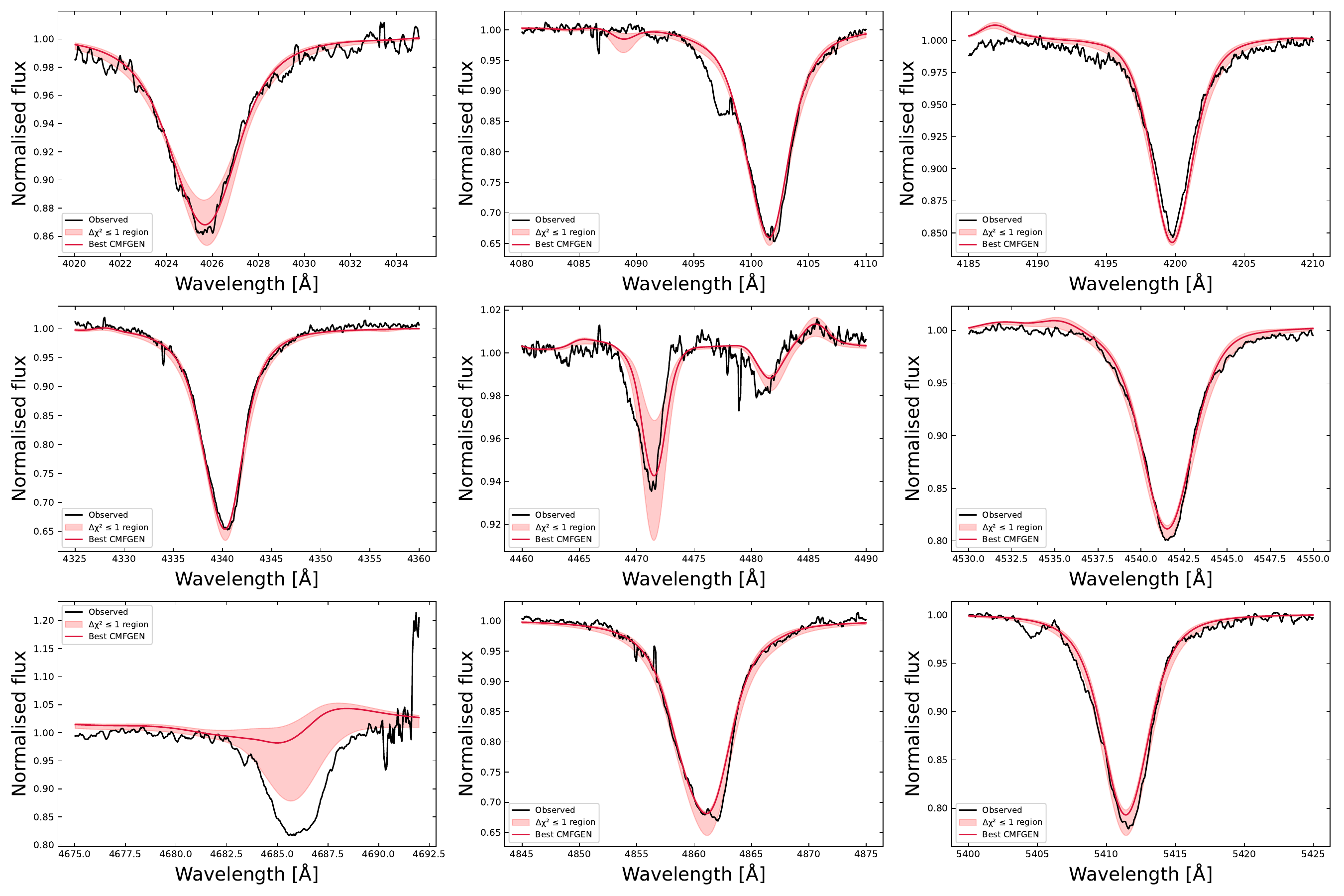}
	\vskip0.5ex
	
	\begin{minipage}{16cm}
		\caption{%
			Comparison between the modified disentangled spectrum of the O5 star (black) and the best-fit CMFGEN model (red) in several spectral regions.
			The C, N, and O surface abundances were kept fixed at their solar values.
			The light red region indicates the range spanned by the synthetic spectra corresponding to models lying within the 1$\sigma$ confidence interval of the best-fit solution.%
			\label{fig:CMFGEN_Ostar}}
	\end{minipage}
\end{sidewaysfigure}

For the O5 star, we find an effective temperature of $T_{\text{eff}} = 41400~(2000)$\,K (defined at the Rosseland optical depth $\tau_{\text{Ross}} = 2/3$), a surface gravity of $\log\,g = 3.85~(0.05)$\,cgs, a $\log(L/L_{\odot}) = 5.72~(0.08)$, and a microturbulent velocity of $15\,(5)$\,km~s$^{-1}$.
The projected rotational velocity is $v\,\sin i = 85\,(17)$\,km\,s$^{-1}$ and the macroturbulent velocity is $v_{\text{mac}} = 95\,(33)$\,km\,s$^{-1}$.
Figure~\ref{fig:CMFGEN_Ostar} shows the comparison between the best-fit CMFGEN model (red), the models within the 1$\sigma$ confidence interval (light red), and the modified disentangled spectrum of the O5 star (black).
 
The abundances of the metals were kept fixed at their solar values, as detailed abundance analysis is beyond the scope of this paper.
A~discrepancy is also present for the \ion{He}{ii}\,$\lambda$4686 line which is known to be sensitive to wind properties.

\subsection{The WN star}%
\label{ssec:WNstar}
CMFGEN was also used to model the modified disentangled spectrum of the WN star.
However, we carried out a preliminary by-eye fit, as the large number of physical parameters makes a fully automated approach difficult.
A~more robust atmosphere analysis of the individual components will be performed in Deshmukh et al.\ (in prep).
The fitting process is further complicated by discrepancies between the synthetic and observed continua.

\subsubsection{Atmosphere modelling}%
\label{sssec:WNatm}
The effective temperature was estimated from the ionisation balance of the nitrogen lines (\ion{N}{iii}/\ion{N}{iv}/\ion{N}{v}).
For most WR models, the value of $\log\,g$ has no significant impact on the synthetic spectrum, which is formed predominantly in the wind.
We also varied the mass-loss rate, while  keeping the surface abundances of nitrogen and carbon fixed at their solar values.

\begin{figure*}
	\centering
	\includegraphics[width=\textwidth]{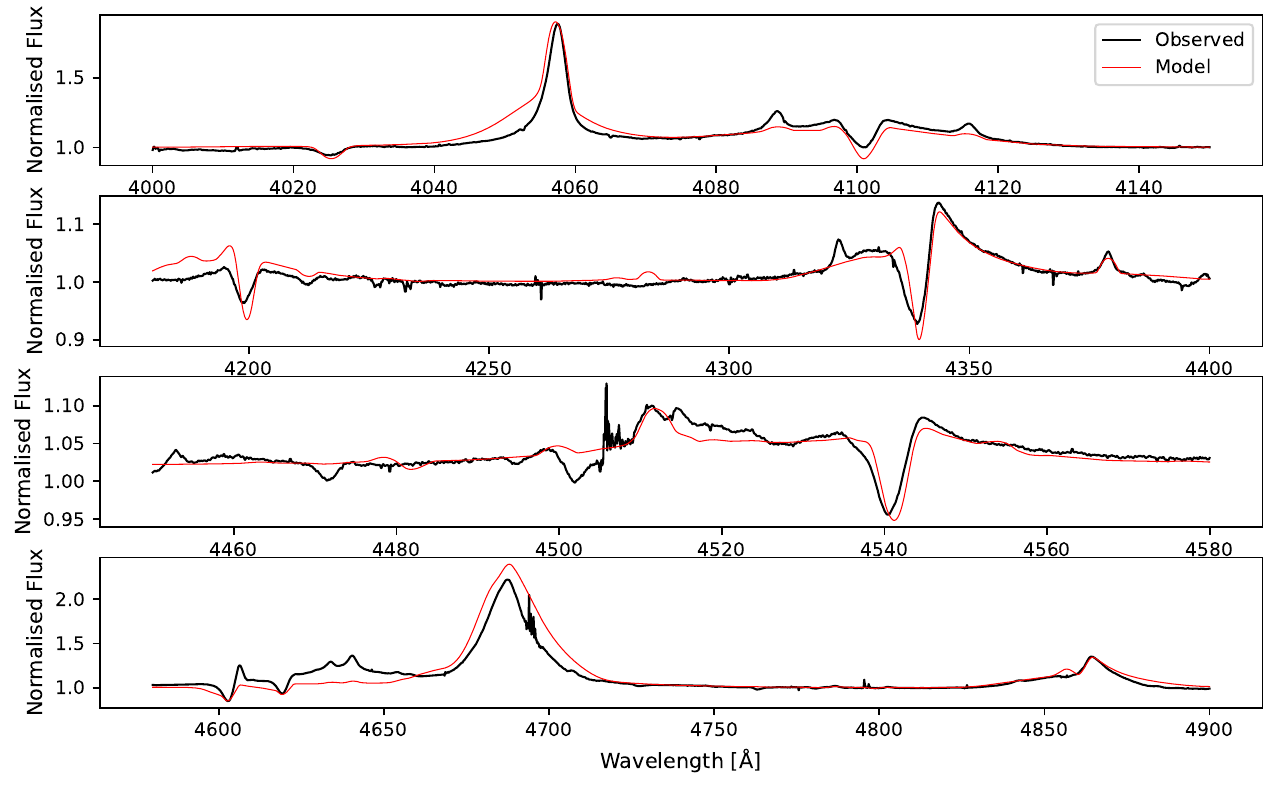}
	\vskip1ex
	
	\begin{minipage}{12cm}
		\caption{%
			Comparison between the modified disentangled spectrum of the WN star (black) and the best-fit CMFGEN model (red) in selected spectral regions.%
			\label{fig:CMFGEN_WNstar}}
	\end{minipage}
\end{figure*}

We find that the WN6ha star has an effective temperature $T_{*}$(WN6ha) of approximatively $45\,000\,(2000)$\,K, a mass-loss rate of about
$\log(\dot{M}/(\mbox{$M_{\odot}$\,yr$^{-1}$})) = -5.1\,(0.5)$,
and a terminal wind speed $v_{\infty} = 2500\,(100)$\,\mbox{km\,s$^{-1}$}.
Finally, we fixed the luminosity to a typical value of $\log(L/L_{\odot}) = 6.0$, as some uncertainties remain regarding this parameter (see Sect.~\ref{ssec:brightnessratio} for further discussion).
We also fixed the clumping factor to 0.1 and the exponent of the $\beta$-velocity law to 1.0.
In this sense, the fit should be regarded as preliminary.
It is shown in Fig.~\ref{fig:CMFGEN_WNstar}.
%
Alternatively, we may estimate the consequences of adopting a higher luminosity of $\log(L/L_{\odot}) = 6.3$.
We find that an increase of 0.3\,dex in luminosity has almost no effect on the synthetic spectrum, provided that the mass-loss rate is increased by 0.2--0.3\,dex.
No change of the temperature is required.
The luminosity and mass-loss rate therefore have a correlated impact, at least in the relevant region of parameter space.
This further underlines the need for accurate brightness ratios.

The disentangled spectrum of the WN6ha star shows the presence of the \ion{He}{i}\,$\lambda$4471 line (see Figs.~\ref{fig:CMFGEN_WNstar} and \ref{fig:figdisWtotal}).
Given the temperature estimated from the CMFGEN fit, no \ion{He}{i}\,$\lambda$4471 line should be present.
This strengthens the conclusion that the spectrum of WR\,25 is composed not of two but of three objects, including an additional OB component cooler than the O5 secondary star.
According to our analysis, this object must either be gravitationally bound to the system or lie along its line of sight.

\subsubsection{The potential intruder}%
\label{sssec:intruder}
We attempt to characterise the nature of the potential third object.
To this end, we measured the equivalent width (EW) of the \ion{He}{i}\,$\lambda$4471 line in the observed spectrum taken at orbital phase 0.98 (see Sect.~\ref{ssec:ospec4471}).
The \ion{He}{i}\,$\lambda$4471 lines of the O5 star and of the intruder are expected to be diluted by the WN star in the same way; therefore, the ratio of their EWs is preserved.
Assuming that the blue-shifted absorption profile originates from a third object and that the red-shifted one arises from the O5 star, we measure EWs of 0.054\,{\AA} and 0.023\,{\AA} respectively, yielding an absolute ratio (O5 over intruder) of 0.41.

\begin{figure*}
	\centering
	\includegraphics[width=1\textwidth]{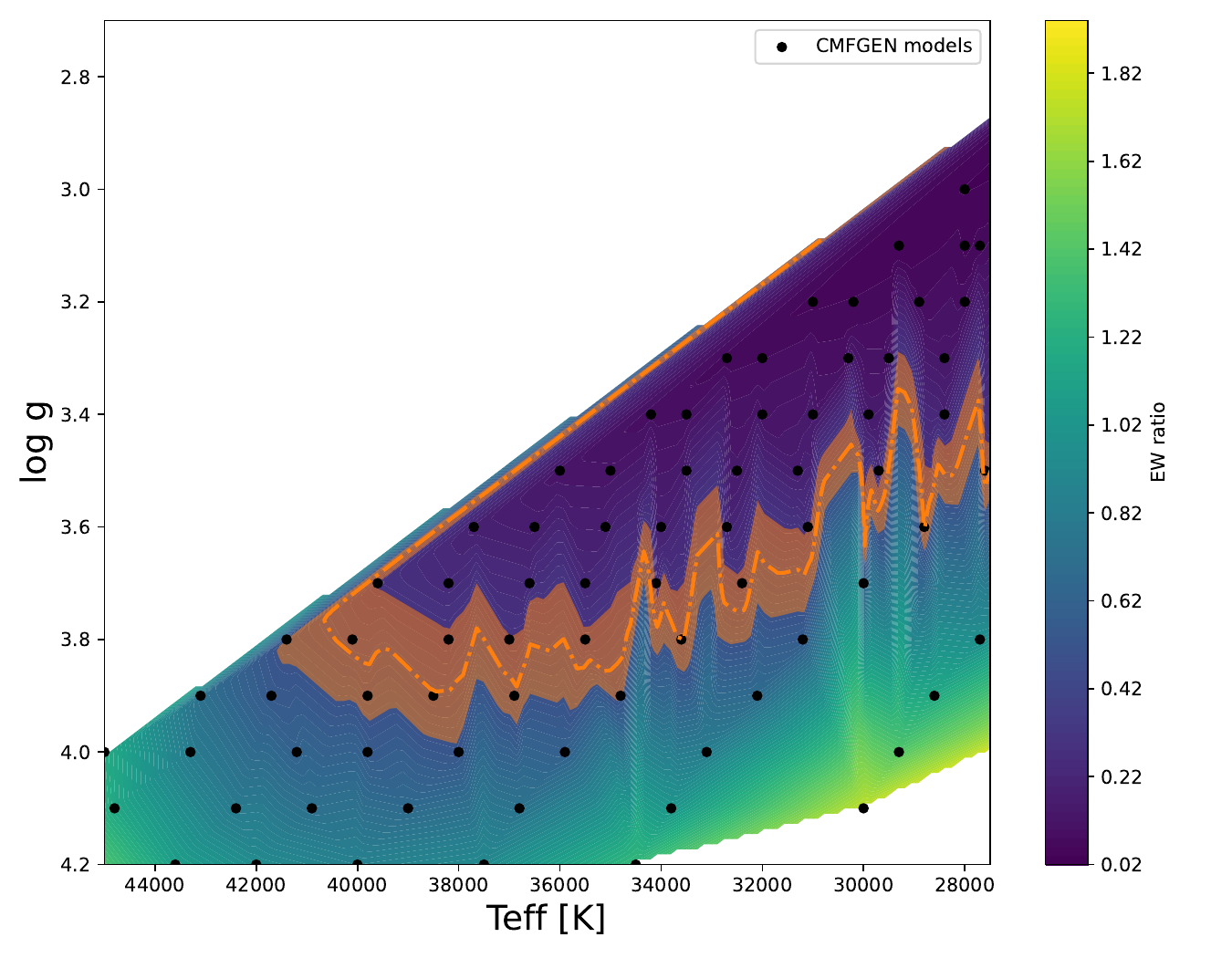}
	
	\begin{minipage}{12cm}
		\caption{%
			Constraint on the effective temperature of the third object based on the \ion{He}{i}\,$\lambda$4471 equivalent-width ratio.
			The orange region shows CMFGEN models whose predicted EW ratios between the O5 component and the intruder reproduce the observed ratio of 0.41 within uncertainties.%
			\label{fig:thirdstar_teff}}
	\end{minipage}
\end{figure*}

This ratio is then multiplied by the dilution ratio between the O5 and the third object.
This dilution is not known, and depends on the nature of the third object.
We tentatively estimated it by simulating the fluxes in the $V$ band of two black bodies with temperatures corresponding to those of the O5 star and the third object.
For each model in our grid (tentatively representing the third star), the dilution ratio was estimated and applied to the EW ratio of the two \ion{He}{i}\,$\lambda$4471 lines.
We then compared this result with the absolute EW ratios computed between the best-fit CMFGEN model for the O5 star and all synthetic spectra in the CMFGEN grid.
These ratios must agree within the uncertainties.
A~conservative uncertainty of 0.1 is adopted to account for potential differences between the estimated dilution ratio and the true dilution ratio of the two stars.
All models for which the absolute ratios are consistent with 0.41 are selected.
In Fig.~\ref{fig:thirdstar_teff}, the orange region corresponds to the set of models satisfying this criterion (the orange diagonal line is an artefact, and must not be considered physical).

Assuming that the intruder is coeval with the primary and secondary components (i.e., that it is a main-sequence star or a giant), its effective temperature is estimated to lie between 34\,000 and 41\,000\,K, corresponding to a spectral type between O8 and O6.
We implicitly assumed that the intruder is located at the same distance as the binary system, which remains to be demonstrated.
Here again, an independent determination of the brightness ratios is a crucial next step.

\subsection{Comments}\label{ssec:comments}
The CMFGEN analysis was performed on the modified disentangled spectra, assuming a brightness ratio 0.26/0.74.
This value is certainly not fully secure, and in this sense, the results should be regarded as preliminary.
However, we consider the results of the analysis of the O5 spectrum to be reasonably robust.
The derived temperature is in good agreement with the inferred spectral type, which is determined independently of the adopted brightness ratio.

The analysis case of the WN component is more problematic, as the disentangled spectrum appears to exhibit \ion{He}{i} lines that are not compatible with the temperature expected for a WN6ha star.
This suggests the presence of a third object, for which we have neither much information nor any secure knowledge of its impact on the brightness ratio.
We just demonstrated that this object is cooler than the two components of the main binary system.
Owing to the lack of information on the contributions of each of the three objects to the total brightness, a more detailed analysis, such as a three-component disentangling, is beyond the scope of the present paper.

\section{Conclusion}\label{sec:conclusion}
We acquired new high-resolution optical spectroscopic data of the binary star WR\,25.
These new observations, sometimes combined with older ones, were carefully analysed and improved the phase coverage of the 208\,d period exhibited by the RV variations of the WN star, first discovered by \citet{2006A&A...460..777G}.
The data allowed us to derive new SB1 ephemerides that are significantly more accurate than the preliminary orbital solution proposed by \citet{2006A&A...460..777G}.
The resulting orbital parameters are given in the fourth column of Table~\ref{tabelement}.
We derive a semi-amplitude of $K_{\text{WN}} = 53.18\,(0.82)$\,km\,s$^{-1}$.
These results are based on a few selected lines that are expected to provide an adequate representation of the orbital motion of the WN component.

A~fainter contribution to the WR\,25 spectrum is present and can be attributed to an O~star, as first pointed out by \citet{2008RMxAC..33...91G}.
In the present work, we identified this additional component in several spectroscopic lines and provided definitive proof that it originates from an O~star gravitationally bound to the WN component.
Unfortunately, the \mbox{O-star} contribution is not easily detectable at all orbital phases and is often blended with the absorption lines of the WN component.
To measure and constrain the RV semi-amplitude of the O~star, we applied spectral disentangling techniques that provided $K_{\text{O}} = 107.36\,(19.24)$\,km\,s$^{-1}$ and thus a mass ratio of 2.02\,(0.36).
These values remain relatively uncertain.
Taking the mass ratio at face value and assuming that the O~star is a canonical star following the calibration of \citet{2005A&A...436.1049M}, we derived a mass for the WN component in the range 75--84\,$M_{\odot}$, and certainly above 55\,$M_{\odot}$.
This confirms that WR\,25 is, as suspected, a very massive star similar to the class of WNLh objects.
After completion of the present work, a broadly comparable SB2 solution was also obtained by applying a disentangling technique to the available FEROS spectra.
This result will be presented in a forthcoming paper (in preparation) describing the SB2 orbital solutions from the OWN Survey \citep[see, e.g.,][for details of the project]{2017IAUS..329...89B}.

From the disentangled spectrum of the WN component, we confirm the WN6ha classification identification.
We demonstrated that the absorption component of the H$\beta$ P-Cygni profile observed in WR\,25 during most of the time is not intrinsic to the WN star, but instead originates from the \mbox{O-star} companion.
We therefore argue against the O2.5If$^{*}$/WN6ha classification proposed by \citet{2011MNRAS.416.1311C} for WR\,25.
This conclusion relies primarily on our use of high-resolution, high-S/N spectra, 

WR\,25 is often regarded as a weak-line WN star with the highest known hydrogen content among the Galactic WNLh stars and exhibiting absorption in the upper Balmer lines.
We confirmed the presence of numerous absorption lines, many of which are blended with those of the \mbox{O-star} companion.
We also argued that the relative faintness of the emission lines in the normalised spectra is primarily due to dilution by the O~companion.
Our high-S/N data allowed us to detect, and subsequently confirm through spectral disentangling, the presence of several weak \ion{N}{iii} lines that are also observed in WR\,24 and WR\,78 (see Sect.~\ref{ssec:wrspec_DS1}).
The spectrum of the WN component appears to exhibit a \ion{He}{i}\,$\lambda$4471 line, although this feature could in fact originate from an intruder (Sects.~\ref{ssec:ospec4471} and \ref{sssec:intruder}).
Using a CMFGEN analysis of the WN spectrum, we derived physical parameters for the WN6ha object (see Table~\ref{tab:cmfgen} for the temperature).
The difficulty in reproducing the \ion{He}{i} lines in the disentangled spectrum of the WN component suggests that an additional, probably late \mbox{O~type}, star contributes to the composite spectrum of WR\,25.
\begin{table}
	\centering
	\begin{minipage}{84mm}
		\caption{%
			Physical parameters of the WN6ha and O5 stars and their $1\sigma$ errors.%
			\label{tab:cmfgen}}
	\end{minipage}
	\vskip1ex
	
	\begin{tabular}{ccc}
		\toprule
		\textbf{Parameter}              & \textbf{WN6ha} & \textbf{O5}  \\
		\midrule
		$T_{*}$ [K]                     & 45000 (2000)   & ---          \\
		$T_{\text{eff}}$ [K]            & ---            & 41400 (2000) \\
		$\log\,g$ [cgs]                 & ---            & 3.85 (0.05)  \\
		$v\,\sin i$ [km\,s$^{-1}$]      & ---            & 85\,(17)     \\
		$v_{\text{mac}}$ [km\,s$^{-1}$] & ---            & 95\,(33)     \\
		\bottomrule
	\end{tabular}
\end{table}

The gravitationally bound companion was found to be of spectral type O5(f$^{+}$).
Its luminosity class is either main-sequence (V) or giant (III), but certainly not a supergiant.
A~detailed CMFGEN analysis yielded the physical parameters listed in Table~\ref{tab:cmfgen}.

The absolute magnitude of WR\,25, and consequently its luminosity, remain difficult to determine.
In the \emph{Gaia} era, this uncertainty is dominated by the poorly constrained interstellar extinction.
This stems from the well-established inconsistency between the extinction inferred from the blue/UV domain and derived from the red/NIR domain \citep[see, e.g., Sect.~4.2 of][]{1995A&A...293..427C}.

An improved understanding of WR\,25 will certainly require a substantial observational effort.
We identified two main avenues for future progress.
First, we found only one secure spectral line of the O~star that appears suitable for deriving a precise orbital motion, namely \ion{O}{iii}\,$\lambda$5592.
This line is present in the \mbox{O-star} spectrum and lies in a spectral region where the WN component exhibits almost no features.
However, this line is faint and has a depth of only about 1{\%} of the continuum.
Its exploitation will therefore require spectra with even higher S/N ratios than those currently available, while maintaining the same spectral resolution.
The search for additional lines of this kind, possibly even in other wavelength domains, could also prove beneficial.
Acquiring extremely high-S/N data over the 208\,d orbital period of the system represents a genuine observational challenge.
Second, a precise determination of the brightness ratio is still lacking.
Such a measurement is essential to improve the luminosity estimates of both components, and to better quantify line-dilution effects.
This will also allow us to assess the intrinsic luminosities of the WN emission lines, beyond spectrophotometric analyses.
The projected separation of both objects on the sky is probably slightly larger than 1\,mas, making long-baseline interferometry the only technique currently capable of directly addressing this issue.
Although this lies at the limit of present-day capabilities, such observations could yield the orbital inclination of the system (provided that the close environment is not excessively crowded) and thus the absolute masses of the components.
A~first exploratory observation will be reported by Deshmukh et al.\ (in preparation).

\begin{acknowledgments}
The authors are thankful to the two anonymous referees who contributed to the improvement of the paper.
This work is based on data collected at the European Southern Observatory (La Silla, Chile) and at the Las Campanas Observatory (Chile), including one observation at the Magellan II (Clay) \mbox{6.5-m} Telescope.
The raw data were processed using the \textsc{midas} and \textsc{iraf} reduction packages.
This study used \textsc{iraf} versions 2.13, 2.14, 2.15 and 2.16, which were developed at the National Optical Astronomy Observatories (NOAO, now integrated into the National Optical-Infrared Astronomy Research Laboratory, NOIRLab).
These facilities, as well as NSF NOIRLab, are operated by the Association of Universities for Research in Astronomy, Inc.\ (AURA) under a cooperative agreement with the U.S.\ National Science Foundation (NSF).
The first author would like to acknowledge discussions with Pascal Quinet about atomic transitions and related databases.
He is also greatly indebted to Alain Detal and Francine Mélen for various forms of support related to computing.
The Belgian Fund for Scientific Research---F.R.S.-FNRS has provided financial support that was essential for this project.
\end{acknowledgments}

\begin{furtherinformation}

\begin{authorcontributions}
The spectroscopic observational campaign was organised by EG on the European Southern Observatory side and by RG, NM and RB on the Las Campanas Observatory side.
HS performed the basic FEROS reduction, as well as the global continuum normalisation.
RG performed the basic reduction of the Las Campanas data.
LM contributed to the spectral disentangling and performed the detailed CMFGEN analysis.
EG, RG and LM are responsible for the detailed analysis.
EG wrote the original draft of the paper and directed the review and editing of the final version, with input from RG, LM, NM and HS.
\end{authorcontributions}

\begin{conflictsofinterest}
The authors declare that there is no conflict of interest.
\end{conflictsofinterest}

\begin{orcids}
	\orcid{0000-0002-5536-5993}{Eric}{Gosset}
	\orcid{0000-0002-5227-9627}{Roberto}{Gamen}
	\orcid{0000-0003-0688-7987}{Laurent}{Mahy}
	\orcid{0000-0003-2535-3091}{Nidia}{Morrell}
	\orcid{0000-0001-6656-4130}{Hugues}{Sana}
	\orcid{0000-0003-1086-1579}{Rodolfo~H.}{Barbá}
\end{orcids}

\end{furtherinformation}

\providecommand{\noopsort}[1]{}

\clearpage

\counterwithin{figure}{section}
\counterwithin{table}{section}
\begin{appendix}

\section{Journal of the Observations}\label{sec:appA}
This appendix presents the journal of the observations directly associated with the present work.
The four tables correspond to the four data sets (I, II, III, and IV).
For each observation, we provide the date, the Universal Time (UT) at the start of the exposure (except for Table~\ref{tabA3}), the Heliocentric Julian Day at mid-exposure, the exposure time and the orbital phase computed using the adopted period $P$ and phase-zero time $T_{0}$ from the fourth column of Table~\ref{tabelement}.
\begin{table*}[h!]
	\centering
	\begin{minipage}{127mm}
		\centering
		\caption{%
			Journal of the observations (data set~I) including exposure time.%
			\label{tabA1}}
	\end{minipage}
	\vskip1ex
	
	\begin{tabular}{cccr@{~~~~~}cc}
		\toprule
		\textbf{Date}
		  & \textbf{UT}
		    & \textbf{HJD(2400000+)}
		      & \multicolumn{1}{c}{\hskip-1ex\textbf{Exp.~}}
		        & \textbf{Phase}
		          & \textbf{Global} \\
		\textbf{yyyy-mm-dd}
		  & \textbf{hh:mm:ss}
		    & \textbf{(mid-exposure)}
		      & \multicolumn{1}{c}{\hskip-1ex\textbf{[s]}}
		        & & \textbf{(Y/N)} \\ 
		\midrule
		2006-03-02 & 02:33:41 & 53796.6176 & 1500 & 0.582 & Y \\
		2006-03-04 & 02:30:54 & 53798.6158 & 1500 & 0.592 & Y \\
		2006-03-06 & 03:49:50 & 53800.6706 & 1500 & 0.602 & Y \\
		2006-04-08 & 00:23:53 & 53833.5275 & 1380 & 0.760 & Y \\
		2006-04-10 & 02:23:22 & 53835.6105 & 1380 & 0.770 & Y \\
		2006-04-12 & 01:33:33 & 53837.5759 & 1380 & 0.779 & Y \\
		2006-05-04 & 23:40:24 & 53860.4978 & 1500 & 0.890 & Y \\
		2006-05-08 & 23:18:56 & 53864.4829 & 1500 & 0.909 & Y \\
		2006-05-13 & 02:49:37 & 53868.6284 & 1380 & 0.929 & Y \\
		2006-06-10 & 23:00:52 & 53897.4685 & 1380 & 0.068 & Y \\
		2006-06-12 & 23:34:10 & 53899.4916 & 1380 & 0.078 & Y \\
		2006-06-15 & 01:52:07 & 53901.5873 & 1380 & 0.088 & Y \\
		2006-06-26 & 01:11:24 & 53912.5575 & 1200 & 0.141 & N \\
		2006-07-01 & 23:45:53 & 53918.4988 & 1380 & 0.169 & Y \\
		2006-12-03 & 07:18:50 & 54072.8108 & 1380 & 0.912 & N \\
		2006-12-15 & 05:50:43 & 54084.7501 & 1380 & 0.970 & Y \\
		2006-12-23 & 07:40:45 & 54092.8269 & 1380 & 0.009 & Y \\
		2006-12-25 & 07:19:46 & 54094.8124 & 1380 & 0.018 & Y \\
		2007-05-27 & 02:15:50 & 54247.6017 &  900 & 0.754 & Y \\
		2009-03-20 & 00:13:15 & 54910.5224 & 1800 & 0.947 & N \\
		2009-03-21 & 00:16:08 & 54911.5244 & 1800 & 0.952 & N \\
		2009-03-22 & 00:04:23 & 54912.5162 & 1800 & 0.956 & N \\
		2009-03-22 & 23:43:46 & 54913.5019 & 1800 & 0.961 & Y \\
		2009-03-23 & 23:57:43 & 54914.5116 & 1800 & 0.966 & N \\
		2009-03-26 & 00:39:15 & 54916.5405 & 1800 & 0.976 & Y \\
		2009-03-28 & 01:46:49 & 54918.5875 & 1800 & 0.986 & Y \\
		2009-03-31 & 01:56:07 & 54921.5940 & 1800 & 0.000 & Y \\
		\bottomrule\addlinespace[\belowrulesep]
		\multicolumn{6}{p{123mm}}{\footnotesize
			{\bf{Notes:}} A `Y' in the last column indicates that the global version of the spectrum exists.}
	\end{tabular}
\end{table*} 
%
%
%
\begin{table}
\centering
\caption{%
Journal of the observations (data set~II).%
\label{tabA2}}
\vskip1ex

\begin{tabular}{ccccc}
\toprule
\textbf{Date}
  & \textbf{UT}
    & \textbf{HJD(2400000+)}
      & \textbf{Exp.}
        & \textbf{Phase} \\
 \textbf{yyyy-mm-dd}
  & \textbf{hh:mm:ss}
    & \textbf{(mid-exposure)}
      & \textbf{[s]}
        &   \\ 
\midrule
 2007-02-13 & 07:00:30 & 54144.8064 & 2220 & 0.259 \\
\bottomrule
\end{tabular}
\end{table} 
%
%
%
\begin{table}
	\centering
	\caption{%
		Journal of the observations (data set~III).%
		\label{tabA3}}
	\vskip1ex
	
	\begin{tabular}{cccrc}
		\toprule
		\textbf{Date}
		  & \textbf{UT}
		    & \textbf{HJD(2400000+)}
		      & \multicolumn{1}{c}{\textbf{Exp.}}
		        & \textbf{Phase} \\
		\textbf{yyyy-mm-dd}
		  & \textbf{hh:mm:ss}
		    & \textbf{(mid-exposure)}
	          & \multicolumn{1}{c}{\textbf{[s]}}
		        & \\ 
		\midrule
		2001-05-08 & 03:15:25 & 52037.6384 &  600 & 0.111 \\
		2001-05-09 & 01:59:50 & 52038.5859 &  600 & 0.115 \\
		2001-05-10 & 02:46:04 & 52039.6180 &  600 & 0.120 \\
		2001-05-11 & 02:52:19 & 52040.6223 &  900 & 0.125 \\
		2002-03-02 & 03:35:12 & 52335.6516 &  600 & 0.546 \\
		2002-03-04 & 04:23:33 & 52337.6853 &  800 & 0.556 \\
		2002-03-05 & 03:12:22 & 52338.6359 &  800 & 0.560 \\
		2002-03-06 & 03:16:48 & 52339.6391 &  900 & 0.565 \\
		2002-04-19 & 00:05:50 & 52383.5071 & 1800 & 0.777 \\
		2003-05-23 & 00:09:11 & 52782.5087 & 1800 & 0.698 \\
		2003-05-23 & 23:53:07 & 52783.4975 & 1800 & 0.703 \\
		2003-05-24 & 23:46:28 & 52784.4929 & 1800 & 0.708 \\
		2004-05-06 & 00:10:43 & 53131.5101 & 1500 & 0.379 \\
		2004-05-08 & 02:27:23 & 53133.6051 & 1500 & 0.389 \\
		2004-05-09 & 23:35:06 & 53135.4854 & 1200 & 0.398 \\
		\bottomrule\addlinespace[\belowrulesep]
		\multicolumn{5}{p{110mm}}{\footnotesize
			{\textbf{Notes:}} The UT values in this table correspond to the mid-exposure.}
	\end{tabular}
\end{table} 
%
\clearpage
%
\begin{table}
	\centering
	\begin{minipage}{160mm}
		\centering
		\caption{%
			Journal of the observations (data set~IV).%
			\label{tabA4}}
	\end{minipage}
	\vskip1ex
	
	\begin{tabular}{ccccc}
		\toprule
		\textbf{Date}
		  & \textbf{UT}
		    & \textbf{HJD(2400000+)}
		      & \textbf{Exp.}
		        & \textbf{Phase} \\
		\textbf{yyyy-mm-dd}
		  & \textbf{hh:mm:ss}
		    & \textbf{(mid-exposure)}
		      & \textbf{[s]}
		        & \\ 
		\midrule
		2006-05-18 & 01:07:34 & 53873.5529 & 600 & 0.953 \\
		2006-05-19 & 01:12:50 & 53874.5645 & 900 & 0.958 \\
		2006-05-20 & 00:49:23 & 53875.5402 & 600 & 0.962 \\
		2006-05-21 & 01:27:05 & 53876.5675 & 800 & 0.967 \\
		2006-05-21 & 01:45:45 & 53876.5781 & 400 & 0.967 \\
		2006-05-22 & 00:37:32 & 53877.5319 & 600 & 0.972 \\
		2006-05-22 & 02:36:22 & 53877.6157 & 800 & 0.972 \\
		2006-05-23 & 00:21:19 & 53878.5206 & 600 & 0.977 \\
		2006-05-23 & 00:49:24 & 53878.5401 & 600 & 0.977 \\
		2006-05-23 & 01:20:51 & 53878.5619 & 600 & 0.977 \\
		2007-01-02 & 08:06:17 & 54102.8424 & 900 & 0.057 \\
		2007-01-02 & 08:23:54 & 54102.8529 & 600 & 0.057 \\
		\bottomrule
	\end{tabular}
\end{table} 


\vspace*{10mm}

\section{\texorpdfstring{Radial Velocities for the \ion{N}{iv}\,$\boldsymbol\lambda$4058 Line}{Radial Velocities for the N IV lambda 4058 Line}}%
\label{sec:appB}
This appendix presents the material related to the radial velocities measured from the \ion{N}{iv}\,$\lambda$4058 line.
We also present an extended version of Fig.~\ref{fig:fgcorr4058add} that includes, in addition, the data from data sets II and III.
The overall agreement between the various measurements supports the idea that the two fitted Gaussians provide a suitable representation of the total line profile, whose shape remains remarkably stable throughout the orbital cycle despite the relative orbital motion.
\begin{figure}[b]
	\centering
	\includegraphics[width=0.67\textwidth]{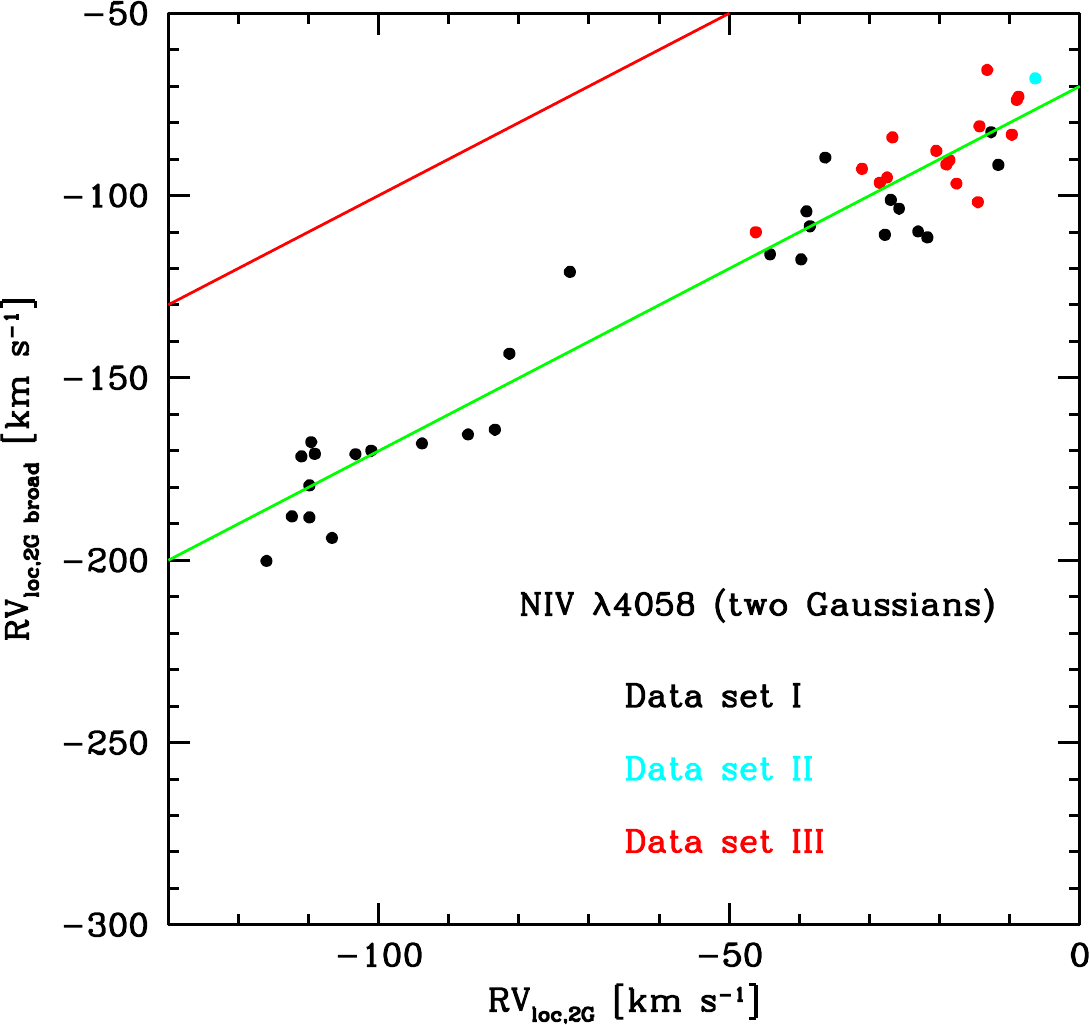}
	\begin{minipage}{12cm}
		\centering
		\caption{%
			Comparison between the RVs of Gaussian~1 with those of Gaussian~2, which fits the broader component.
			This figure extends Fig.~\ref{fig:fgcorr4058add} by including, in addition to data set~I, the measurements from data sets II and III.%
			\label{fig:fig6tout}}
	\end{minipage}
\end{figure}
%
%
%
\begin{sidewaystable}
\centering
\caption{%
	Radial velocities [km\,s$^{-1}$] for the \ion{N}{iv} $\lambda$4058 line (data set~I).%
	\label{tabB1}}
\vskip1ex

\begin{tabular}{cccccccc}
\toprule
\textbf{HJD(2400000+)}
  & \textbf{Phase}
    & \textbf{RV$_{\mathbf{loc}}$}
      & \multicolumn{2}{c}{\textbf{RV$_{\mathbf{loc}}$}}
          & \textbf{RV$_{\mathbf{glo}}$}
      	    & \multicolumn{2}{c}{\textbf{RV$_{\mathbf{glo}}$}} \\
  &
    & \textbf{One Gaussian}
      & \textbf{Gaussian~1}
        & \textbf{Gaussian~2}
          & \textbf{One Gaussian}
            & \textbf{Gaussian~1}
              & \textbf{Gaussian~2} \\ 
\midrule
53796.6176 & 0.582 & $~-29.33$ & $-21.75$ & $-111.44$ & $-29.80$ & $-18.40$ & $~-80.71$ \\
53798.6158 & 0.592 & $~-33.69$ & $-26.96$ & $-101.17$ & $-33.78$ & $-23.53$ & $~-83.30$ \\
53800.6706 & 0.602 & $~-32.91$ & $-23.07$ & $-109.83$ & $-33.05$ & $-19.43$ & $~-91.13$ \\
53833.5275 & 0.760 & $~-47.27$ & $-38.97$ & $-104.33$ & $-45.97$ & $-36.12$ & $-101.60$ \\
53835.6105 & 0.770 & $~-50.38$ & $-44.17$ & $-116.09$ & $-49.49$ & $-42.19$ & $-100.48$ \\
53837.5759 & 0.779 & $~-46.38$ & $-39.74$ & $-117.50$ & $-47.88$ & $-35.10$ & $~-98.87$ \\
53860.4978 & 0.890 & $~-77.54$ & $-72.72$ & $-120.90$ & $-77.23$ & $-70.46$ & $-136.19$ \\
53864.4829 & 0.909 & $~-90.53$ & $-83.40$ & $-164.16$ & $-90.75$ & $-81.14$ & $-146.17$ \\
53868.6284 & 0.929 & $-100.68$ & $-93.80$ & $-167.97$ & $-99.56$ & $-91.47$ & $-156.54$ \\
53897.4685 & 0.068 & $~-46.50$ & $-38.50$ & $-108.39$ & $-45.34$ & $-36.99$ & $~-95.64$ \\
53899.4916 & 0.078 & $~-35.61$ & $-27.79$ & $-110.75$ & $-34.19$ & $-25.72$ & $~-92.42$ \\
53901.5873 & 0.088 & $~-33.53$ & $-25.78$ & $-103.56$ & $-33.46$ & $-24.36$ & $~-84.03$ \\
53912.5575 & 0.141 & $~-20.52$ & $-12.66$ & $~-82.59$ &   ---    &   ---    &   ---    \\
\bottomrule
\end{tabular}
\end{sidewaystable}
\addtocounter{table}{-1}
\begin{sidewaystable}
\centering
\caption{\emph{continued}}
\vskip1ex

\begin{tabular}{cccccccc}
\toprule
\textbf{HJD(2400000+)}
  & \textbf{Phase}
    & \textbf{RV$_{\mathbf{loc}}$}
      & \multicolumn{2}{c}{\textbf{RV$_{\mathbf{loc}}$}}
          & \textbf{RV$_{\mathbf{glo}}$}
      	    & \multicolumn{2}{c}{\textbf{RV$_{\mathbf{glo}}$}} \\
  &
    & \textbf{One Gaussian}
      & \textbf{Gaussian~1}
        & \textbf{Gaussian~2}
          & \textbf{One Gaussian}
            & \textbf{Gaussian~1}
              & \textbf{Gaussian~2} \\ 
\midrule
53918.4988 & 0.169 & $~-19.39$ & $~-11.63$ & $~-91.60$ & $~-19.28$ & $~~-9.29$ & $~-73.02$ \\
54072.8108 & 0.912 & $~-90.43$ & $~-81.34$ & $-143.32$ &    ---    &    ---    &    ---    \\
54084.7501 & 0.970 & $-124.57$ & $-115.97$ & $-200.20$ & $-124.17$ & $-112.61$ & $-181.67$ \\
54092.8269 & 0.009 & $-107.67$ & $-101.04$ & $-169.92$ & $-107.13$ & $~-99.50$ & $-155.90$ \\
54094.8124 & 0.018 & $~-95.71$ & $~-87.24$ & $-165.51$ & $~-95.53$ & $~-84.46$ & $-154.80$ \\
54247.6017 & 0.754 & $~-41.91$ & $~-36.31$ & $~-89.57$ & $~-41.74$ & $~-34.26$ & $~-84.13$ \\
54910.5224 & 0.947 & $-114.22$ & $-103.28$ & $-170.89$ &    ---    &    ---    &    ---    \\
54911.5244 & 0.952 & $-116.74$ & $-109.09$ & $-170.81$ &    ---    &    ---    &    ---    \\
54912.5162 & 0.956 & $-117.56$ & $-109.61$ & $-167.58$ &    ---    &    ---    &    ---    \\
54913.5019 & 0.961 & $-115.38$ & $-106.63$ & $-193.90$ & $-114.36$ & $-104.52$ & $-170.02$ \\
54914.5116 & 0.966 & $-116.74$ & $-109.86$ & $-179.47$ &    ---    &    ---    &    ---    \\
54916.5405 & 0.976 & $-120.95$ & $-112.36$ & $-187.98$ & $-120.65$ & $-109.67$ & $-175.76$ \\
54918.5875 & 0.986 & $-119.01$ & $-109.85$ & $-188.23$ & $-118.57$ & $-107.82$ & $-171.83$ \\
54921.5940 & 0.000 & $-117.63$ & $-110.99$ & $-171.53$ & $-116.75$ & $-108.66$ & $-160.43$ \\
\bottomrule
\end{tabular}
\end{sidewaystable}
%
%
%
\begin{table}
\centering
\begin{minipage}{105mm}
\caption{%
Radial velocities [km\,s$^{-1}$] for the \ion{N}{iv}\,$\lambda$4058 line (data sets II and III).%
\label{tabB2}}
\end{minipage}
\vskip1ex

\begin{tabular}{ccccc}
\toprule
\textbf{HJD(2400000+)}
  & \textbf{Phase}
    & \textbf{RV\textsubscript{loc}} & 
\multicolumn{2}{c}{\textbf{RV\textsubscript{loc}}} \\
  &
    & \textbf{One Gauss.}
      & \textbf{1} & \textbf{2} \\ 
\midrule
 54144.8064 & 0.259 & $-15.99$ & $~-6.35$ & $~-67.87$ \\
  & & & & \\
 52037.6384 & 0.111 & $-13.22$ & $-13.22$ & $~-65.56$ \\
 52038.5859 & 0.115 & $-18.59$ & $-18.59$ & $~-90.30$ \\
 52039.6180 & 0.120 & $-19.04$ & $-19.04$ & $~-91.49$ \\
 52040.6223 & 0.125 & $-14.53$ & $-14.53$ & $-101.78$ \\
 52335.6516 & 0.546 & $-14.30$ & $-14.30$ & $~-80.98$ \\
 52337.6853 & 0.556 & $-26.72$ & $-26.72$ & $~-84.06$ \\
 52338.6359 & 0.560 & $-17.59$ & $-17.59$ & $~-96.70$ \\
 52339.6391 & 0.565 & $-20.46$ & $-20.48$ & $~-87.74$ \\
 52383.5071 & 0.777 & $-46.21$ & $-46.21$ & $-110.04$ \\
 52782.5087 & 0.698 & $-31.06$ & $-31.06$ & $~-92.66$ \\
 52783.4975 & 0.703 & $-28.52$ & $-28.52$ & $~-96.48$ \\
 52784.4929 & 0.708 & $-27.49$ & $-27.49$ & $~-95.03$ \\
 53131.5101 & 0.379 & $~-9.70$ & $~-9.70$ & $~-83.27$ \\ 
 53133.6051 & 0.389 & $~-8.75$ & $~-8.75$ & $~-72.87$ \\
 53135.4854 & 0.398 & $~-9.00$ & $~-9.01$ & $~-73.78$ \\
\bottomrule
\end{tabular}
\end{table}
%
%
%
\begin{table}
\centering
\begin{minipage}{111mm}
\centering
\caption{%
Radial velocities [km\,s$^{-1}$] for the \ion{N}{iv}\,$\lambda$4058 line (data set~IV).%
\label{tabB3}}
\end{minipage}
\vskip1ex

\begin{tabular}{ccccc}
\toprule
\textbf{HJD(2400000+)}
  & \textbf{Phase}
    & \textbf{RV\textsubscript{loc}}
      & \multicolumn{2}{c}{\textbf{RV\textsubscript{loc}}} \\
  & & \textbf{One Gauss.}
      & \textbf{1} & \textbf{2} \\ 
\midrule
 53873.5529 & 0.953 & $-115.18$ & $-109.95$ & $-302.43$ \\
 53874.5645 & 0.958 & $-119.71$ & $~-98.47$ & $-243.40$ \\
 53875.5402 & 0.962 & $-119.38$ & $-106.57$ & $-265.07$ \\
 53876.5675 & 0.967 & $-124.06$ & $-105.20$ & $-221.95$ \\
 53876.5781 & 0.967 & $-124.38$ & $-108.23$ & $-238.55$ \\
 53877.5319 & 0.972 & $-125.01$ & $-107.00$ & $-242.54$ \\
 53877.6157 & 0.972 & $-124.23$ & $-111.53$ & $-260.73$ \\
 53878.5206 & 0.977 & $-124.38$ & $-108.93$ & $-154.05$\makebox[0pt][l]{:} \\
 53878.5401 & 0.977 & $-122.28$ & $-113.84$ & $-279.68$ \\
 53878.5619 & 0.977 & $-122.68$ & $-113.49$ & $-290.96$ \\
 54102.8424 & 0.057 & $~-49.83$ & $~-29.07$ & $-125.13$ \\
 54102.8529 & 0.057 & $~-50.45$ & $~-28.31$ & $-122.53$ \\
\bottomrule
\end{tabular}
\end{table} 

\clearpage

\section{Radial Velocities of the WN Component for the Selected Spectral Lines}%
\label{sec:appC}
This appendix presents the ``gold sample'' of measured radial velocities for the selected subset of spectral lines that are intended to best represent the relative orbital motion of the WN star.

\begin{table}[h!]
	\centering
	\begin{sideways}
		\begin{minipage}{200mm}
			\centering
			\begin{minipage}{160mm}
				\caption{%
					Radial velocities RV$_{\text{loc}}$ [km\,s$^{-1}$] of the WN component in the selected spectral lines.%
					\label{tabC}}
			\end{minipage}
			\vskip1ex
			
			\begin{tabular}{ccr@{.}lr@{.}lr@{.}lr@{.}lr@{.}lr@{.}lr@{.}lr@{.}l}
				\toprule
				\textbf{HJD(2400000+)}
				  & \textbf{Phase}
				    & \multicolumn{2}{c}{\textbf{\ion{N}{iv}}}
				      & \multicolumn{2}{c}{\textbf{\ion{N}{v}}}
				        & \multicolumn{2}{c}{\textbf{\ion{N}{iv}}}
				          & \multicolumn{2}{c}{\textbf{\ion{N}{iv}}}
				            & \multicolumn{2}{c}{\textbf{\ion{N}{v}}}
				              & \multicolumn{2}{c}{\textbf{\ion{N}{iv}}}
				                & \multicolumn{2}{c}{\textbf{\ion{Si}{iv}}}
				                  & \multicolumn{2}{c}{\textbf{Mean}} \\
				  & & \multicolumn{2}{c}{$\boldsymbol\lambda$\textbf{4058}}
				      & \multicolumn{2}{c}{$\boldsymbol\lambda$\textbf{4604}}
				        & \multicolumn{2}{c}{$\boldsymbol\lambda$\textbf{6220}}
				          & \multicolumn{2}{c}{$\boldsymbol\lambda$\textbf{7110}}
				            & \multicolumn{2}{c}{$\boldsymbol\lambda$\textbf{4945}}
				              & \multicolumn{2}{c}{$\boldsymbol\lambda$\textbf{5737}}
				                & \multicolumn{2}{c}{$\boldsymbol\lambda$\textbf{4089}} \\
				\midrule
				 53796.6176 & 0.582 &  $-21$&75 & 162&93 &   $-5$&36 &  $-18$&12 &    26&79 &  $-15$&49 &  $-19$&14 &    11&83 \\
				 53798.6158 & 0.592 &  $-26$&96 & 158&56 &  $-12$&34 &  $-22$&98 &    24&18 &  $-20$&13 &  $-20$&38 &     7&56 \\
				 53800.6706 & 0.602 &  $-23$&07 & 156&90 &   $-0$&24 &  $-20$&75 &    35&07 &  $-17$&64 &  $-17$&74 &    12&21 \\
				 53833.5275 & 0.760 &  $-38$&97 & 150&80 &  $-11$&17 &  $-35$&25 &    21&15 &  $-30$&97 &  $-32$&48 &  $-0$&56 \\
				 53835.6105 & 0.770 &  $-44$&17 & 145&13 &  $-25$&71 &  $-41$&11 &     9&39 &  $-33$&42 &  $-32$&77 &  $-7$&10 \\
				 53837.5759 & 0.779 &  $-39$&74 & 149&59 &  $-12$&41 &  $-36$&54 &    12&02 &  $-31$&31 &  $-28$&30 &  $-1$&96 \\
				 53860.4978 & 0.890 &  $-72$&72 & 115&07 &  $-56$&85 &  $-65$&85 & $-32$&90 &  $-63$&47 &  $-58$&51 & $-37$&47 \\
				 53864.4829 & 0.909 &  $-83$&40 &  99&18 &  $-72$&08 &  $-75$&24 & $-34$&85 &  $-71$&60 &  $-74$&64 & $-48$&53 \\
				 53868.6284 & 0.929 &  $-93$&80 &  94&09 &  $-77$&54 &  $-84$&73 & $-37$&63 &  $-79$&85 &  $-89$&08 & $-56$&51 \\
				 53897.4685 & 0.068 &  $-38$&50 & 146&78 &  $-22$&75 &  $-37$&22 &    14&75 &  $-33$&18 &  $-30$&28 &  $-3$&92 \\
				 53899.4916 & 0.078 &  $-27$&79 & 161&80 &    $0$&97 &  $-24$&43 &    30&24 &  $-21$&07 &  $-19$&21 &    10&49 \\
				 53901.5873 & 0.088 &  $-25$&78 & 159&00 &   $-7$&38 &  $-23$&97 &    31&46 &  $-19$&83 &  $-14$&66 &    10&25 \\
				 53912.5575 & 0.141 &  $-12$&66 & 173&15 &    $4$&46 &  $-11$&41 &    44&98 &   $-5$&21 &   $-6$&75 &    22&79 \\
				 53918.4988 & 0.169 &  $-11$&63 & 177&91 &   $15$&74 &   $-7$&28 &    46&64 &   $-1$&42 &   $-5$&43 &    26&78 \\
				 54072.8108 & 0.912 &  $-81$&34 & 103&92 &  $-70$&78 &  $-77$&89 & $-30$&84 &  $-73$&45 &  $-75$&59 & $-47$&58 \\
				\bottomrule
			\end{tabular}
		\end{minipage}
	\end{sideways}
\end{table}
\addtocounter{table}{-1}
\begin{sidewaystable}
	\centering
	\begin{minipage}{160mm}
		\caption{\emph{continued}}
	\end{minipage}
	\vskip1ex
	
	\begin{tabular}{ccr@{.}lr@{.}lr@{.}lr@{.}lr@{.}lr@{.}lr@{.}lr@{.}l}
		\toprule
		\textbf{HJD(2400000+)}
		  & \textbf{Phase}
		    & \multicolumn{2}{c}{\textbf{\ion{N}{iv}}}
		      & \multicolumn{2}{c}{\textbf{\ion{N}{v}}}
		        & \multicolumn{2}{c}{\textbf{\ion{N}{iv}}}
		          & \multicolumn{2}{c}{\textbf{\ion{N}{iv}}}
		            & \multicolumn{2}{c}{\textbf{\ion{N}{v}}}
		              & \multicolumn{2}{c}{\textbf{\ion{N}{iv}}}
		                & \multicolumn{2}{c}{\textbf{\ion{Si}{iv}}}
		                  & \multicolumn{2}{c}{\textbf{Mean}} \\
		  & & \multicolumn{2}{c}{$\boldsymbol\lambda$\textbf{4058}}
		      & \multicolumn{2}{c}{$\boldsymbol\lambda$\textbf{4604}}
		        & \multicolumn{2}{c}{$\boldsymbol\lambda$\textbf{6220}}
		          & \multicolumn{2}{c}{$\boldsymbol\lambda$\textbf{7110}}
		            & \multicolumn{2}{c}{$\boldsymbol\lambda$\textbf{4945}}
		              & \multicolumn{2}{c}{$\boldsymbol\lambda$\textbf{5737}}
		                & \multicolumn{2}{c}{$\boldsymbol\lambda$\textbf{4089}} \\
		\midrule
		54084.7501 & 0.970 & $-115$&97 &  66&66 & $-101$&86 & $-111$&83 & $-60$&97 & $-101$&72 & $-113$&13 & $-80$&84 \\
		54092.8269 & 0.009 & $-101$&04 &  84&67 &  $-87$&78 &  $-97$&69 & $-37$&06 &  $-84$&94 &  $-93$&92 & $-63$&55 \\
		54094.8124 & 0.018 &  $-87$&24 & 101&46 &  $-75$&61 &  $-78$&90 & $-40$&42 &  $-76$&06 &  $-79$&99 & $-51$&97 \\
		54247.6017 & 0.754 &  $-36$&31 & 148&98 &   $-6$&52 &  $-32$&84 &    25&51 &  $-27$&52 &  $-26$&69 &     2&51 \\
		54910.5224 & 0.947 & $-103$&28 &  83&77 &  $-90$&21 &  $-96$&14 & $-46$&86 &  $-93$&11 & $-104$&48 & $-68$&20 \\
		54911.5244 & 0.952 & $-109$&09 &  77&50 &  $-94$&01 & $-101$&21 & $-43$&42 &  $-95$&91 & $-106$&61 & $-71$&40 \\
		54912.5162 & 0.956 & $-109$&61 &  72&95 &  $-93$&53 & $-102$&67 & $-46$&99 &  $-98$&76 & $-101$&40 & $-72$&44 \\
		54913.5019 & 0.961 & $-106$&63 &  74&11 &  $-94$&30 & $-107$&31 & $-63$&43 &  $-98$&14 &  $-99$&35 & $-74$&59 \\
		54914.5116 & 0.966 & $-109$&86 &  77&72 &  $-93$&27 & $-108$&42 & $-48$&13 &  $-94$&61 & $-104$&99 & $-72$&66 \\
		54916.5405 & 0.976 & $-112$&36 &  72&98 &  $-96$&75 & $-109$&02 & $-67$&04 &  $-97$&68 & $-101$&91 & $-76$&98 \\
		54918.5875 & 0.986 & $-109$&85 &  81&75 &  $-92$&16 & $-104$&83 & $-61$&91 &  $-94$&66 & $-108$&73 & $-73$&92 \\
		54921.5940 & 0.000 & $-110$&99 &  76&13 &  $-92$&74 & $-109$&52 & $-59$&66 &  $-96$&38 & $-104$&70 & $-74$&99 \\[2ex]
		54144.8064 & 0.259 &   $-6$&35 & 176&90 &     20&56 &   $-5$&01 &    45&18 &     10&47 &   $-2$&15 &    30&36 \\
		\bottomrule 
	\end{tabular}
\end{sidewaystable}
\addtocounter{table}{-1}
\begin{sidewaystable}
	\centering
	\begin{minipage}{160mm}
		\caption{\emph{continued}}
	\end{minipage}
	\vskip1ex
	
	\begin{tabular}{ccr@{.}lr@{.}lr@{.}lr@{.}lr@{.}lr@{.}lr@{.}lr@{.}l}
		\toprule
		\textbf{HJD(2400000+)}
		  & \textbf{Phase}
		    & \multicolumn{2}{c}{\textbf{\ion{N}{iv}}}
		      & \multicolumn{2}{c}{\textbf{\ion{N}{v}}}
		        & \multicolumn{2}{c}{\textbf{\ion{N}{iv}}}
		          & \multicolumn{2}{c}{\textbf{\ion{N}{iv}}}
		            & \multicolumn{2}{c}{\textbf{\ion{N}{v}}}
		              & \multicolumn{2}{c}{\textbf{\ion{N}{iv}}}
		                & \multicolumn{2}{c}{\textbf{\ion{Si}{iv}}}
		                  & \multicolumn{2}{c}{\textbf{Mean}} \\
		  & & \multicolumn{2}{c}{$\boldsymbol\lambda$\textbf{4058}}
		      & \multicolumn{2}{c}{$\boldsymbol\lambda$\textbf{4604}}
		        & \multicolumn{2}{c}{$\boldsymbol\lambda$\textbf{6220}}
		          & \multicolumn{2}{c}{$\boldsymbol\lambda$\textbf{7110}}
		            & \multicolumn{2}{c}{$\boldsymbol\lambda$\textbf{4945}}
		              & \multicolumn{2}{c}{$\boldsymbol\lambda$\textbf{5737}}
		                & \multicolumn{2}{c}{$\boldsymbol\lambda$\textbf{4089}} \\
		\midrule
		52037.6384 & 0.111 &  $-13$&22 & 141&54 &  $-6$&10 &  $-16$&89 &    37&16 & $-17$&76 &  $-18$&59 &    11&30 \\
		52038.5859 & 0.115 &  $-18$&59 & 168&65 &     4&51 &  $-19$&72 &    41&66 & $-13$&61 &  $-10$&86 &    17&85 \\
		52039.6180 & 0.120 &  $-19$&04 & 162&66 & $-10$&38 &  $-23$&91 &    45&88 & $-16$&76 &  $-18$&14 &    13&32 \\
		52040.6223 & 0.125 &  $-14$&53 & 166&51 &  $-3$&98 &  $-16$&27 &    41&61 & $-13$&48 &   $-4$&68 &    18&30 \\
		52335.6516 & 0.546 &  $-14$&30 & 152&13 &  $-7$&46 &  $-16$&57 &    21&81 & $-18$&99 &  $-20$&61 &     9&85 \\
		52337.6853 & 0.556 &  $-26$&72 & 153&48 & $-10$&36 &  $-13$&62 &    33&80 & $-20$&10 &  $-17$&85 &    10&22 \\
		52338.6359 & 0.560 &  $-17$&59 & 165&33 &     1&75 &  $-11$&74 &    34&21 & $-14$&69 &  $-16$&49 &    16&25 \\
		52339.6391 & 0.565 &  $-20$&48 & 144&16 &  $-8$&73 &  $-13$&49 &    21&67 & $-15$&21 &  $-11$&85 &     9&86 \\
		52383.5071 & 0.777 &  $-46$&21 & 137&29 & $-29$&44 &  $-44$&09 &     5&68 & $-40$&73 &  $-38$&43 & $-11$&86 \\
		52782.5087 & 0.698 &  $-31$&06 & 147&90 & $-16$&37 &  $-30$&99 &    20&0: & $-24$&51 &  $-18$&77 &     2&73 \\
		52783.4975 & 0.703 &  $-28$&52 & 137&29 & $-14$&16 &  $-28$&09 &    19&63 & $-19$&79 &  $-18$&77 &     2&93 \\
		52784.4929 & 0.708 &  $-27$&49 & 140&13 &  $-2$&10 &  $-30$&09 &    17&58 & $-21$&25 &  $-21$&25 &     4&07 \\
		53131.5101 & 0.379 &   $-9$&70 & 150&32 &  $-0$&44 &   $-9$&85 &    32&04 &  $-3$&79 &   $-2$&81 &    18&39 \\ 
		53133.6051 & 0.389 &   $-8$&75 & 171&59 &     0&52 &      2&26 &    42&55 &  $-1$&52 &   $-0$&58 &    25&57 \\
		53135.4854 & 0.398 &   $-9$&01 & 179&11 &     1&25 &   $-9$&78 &    43&64 &  $-2$&96 &   $-0$&19 &    25&00 \\
		\bottomrule\addlinespace[\belowrulesep]
		\multicolumn{18}{p{175mm}}{\footnotesize
			{\bf{Notes:}} except for the \ion{N}{iv}\,$\lambda$4058 line, the RVs have an arbitrary zero-point (see Sect.~\ref{ssec:wrspec_DS1} \emph{et seq.}).
			The `Mean' column is a combination of the other columns that is shifted to obtain a $\gamma_{\text{WN}}$ of zero.}
	\end{tabular}
\end{sidewaystable}

\clearpage

\section{Radial Velocities of the O Component}\label{sec:appD}
This appendix contains the material related to the RVs used to directly trace the motion of the \mbox{O-star} companion.
Table~\ref{tabD1} lists the radial velocities measured from the few spectral lines for which the \mbox{O-star} contribution can be clearly separated from that of the WN component owing to significant deblending.
These RVs correspond to a small, limited region of values for the phases around 0.98.
Table~\ref{tabD2} presents the RVs derived from the \ion{He}{i}\,$\lambda$4471 line after subtraction of a line assumed to be constant in radial velocity, depth, and width.
This line was initially attributed to the WN component but could in fact originate from the intruder (the third spectrum).
\begin{table}[h]
\centering
\begin{minipage}{160mm}
\centering
\caption{%
Directly measured radial velocities for the O~star.%
\label{tabD1}}
\end{minipage}
\vskip1ex

\begin{tabular}{cccc}
\toprule
\textbf{HJD}
  & \textbf{RV}
    & \textbf{RV}
      & \textbf{RV} \\
\textbf{(2400000+)}
  & \textbf{\ion{He}{i}\,$\boldsymbol\lambda$4471}
    & \textbf{\ion{He}{ii}\,$\boldsymbol\lambda$4542}
      & \textbf{\ion{N}{iv}\,$\boldsymbol\lambda\boldsymbol\lambda$5204-5205} \\
  & \textbf{[km\,s\textsuperscript{--1}]}
    & \textbf{[km\,s\textsuperscript{--1}]}
      & \textbf{[km\,s\textsuperscript{--1}]} \\
\midrule
 53868.6284 & ---    & ---    & 117.52 \\
 54084.7501 & 179.31 & 170.18 & 180.63 \\
 54092.8269 & 130.89 & 144.48 & 137.78 \\
 54910.5224 & 149.09 & 162.89 & 146.23 \\
 54911.5244 & 145.00 & 157.14 & 161.15 \\
 54912.5162 & 150.29 & 158.44 & 157.88 \\
 54913.5019 & 169.45 & 165.02 & 174.95 \\
 54914.5116 & 169.87 & 178.07 & 170.48 \\
 54916.5405 & 171.48 & 174.96 & 188.29 \\
 54918.5875 & 180.97 & 190.96 & 181.13 \\
 54921.5940 & 160.05 & 167.77 & 161.19 \\
\bottomrule
\end{tabular}
\end{table}
%
\begin{table}
	\centering
	\begin{minipage}{131mm}
	\caption{%
		Radial velocities for the \ion{He}{i}\,$\lambda$4471 line measured from spectra cleaned of the main line, initially attributed to the WN component and assumed to be constant in radial velocity, depth, and width.%
		\label{tabD2}}
	\end{minipage}
	\vskip2ex
	
	\begin{tabular}{cr@{\hskip4.7ex}}
		\toprule
		\textbf{HJD(2400000+)}
		  & \multicolumn{1}{c}{\textbf{Radial velocity}} \\
		    & \multicolumn{1}{c}{\textbf{[km\,s\textsuperscript{--1}]}} \\
		\midrule
		52038.5859 & $-35.53$ \\
		52039.6180 &   $3.52$ \\
		52040.6223 & $-14.58$ \\
		52335.6516 &   $1.85$ \\
		52337.6853 &  $13.10$ \\
		52338.6359 & $-38.73$ \\
		52339.6391 & $-33.72$ \\
		52383.5071 &  $-6.99$ \\
		52782.5087 & $-18.47$ \\
		52783.4975 & $-46.82$ \\
		53135.4854 & $-37.23$ \\
		53796.6176 & $-25.59$ \\
		53798.6158 & $-30.55$ \\
		53800.6706 & $-18.57$ \\
		53833.5275 &   $5.35$ \\
		53835.6105 &   $7.30$ \\
		53837.5759 &  $17.43$ \\
		53864.4829 &  $83.18$ \\
		53897.4685 &   $5.76$ \\
		\bottomrule
	\end{tabular}%
	\hskip1em
	\begin{tabular}{cr@{\hskip4.7ex}}
		\toprule
		\textbf{HJD(2400000+)}
		  & \multicolumn{1}{c}{\textbf{Radial velocity}} \\
		    & \multicolumn{1}{c}{\textbf{[km\,s\textsuperscript{--1}]}} \\
		\midrule
		53899.4916 & $-11.17$ \\
		53901.5873 &  $-4.12$ \\
		53912.5575 & $-35.68$ \\
		53918.4988 & $-37.22$ \\
		54072.8108 &  $76.90$ \\
		54084.7501 & $167.04$ \\
		54092.8269 & $112.77$ \\
		54094.8124 &  $92.16$ \\
		54144.8064 & $-35.86$ \\
		54247.6017 &   $5.71$ \\
		54910.5224 & $149.08$ \\
		54911.5244 & $133.10$ \\
		54912.5162 & $137.11$ \\
		54913.5019 & $153.59$ \\
		54914.5116 & $160.49$ \\
		54916.5405 & $160.31$ \\
		54918.5875 & $169.60$ \\
		54921.5940 & $143.55$ \\
		 \\
		\bottomrule
	\end{tabular}
\end{table} 

\clearpage

\section{Disentangled Spectra}\label{sec:appE}
In this appendix, we present the disentangled spectrum of the secondary O5 component and that of the primary WN component, although the latter may be affected by the presence of a third, fainter object.
Figure~\ref{fig:figdisOtotal} shows an excerpt of the \mbox{O5-star} spectrum together with the identification of the main spectral lines.
The disentangled spectrum exhibits features with amplitudes well below the 1{\%} level in continuum units.
We recall that the disentangling procedure was performed assuming a brightness ratio of 0.5/0.5, a choice that is convenient for interpretation but is certainly not the correct value.

Beyond the usual lines expected to be present in an O5 star and those discussed in Sect.~\ref{ssec:disentanglingspectra}, it is noteworthy that the spectrum displays faint but reliable detections of the \ion{N}{v}\,$\lambda \lambda$4604-4619 doublet in absorption as well as the \ion{O}{iii}\,$\lambda$5592 absorption line.
The \ion{S}{iv}\,$\lambda$4486 and \ion{S}{iv}\,$\lambda$4504 lines are also present \citep{2001ASPC..242..229W}.
They appear to exhibit P-Cygni profiles, although the emission component of \ion{S}{iv}\,$\lambda$4504 is affected by a disentangling artefact and the possible absorption component is blended with a DIB.
The P-Cygni profile of \ion{S}{iv}\,$\lambda$4486 is consistent with the synthetic spectrum, as can be seen in Fig.~\ref{fig:CMFGEN_Ostar} (Sect.~\ref{ssec:Ostar}).
The line \ion{C}{iii}\,$\lambda$5696 is also present as a weak emission feature, while the \ion{C}{iv}\,$\lambda \lambda$5801-5812 lines are visible in absorption.
The \ion{N}{iii}\,$\lambda$4514 line is clearly absent, supporting the idea that the O5 does not belong to the supergiant luminosity class \citep{1988A&AS...76..427M}.
It is worth noting that artefacts of the disentangling procedure affecting constant spectral features are visible in the vicinity of the nebular lines [\ion{O}{iii}]\,$\lambda$5007 and [\ion{O}{iii}]\,$\lambda$4959, which are visible in the individual spectra.

Figure~\ref{fig:figdisWtotal} presents the disentangled spectrum of the WN component.
In addition to the usual emission lines of a WN6ha object (see Sect.~\ref{sec:wrspec}), a few weak lines appear to move with the WN component, in particular the \ion{N}{iii}\,$\lambda$4321, \ion{N}{iii}\,$\lambda$4842, and \ion{N}{iii}\,$\lambda$4882-4884 lines.
The position at rest of the line labelled \ion{N}{iii}\,$\lambda$4321 in the disentangled WN spectrum is actually at 4322.8\,{\AA}, which may cast some doubt on the identification.
The \ion{C}{iii}\,$\lambda$5696 line appears also to be present as a weak emission feature.
This line therefore seems to be present in both objects, although the relatively small RV variations of the WN star mean that it could instead originate from the possible third object.
The \ion{He}{i}\,$\lambda$4471 and \ion{He}{i}\,$\lambda$5876 absorption lines are also present and they are difficult to explain (see Sect.~\ref{sec:CMFGEN}).
These features could likewise be associated with the possible third object.
\begin{figure}
	\centering
	\includegraphics[width=0.67\textwidth]{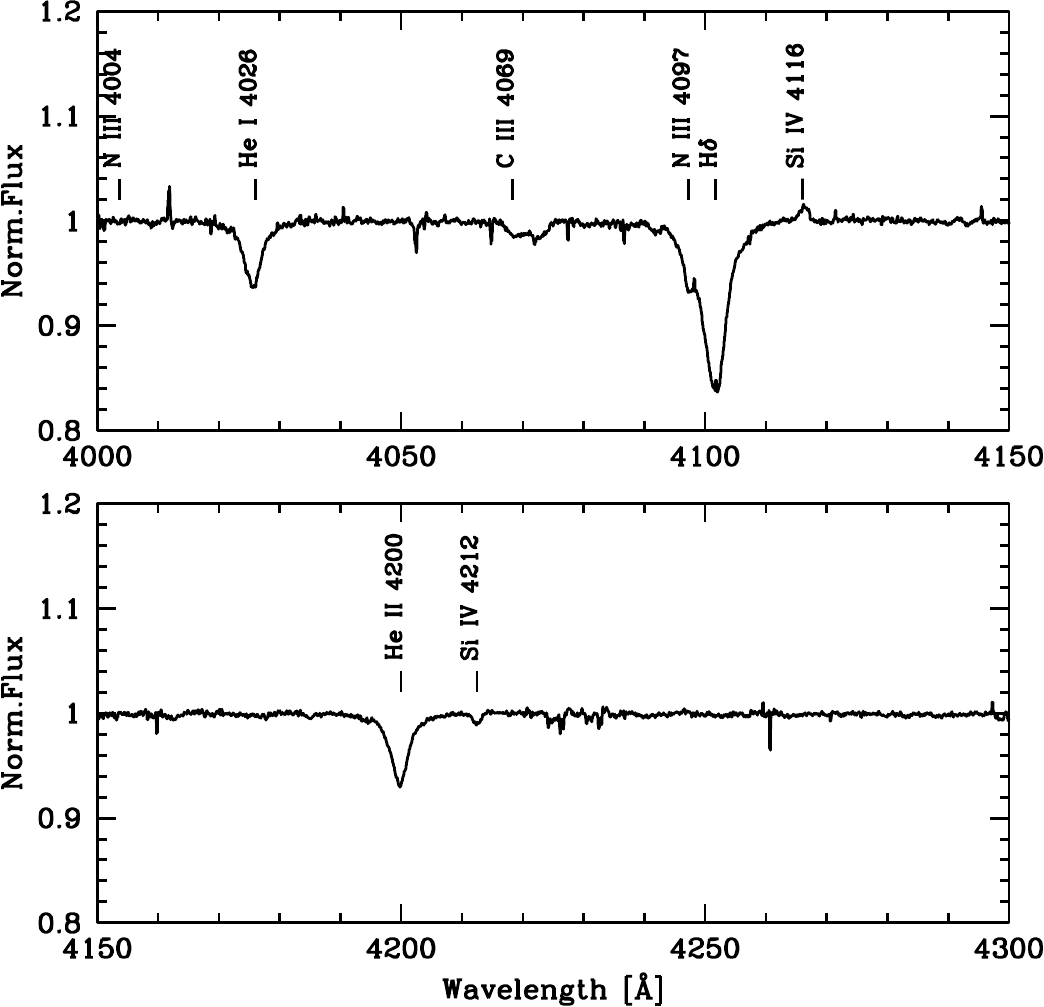} 
	\includegraphics[width=0.67\textwidth]{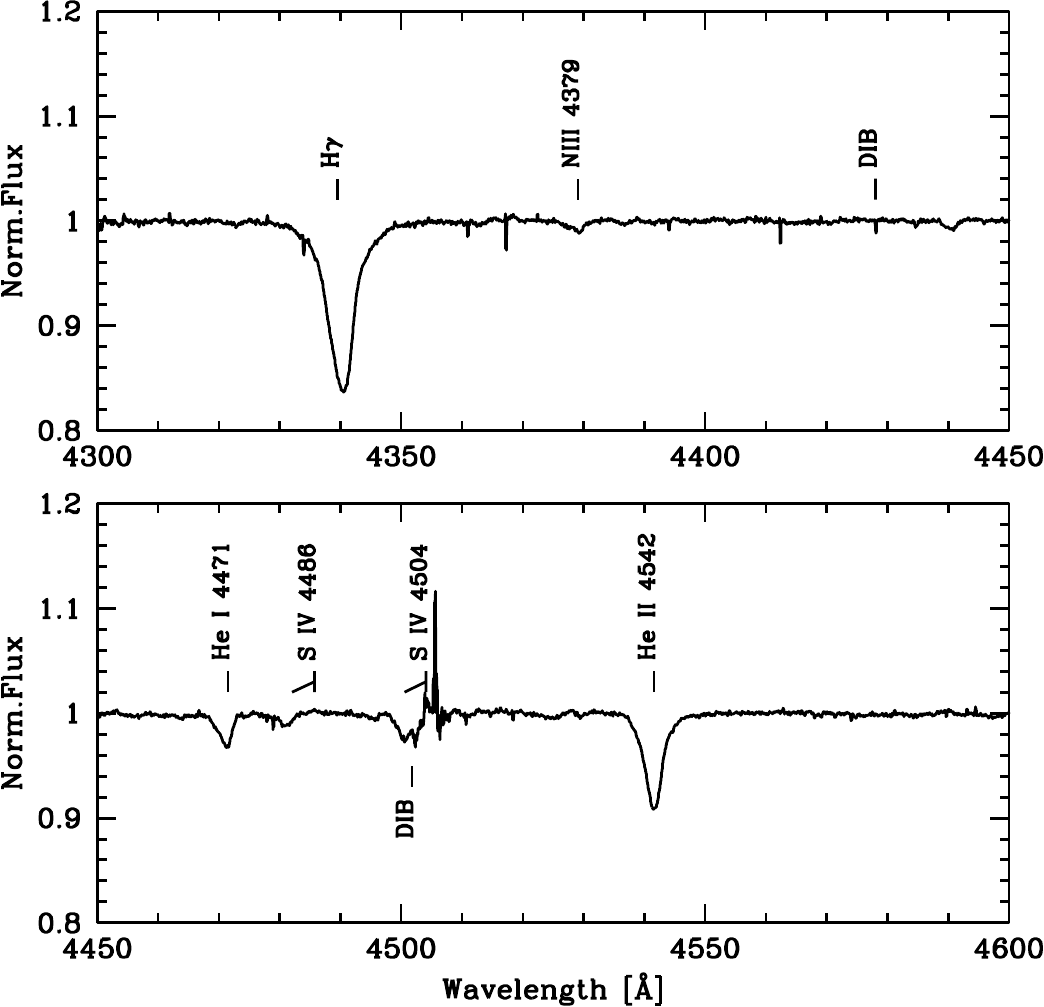}
	\vskip1ex
	
	\begin{minipage}{12cm}
		\caption{%
		Disentangled spectrum of the O5 star, with identifications of the main lines.%
		\label{fig:figdisOtotal}}
	\end{minipage}
\end{figure}
%
\begin{figure}
	\centering
	\includegraphics[width=0.67\textwidth]{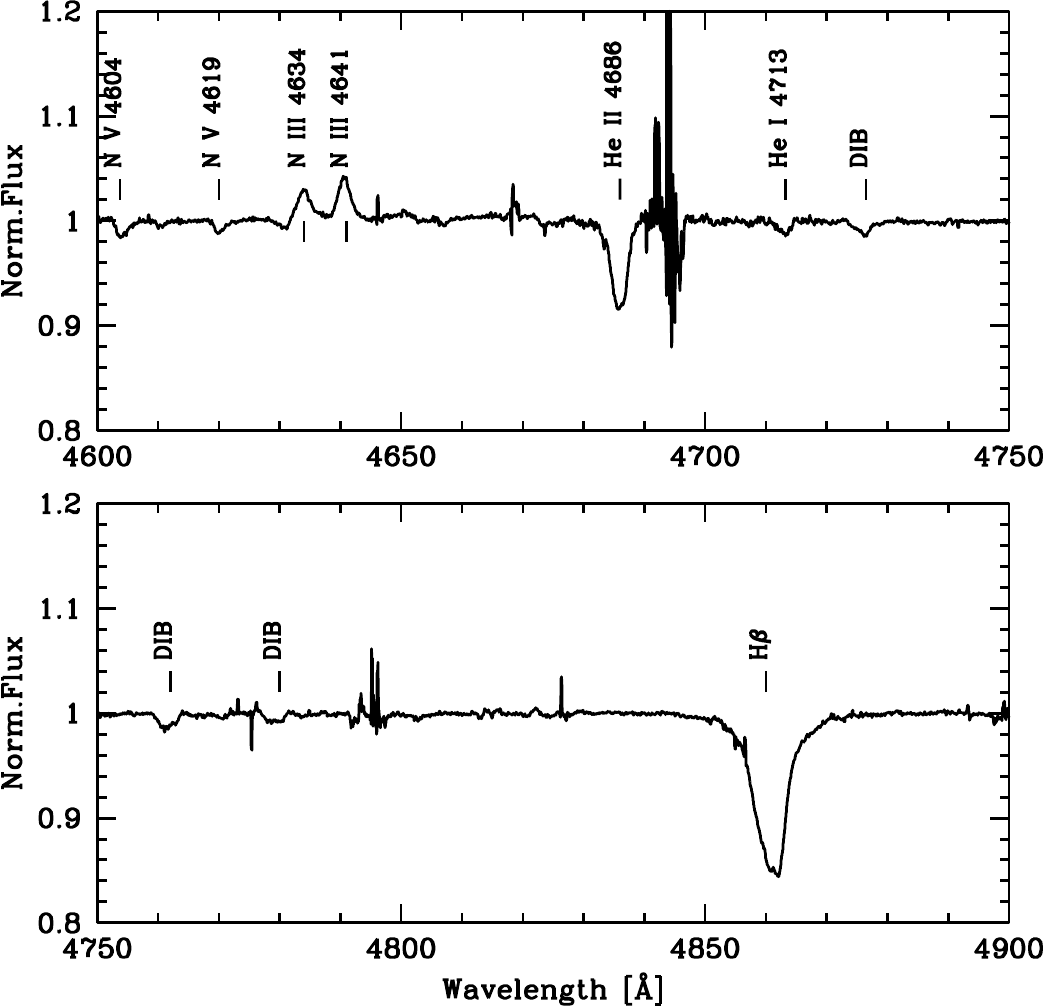}
	\includegraphics[width=0.67\textwidth]{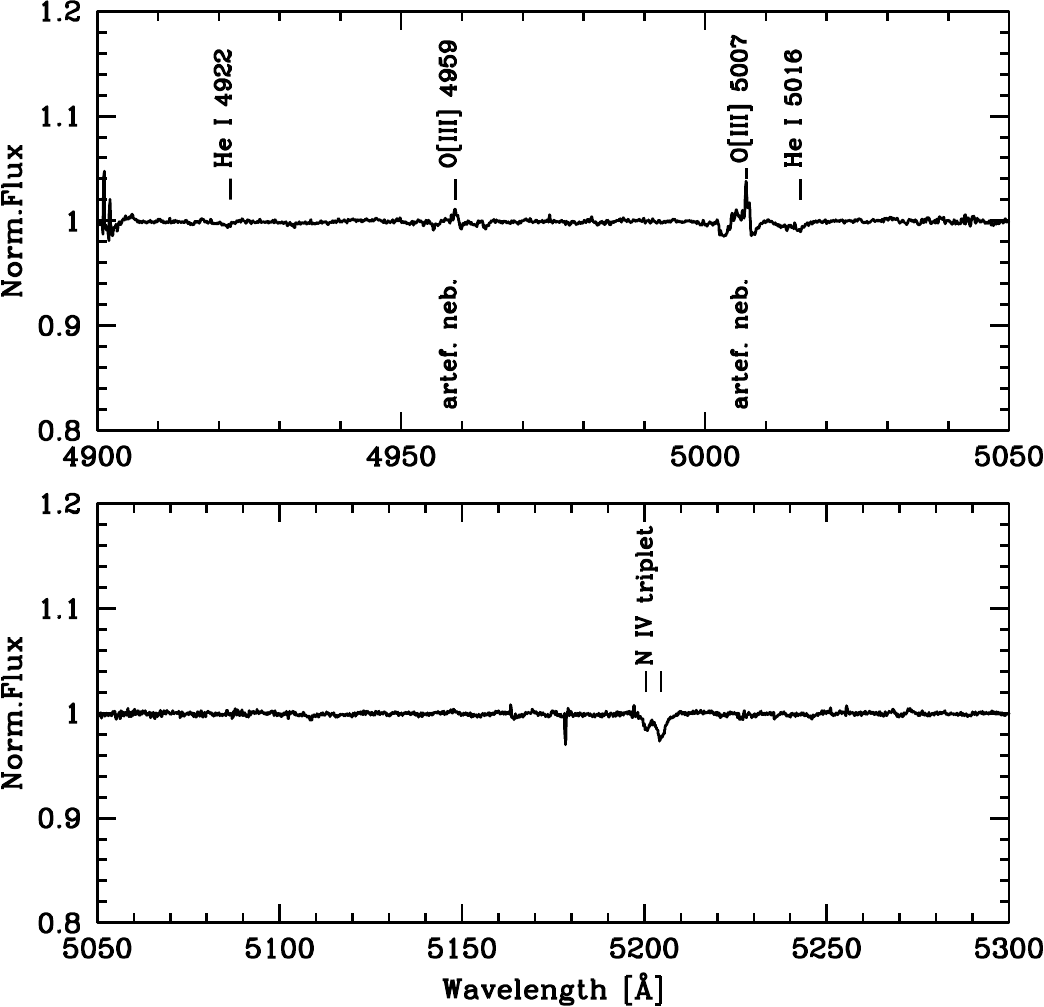}
	\vskip1ex
	
	\begin{minipage}{12cm}
		\addtocounter{figure}{-1}
		\caption{\emph{continued.}}
	\end{minipage}
\end{figure}
%
\begin{figure}
	\centering
	\includegraphics[width=0.67\textwidth]{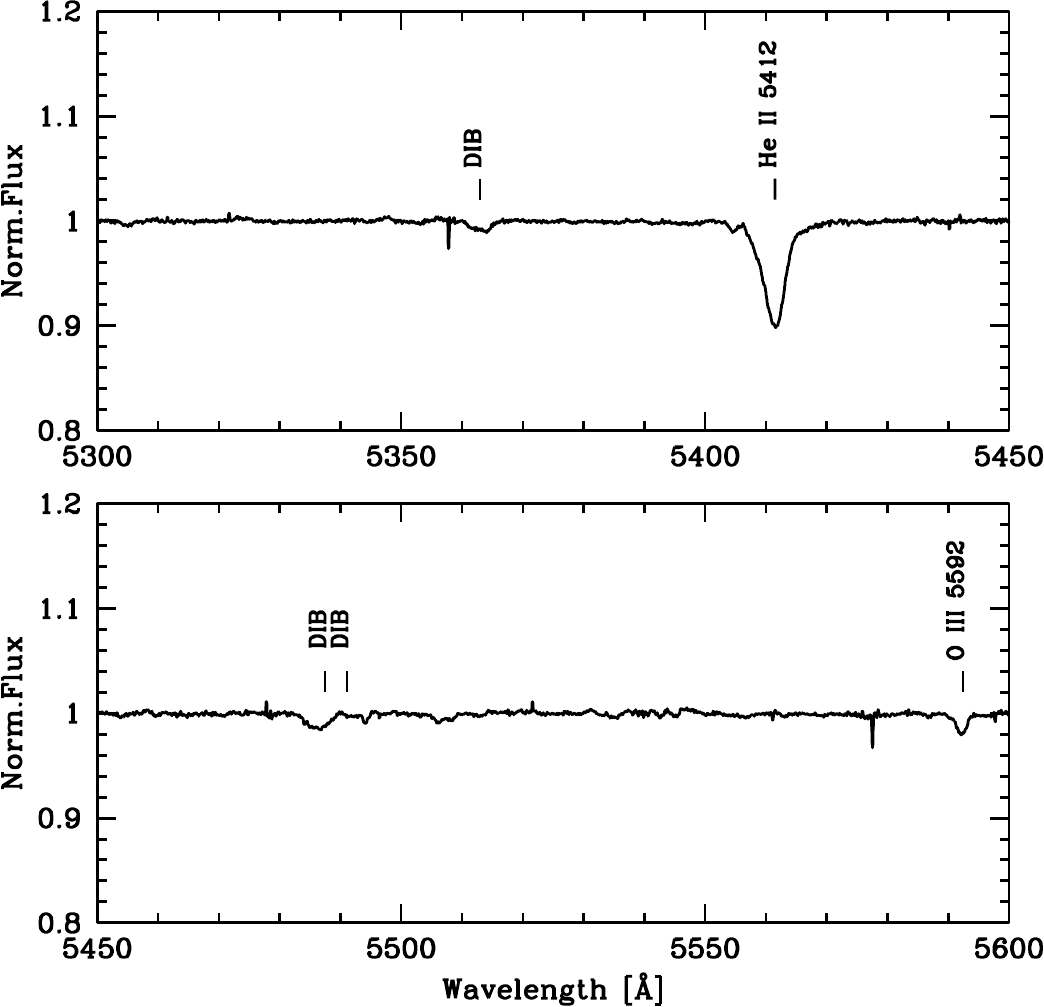}
	\includegraphics[width=0.67\textwidth]{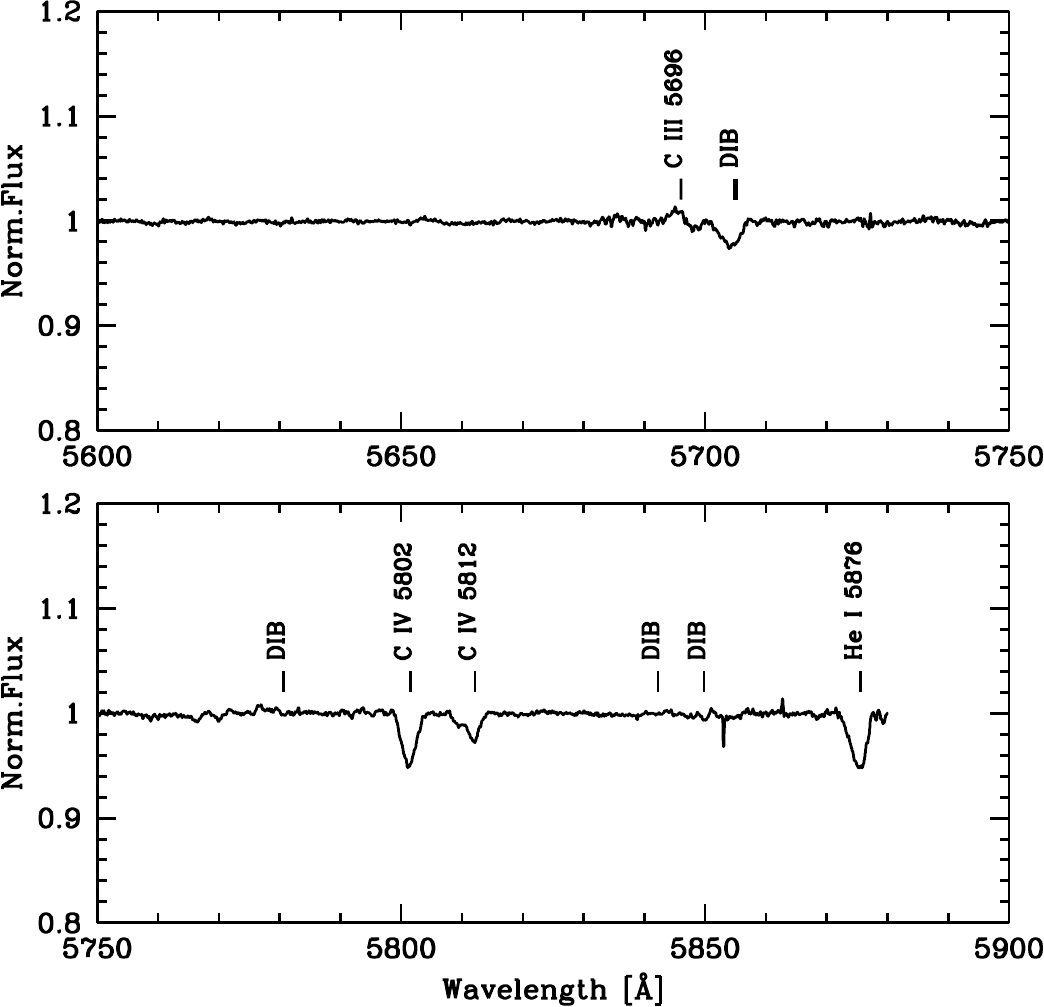}
	\vskip1ex
	
	\begin{minipage}{12cm}
		\addtocounter{figure}{-1}
		\caption{\emph{continued.}}
	\end{minipage}
\end{figure}
%
\begin{figure}
	\centering
	\includegraphics[width=0.67\textwidth]{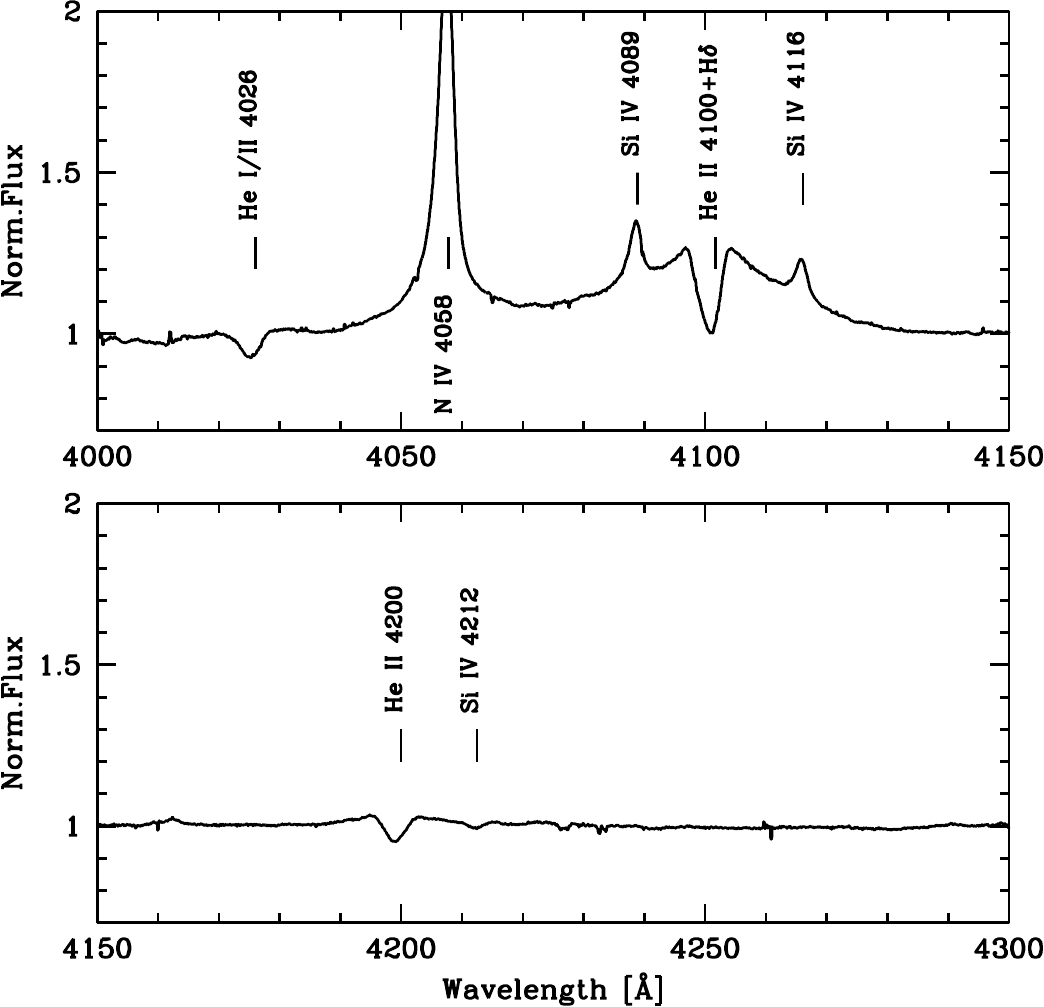} 
	\includegraphics[width=0.67\textwidth]{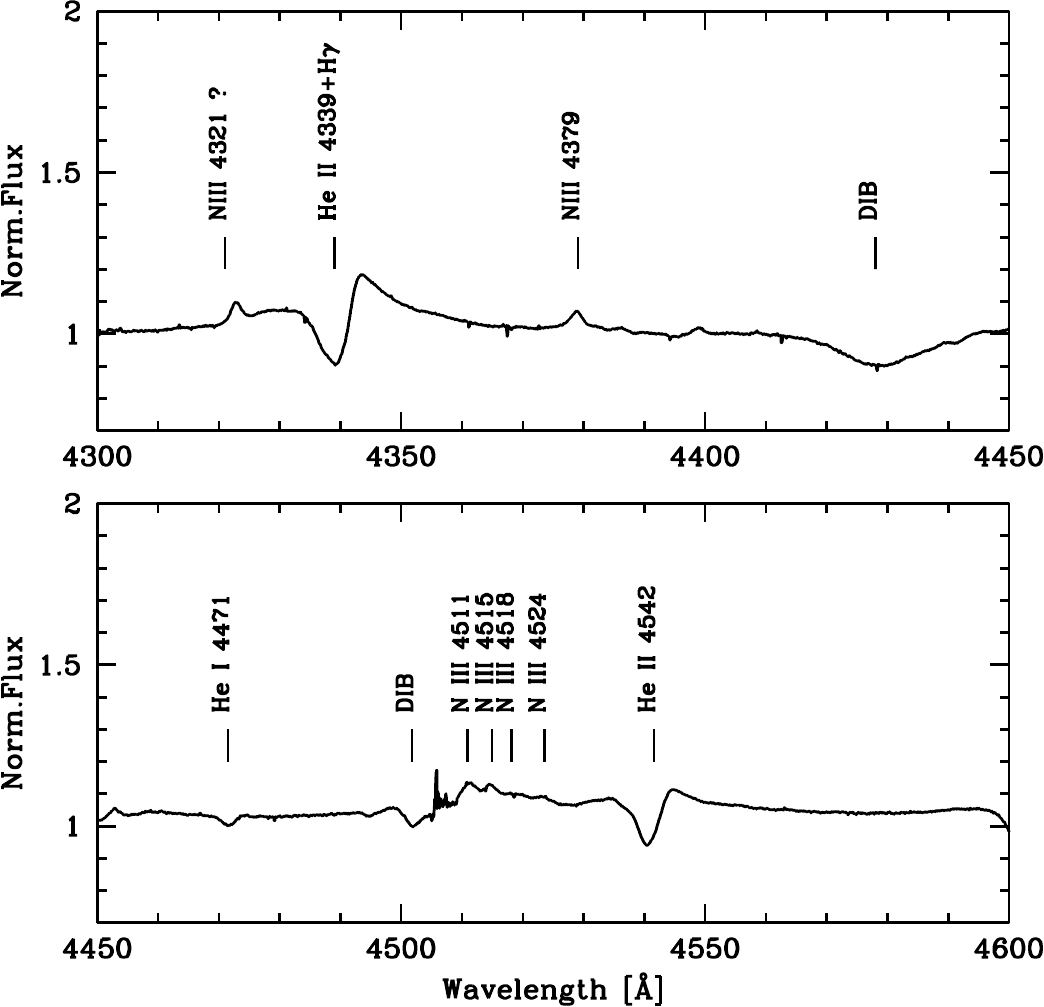}
	\vskip1ex
	
	\begin{minipage}{12cm}
		\caption{%
		Disentangled spectrum of the WN6ha star, with identifications of the main lines.%
		\label{fig:figdisWtotal}}
	\end{minipage}
\end{figure}
%
\begin{figure}
	\centering
	\includegraphics[width=0.67\textwidth]{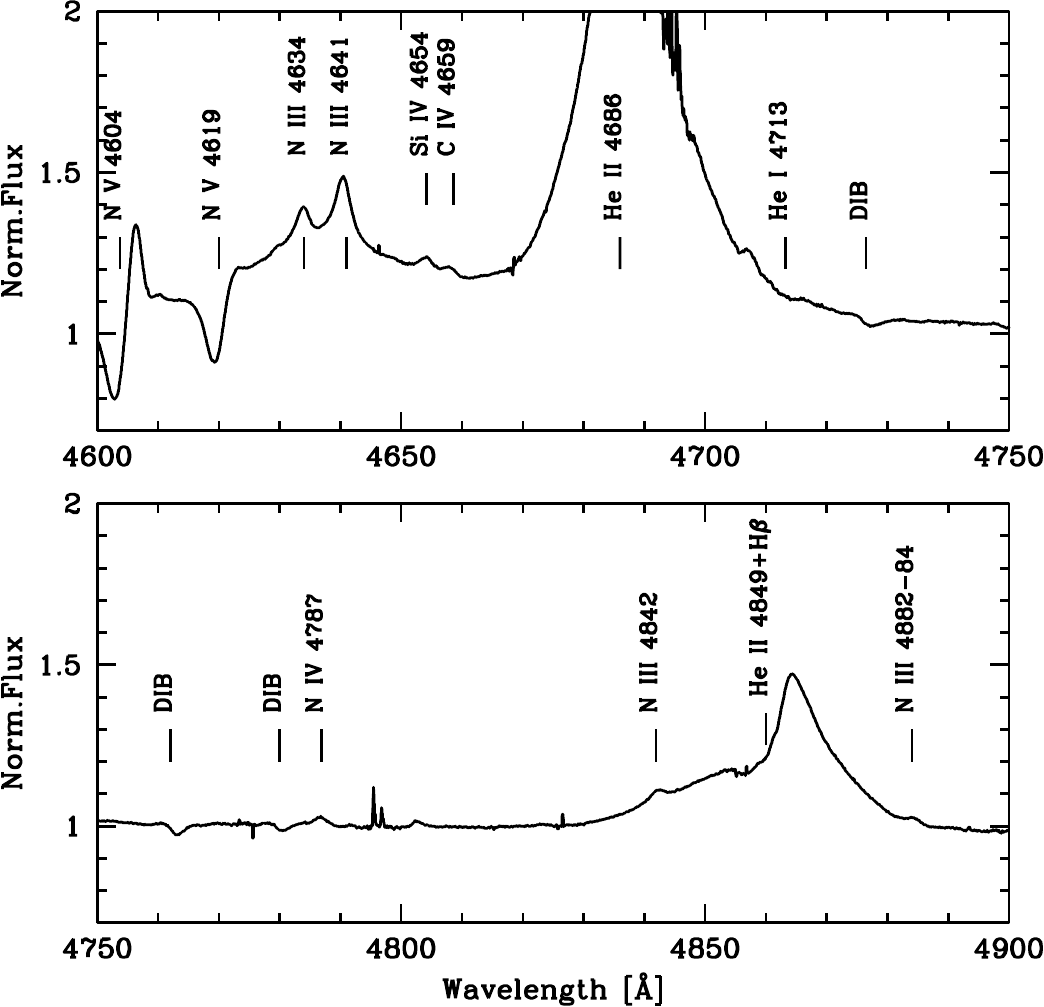}
	\includegraphics[width=0.67\textwidth]{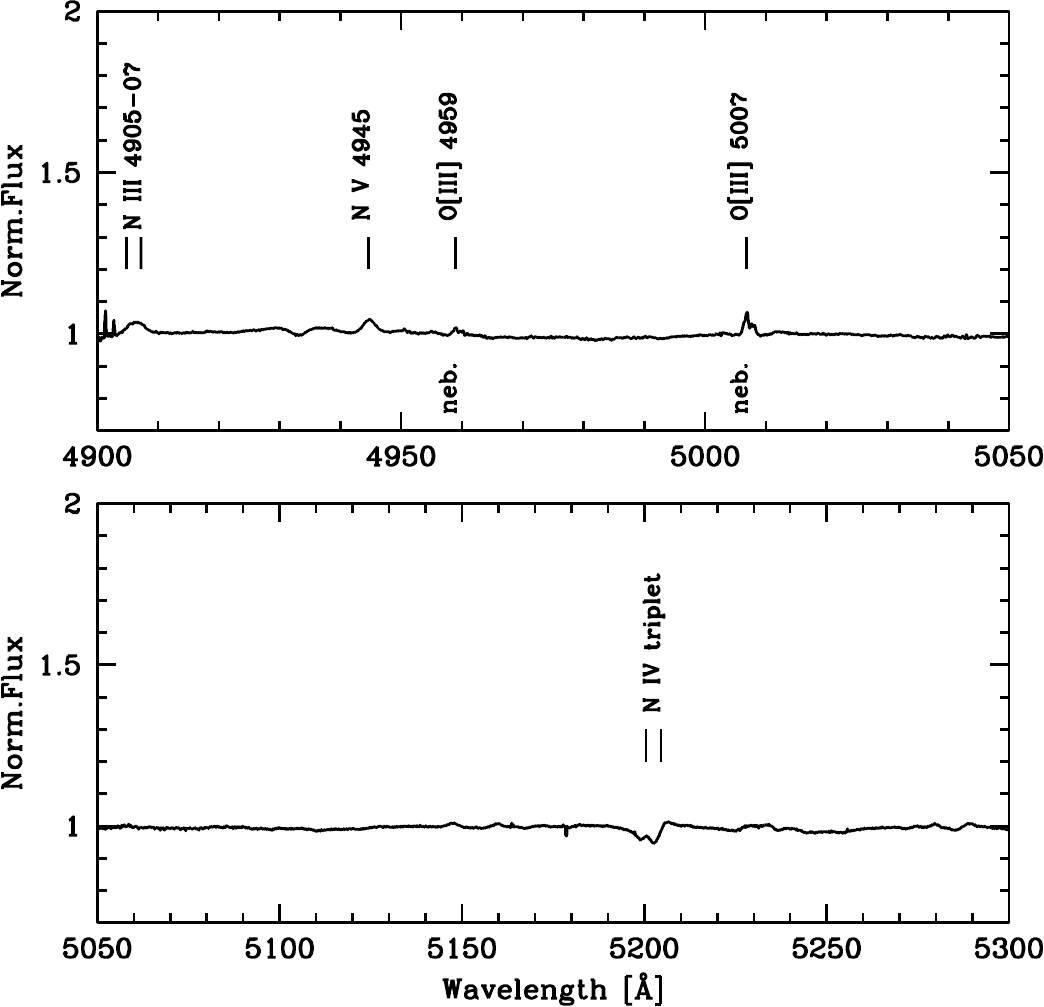}
	\vskip1ex
	
	\begin{minipage}{12cm}
		\addtocounter{figure}{-1}
		\caption{\emph{continued.}}
	\end{minipage}
\end{figure}
%
\begin{figure}
	\centering
	\includegraphics[width=0.67\textwidth]{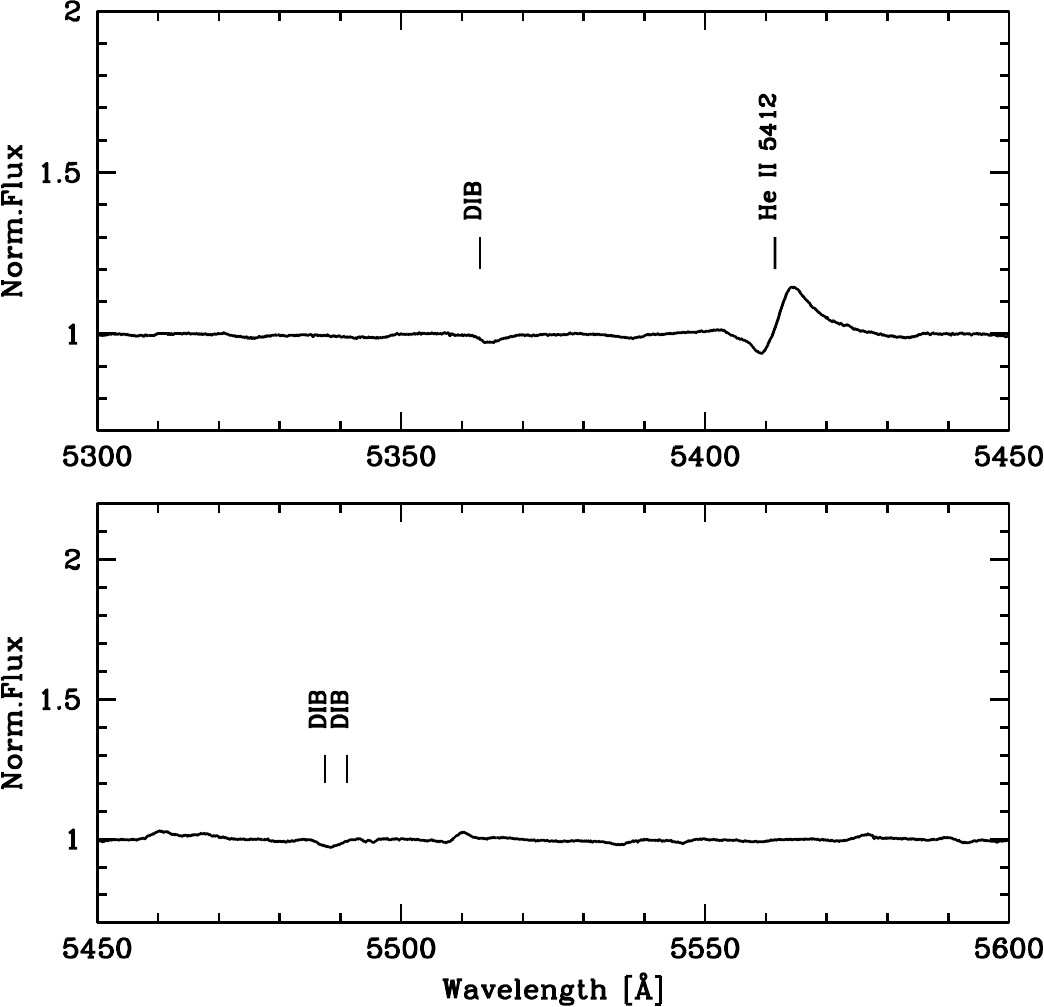}
	\includegraphics[width=0.67\textwidth]{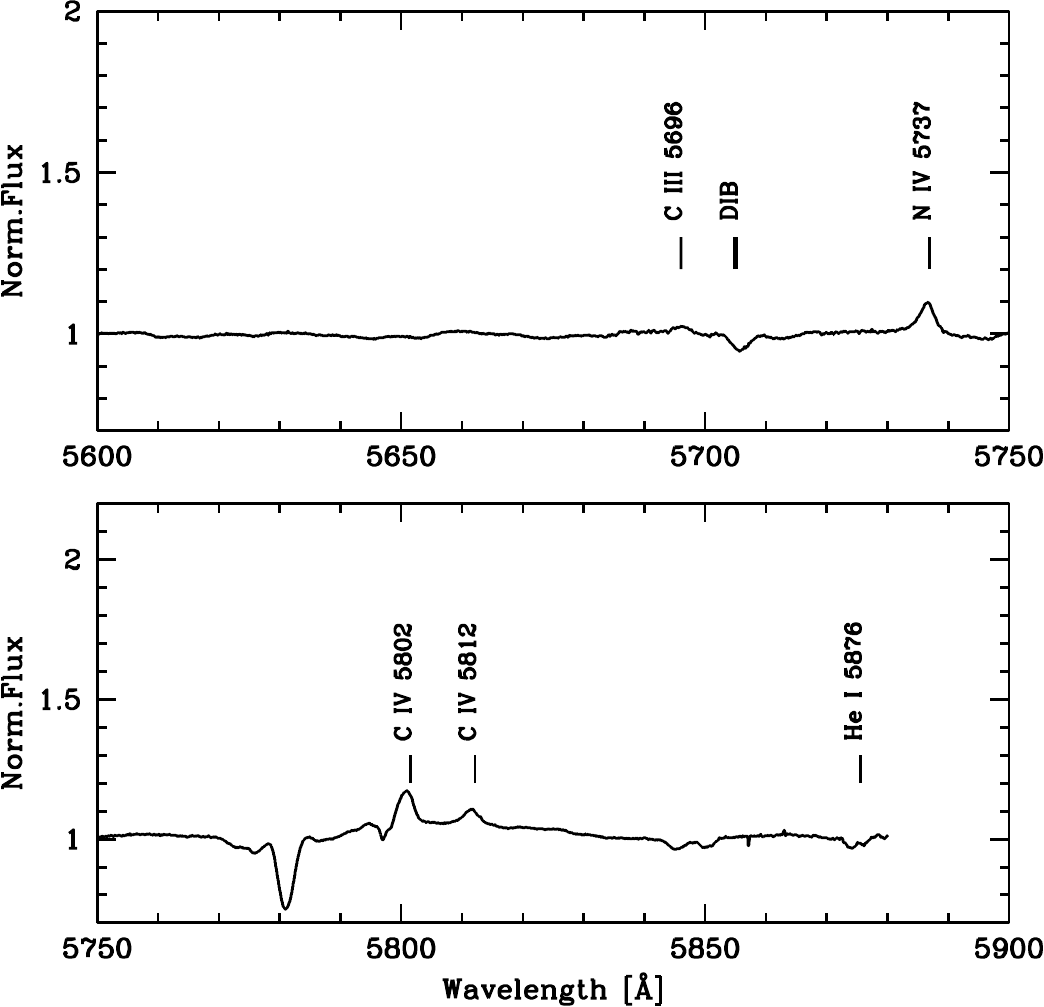}
	\vskip1ex
	
	\begin{minipage}{12cm}
		\addtocounter{figure}{-1}
		\caption{\emph{continued.}}
	\end{minipage}
\end{figure}
%
%
%
\end{appendix}
\end{document}